\documentclass[onecolumn,prx,superscriptaddress,longbibliography,nofootinbib]{revtex4-2}

\usepackage[dvips]{graphicx} 
\usepackage{amsfonts,amscd,amsmath,amsthm,amssymb}
\usepackage{rotating}
\usepackage{enumerate}
\usepackage{epsfig}
\usepackage{subcaption}
\usepackage{xcolor}
\usepackage[colorlinks = true]{hyperref}
\usepackage{physics}
\usepackage{epstopdf}
\usepackage{framed}
\usepackage{threeparttable,booktabs,multirow}
\usepackage{color}
\usepackage{comment}
\usepackage[ruled,vlined]{algorithm2e}
\usepackage[most]{tcolorbox}

\graphicspath{{./figure/}}

\usepackage{tikz}
\usetikzlibrary{tikzmark, calc, fit, positioning}
\usetikzlibrary{shapes,arrows,arrows.meta}
\usetikzlibrary{quantikz}

\newtheorem{lemma}{Lemma}

\newtheorem{observation}{Observation}

\newtcolorbox[auto counter]{mybox}[2][]{
	enhanced,
	breakable,
	colback=blue!5!white,
	colframe=blue!75!black,
	fonttitle=\bfseries,
	title=Box \thetcbcounter: #2,#1
}

\newcommand\qOutput[3]{\draw[fill=#3] (#1,#2) +(0,-2) -- +(3,0) -- +(0,2) -- +(0,-2);}
\newcommand{\red}[1]{{\color{red} #1}}
\newcommand{\blue}[1]{{\color{blue} #1}}
\DeclareMathOperator{\oplusIs}{\mathord{\oplus\!\!=}}

\begin{document}
	\preprint{APS/123-QED}
	
	\title{Resource analysis of Shor's elliptic curve algorithm with an improved quantum adder on a two-dimensional lattice}
	\author{Quan Gu}
	\thanks{All the authors contributed equally to this work}
	\author{Han Ye}
	\thanks{All the authors contributed equally to this work}
	\author{Junjie Chen}
	\thanks{All the authors contributed equally to this work}
	\author{Xiongfeng Ma}
	\email{xma@tsinghua.edu.cn}
	\affiliation{Center for Quantum Information, Institute for Interdisciplinary Information Sciences, Tsinghua University, Beijing, 100084 China}

	\begin{abstract}
	Quantum computers have the potential to break classical cryptographic systems by efficiently solving problems such as the elliptic curve discrete logarithm problem using Shor's algorithm. While resource estimates for factoring-based cryptanalysis are well established, comparable evaluations for Shor's elliptic curve algorithm under realistic architectural constraints remain limited. In this work, we propose a carry-lookahead quantum adder that achieves Toffoli depth $\log n + \log\log n + O(1)$ with only $O(n)$ ancillas, matching state-of-the-art performance in depth while avoiding the prohibitive $O(n\log n)$ space overhead of existing approaches. Importantly, our design is naturally compatible with the two-dimensional nearest-neighbor architectures and introduce only a constant-factor overhead. Further, we perform a comprehensive resource analysis of Shor's elliptic curve algorithm on two-dimensional lattices using the improved adder. By leveraging dynamic circuit techniques with mid-circuit measurements and classically controlled operations, our construction incorporates the windowed method, Montgomery representation, and quantum tables, and substantially reduces the overhead of long-range gates. For cryptographically relevant parameters, we provide precise resource estimates. In particular, breaking the NIST P-256 curve, which underlies most modern public-key infrastructures and the security of Bitcoin, requires about $4300$ logical qubits and logical Toffoli fidelity about $10^{-9}$. These results establish new benchmarks for efficient quantum arithmetic and provide concrete guidance toward the experimental realization of Shor's elliptic curve algorithm.
	\end{abstract}

	\maketitle
	
	
	\section{Introduction}\label{sec:intro}
	
	Quantum algorithms, such as Shor's algorithm~\cite{shor1997polynomial}, Grover's search~\cite{grover1996afast}, the HHL algorithm~\cite{harrow2009quantum}, and Hamiltonian simulation~\cite{berry2015hamiltonian}, offer potential speedups over their classical counterparts. Tremendous efforts have been devoted to their experimental realization across a variety of physical platforms~\cite{arute2019quantum, zhong2020jiuzhang, moses2023racetrack}. However, noise and decoherence make demonstrating quantum advantage with noisy intermediate-scale quantum (NISQ) devices challenging~\cite{preskill2018quantum}. As recent works suggest that classically hard problem instances remain beyond the reach of NISQ algorithms~\cite{aharonov1996limitations, yan2023limitations}, it becomes essential to explore the precise resource requirements for implementing large-scale quantum algorithms under realistic hardware assumptions.

	One of the most prominent cryptographic applications of quantum algorithms is Shor's solution to the elliptic curve discrete logarithm problem. The hardness of this problem underpins widely deployed cryptographic systems such as key agreement~\cite{diffie1976new}, digital signatures~\cite{elgamal1985public, johnson2001elliptic}, and pseudorandom generators~\cite{Dubal2013pseudorandom}, with notable implementations including the NIST P-256, P-384, and P-521 curves~\cite{pub2000digital}. In particular, Bitcoin relies on the secp256k1 elliptic curve to ensure the security of its transactions~\cite{qu1999sec}. An efficient algorithm for solving the elliptic-curve discrete logarithm problem would put at risk any bitcoins whose public keys have been revealed, as well as funds in transactions that have been broadcast to the network but not yet confirmed on the blockchain, which may worth trillions of dollars~\cite{aggarwal2017quantum}. The security of systems ranging from internet protocols to blockchain technologies relies directly on the conjectured classical intractability of this problem~\cite{Satoshi2009bitcoin}. Although believed to be hard classically, Shor's algorithm can solve the elliptic curve discrete logarithm problem in polynomial time~\cite{shor1997polynomial, proos2004shors}. While the resource demands for Shor's factoring algorithm have been extensively analyzed~\cite{fowler2012surface, gorman2017quantum, Gidney2021howtofactorbit, ha2022resource, gidney2025factor}, systematic studies for Shor's elliptic curve algorithm remain comparatively limited~\cite{roetteler2017quantum, haner2020improved}.

	Any realistic analysis must account for architectural constraints. Leading quantum platforms such as superconducting qubits~\cite{arute2019quantum} and topological qubits~\cite{fowler2012surface} are naturally restricted to two-dimensional (2D) nearest-neighbor layouts. This geometry strikes a balance between scalability, fabrication feasibility, and compatibility with error correction, but it also imposes significant routing and depth overheads for nonlocal gates. Since Shor's elliptic curve algorithm relies heavily on modular arithmetic, these architectural constraints make the efficiency of arithmetic subroutines, especially quantum adders, central to any realistic resource estimate.

	At the core of modular arithmetic, quantum adders serve as one of the most fundamental primitives in quantum computing. Optimizing the trade-off between circuit depth and ancilla overhead is especially crucial for the NISQ devices, where the number of available qubits is limited and circuit depth is constrained by decoherence and uncorrectable noise. A shallower adder not only reduces execution time but also mitigates cumulative noise, while minimizing ancilla qubits enhances scalability and physical realizability on current hardware platforms. Ripple-carry adders achieve linear Toffoli depth with minimal ancilla overhead~\cite{cuccaro2004new, gidney2018halving}, whereas carry-lookahead adders attain logarithmic depth at the cost of linear or larger ancillary space. As key references for evaluating the trade-off between depth and ancilla usage, the most well-known carry-lookahead adder achieves a depth of $2\log n + O(1)$ with $O(n)$ ancillas~\cite{draper2004logarithmic}, while the best known depth of $\log n + 1$ has recently been obtained at the expense of $O(n\log n)$ ancillas~\cite{wang2024optimal}.

	In this work, we propose a new carry-lookahead quantum adder that achieves Toffoli depth $\log n + \log\log n + O(1)$ with only $O(n)$ ancillas, achieving state-of-the-art performance of depth while avoiding the prohibitive space overhead of prior designs. Our construction is naturally compatible with two-dimensional nearest-neighbor architectures with only constant-factor overhead in this setting. A comparison of the resource costs of our in-place adder with previous works is shown in Table~\ref{table:adder-comparison}.

	\begin{table}[ht]
		\caption{Comparison of our in-place quantum adder with previous works}\label{table:adder-comparison}
		\begin{threeparttable}
    \begin{tabular}{cccc}
        \toprule
        \bf Adder & \bf qubit number & \bf Toffoli depth & \bf Toffoli number \\ 
        \midrule
        Brent-Kung tree~\cite{draper2004logarithmic}\tnote{1} & $5n+O(1)$                   & $4\log n +O(1)$              & $10n+O(1)$         \\ 
        Sklansky tree~\cite{wang2024optimal}\tnote{2}   & $\frac{1}{2}n\log n+3n+O(1)$ & $2\log n+O(1)$               & $2n\log n+2n+O(1)$ \\ 
        Our work             & $6n+O(1)$                   & $2\log n +2\log\log n +O(1)$ & $14n+O(1)$         \\ 
        \bottomrule
    \end{tabular}
    \begin{tablenotes}
        \footnotesize
        \item[1] In the original paper, it costs $4n+O(1)$ bits. However, in order to achieve a dynamic Toffoli gate with $T$-depth $1$, one additional column is needed.
        \item[2] We made some quick improvements on the qubit numbers compared to the original paper.
    \end{tablenotes}
\end{threeparttable}
	\end{table}

\begin{figure}[h]
		\centering
		\captionsetup{justification=raggedright,singlelinecheck=false}
		\includegraphics[width=0.7\textwidth]{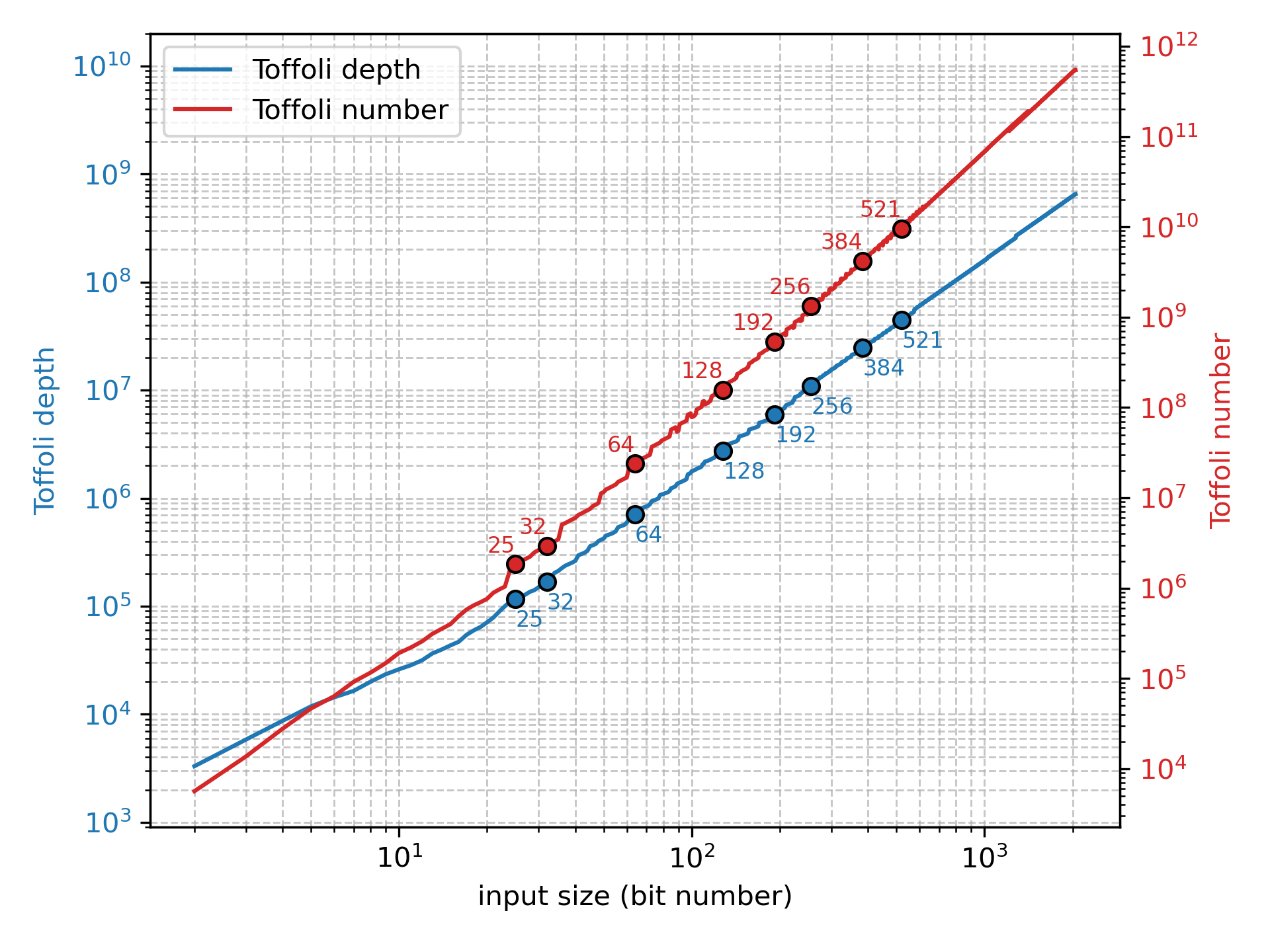}
		\caption{The relationship between the Toffoli depth, Toffoli number, and the input size $n$ (i.e., the bit length of the elliptic curve). These metrics are evaluated by simulating our circuit for $n$ values from $2$ to $2048$. Minor fluctuations in the Toffoli number appear when the window size increases by one.}\label{figure:circuit-parameter-plot}
	\end{figure}

	Building on this improved adder, we present a comprehensive resource analysis of Shor's elliptic curve algorithm compiled for two-dimensional nearest-neighbor lattices. Our construction integrates several key algorithmic ingredients, including the windowed trick~\cite{haner2020improved}, Montgomery representation~\cite{montgomery1985modular}, and quantum tables~\cite{babbush2018encoding}, while leveraging dynamic circuit techniques to reduce the cost of nonlocal operations. Dynamic circuits, incorporating mid-circuit measurements, feedforward, and classically controlled operations, enable polynomial overheads for long-range gates to be traded for constant spatial overhead. Beyond their foundational role in fault-tolerant quantum error correction~\cite{Shor1995code, gottesman1997stabilizer}, dynamic circuits have also demonstrated broad utility in state preparation~\cite{Piroli2021adaptive, Yan2025Variational}, measurement-based quantum computing~\cite{Raussendorf2003MBQC, Briegel2009MBQC}, and gate compilation~\cite{Takahashi2016collapse}. By exploiting this capability, our analysis achieves dramatically reduced depth in 2D layout. We provide explicit circuit designs and quantify resource costs in terms of circuit depth, qubit number, and logical gate counts. We summarize the exact costs for input sizes of $25$, $32$, $192$, $256$, $384$, and $521$, and also the leading-order terms for general $n$ in Table~\ref{table:resource-estimate}. A figure illustrating the relation between Toffoli number, Toffoli depth and input size is shown in Fig.~\ref{figure:circuit-parameter-plot}. Specifically, for cryptographically relevant security levels, we estimate that solving a single instance of the NIST P-256 curve, whose hardness underlies the security of Bitcoin and other major blockchain systems, would require at least $4300$ logical qubits and Toffoli-gate fidelity better than $10^{-9}$.

	\begin{table}[ht]
		\caption{Resource estimation of Shor's algorithm for the elliptic curve discrete logarithm problem. The last row shows the leading terms.}\label{table:resource-estimate}
		\begin{threeparttable}
    \begin{tabular}{*{7}{c}}
        \toprule
        \bf Input size & \bf qubit number & \bf CNOT depth & \bf CNOT count & \bf Toffoli depth & \bf Toffoli count \\
        \midrule
        25                       & $425$                 & $2.12 \times 10^{5}$                                  & $3.55 \times 10^{6}$                                 & $1.17 \times 10^{5}$   & $1.84 \times 10^{6}$       \\
        32                       & $544$                 & $3.69 \times 10^{5}$                                  & $7.86 \times 10^{6}$                                 & $1.68 \times 10^{5}$   & $2.94 \times 10^{6}$       \\    
        192                      & $3264$                & $1.11 \times 10^{7}$                                  & $1.30 \times 10^{9}$                                 & $5.95 \times 10^{6}$   & $5.32 \times 10^{8}$       \\
        256                      & $4352$                & $1.89 \times 10^{7}$                                  & $2.89 \times 10^{9}$                                 & $1.09 \times 10^{7}$   & $1.33 \times 10^{9}$       \\
        384                      & $7528$                & $4.41 \times 10^{7}$                                  & $9.99 \times 10^{9}$                                & $2.47 \times 10^{7}$   & $4.20 \times 10^{9}$      \\
        521                      & $8857$               & $8.58 \times 10^{7}$                                  & $2.67 \times 10^{10}$                                & $4.44 \times 10^{7}$   & $9.51 \times 10^{9}$        \\
        $n$                      & $17n$                 & $96n^2 + 240\frac{n^2}{\log\log n}$ & $80 \frac{n^3}{\log\log n} + 1490\frac{n^3}{\log n}$ & $112n^2 + 80\frac{n^2}{\log\log n}$               & $1144\frac{n^3}{\log n}+280\frac{n^3}{\log n\log\log n}$                \\
        \bottomrule
    \end{tabular}
\end{threeparttable}
	\end{table}

	This paper is organized as follows.
	Section~\ref{sec:preliminaries} reviews the necessary background, including the elliptic curve discrete logarithm problem, Shor's elliptic curve algorithm, the 2D layout model, and the concept of dynamic circuits.
	Section~\ref{sec:improvedadder} presents our improved quantum adder and analyzes its performance.
	Section~\ref{sec:circuit-analysis} provides a detailed resource analysis of Shor's elliptic curve algorithm on a 2D lattice.
	Section~\ref{sec:discussion} discusses the implications of our results and outlines possible directions for future research.

	The appendices provide additional technical details.
	Appendix~\ref{sec:toffoli-gate} describes the construction of the Toffoli gate, a key component of our circuit.
	Appendix~\ref{sec:early-carry-bit} reviews two early designs of carry-lookahead adders that form the basis of our improved adder.
	Appendix~\ref{sec:circuit-construction} details the implementation of several important arithmetic subroutines used in our analysis.
	Appendix~\ref{sec:resource-cost} lists the explicit calculations of the resource costs.
	Appendix~\ref{sec:other-circuit} presents the construction of other essential circuit components.
	Finally, Appendix~\ref{sec:single-layer} demonstrates the implementation of our circuit on a single-layer layout as an alternative to the bi-layer design used in the main text.

	\section{Preliminaries}\label{sec:preliminaries}
	In this section, we briefly introduce preliminary knowledge, including the formal definition of elliptic curve discrete logarithm problem, Shor's algorithm to solve it, the layout of a 2D system and the concept and usage of dynamic circuits.

	\subsection{Elliptic curve discrete logarithm problem}
	We only provide a concise overview of the elliptic curve discrete logarithm problem here, and leave the details of elliptic curve digital signature algorithms to Ref.~\cite{johnson2001elliptic}. An elliptic curve can be defined over finite fields $\mathbb{F}_p$, where $p$ is a prime. All subsequent arithmetic operations are thus performed within the finite field $\mathbb{F}_p$: the operations ``$+$'', ``$-$'', and ``$\times$'' are understood modulo $p$, and the division ``$a/b$'' is defined by $a\times b^{-1}$, where $b^{-1}$ denotes the multiplicative inverse of $b$ in $\mathbb{F}_p$.

	Formally, an elliptic curve in short Weierstrass form $\mathsf{E}$ over $\mathbb{F}_p$ is defined as the set of points $P=(x,y)$ satisfying the constraint $y^2=x^3+ax+b$, where constants $a,b\in\mathbb{F}_p$ should fulfill the non-singularity condition $4a^3+27b^3\not\equiv0$. To construct an Abelian group based on the points on $\mathsf{E}$, an identity element $O$ and the point addition should be defined. Define the point $O$ as the point at infinity, which cannot be represented in $\mathbb{F}_p^2$. The addition operation on points of the elliptic curve $\mathsf{E}$ is then defined as follows. Let $P_1, P_2\in \mathsf{E}$ and $P_3 = P_1 + P_2$. If either $P_1 = O$ or $P_2 = O$, the addition is simply defined by $P_3 = P_2$ or $P_3 = P_1$, respectively. If both $P_1$ and $P_2$ are distinct from $O$, we write their coordinates explicitly as $P_1(x_1,y_1)$ and $P_2(x_2,y_2)$. If $x_1=x_2$ and $y_1=-y_2$, we let $P_3=O$ and say that $P_2$ is the inverse of $P_1$, denoted $P_1=-P_2$. Otherwise, $P_3(x_3,y_3)$ can be computed via the following equation
	\begin{equation}\label{equ:point-addition}
		(x_3,y_3)=(\lambda^2-x_1-x_2,-y_2-\lambda(x_3-x_2)),
	\end{equation}
	where the slope
	\begin{equation}
		\lambda = \begin{cases}
			\frac{y_1-y_2}{x_1-x_2}, & \text{if } P_1 \neq P_2, \\
			\frac{3x_1^2+a}{2y_1}, & \text{if } P_1 = P_2.
		\end{cases}
	\end{equation}
	We still use $\mathsf{E}$ to represent the Abelian group. The scalar multiplication between $m \in \mathbb{F}_p$ and $P\in \mathsf{E}$ is defined naturally by repeated addition as
	\begin{equation}
		mP=\underbrace{P+P+\cdots+P}_m
	\end{equation}
	when $m \neq 0$, and $0P=O$.

	Suppose the cardinality of the group $|\mathsf{E}|$ has a large prime factor $q$. One can find a cyclic subgroup of $\mathsf{E}$ of order $q$, denoted as $\{O,G,2G,\ldots,(q-1)G\}$ with $qG=O$. Specifically, $|\mathsf{E}|=q$ typically endows the elliptic curve with advantageous cryptographic properties. However, this constraint is not necessary for the analysis presented here.

	In elliptic curve digital signature algorithms, each user possesses a private key $d \in [1,q-1]$ and a corresponding public key defined as $Q = dG$~\cite{johnson2001elliptic}. The security of these algorithms relies on the computational difficulty of deriving the private key $d$ from the known generator point $G$ and the public key $Q$. That is, given two points $G,Q\in\mathsf{E}$, there is no known classical polynomial-time algorithm for computing $d\in\mathbb{F}_p$ such that $dG=Q$. Successfully solving for $d$ would compromise the encryption scheme by enabling an adversary to forge signatures. This computationally challenging task is widely recognized as the elliptic curve discrete logarithm problem.

	\subsection{Shor's elliptic curve Algorithm}\label{subsec:shor-algorithm}
	Although certain specialized classical algorithms can efficiently attack elliptic curves with particular structural properties, there currently exists no general classical algorithm capable of solving the elliptic curve discrete logarithm problem with a time complexity better than $O(\sqrt{q})$~\cite{johnson2001elliptic}. Consequently, the elliptic curve discrete logarithm problem is widely considered unbreakable by classical computers.

	However, Shor's quantum algorithm provides a general and efficient quantum solution to the elliptic curve discrete logarithm problem. Its procedure can be summarized as follows~\cite{nielsen2010quantum}.
	\begin{equation}
		\begin{aligned}
			&\ket{\mathbf{0}}\ket{\mathbf{0}}\ket{\mathbf{0}}&\textrm{initial state}\\
			\to&\frac{1}{2^n}\sum_{x,y=0}^{2^n-1}\ket{x}\ket{y}\ket{\mathbf{0}}&\textrm{apply H gates on the first two registers}\\
			\to&\frac{1}{2^n}\sum_{x,y=0}^{2^n-1}\ket{x}\ket{y}\ket{xG+yQ}&\textrm{apply controlled point addition}\\
			=&\frac{1}{2^n}\sum_{x,y=0}^{2^n-1}\ket{x}\ket{y}\ket{(x+yd)G}\\
			=&\frac{1}{2^n\sqrt{q}}\sum_{l=0}^{q-1}\sum_{x,y=0}^{2^n-1}e^{2\pi i(x+yd)l/q}\ket{x}\ket{y}\ket{\psi(l)}\\
			=&\frac{1}{2^n\sqrt{q}}\sum_{l=0}^{q-1}\left(\sum_{x=0}^{2^n-1}e^{2\pi ilx/q}\ket{x}\right)\left(\sum_{y=0}^{2^n-1}e^{2\pi idly/q}\ket{y}\right)\ket{\psi(l)}\\
			\to&\frac{1}{\sqrt{q}}\sum_{l=0}^{q-1}\ket{\widetilde{l/q}}\ket{\widetilde{dl/q}}\ket{\psi(l)}&\textrm{apply inverse QFT on the first two registers}\\
			\to&\ket{\widetilde{l/q}}\ket{\widetilde{dl/q}}&\textrm{measure the first two registers}\\
		\end{aligned}
	\end{equation}
	Here, the equation in the fifth line is based on the quantum Fourier transform (QFT), where $\ket{\psi(l)}$ is defined as $\ket{\psi(l)}=\frac{1}{\sqrt{q}}\sum_{j=0}^{q-1}e^{-2\pi i lj/q}\ket{jG}$. After applying the inverse quantum Fourier transform, the state $\sum_{x=0}^{2^n-1}e^{2\pi ilx/q}\ket{x}$ is transformed to $\ket{\widetilde{l/q}}$, which closely approximates the computational basis state $\ket{l/q}$. $\ket{\widetilde{dl/q}}$ is obtained and near to $\ket{dl/q}$ in the same way. Therefore, if we measure the first two registers on the computational basis, the outcome of the second register should be $d$ times the outcome of the first register with high probability. A simple diversity will give $d$ and the problem can be solved in polynomial time and space of $n=\Theta(\log{d})$.

	The main computational primitives in Shor's algorithm are controlled point addition and QFT\@. While the circuit of QFT is well established~\cite{shor1997polynomial}, there needs additional design for realizing controlled point addition $\ket{x}\ket{P_1}\ket{P_2}\mapsto\ket{x}\ket{P_1}\ket{P_2+xP_1}$. Note that the core issue is to design the point addition $\ket{P_1}\ket{P_2}\mapsto\ket{P_1}\ket{P_2+P_1}$, since the control operation can be realized by adding a quantum control on each gate in the original circuit. Each point $\ket{P}$ should be represented as $\ket{x}\ket{y}$. One can use the classical algorithm for point addition based on Eq.~\eqref{equ:point-addition} and replace the corresponding classical operations with reversible quantum operations. Here, ``reversible'' means that we cannot erase any information during the computation process. As shown in Algorithm~\ref{algorithm:point-addition}, we need quantum circuits for modular addition $\ket{x_1}\ket{x_2}\mapsto\ket{x_1-x_2}\ket{x_2}$, modular multiplication $\ket{x}\ket{y}\ket{\lambda}\mapsto\ket{x}\ket{y\oplus \lambda\cdot x}\ket{\lambda}$, and modular division $\ket{x}\ket{y}\ket{\lambda}\mapsto\ket{x}\ket{y}\ket{\lambda\oplus \frac{y}{x}}$, which will be discussed in Appendix~\ref{sec:circuit-construction}.

	\begin{algorithm}[ht]
		\SetKwInOut{Input}{Input}
		\SetKwInOut{Output}{Output}
		
		\Input{$x_1, x_2, y_1, y_2$}
		\Output{$x_3, y_3$}
		\BlankLine{}
		$x \gets x_1, y \gets y_1$ \;
		$x \gets x-x_2, y \gets y-y_2$ \tcp*{$x=x_1-x_2,y=y_1-y_2$}
		$\lambda \gets \frac{y}{x}$ \tcp*{$\lambda=\frac{y_1-y_2}{x_1-x_2}$}
		$y \gets y \oplus (\lambda \cdot x)$ \tcp*{$\oplus$ for bit-wise XOR\@; $y=0$}
		$x \gets x + 3x_2 - \lambda^2$ \tcp*{$x=x_1+2x_2-\lambda^2=x_2-x_3$}
		$y \gets \lambda\cdot x$ \tcp*{$y=\lambda\cdot(x_2-x_3)=y_2+y_3$}
		$\lambda \gets \lambda\oplus \frac{y}{x}$  \tcp*{$\oplus$ for bit-wise XOR\@; $\lambda=0$}
		$x \gets x_2-x, y \gets y-y_2$ \tcp*{$x=x_3,y=y_3$}
		\KwRet{$x,y$}
		\caption{(reversible) classical algorithm for point addition~\cite{roetteler2017quantum}}\label{algorithm:point-addition}
	\end{algorithm}

	\subsection{2D Layout}\label{subsec:2d-layout}
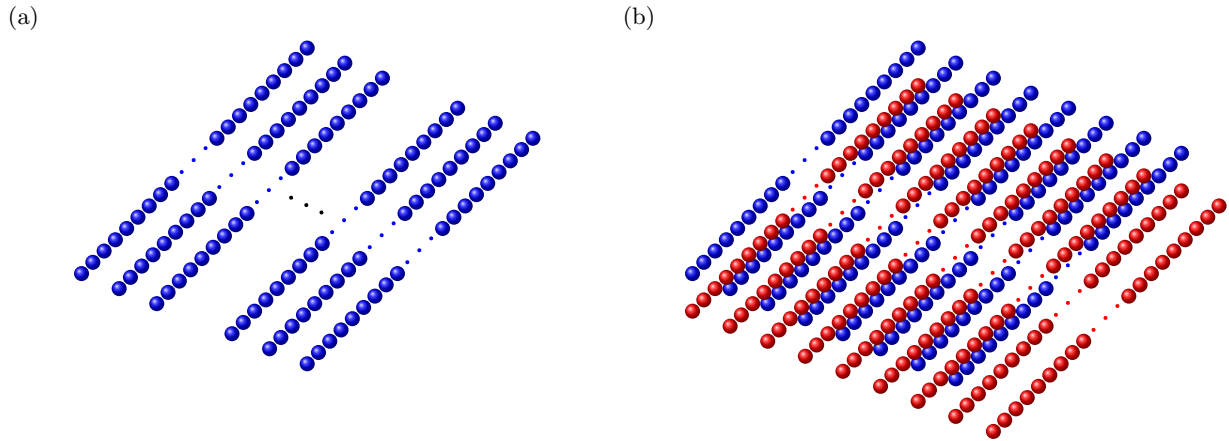
\begin{figure}[ht]
		\captionsetup{justification=raggedright,singlelinecheck=false}
		\begin{subfigure}[t]{0.45\textwidth}
			\caption{}\label{figure:qubit-layout-single}
			\centering
			\begin{tikzpicture}
    \foreach \x in {0,1,2,4,5,6} {
        \foreach \i in {0,0.15,...,1.2} {
            \shade[ball color=blue] (\x*0.5 + \i, \x*-0.2 + \i) circle (0.1);
        }
        \foreach \i in {1.8,1.95,...,3} {
            \shade[ball color=blue] (\x*0.5 + \i, \x*-0.2 + \i) circle (0.1);
        }
        \node at (\x*0.5 + 1.53, \x*-0.2 + 1.53) {\begin{turn}{45}\blue{\Large$\cdots$}\end{turn}};
    }
    \node at (3.02, 0.88) {\begin{turn}{-26.5}\Large$\cdots$\end{turn}};
\end{tikzpicture}
		\end{subfigure}
		\begin{subfigure}[t]{0.5\textwidth}
			\caption{}\label{figure:qubit-layout-bilayer}
			\centering
			\begin{tikzpicture}
    \foreach \x in {0,1,2,3,4,5,6,7} {
        \foreach \i in {0,0.15,...,1.2} {
            \shade[ball color=blue] (\x*0.5 + \i, \x*-0.2 + \i) circle (0.1);
        }
        \foreach \i in {1.8,1.95,...,3} {
            \shade[ball color=blue] (\x*0.5 + \i, \x*-0.2 + \i) circle (0.1);
        }
        \node at (\x*0.5 + 1.53, \x*-0.2 + 1.53) {\begin{turn}{45}\blue{\Large$\cdots$}\end{turn}};
    }
    \foreach \x in {0,1,2,3,4,5,6,7,8} {
        \foreach \i in {0,0.15,...,1.2} {
            \shade[ball color=red] (\x*0.5 + \i, \x*-0.2 - 0.5 + \i) circle (0.1);
        }
        \foreach \i in {1.8,1.95,...,3} {
            \shade[ball color=red] (\x*0.5 + \i, \x*-0.2 - 0.5 + \i) circle (0.1);
        }
        \node at (\x*0.5 + 1.53, \x*-0.2 - 0.5 + 1.53) {\begin{turn}{45}\red{\Large$\cdots$}\end{turn}};
    }
\end{tikzpicture}
		\end{subfigure}
		\caption{Two diagrams illustrating the qubit layout.\ (a) The diagram for the single-layer layout, each ball representing a qubit, and each line consists of $n$ qubits.\ (b) The diagram for the bi-layer layout, each ball representing a qubit, and each line consists of $n$ qubits. Blue balls stand for the qubits in the upper layer, red balls stand for those in the lower layer.}\label{figure:qubit-layout}
	\end{figure}

	Two-dimensional (2D) qubit layouts are widely favored in quantum computing, as they offer the minimal geometry required for implementing high-threshold topological codes while remaining compatible with leading physical platforms, most notably superconducting architectures. In this work, we introduce two 2D layouts tailored for Shor's algorithm, each comprising $17n$ qubits. The first is a single-layer $n \times 17$ lattice, referred to as the ``single-layer layout'', and the second is a bi-layer $n \times 9 \times 2$ lattice, referred to as the ``bi-layer layout''. Among them, $2n$ qubits are to store the result of point addition, $13n$ qubits are for intermediate computations, and the other $2n$ qubits are ancillas. For algorithmic convenience, we permit a slight extension in the column size, such as $n + \log n$, rather than fixing it exactly at $n$. The two layouts are illustrated in Fig.~\ref{figure:qubit-layout}. The bi-layer layout enables more efficient execution of nearest-neighbor gates, whereas the single-layer layout requires gate teleportation or swap operations, incurring only constant overhead. In the main text, we focus on circuit design based on the bi-layer layout, while additional details of the single-layer layout are provided in Appendix~\ref{sec:single-layer}.

	To manipulate $n$-qubit registers (e.g., applying arithmetic operations such as addition or multiplication), we organize the qubits vertically into columns containing $n$ qubits each. Columns of idle qubits, that is, qubits in state $\ket{0}$, can be employed by dynamic circuits to facilitate the implementation of circuits requiring long-range gate interactions.

	Throughout our analysis, certain multi-qubit gates inevitably act on qubits that are not nearest neighbors. To quantify this aspect, we define the gate interaction distance as the maximum Manhattan distance between any two qubits involved in a gate on the lattice, where the Manhattan distance between two qubits located at positions $(x_1, y_1)$ and $(x_2, y_2)$ is given by $|x_1 - x_2| + |y_1 - y_2|$. For practical quantum computing implementations, gate interaction distances of gate bricks used for compilation should remain constant, since it only introduces a constant overhead to decompose them into nearest-neighbor two-qubit gates. While we retain the multi-qubit representation for clarity and conciseness, the gate interaction distance provides a direct measure of the resource overhead incurred when compiling abstract gates into hardware-compatible nearest-neighbor interactions, and is thus an essential consideration for practical implementations.

	\subsection{Dynamic Circuits}\label{subsec:dynamic circuits}
	Dynamic quantum circuits, which integrate unitary gates, mid-circuit measurements, classical post-processing of measurement outcomes, and classically controlled unitary operations conditioned on the results of classical computing, offer computational capabilities beyond those of purely unitary circuits. They have been shown to provide advantages in quantum state preparation~\cite{Piroli2021adaptive, Yan2025Variational, Malz2024MPS, piroli2024approximatingmanybodyquantumstates} as well as in computational tasks~\cite{Takahashi2016collapse, Raussendorf2003MBQC, Briegel2009MBQC}. A key feature of dynamic circuits is the ability to trade temporal complexity for spatial complexity, which can be accommodated by idle ancillary qubits. Leveraging dynamic circuits, it is possible to implement long-range quantum gates and generate long-range entangled states with substantially reduced depth, even under nearest-neighbor constraints. In this work, we focus on two fundamental dynamic circuits: the long-range Toffoli gate and unbounded GHZ-state generation~\cite{B_umer_2024}, which we discuss in detail below.

\begin{figure}[ht]
		\captionsetup{justification=raggedright,singlelinecheck=false}
		\begin{subfigure}[t]{0.5\textwidth}
			\caption{}\label{figure:Toffoli-circuit-middle}
			\centering
			\begin{tikzpicture}[scale=1.000000,x=1pt,y=1pt]
\filldraw[color=white] (0.000000, -7.500000) rectangle (129.000000, 157.500000);
\draw[color=black] (0.000000,150.000000) -- (129.000000,150.000000);
\draw[color=black] (0.000000,150.000000) node[left] {$\ket{x}$};
\draw[color=black] (0.000000,135.000000) -- (129.000000,135.000000);
\draw[color=black] (0.000000,135.000000) node[left] {$\ket{0}$};
\draw[color=black] (0.000000,120.000000) node[anchor=mid east] {$\vdots$};
\draw[color=black] (0.000000,105.000000) -- (129.000000,105.000000);
\draw[color=black] (0.000000,105.000000) node[left] {$\ket{0}$};
\draw[color=black] (0.000000,90.000000) -- (117.000000,90.000000);
\draw[color=black] (117.000000,89.500000) -- (129.000000,89.500000);
\draw[color=black] (117.000000,90.500000) -- (129.000000,90.500000);
\draw[color=black] (0.000000,90.000000) node[left] {$\ket{0}$};
\draw[color=black] (0.000000,75.000000) -- (129.000000,75.000000);
\draw[color=black] (0.000000,75.000000) node[left] {$\ket{z}$};
\draw[color=black] (0.000000,60.000000) -- (117.000000,60.000000);
\draw[color=black] (117.000000,59.500000) -- (129.000000,59.500000);
\draw[color=black] (117.000000,60.500000) -- (129.000000,60.500000);
\draw[color=black] (0.000000,60.000000) node[left] {$\ket{0}$};
\draw[color=black] (0.000000,45.000000) -- (129.000000,45.000000);
\draw[color=black] (0.000000,45.000000) node[left] {$\ket{0}$};
\draw[color=black] (0.000000,30.000000) node[anchor=mid east] {$\vdots$};
\draw[color=black] (0.000000,15.000000) -- (129.000000,15.000000);
\draw[color=black] (0.000000,15.000000) node[left] {$\ket{0}$};
\draw[color=black] (0.000000,0.000000) -- (129.000000,0.000000);
\draw[color=black] (0.000000,0.000000) node[left] {$\ket{y}$};
\draw (9.000000,150.000000) -- (9.000000,0.000000);
\begin{scope}
\draw[fill=white] (9.000000, 75.000000) circle(3.000000pt);
\clip (9.000000, 75.000000) circle(3.000000pt);
\draw (6.000000, 75.000000) -- (12.000000, 75.000000);
\draw (9.000000, 72.000000) -- (9.000000, 78.000000);
\end{scope}
\filldraw (9.000000, 150.000000) circle(1.500000pt);
\filldraw (9.000000, 0.000000) circle(1.500000pt);
\draw[fill=white,color=white] (24.000000, -6.000000) rectangle (39.000000, 156.000000);
\draw (31.500000, 75.000000) node {$=$};
\draw (54.000000,150.000000) -- (54.000000,90.000000);
\begin{scope}
\draw[fill=white] (54.000000, 90.000000) circle(3.000000pt);
\clip (54.000000, 90.000000) circle(3.000000pt);
\draw (51.000000, 90.000000) -- (57.000000, 90.000000);
\draw (54.000000, 87.000000) -- (54.000000, 93.000000);
\end{scope}
\filldraw (54.000000, 150.000000) circle(1.500000pt);
\draw (54.000000,60.000000) -- (54.000000,0.000000);
\begin{scope}
\draw[fill=white] (54.000000, 60.000000) circle(3.000000pt);
\clip (54.000000, 60.000000) circle(3.000000pt);
\draw (51.000000, 60.000000) -- (57.000000, 60.000000);
\draw (54.000000, 57.000000) -- (54.000000, 63.000000);
\end{scope}
\filldraw (54.000000, 0.000000) circle(1.500000pt);
\draw (72.000000,90.000000) -- (72.000000,60.000000);
\begin{scope}
\draw[fill=white] (72.000000, 75.000000) circle(3.000000pt);
\clip (72.000000, 75.000000) circle(3.000000pt);
\draw (69.000000, 75.000000) -- (75.000000, 75.000000);
\draw (72.000000, 72.000000) -- (72.000000, 78.000000);
\end{scope}
\filldraw (72.000000, 90.000000) circle(1.500000pt);
\filldraw (72.000000, 60.000000) circle(1.500000pt);
\begin{scope}
\draw[fill=white] (93.000000, 90.000000) +(-45.000000:8.485281pt and 8.485281pt) -- +(45.000000:8.485281pt and 8.485281pt) -- +(135.000000:8.485281pt and 8.485281pt) -- +(225.000000:8.485281pt and 8.485281pt) -- cycle;
\clip (93.000000, 90.000000) +(-45.000000:8.485281pt and 8.485281pt) -- +(45.000000:8.485281pt and 8.485281pt) -- +(135.000000:8.485281pt and 8.485281pt) -- +(225.000000:8.485281pt and 8.485281pt) -- cycle;
\draw (93.000000, 90.000000) node {$H$};
\end{scope}
\begin{scope}
\draw[fill=white] (93.000000, 60.000000) +(-45.000000:8.485281pt and 8.485281pt) -- +(45.000000:8.485281pt and 8.485281pt) -- +(135.000000:8.485281pt and 8.485281pt) -- +(225.000000:8.485281pt and 8.485281pt) -- cycle;
\clip (93.000000, 60.000000) +(-45.000000:8.485281pt and 8.485281pt) -- +(45.000000:8.485281pt and 8.485281pt) -- +(135.000000:8.485281pt and 8.485281pt) -- +(225.000000:8.485281pt and 8.485281pt) -- cycle;
\draw (93.000000, 60.000000) node {$H$};
\end{scope}
\draw (116.500000,150.000000) -- (116.500000,90.000000);
\draw (117.500000,150.000000) -- (117.500000,90.000000);
\begin{scope}
\draw[fill=white] (117.000000, 150.000000) +(-45.000000:8.485281pt and 8.485281pt) -- +(45.000000:8.485281pt and 8.485281pt) -- +(135.000000:8.485281pt and 8.485281pt) -- +(225.000000:8.485281pt and 8.485281pt) -- cycle;
\clip (117.000000, 150.000000) +(-45.000000:8.485281pt and 8.485281pt) -- +(45.000000:8.485281pt and 8.485281pt) -- +(135.000000:8.485281pt and 8.485281pt) -- +(225.000000:8.485281pt and 8.485281pt) -- cycle;
\draw (117.000000, 150.000000) node {$Z$};
\end{scope}
\filldraw (117.000000, 90.000000) circle(1.500000pt);
\draw[fill=white] (111.000000, 84.000000) rectangle (123.000000, 96.000000);
\draw[very thin] (117.000000, 90.600000) arc (90:150:6.000000pt);
\draw[very thin] (117.000000, 90.600000) arc (90:30:6.000000pt);
\draw[->,>=stealth] (117.000000, 84.600000) -- +(80:10.392305pt);
\draw (116.500000,60.000000) -- (116.500000,0.000000);
\draw (117.500000,60.000000) -- (117.500000,0.000000);
\begin{scope}
\draw[fill=white] (117.000000, 0.000000) +(-45.000000:8.485281pt and 8.485281pt) -- +(45.000000:8.485281pt and 8.485281pt) -- +(135.000000:8.485281pt and 8.485281pt) -- +(225.000000:8.485281pt and 8.485281pt) -- cycle;
\clip (117.000000, 0.000000) +(-45.000000:8.485281pt and 8.485281pt) -- +(45.000000:8.485281pt and 8.485281pt) -- +(135.000000:8.485281pt and 8.485281pt) -- +(225.000000:8.485281pt and 8.485281pt) -- cycle;
\draw (117.000000, 0.000000) node {$Z$};
\end{scope}
\filldraw (117.000000, 60.000000) circle(1.500000pt);
\draw[fill=white] (111.000000, 54.000000) rectangle (123.000000, 66.000000);
\draw[very thin] (117.000000, 60.600000) arc (90:150:6.000000pt);
\draw[very thin] (117.000000, 60.600000) arc (90:30:6.000000pt);
\draw[->,>=stealth] (117.000000, 54.600000) -- +(80:10.392305pt);
\draw[color=black] (129.000000,150.000000) node[right] {$\ket{x}$};
\draw[color=black] (129.000000,135.000000) node[right] {$\ket{0}$};
\draw[color=black] (129.000000,120.000000) node[anchor=mid west] {$\vdots$};
\draw[color=black] (129.000000,105.000000) node[right] {$\ket{0}$};
\draw[color=black] (129.000000,75.000000) node[right] {$\ket{z\oplus x\cdot y}$};
\draw[color=black] (129.000000,45.000000) node[right] {$\ket{0}$};
\draw[color=black] (129.000000,30.000000) node[anchor=mid west] {$\vdots$};
\draw[color=black] (129.000000,15.000000) node[right] {$\ket{0}$};
\draw[color=black] (129.000000,0.000000) node[right] {$\ket{y}$};
\end{tikzpicture}
		\end{subfigure}
		\begin{subfigure}[t]{0.45\textwidth}
			\caption{}\label{figure:Toffoli-layout}
			\centering
			\begin{tikzpicture}
    \draw[line width=10pt,draw=blue!30,fill=blue!30,line join=round,line cap=round] (2.4,-2.4) |- (3.2,-3.2) -- cycle;
    \foreach \x in {0,3,7} {
        \draw[fill=red] (\x*0.8,0) circle (0.1);
        \draw[fill=red] (\x*0.8,-2.4) circle (0.1);
        \draw[fill=red] (\x*0.8,-5.2) circle (0.1);
    }
    \foreach \x in {1,2,4,5,6} {
        \draw[fill=black] (\x*0.8,0) circle (0.1);
        \draw[fill=black] (\x*0.8,-2.4) circle (0.1);
        \draw[fill=black] (\x*0.8,-5.2) circle (0.1);
    }
    \foreach \x in {0,1,2,5,6,7} {
        \draw[fill=white] (\x*0.8,-0.8) circle (0.1);
        \draw[fill=white] (\x*0.8,-3.2) circle (0.1);
        \draw[fill=white] (\x*0.8,-6) circle (0.1);
    }
    \foreach \x in {3,4} {
        \draw[fill=blue] (\x*0.8,-0.8) circle (0.1);
        \draw[fill=blue] (\x*0.8,-3.2) circle (0.1);
        \draw[fill=blue] (\x*0.8,-6) circle (0.1);
    }
    \draw[-Latex] (-0.2,-0.1) |- (2.3,-1) node[pos=0.75,below] {long-range CNOT};
    \draw[-Latex] (5.8,-0.1) |- (3.3,-1) node[pos=0.75,below] {long-range CNOT};

    \draw[-Latex,thick] (2.8,-1.2) -- (2.8,-2);
    \node at (2.8,-3.6) {local Toffoli};
    \draw[-Latex,thick] (2.8,-4) -- (2.8,-4.8);

    \draw[-Latex] (2.3,-6.2) -| (-0.2,-5.3) node[pos=0.25,below] {uncomputation};
    \draw[-Latex] (3.3,-6.2) -| (5.8,-5.3) node[pos=0.25,below] {uncomputation};
\end{tikzpicture}
		\end{subfigure}
		\begin{subfigure}[t]{0.5\textwidth}
			\caption{}\label{figure:Toffoli-circuit-edge}
			\centering
			\begin{tikzpicture}[scale=1.000000,x=1pt,y=1pt]
\filldraw[color=white] (0.000000, -7.500000) rectangle (165.000000, 157.500000);
\draw[color=black] (0.000000,150.000000) -- (165.000000,150.000000);
\draw[color=black] (0.000000,150.000000) node[left] {$\ket{x}$};
\draw[color=black] (0.000000,135.000000) -- (165.000000,135.000000);
\draw[color=black] (0.000000,135.000000) node[left] {$\ket{0}$};
\draw[color=black] (0.000000,120.000000) node[anchor=mid east] {$\vdots$};
\draw[color=black] (0.000000,105.000000) -- (165.000000,105.000000);
\draw[color=black] (0.000000,105.000000) node[left] {$\ket{0}$};
\draw[color=black] (0.000000,90.000000) -- (153.000000,90.000000);
\draw[color=black] (153.000000,89.500000) -- (165.000000,89.500000);
\draw[color=black] (153.000000,90.500000) -- (165.000000,90.500000);
\draw[color=black] (0.000000,90.000000) node[left] {$\ket{0}$};
\draw[color=black] (0.000000,75.000000) -- (165.000000,75.000000);
\draw[color=black] (0.000000,75.000000) node[left] {$\ket{y}$};
\draw[color=black] (0.000000,60.000000) -- (165.000000,60.000000);
\draw[color=black] (0.000000,60.000000) node[left] {$\ket{0}$};
\draw[color=black] (0.000000,45.000000) -- (165.000000,45.000000);
\draw[color=black] (0.000000,45.000000) node[left] {$\ket{0}$};
\draw[color=black] (0.000000,30.000000) node[anchor=mid east] {$\vdots$};
\draw[color=black] (0.000000,15.000000) -- (165.000000,15.000000);
\draw[color=black] (0.000000,15.000000) node[left] {$\ket{0}$};
\draw[color=black] (0.000000,0.000000) -- (165.000000,0.000000);
\draw[color=black] (0.000000,0.000000) node[left] {$\ket{z}$};
\draw (9.000000,150.000000) -- (9.000000,0.000000);
\begin{scope}
\draw[fill=white] (9.000000, 0.000000) circle(3.000000pt);
\clip (9.000000, 0.000000) circle(3.000000pt);
\draw (6.000000, 0.000000) -- (12.000000, 0.000000);
\draw (9.000000, -3.000000) -- (9.000000, 3.000000);
\end{scope}
\filldraw (9.000000, 150.000000) circle(1.500000pt);
\filldraw (9.000000, 75.000000) circle(1.500000pt);
\draw[fill=white,color=white] (24.000000, -6.000000) rectangle (39.000000, 156.000000);
\draw (31.500000, 75.000000) node {$=$};
\draw (54.000000,150.000000) -- (54.000000,90.000000);
\begin{scope}
\draw[fill=white] (54.000000, 90.000000) circle(3.000000pt);
\clip (54.000000, 90.000000) circle(3.000000pt);
\draw (51.000000, 90.000000) -- (57.000000, 90.000000);
\draw (54.000000, 87.000000) -- (54.000000, 93.000000);
\end{scope}
\filldraw (54.000000, 150.000000) circle(1.500000pt);
\draw (72.000000,90.000000) -- (72.000000,60.000000);
\begin{scope}
\draw[fill=white] (72.000000, 60.000000) circle(3.000000pt);
\clip (72.000000, 60.000000) circle(3.000000pt);
\draw (69.000000, 60.000000) -- (75.000000, 60.000000);
\draw (72.000000, 57.000000) -- (72.000000, 63.000000);
\end{scope}
\filldraw (72.000000, 90.000000) circle(1.500000pt);
\filldraw (72.000000, 75.000000) circle(1.500000pt);
\draw (90.000000,60.000000) -- (90.000000,0.000000);
\begin{scope}
\draw[fill=white] (90.000000, 0.000000) circle(3.000000pt);
\clip (90.000000, 0.000000) circle(3.000000pt);
\draw (87.000000, 0.000000) -- (93.000000, 0.000000);
\draw (90.000000, -3.000000) -- (90.000000, 3.000000);
\end{scope}
\filldraw (90.000000, 60.000000) circle(1.500000pt);
\draw (108.000000,90.000000) -- (108.000000,60.000000);
\begin{scope}
\draw[fill=white] (108.000000, 60.000000) circle(3.000000pt);
\clip (108.000000, 60.000000) circle(3.000000pt);
\draw (105.000000, 60.000000) -- (111.000000, 60.000000);
\draw (108.000000, 57.000000) -- (108.000000, 63.000000);
\end{scope}
\filldraw (108.000000, 90.000000) circle(1.500000pt);
\filldraw (108.000000, 75.000000) circle(1.500000pt);
\begin{scope}
\draw[fill=white] (129.000000, 90.000000) +(-45.000000:8.485281pt and 8.485281pt) -- +(45.000000:8.485281pt and 8.485281pt) -- +(135.000000:8.485281pt and 8.485281pt) -- +(225.000000:8.485281pt and 8.485281pt) -- cycle;
\clip (129.000000, 90.000000) +(-45.000000:8.485281pt and 8.485281pt) -- +(45.000000:8.485281pt and 8.485281pt) -- +(135.000000:8.485281pt and 8.485281pt) -- +(225.000000:8.485281pt and 8.485281pt) -- cycle;
\draw (129.000000, 90.000000) node {$H$};
\end{scope}
\draw (152.500000,150.000000) -- (152.500000,90.000000);
\draw (153.500000,150.000000) -- (153.500000,90.000000);
\begin{scope}
\draw[fill=white] (153.000000, 150.000000) +(-45.000000:8.485281pt and 8.485281pt) -- +(45.000000:8.485281pt and 8.485281pt) -- +(135.000000:8.485281pt and 8.485281pt) -- +(225.000000:8.485281pt and 8.485281pt) -- cycle;
\clip (153.000000, 150.000000) +(-45.000000:8.485281pt and 8.485281pt) -- +(45.000000:8.485281pt and 8.485281pt) -- +(135.000000:8.485281pt and 8.485281pt) -- +(225.000000:8.485281pt and 8.485281pt) -- cycle;
\draw (153.000000, 150.000000) node {$Z$};
\end{scope}
\filldraw (153.000000, 90.000000) circle(1.500000pt);
\draw[fill=white] (147.000000, 84.000000) rectangle (159.000000, 96.000000);
\draw[very thin] (153.000000, 90.600000) arc (90:150:6.000000pt);
\draw[very thin] (153.000000, 90.600000) arc (90:30:6.000000pt);
\draw[->,>=stealth] (153.000000, 84.600000) -- +(80:10.392305pt);
\draw[color=black] (165.000000,150.000000) node[right] {$\ket{x}$};
\draw[color=black] (165.000000,135.000000) node[right] {$\ket{0}$};
\draw[color=black] (165.000000,120.000000) node[anchor=mid west] {$\vdots$};
\draw[color=black] (165.000000,105.000000) node[right] {$\ket{0}$};
\draw[color=black] (165.000000,75.000000) node[right] {$\ket{y}$};
\draw[color=black] (165.000000,60.000000) node[right] {$\ket{0}$};
\draw[color=black] (165.000000,45.000000) node[right] {$\ket{0}$};
\draw[color=black] (165.000000,30.000000) node[anchor=mid west] {$\vdots$};
\draw[color=black] (165.000000,15.000000) node[right] {$\ket{0}$};
\draw[color=black] (165.000000,0.000000) node[right] {$\ket{z \oplus x \cdot y}$};
\end{tikzpicture}
		\end{subfigure}
		\begin{subfigure}[t]{0.45\textwidth}
			\caption{}\label{figure:Toffoli-long-range-CNOT}
			\centering
			\input{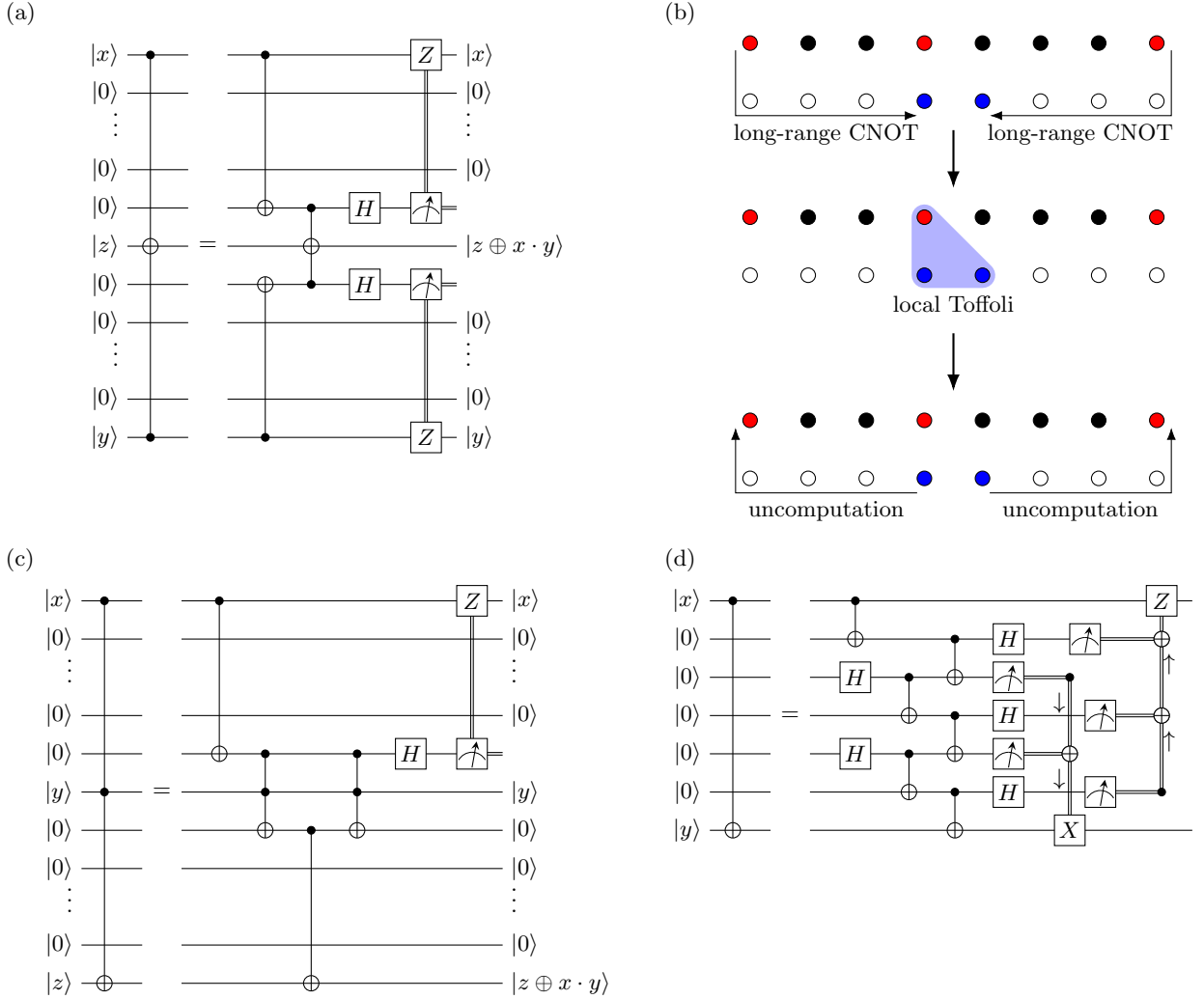}
		\end{subfigure}
		\caption{Implementation of the long-range Toffoli gate using dynamic circuits.\ (a) An example of the circuit for implementing a long-range Toffoli gate with the output qubit in the middle.\ (b) The layout of ancillary qubits for the circuit in (a). We use red circles to represent the data qubits to implement Toffoli, black circles to represent the unused data qubits, white circles to represent the ancillary qubits initialized in $\ket{0}$, and blue circles to represent ancillas storing intermediate data.\ (c) An example of the circuit for implementing the long-range Toffoli gate with the output qubit on the edge. Notably, the second Toffoli gate is used for uncomputing the ancillary qubit, which can be designed with zero $T$-depth, as explained in Appendix~\ref{sec:toffoli-gate}. Therefore, the two circuits shown by (a) and (c) have the same $T$-depth.\ (d) An example of implementing a long-range CNOT using dynamic circuit with a depth of 6.}\label{figure:long-range Toffoli}
	\end{figure}

	Suppose we wish to implement a long-range Toffoli gate on three target qubits arranged along the same chain. Since data qubits cannot serve as ancillas or be directly measured during the process, we introduce an adjacent chain of ancillary qubits, each initialized to $\ket{0}$. As illustrated in Figs.~\ref{figure:Toffoli-circuit-middle} and~\ref{figure:Toffoli-circuit-edge}, long-range CNOTs, which are enabled by dynamic circuits, can be combined to realize the desired long-range Toffoli gate. The explicit circuit construction is provided in Figs.~\ref{figure:Toffoli-layout} and~\ref{figure:Toffoli-long-range-CNOT}. During the protocol, every ancilla qubit is measured once and reset to $\ket{0}$, allowing their reuse in subsequent dynamic circuits.

\begin{figure}[ht]
		\captionsetup{justification=raggedright,singlelinecheck=false}
		\centering
		\begin{subfigure}[t]{0.6\textwidth}
			\caption{}\label{figure:GHZ-layout}
			\centering
			\input{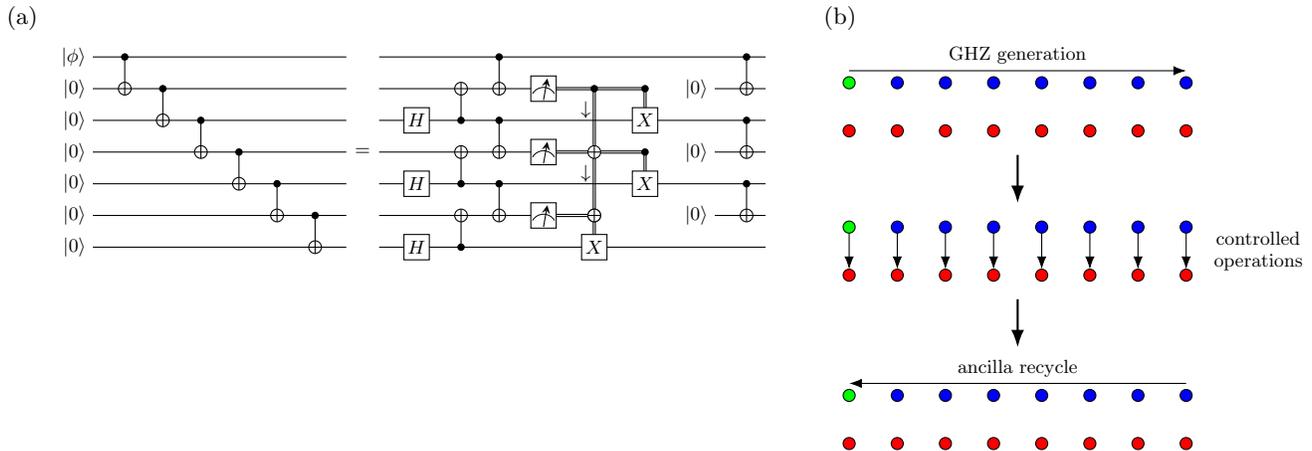}
		\end{subfigure}
		\begin{subfigure}[t]{0.38\textwidth}
			\caption{}\label{figure:GHZ-circuit}
			\centering
			\begin{tikzpicture}[scale=0.8,transform shape]
    \draw[fill=green] (0,0) circle (0.1);
    \draw[fill=green] (0,-2.4) circle (0.1);
    \draw[fill=green] (0,-5.2) circle (0.1);
    \foreach \x in {1,2,...,7} {
        \draw[fill=blue] (\x*0.8,0) circle (0.1);
        \draw[fill=blue] (\x*0.8,-2.4) circle (0.1);
        \draw[fill=blue] (\x*0.8,-5.2) circle (0.1);
    }
    \foreach \x in {0,1,...,7} {
        \draw[fill=red] (\x*0.8,-0.8) circle (0.1);
        \draw[fill=red] (\x*0.8,-3.2) circle (0.1);
        \draw[-Latex] (\x*0.8,-2.5) -- (\x*0.8,-3.1);
        \draw[fill=red] (\x*0.8,-6) circle (0.1);
    }

    \draw[-Latex] (0,0.2) -- (5.6,0.2) node[pos=0.5,above] {GHZ generation};
    \draw[-Latex,thick] (2.8,-1.2) -- (2.8,-2);
    \draw[-Latex,thick] (2.8,-3.6) -- (2.8,-4.4);
    \draw[-Latex] (5.6,-5) -- (0,-5) node[pos=0.5,above] {ancilla recycle};
    \node[align=center] at (6.8,-2.8) {controlled\\operations};
\end{tikzpicture}
		\end{subfigure}
		\caption{Implementation of GHZ state generation.\ (a) Two equivalent circuits for implementing GHZ state generation.\ (b) Implement parallel controlled gates using GHZ state generation. We use red circles to represent the data qubits to implement controlled gates, green circles to represent the control qubit, and blue circles to represent ancillas storing intermediate data.}\label{figure:GHZ}
	\end{figure}

	Another essential dynamic-circuit primitive in our circuit is unbounded GHZ state generation, which achieves the generalized quantum state ``copying'' operation:
	\begin{align}
		a_0\ket{\psi_0}\ket{0}+a_1\ket{\psi_1}\ket{1}\mapsto a_0\ket{\psi_0}\ket{0}^{\otimes n}+a_1\ket{\psi_1}\ket{1}^{\otimes n}.
	\end{align}
	The preparation of such states with dynamic circuits of depth 6 and $n-1$ ancillary qubits is well established; a representative implementation is shown in Fig.~\ref{figure:GHZ-layout}~\cite{B_umer_2024}.

	Since the global phase is uniformly shared among all qubits in a GHZ state, the choice of control qubit in a controlled operation is irrelevant. Consequently, when multiple controlled gates share the same control qubit, one can first prepare a sufficiently large GHZ state and assign different qubits within it as individual controls, enabling all controlled gates to be executed in parallel, as illustrated in Fig.~\ref{figure:GHZ-circuit}. To recycle the ancillary qubits, any $n-1$ qubits of the GHZ state are measured in the $X$ basis, and an appropriate corrective operation ($\mathbb{I}$ or $Z$) is applied to the remaining qubit depending on the measurement outcomes. On a two-dimensional lattice, this procedure can still be realized with constant-depth local gates and efficient classical post-processing, provided the ancillary qubits are placed in a column adjacent to the data qubits.
	
	\section{Improved Quantum Adder}\label{sec:improvedadder}
	Efficient quantum addition is a fundamental building block for implementing arithmetic-intensive algorithms such as Shor's factoring and discrete logarithm algorithms. The performance of quantum adders directly impacts both the time and space complexity of these algorithms, since addition operations dominate the arithmetic subroutines. Improving the depth of quantum adders reduces overall circuit runtime and mitigates decoherence, while reducing the ancillary qubit overhead is critical for scalability on near-term and fault-tolerant architectures. Consequently, the design of quantum adders that simultaneously optimize time and space resources plays a central role in advancing the practical feasibility of large-scale quantum algorithms.  
	
	The classical adder computes the summation of two $n$-bit numbers $x$ and $y$. The quantum adder implements a corresponding unitary $U$ that produces $\ket{s}=\ket{x+y}$ given $\ket{x}\ket{y}$. Typically, there are two types of quantum adders: the out-place adder $U\ket{x}\ket{y}\ket{0}=\ket{x}\ket{y}\ket{x+y}$ and the in-place adder $U\ket{x}\ket{y}=\ket{x}\ket{x + y}$. We will only discuss in-place adder in the main text and leave the discussion of out-place adder and other variants at Appendix~\ref{subsec:variants-addition}. Our adder design is based on reversible classical adders, whose essential task is to compute the carry bits $c$. Let $x_i$ and $y_i$ denote the $i$-th bits of $x$ and $y$ respectively, and let $c_i$ denote the $i$-th carry bit. The final sum bits can be obtained in parallel given the carry bits as $s_i=x_i \oplus y_i \oplus c_i$.
	
	Naively, the carry bits can be computed using a recursion equation $c_i = x_i y_i + x_i c_{i-1} + y_i c_{i-1}$. However, a direct implementation requires circuit depth $O(n)$, since the information from the data bits is propagated sequentially. In contrast, parallelization can be exploited to reduce the depth. By pre-processing the input bits $x$ and $y$, one can enable parallel evaluation of carry bits, which is precisely the idea underlying classical carry-lookahead adders. To this end, the propogation state $p[i, j]$ and generation state $g[i, j]$ is defined by the recursion equations 
	\begin{equation}
		\begin{gathered}
			p[i, j] = p[i, k] \cdot p[k, j], \\
			g[i, j] = g[i, k] \oplus (p[i, k] \cdot g[k, j]).
		\end{gathered}
		\label{eq:pg3}
	\end{equation}
	for $i<k<j$. The initial values give $p[i,i+1] = x_i \oplus y_i$ and $g[i,i+1] = x_i \cdot y_i$, and the carry bits are given by $c_j=g[0,j]$. The problem of solving the former recursion equation is known as ``prefix sum problem''~\cite{blelloch1990prefix}. Solving this problem is the key part of designing quantum adder.
	
	Various classical parallel prefix sum algorithms provide different trade-offs between time and space, which directly influence the design of quantum adders. The most well-known quantum carry-lookahead adder~\cite{draper2004logarithmic} computes the carry bits using a Brent-Kung tree~\cite{brent1982regular}, achieving depth $2\log n + O(1)$ with $O(n)$ ancillas. By contrast, a more recent carry-lookahead construction based on the Sklansky tree~\cite{sklansky2009conditional, wang2024optimal} reduces the depth to $\log n + 1$ but at the expense of $O(n \log n)$ ancillas. In this section, we introduce an improved carry-lookahead quantum adder that achieves Toffoli depth $\log n + \log\log n + O(1)$ with only $O(n)$ ancillas, thereby combining the advantages of both approaches.
	
	\subsection{Carry Bit Computing}
	Before introducing our algorithm for carry bit computing, we briefly introduce two previous methods~\cite{draper2004logarithmic,wang2024optimal}. Intuitively, Draper's method~\cite{draper2004logarithmic} utilizes a Brent-Kung tree structure~\cite{brent1982regular}, computing the value corresponding to the qubits located at the positions of powers of two in the first $\log n$ steps, and compute the other values in the remaining $\log n$ steps. Wang's method~\cite{wang2024optimal}, on the other hand, employs a Sklansky tree~\cite{sklansky2009conditional}, updating the value of all qubits whose $i$-th bit of their locations are 1 at the $i$-th step. A demonstration of these two methods is shown in Fig.~\ref{fig:carry-draper} and~\ref{fig:carry-wang}, and the details of them can be found in Appendix~\ref{sec:early-carry-bit}.
	
	Our construction combines the advantages of Brent-Kung and Sklansky style carry-lookahead schemes within a windowed framework. The $n$ qubits are partitioned into $m=\lceil n/k \rceil$ blocks of size $k=\lfloor \log n \rfloor$ (with the final block truncated if necessary). In the ``block phase'', each $k$-qubit block is processed in parallel using a Brent-Kung tree to compute $p[ik,ik+j]$ and $g[ik,ik+j]$ for $0 \leq i < m$ and $1 \leq j \leq k$. In the ``inter-block phase'', the block outputs $p[ik,(i+1)k]$ and $g[ik,(i+1)k]$ are combined using a Sklansky tree, yielding the higher-level carries $g[0,ik]$. In the merge phase, these intermediate results are integrated with the block-level values to obtain $g[0,ik+j]$. Finally, all auxiliary values are uncomputed. The full procedure is summarized in Algorithm~\ref{algorithm:carry-our}, and a schematic illustration is provided in Fig.~\ref{fig:carry-ours}.
	
	\begin{algorithm}[ht]
		\SetKwInOut{Input}{Input}
		\SetKwInOut{Output}{Output}
		\SetKw{for}{for}
		\SetKwProg{Fn}{Function}{:}{}
		\SetKwFunction{draperIn}{ComputeCarryDraperPGCompute}
		\SetKwFunction{draperOut}{ComputeCarryDraperPUncompute}
		\SetKwFunction{wang}{ComputeCarryWang}
		\SetKwFunction{our}{ComputeCarryOur}
		
		\LinesNumbered{}
		\Fn{\our{$P_0;G;P_1,P_2,a$}}{
			\Input{$n$-qubit $P_0$ storing $p[i-1,i]$; $n$-qubit $G$ storing $g[i-1,i]$; $n$-qubit ancilla $P_1$, $n$-qubit ancilla $P_2$, $n$-qubit ancilla $a$, all initialized to 0.}
			\Output{$P_0,P_1,P_2,a$ remains the same; $G$ holds carry bits $c_i=g[0,i]$.}
			$k\gets\lfloor\log n\rfloor, m\gets\left\lceil\frac{n}{k}\right\rceil$\;
			\For{$i\gets0,\ldots,m-1$}{
				\draperIn{$P_0[ik,(i+1)k];G[ik,(i+1)k];P_1[ik,(i+1)k],P_2[ik,(i+1)k];a[2ik,2(i+1)k]$}\tcp*{set $P_1[ik+j],G[ik+j]$ to $p[ik,ik+j],g[ik,ik+j]$}
			}
			\wang{$\{P_1[k],P_1[2k],\ldots,P_1[n]\};\{G_1[k],G_1[2k],\ldots,G_1[n]\};a$}\tcp*{set $G[ik]$ to $g[0,ik]$}
			\For{$i\gets0,\ldots,m-1$}{
				\For{$j\gets1,\ldots,\min\{k,n-ik\}$}{
					$G[ik+j]\oplusIs G[ik]\cdot P_1[ik+j]$\tcp*{set $G[ik+j]$ to $g[0,ik+j]$}
				}
			}
			\For{$i\gets0,\ldots,m-1$}{
				\draperOut{$P_0[ik,(i+1)k];;P_1[ik,(i+1)k],P_2[ik,(i+1)k];a[ik,(i+1)k]$}\tcp*{recover $P_1, P_2$ to 0}
			}
		}
		\caption{our algorithm for computing carry bits.}\label{algorithm:carry-our}
	\end{algorithm}
	
	\begin{figure}
		\captionsetup{justification=raggedright,singlelinecheck=false}
		\begin{subfigure}[t]{0.45\textwidth}
			\caption{}\label{fig:carry-draper}
			\centering
			\begin{tikzpicture}
    \foreach \x in {1,...,8} {
        \pgfmathtruncatemacro{\xbf}{\x-1}
        \node at (8-\x, 0.5) {\x};
        \draw[-Latex, dashed] (8-\x, 0) -- (8-\x, -4.5);

        \node (a\x) [draw, shape=circle, minimum size=0.4cm, inner sep=0pt, fill=blue!20] at (8-\x, 0) {\xbf};

        \node at (8-\x, -5) {$c_{\x}$};
    }
    \foreach \x in {2,4,6,8} {
        \pgfmathtruncatemacro{\xbf}{\x-2}
        \pgfmathtruncatemacro{\xedge}{\x-1}
        \node (b\x) [draw, shape=circle, minimum size=0.4cm, inner sep=0pt, fill=blue!20] at (8-\x, -0.8) {\xbf};
        \draw[-Latex] (a\xedge) -- (b\x);
    }
    \foreach \x in {4,8} {
        \pgfmathtruncatemacro{\xbf}{\x-4}
        \pgfmathtruncatemacro{\xedge}{\x-2}
        \node (c\x) [draw, shape=circle, minimum size=0.4cm, inner sep=0pt, fill=blue!20] at (8-\x, -1.6) {\xbf};
        \draw[-Latex] (b\xedge) -- (c\x);
    }
    \node (d8) [draw, shape=circle, minimum size=0.4cm, inner sep=0pt, fill=blue!20] at (0, -2.4) {0};
    \draw[-Latex] (c4) -- (d8);

    \node (e6) [draw, shape=circle, minimum size=0.4cm, inner sep=0pt, fill=blue!20] at (2, -3.2) {0};
    \draw[-Latex] (4, -2.4) -- (e6);
    \foreach \x in {3,5,7} {
        \pgfmathtruncatemacro{\xedge}{\x-1}
        \node (f\x) [draw, shape=circle, minimum size=0.4cm, inner sep=0pt, fill=blue!20] at (8-\x, -4) {0};
    }
    \draw[-Latex] (e6) -- (f7);
    \draw[-Latex] (4, -3.2) -- (f5);
    \draw[-Latex] (6, -3.2) -- (f3);
\end{tikzpicture}
		\end{subfigure}
		\qquad
		\begin{subfigure}[t]{0.45\textwidth}
			\caption{}\label{fig:carry-wang}
			\centering
			\begin{tikzpicture}
    \foreach \x in {1,...,8} {
        \pgfmathtruncatemacro{\xbf}{\x-1}
        \node at (8-\x, 0.5) {\x};
        \draw[-Latex, dashed] (8-\x, 0) -- (8-\x, -2.9);

        \node (a\x) [draw, shape=circle, minimum size=0.4cm, inner sep=0pt, fill=blue!20] at (8-\x, 0) {\xbf};

        \node at (8-\x, -3.4) {$c_{\x}$};
    }

    \foreach \x in {2,4,6,8} {
        \pgfmathtruncatemacro{\xbf}{\x-2}
        \pgfmathtruncatemacro{\xedge}{\x-1}
        \node (b\x) [draw, shape=circle, minimum size=0.4cm, inner sep=0pt, fill=blue!20] at (8-\x, -0.8) {\xbf};
        \draw[-Latex] (a\xedge) -- (b\x);
    }

    \foreach \x in {7,8} {
        \node (c\x) [draw, shape=circle, minimum size=0.4cm, inner sep=0pt, fill=blue!20] at (8-\x, -1.6) {4};
    }
    \foreach \x in {3,4} {
        \node (c\x) [draw, shape=circle, minimum size=0.4cm, inner sep=0pt, fill=blue!20] at (8-\x, -1.6) {0};
    }
    \draw[-Latex] (b6) -- (c7);
    \draw[-Latex] (1.5, -1.2) -- ++(-1,0) -- (c8);
    \draw[-Latex] (b2) -- (c3);
    \draw[-Latex] (5.5, -1.2) -- ++(-1,0) -- (c4);

    \foreach \x in {5,6,7,8} {
        \node (d\x) [draw, shape=circle, minimum size=0.4cm, inner sep=0pt, fill=blue!20] at (8-\x, -2.4) {0};
    }
    \draw[-Latex] (c4) -- (d5);
    \draw[-Latex] (3.5, -2) -- ++(-1,0) -- (d6);
    \draw[-Latex] (2.5, -2) -- ++(-1,0) -- (d7);
    \draw[-Latex] (1.5, -2) -- ++(-1,0) -- (d8);

\end{tikzpicture}
		\end{subfigure}
		\begin{subfigure}[t]{\textwidth}
			\caption{}\label{fig:carry-ours}
			\centering
			\begin{tikzpicture}
    \foreach \x in {1,...,16} {
        \pgfmathtruncatemacro{\xbf}{\x-1}
        \node at (16-\x, 0.5) {\x};
        \draw[-Latex, dashed] (16-\x, 0) -- (16-\x, -5.3);

        \node (a\x) [draw, shape=circle, minimum size=0.4cm, inner sep=0pt, fill=blue!20] at (16-\x, 0) {\xbf};

        \node at (16-\x, -5.8) {$c_{\x}$};
    }
    \foreach \x in {2,4,...,16} {
        \pgfmathtruncatemacro{\xbf}{\x-2}
        \pgfmathtruncatemacro{\xedge}{\x-1}
        \node (b\x) [draw, shape=circle, minimum size=0.4cm, inner sep=0pt, fill=blue!20] at (16-\x, -0.8) {\xbf};
        \draw[-Latex] (a\xedge) -- (b\x);
    }
    \foreach \x in {4,8,12,16} {
        \pgfmathtruncatemacro{\xbf}{\x-4}
        \pgfmathtruncatemacro{\xedge}{\x-2}
        \node (c\x) [draw, shape=circle, minimum size=0.4cm, inner sep=0pt, fill=blue!20] at (16-\x, -1.6) {\xbf};
        \draw[-Latex] (b\xedge) -- (c\x);
    }
    \foreach \x in {3,7,11,15} {
        \pgfmathtruncatemacro{\xbf}{\x-3}
        \node (d\x) [draw, shape=circle, minimum size=0.4cm, inner sep=0pt, fill=blue!20] at (16-\x, -2.4) {\xbf};
        \draw[-Latex] (16-\x+1, -1.6) -- (d\x);
    }

    \node (e8) [draw, shape=circle, minimum size=0.4cm, inner sep=0pt, fill=blue!20] at (8, -3.2) {0};
    \draw[-Latex] (12, -2.4) -- (e8);
    \node (e16) [draw, shape=circle, minimum size=0.4cm, inner sep=0pt, fill=blue!20] at (0, -3.2) {8};
    \draw[-Latex] (4, -2.4) -- (e16);

    \node (f12) [draw, shape=circle, minimum size=0.4cm, inner sep=0pt, fill=blue!20] at (4, -4) {0};
    \draw[-Latex] (e8) -- (f12);
    \node (f16) [draw, shape=circle, minimum size=0.4cm, inner sep=0pt, fill=blue!20] at (0, -4) {0};
    \draw[-Latex] (6, -3.6) -- ++(-4,0) -- (f16);

    \foreach \x in {1,2,3} {
        \pgfmathtruncatemacro{\xone}{\x*4+1}
        \pgfmathtruncatemacro{\xtwo}{\x*4+2}
        \pgfmathtruncatemacro{\xthree}{\x*4+3}
        \node (g\xone) [draw, shape=circle, minimum size=0.4cm, inner sep=0pt, fill=blue!20] at (16-\xone, -4.8) {0};
        \node (g\xtwo) [draw, shape=circle, minimum size=0.4cm, inner sep=0pt, fill=blue!20] at (16-\xtwo, -4.8) {0};
        \node (g\xthree) [draw, shape=circle, minimum size=0.4cm, inner sep=0pt, fill=blue!20] at (16-\xthree, -4.8) {0};
        \draw[-Latex] (16-\x*4-0.5, -4.4) -- ++(-1,0) -- (g\xtwo);
        \draw[-Latex] (16-\x*4-1.5, -4.4) -- ++(-1,0) -- (g\xthree);
    }
    \draw[-Latex] (12, -4) -- (g5);
    \draw[-Latex] (8, -4) -- (g9);
    \draw[-Latex] (f12) -- (g13);

    \draw[densely dotted, color=green!50!black] (-0.5, -2.6) -- (16.5, -2.6);
    \draw[densely dotted, color=green!50!black] (-0.5, -4.2) -- (16.5, -4.2);

    \node[align=center] at (15.8, -1.2) {block\\phase};
    \node[align=center] at (15.8, -3.4) {inter-block\\phase};
    \node[align=center] at (15.8, -4.75) {merge\\phase};
\end{tikzpicture}
		\end{subfigure}
		\caption{Demonstration of carry bit computing by solving the prefix sum problem. Each column corresponds to a group of qubits that store the values of $p$ and $g$, while each row represents a layer of Toffoli gates. If the $j$-th column is labeled $i$, it indicates that the values of $p[i,j]$ and $g[i,j]$ are stored in the qubits. Since $p[i,k]$ and $g[i,k]$ can be derived from $p[i,j]$, $g[i,j]$ together with $p[j,k]$, $g[j,k]$, the label of the $k$-th column can be updated from $j$ to $i$ whenever the $j$-th column is already labeled $i$. Proceeding in this way, the process eventually labels every column with $0$, meaning that $g[0,i]$ has been obtained for all $i$. These values correspond exactly to the carry bits.\ (a) Method based on the Brent-Kung tree~\cite{draper2004logarithmic}.\ (b) Method based on the Sklansky tree~\cite{wang2024optimal}.\ (c) Our method combining both approaches.}\label{fig:carry}
	\end{figure}
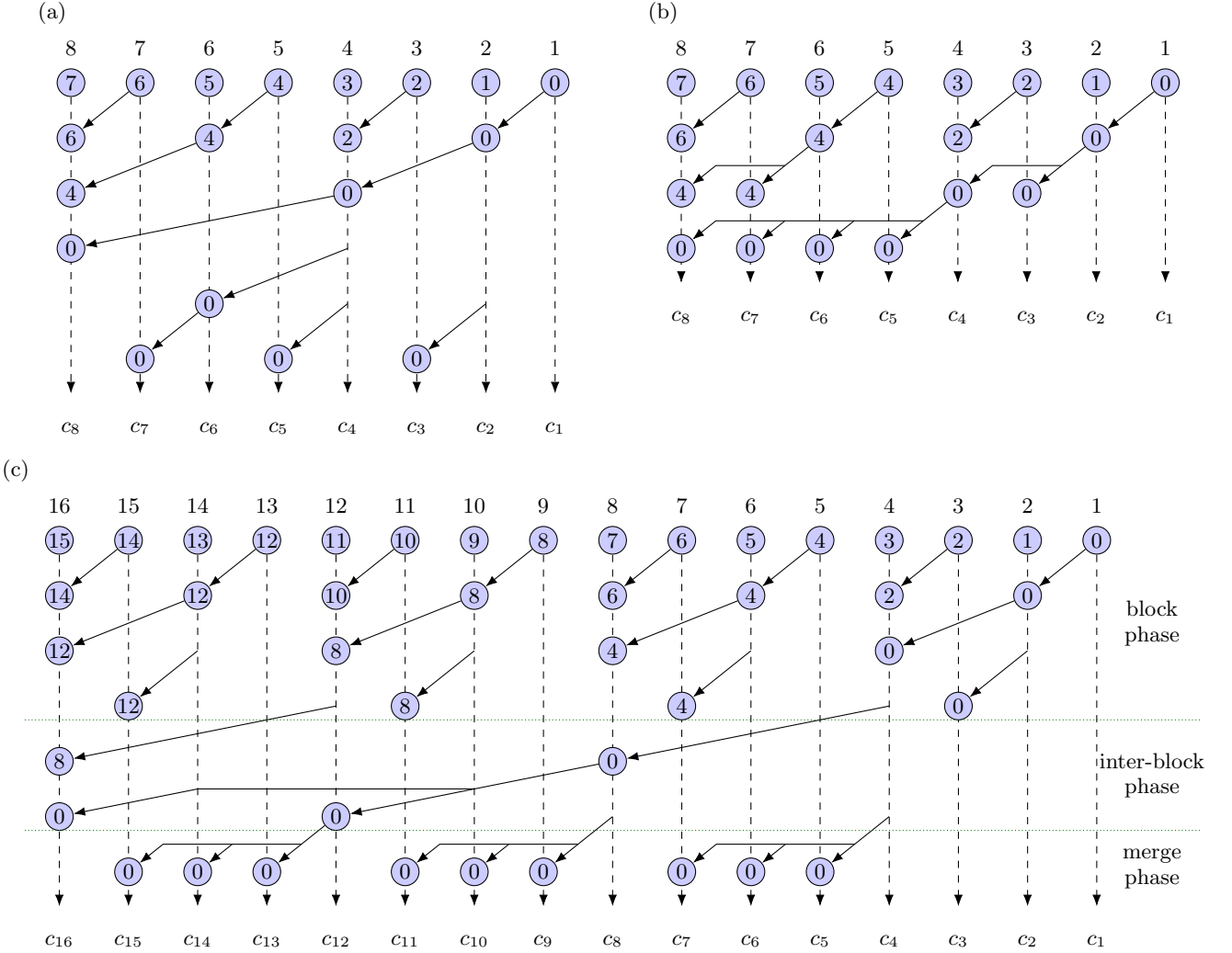
	
	We now analyze the resource cost of the proposed method, where ``$\approx$'' indicates that only leading terms are retained. Here, we use ``I-O qubits'' to denote the the qubits served as both input and output qubits, which are not counted in ``input qubits'' or ``output qubits''. In the block phase, the circuit uses $n$ input qubits, $2n$ output qubits, $n$ I-O qubits, and $n$ ancillas. The Toffoli and CNOT depths are $\approx 2\log k \approx 2\log\log n$, and the Toffoli and CNOT counts are $\approx 4n$. In the inter-block phase, the circuit requires $m$ inputs, $m$ I-O qubits, and $\frac{1}{2}m\log m+m \approx \frac{1}{2}n$ ancillas, with Toffoli and CNOT depths $\approx \log m \approx \log n - \log\log n$ and gate counts $\approx m\log m \approx n - n\frac{\log\log n}{\log n}$. In the merge phase, $k-1$ copies of $G[ik]$ are prepared for each block using GHZ-state generation, costing $m \cdot \frac{3}{2}(k-1)$ CNOT gates with depth $3$ and $(n-m)$ ancillas. The subsequent Toffoli gates are executed simultaneously, requiring an additional $(n-m)$ ancillas (two per Toffoli gate), after which the GHZ states are uncomputed at no extra Toffoli or CNOT cost. Thus, the merge phase uses $2(n-m)$ ancillas in total. This ancilla cost can be reduced to $\frac{2(n-m)}{k}$ at the expense of increasing the Toffoli depth to $k$. In this work, we allocate $n$ ancillas for the merge phase, achieving Toffoli depth two. Overall, the construction involves $n$ input qubits, $n$ I-O qubits, and $3n$ ancillas, for a total of $5n$ qubits. The circuit depth and gate numbers are given by
	\begin{equation}
		\begin{aligned}
			\text{Toffoli depth} &= \left\lfloor \log k \right\rfloor + \left\lfloor \log \frac{2k}{3} \right\rfloor + \left\lceil \log m \right\rceil + 1 = \log n + \log\log n + O(1), \\
			\text{Toffoli number} &= m\cdot (4k-2w(k)-2 \left\lfloor \log k \right\rfloor-2) + 2\sum_{i=0}^{m-1}w(i) + n-m = 6n + O(1), \\
			\text{CNOT depth} &=  \left\lfloor \log k \right\rfloor + \left\lfloor \log \frac{2k}{3} \right\rfloor+2 + \left\lceil \log m \right\rceil+1 + 1 = \log n + \log\log n + O(1), \\
			\text{CNOT number} &= m\cdot (4k-2w(k)-2 \left\lfloor \log k \right\rfloor-2) + 2\sum_{i=0}^{m-1}w(i) + m\cdot \frac{3}{2}(k-1) = \frac{13}{2}n + O(1),
		\end{aligned}
	\end{equation}
	where $w(n) = n - \sum_{i=1}^\infty \left\lfloor \frac{n}{2^i} \right\rfloor$ is the number of ``$1$''s in the binary representation of $n$.
	
	In conclusion, we obtain a quantum carry-bit computation algorithm with Toffoli depth $(1+o(1))\log n$ and only linear ancilla overhead, representing the best known trade-off to date. Building on the carry-bit construction, the in-place quantum adder is straightforward following the framework of~\cite{draper2004logarithmic}, while the out-of-place variant and other extensions are described in Appendix~\ref{subsec:variants-addition}.

	\subsection{In-place Quantum Adder}\label{subsec:inplace-adder}
	Suppose we have successfully computed the carry bits $c$, i.e., $\ket{x}\ket{y}\ket{0}\mapsto\ket{x}\ket{y}\ket{c}$. To obtain an in-place adder $\ket{x}\ket{y}\ket{0}\mapsto\ket{x}\ket{s}\ket{0}$ with $s=x+y$, we first apply CNOT gates to compute $\ket{x}\ket{s}\ket{c}$ using the relation $s_i = x_i \oplus y_i \oplus c_i$. The remaining task is to uncompute the carry register $\ket{c}$, which can be accomplished by a standard trick. For the $n$-bit sum $s=x+y$, let $s'$ denote its bitwise complement, i.e., $s' = 2^n - x - y - 1 = x \oplus y \oplus c \oplus (-1)$, where we use $\oplus$ denotes bitwise XOR for simplicity. Noting that $x + s' = x + (2^n - x - y - 1) = 2^n - y - 1 = y'$, and denoting the carry string generated by $x$ and $s'$ as $c'$, we find
	\begin{equation}
		\begin{gathered}
			x \oplus s' \oplus c' = y', \\
			x \oplus (x \oplus y \oplus c \oplus (-1)) \oplus c' = y \oplus (-1), \\
			c' = c.
		\end{gathered}
	\end{equation}
	Therefore, the carry-bit string generated by $(x,s')$ is identical to that generated by $(x,y)$. Consequently, after computing $s$ and $c$, we may negate $s$ to obtain $s'$, and then recompute the carry bits in reverse to erase $c$. The complete procedure is summarized in Algorithm~\ref{algorithm:in-place-adder}.
	
	For clarity, we denote the Toffoli depth of the carry-bit computation as $d_t$, the Toffoli count as $n_t$, the CNOT depth as $d_c$, the CNOT count as $n_c$, and the ancilla overhead as $s_a$. In our in-place adder, the procedure requires $2n$ input qubits, one output qubit for the leading carry bit (which may be omitted), and $s_a+n$ ancillas. The overall Toffoli depth is $2d_t+2$, with Toffoli count $2n_t+2n-1$, and the CNOT depth is $2d_c+4$, with CNOT count $2n_c+4n-5$. When instantiated with our carry-bit computation scheme, the leading terms of these costs become
	\begin{equation}
		\begin{aligned}
			\text{Toffoli depth} &= 2\log n + 2\log\log n + O(1), \\
			\text{Toffoli number} &= 14n + O(1), \\
			\text{CNOT depth} &= 2\log n + 2\log\log n + O(1), \\
			\text{CNOT number} &= 17n + O(1), \\
			\text{Ancilla} &= 4n + O(1).
		\end{aligned}
	\end{equation}
	
	Moreover, integer subtraction $\ket{x}\ket{y}\mapsto\ket{x}\ket{x-y}$ can be reduced to integer addition combined with two layers of bit flips. Specifically, we can first apply Pauli-$X$ gates to flip all bits of $x$, obtaining $\ket{\bar{x}} = X^{\otimes n}\ket{x} = \ket{2^n - x - 1}$, perform the integer addition, and then flip $\ket{\bar{x}}$ and the outcome bits back. Since $X$ gates are not counted in our resource analysis, integer subtraction is equivalent to integer addition in the following discussion.
	
	\begin{algorithm}[ht]
		\SetKwInOut{Input}{Input}
		\SetKwInOut{Output}{Output}
		\SetKw{for}{for}
		\SetKwProg{Fn}{Function}{:}{}
		\SetKwFunction{InPlaceAdder}{InPlaceAdder}
		\SetKwFunction{ComputeCarry}{ComputeCarry}
		\SetKwFunction{ComputeCarryInv}{$\texttt{ComputeCarry}^{-1}$}
		
		\LinesNumbered{}
		\Fn{\InPlaceAdder{$x;y;c[n];c[1,\ldots, n-1],a$}}{
			\Input{$n$-qubit number $x$; $n$-qubit number $y$; $n$-qubit $s$, initialized to 0; ancilla $a$ (size depends on \ComputeCarry), initialized to 0.}
			\Output{$x,c[1,\ldots, n-1],a$ remain the same; $y$ holds sum bits; $c[n]$ holds the leading carry bit.}
			$c[i+1] \oplusIs x[i]y[i]$ \for{} $i \gets 0 ,\ldots, n-1$\tcp*{set $c[i+1] = g[i,i+1]$}
			$y[i] \oplusIs x[i]$ \for{} $i \gets 0 ,\ldots, n-1$\tcp*{set $y[i] = p[i,i+1]$ for $i\geq1$; $y[0]=s_0$}
			\ComputeCarry{$y;c;a$}\tcp*{set $c[i] = c_i$ for $i\geq 1$}
			$y[i] \oplusIs c[i]$ \for{} $i \gets 1 ,\ldots, n-1$\tcp*{set $y[i] = s_i$}
			$y[i]=-y[i]$ \for{} $i \gets 0 ,\ldots, n-2$\tcp*{negate $y$ to get $s'$}
			$y[i] \oplusIs x[i]$ \for{} $i \gets 1 ,\ldots, n-2$\tcp*{set $y[i]=p'[i,i+1]$ for $s'$ and $x$}
			\ComputeCarryInv{$y;c;a$}\tcp*{set $c[i+1]=g'[i,i+1]=x_i \cdot s'_i$ for $i\leq n-1$}
			$y[i] \oplusIs x[i]$ \for{} $i \gets 1 ,\ldots, n-2$\tcp*{recover $y$ to $s'$}
			$c[i+1] \oplusIs x[i]y[i]$ \for{} $i \gets 0 ,\ldots, n-2$\tcp*{recover $c$ to 0 for $i\leq n-1$}
			$y[i]=-y[i]$ \for{} $i \gets 0 ,\ldots, n-2$\tcp*{recover $y$ to $s$}
		}
		\caption{quantum algorithm for in-place adder.}\label{algorithm:in-place-adder}
	\end{algorithm}

	\section{Circuit Analysis}\label{sec:circuit-analysis}
	In this section, we present the quantum circuits used in our analysis and evaluate their resource requirements. Since the full quantum circuit for Shor's algorithm is highly complex, we begin in Sec.~\ref{subsec:point-addition} with the circuit for controlled point addition, which serves as a fundamental building block for the overall construction. The complete circuit, assembled from these building blocks, is described in Sec.~\ref{subsec:full-circuit}, and its resource requirements are analyzed in Sec.~\ref{subsec:circuit-parameters}.

	\subsection{Point Addition}\label{subsec:point-addition}
\begin{figure}[ht]
		\captionsetup{justification=raggedright,singlelinecheck=false}
		\centering
		\begin{tikzpicture}[scale=1.000000,x=1pt,y=1pt]
\filldraw[color=white] (0.000000, -7.500000) rectangle (416.000000, 52.500000);
\draw[color=black] (0.000000,45.000000) -- (440.000000,45.000000);
\draw[color=black] (0.000000,45.000000) node[left] {$\ket{x_1}$};
\draw[color=black] (0.000000,30.000000) -- (440.000000,30.000000);
\draw[color=black] (0.000000,30.000000) node[left] {$\ket{y_1}$};
\draw[color=black] (0.000000,15.000000) -- (440.000000,15.000000);
\draw[color=black] (0.000000,15.000000) node[left] {$\ket{\lambda=0}$};
\draw[color=black] (0.000000,0.000000) -- (440.000000,0.000000);
\draw[color=black] (0.000000,0.000000) node[left] {$\ket{ctrl}$};
\draw (6.000000, 39.000000) -- (14.000000, 51.000000);
\draw (12.000000, 48.000000) node[right] {$\scriptstyle{n}$};
\draw (6.000000, 24.000000) -- (14.000000, 36.000000);
\draw (12.000000, 33.000000) node[right] {$\scriptstyle{n}$};
\draw (6.000000, 9.000000) -- (14.000000, 21.000000);
\draw (12.000000, 18.000000) node[right] {$\scriptstyle{n}$};
\begin{scope}
\draw[fill=white] (41.000000, 45.000000) +(-45.000000:21.213203pt and 8.485281pt) -- +(45.000000:21.213203pt and 8.485281pt) -- +(135.000000:21.213203pt and 8.485281pt) -- +(225.000000:21.213203pt and 8.485281pt) -- cycle;
\clip (41.000000, 45.000000) +(-45.000000:21.213203pt and 8.485281pt) -- +(45.000000:21.213203pt and 8.485281pt) -- +(135.000000:21.213203pt and 8.485281pt) -- +(225.000000:21.213203pt and 8.485281pt) -- cycle;
\draw (41.000000, 45.000000) node {$-x_2$};
\end{scope}
\draw (41.000000,30.000000) -- (41.000000,0.000000);
\begin{scope}
\draw[fill=white] (41.000000, 30.000000) +(-45.000000:21.213203pt and 8.485281pt) -- +(45.000000:21.213203pt and 8.485281pt) -- +(135.000000:21.213203pt and 8.485281pt) -- +(225.000000:21.213203pt and 8.485281pt) -- cycle;
\clip (41.000000, 30.000000) +(-45.000000:21.213203pt and 8.485281pt) -- +(45.000000:21.213203pt and 8.485281pt) -- +(135.000000:21.213203pt and 8.485281pt) -- +(225.000000:21.213203pt and 8.485281pt) -- cycle;
\draw (41.000000, 30.000000) node {$-y_2$};
\end{scope}
\filldraw (41.000000, 0.000000) circle(1.500000pt);
\draw (83.000000,45.000000) -- (83.000000,0.000000);
\begin{scope}
\draw[fill=white] (83.000000, 30.000000) +(-45.000000:21.213203pt and 29.698485pt) -- +(45.000000:21.213203pt and 29.698485pt) -- +(135.000000:21.213203pt and 29.698485pt) -- +(225.000000:21.213203pt and 29.698485pt) -- cycle;
\clip (83.000000, 30.000000) +(-45.000000:21.213203pt and 29.698485pt) -- +(45.000000:21.213203pt and 29.698485pt) -- +(135.000000:21.213203pt and 29.698485pt) -- +(225.000000:21.213203pt and 29.698485pt) -- cycle;
\draw (83.000000, 30.000000) node {$\verb|div|$};
\end{scope}
\qOutput{98}{15}{black}
\filldraw (83.000000, 0.000000) circle(1.500000pt);
\draw (125.000000,45.000000) -- (125.000000,15.000000);
\begin{scope}
\draw[fill=white] (125.000000, 30.000000) +(-45.000000:21.213203pt and 29.698485pt) -- +(45.000000:21.213203pt and 29.698485pt) -- +(135.000000:21.213203pt and 29.698485pt) -- +(225.000000:21.213203pt and 29.698485pt) -- cycle;
\clip (125.000000, 30.000000) +(-45.000000:21.213203pt and 29.698485pt) -- +(45.000000:21.213203pt and 29.698485pt) -- +(135.000000:21.213203pt and 29.698485pt) -- +(225.000000:21.213203pt and 29.698485pt) -- cycle;
\draw (125.000000, 30.000000) node {$\verb|mul|_F$};
\end{scope}
\qOutput{140}{30}{black}
\begin{scope}
\draw[fill=white] (167.000000, 45.000000) +(-45.000000:21.213203pt and 8.485281pt) -- +(45.000000:21.213203pt and 8.485281pt) -- +(135.000000:21.213203pt and 8.485281pt) -- +(225.000000:21.213203pt and 8.485281pt) -- cycle;
\clip (167.000000, 45.000000) +(-45.000000:21.213203pt and 8.485281pt) -- +(45.000000:21.213203pt and 8.485281pt) -- +(135.000000:21.213203pt and 8.485281pt) -- +(225.000000:21.213203pt and 8.485281pt) -- cycle;
\draw (167.000000, 45.000000) node {$+3x_2$};
\end{scope}
\draw (209.000000,45.000000) -- (209.000000,15.000000);
\begin{scope}
\draw[fill=white] (209.000000, 45.000000) +(-45.000000:21.213203pt and 8.485281pt) -- +(45.000000:21.213203pt and 8.485281pt) -- +(135.000000:21.213203pt and 8.485281pt) -- +(225.000000:21.213203pt and 8.485281pt) -- cycle;
\clip (209.000000, 45.000000) +(-45.000000:21.213203pt and 8.485281pt) -- +(45.000000:21.213203pt and 8.485281pt) -- +(135.000000:21.213203pt and 8.485281pt) -- +(225.000000:21.213203pt and 8.485281pt) -- cycle;
\draw (209.000000, 45.000000) node {$\verb|squ|^-$};
\end{scope}
\begin{scope}
\draw[fill=white] (209.000000, 15.000000) +(-45.000000:21.213203pt and 8.485281pt) -- +(45.000000:21.213203pt and 8.485281pt) -- +(135.000000:21.213203pt and 8.485281pt) -- +(225.000000:21.213203pt and 8.485281pt) -- cycle;
\clip (209.000000, 15.000000) +(-45.000000:21.213203pt and 8.485281pt) -- +(45.000000:21.213203pt and 8.485281pt) -- +(135.000000:21.213203pt and 8.485281pt) -- +(225.000000:21.213203pt and 8.485281pt) -- cycle;
\draw (209.000000, 15.000000) node {$\verb|squ|^-$};
\end{scope}
\qOutput{224}{45}{black}
\draw (251.000000,45.000000) -- (251.000000,15.000000);
\begin{scope}
\draw[fill=white] (251.000000, 30.000000) +(-45.000000:21.213203pt and 29.698485pt) -- +(45.000000:21.213203pt and 29.698485pt) -- +(135.000000:21.213203pt and 29.698485pt) -- +(225.000000:21.213203pt and 29.698485pt) -- cycle;
\clip (251.000000, 30.000000) +(-45.000000:21.213203pt and 29.698485pt) -- +(45.000000:21.213203pt and 29.698485pt) -- +(135.000000:21.213203pt and 29.698485pt) -- +(225.000000:21.213203pt and 29.698485pt) -- cycle;
\draw (251.000000, 30.000000) node {$\verb|mul|^+$};
\end{scope}
\qOutput{266}{30}{black}
\draw (293.000000,45.000000) -- (293.000000,0.000000);
\begin{scope}
\draw[fill=white] (293.000000, 30.000000) +(-45.000000:21.213203pt and 29.698485pt) -- +(45.000000:21.213203pt and 29.698485pt) -- +(135.000000:21.213203pt and 29.698485pt) -- +(225.000000:21.213203pt and 29.698485pt) -- cycle;
\clip (293.000000, 30.000000) +(-45.000000:21.213203pt and 29.698485pt) -- +(45.000000:21.213203pt and 29.698485pt) -- +(135.000000:21.213203pt and 29.698485pt) -- +(225.000000:21.213203pt and 29.698485pt) -- cycle;
\draw (293.000000, 30.000000) node {$\verb|div|^{-1}$};
\end{scope}
\qOutput{308}{15}{black}
\filldraw (293.000000, 0.000000) circle(1.500000pt);
\begin{scope}
\draw[fill=white] (335.000000, 45.000000) +(-45.000000:21.213203pt and 8.485281pt) -- +(45.000000:21.213203pt and 8.485281pt) -- +(135.000000:21.213203pt and 8.485281pt) -- +(225.000000:21.213203pt and 8.485281pt) -- cycle;
\clip (335.000000, 45.000000) +(-45.000000:21.213203pt and 8.485281pt) -- +(45.000000:21.213203pt and 8.485281pt) -- +(135.000000:21.213203pt and 8.485281pt) -- +(225.000000:21.213203pt and 8.485281pt) -- cycle;
\draw (335.000000, 45.000000) node {$-2x_2$};
\end{scope}
\draw (335.000000,30.000000) -- (335.000000,0.000000);
\begin{scope}
\draw[fill=white] (335.000000, 30.000000) +(-45.000000:21.213203pt and 8.485281pt) -- +(45.000000:21.213203pt and 8.485281pt) -- +(135.000000:21.213203pt and 8.485281pt) -- +(225.000000:21.213203pt and 8.485281pt) -- cycle;
\clip (335.000000, 30.000000) +(-45.000000:21.213203pt and 8.485281pt) -- +(45.000000:21.213203pt and 8.485281pt) -- +(135.000000:21.213203pt and 8.485281pt) -- +(225.000000:21.213203pt and 8.485281pt) -- cycle;
\draw (335.000000, 30.000000) node {$-y_2$};
\end{scope}
\filldraw (335.000000, 0.000000) circle(1.500000pt);
\draw (377.000000,45.000000) -- (377.000000,0.000000);
\begin{scope}
\draw[fill=white] (377.000000, 45.000000) +(-45.000000:21.213203pt and 8.485281pt) -- +(45.000000:21.213203pt and 8.485281pt) -- +(135.000000:21.213203pt and 8.485281pt) -- +(225.000000:21.213203pt and 8.485281pt) -- cycle;
\clip (377.000000, 45.000000) +(-45.000000:21.213203pt and 8.485281pt) -- +(45.000000:21.213203pt and 8.485281pt) -- +(135.000000:21.213203pt and 8.485281pt) -- +(225.000000:21.213203pt and 8.485281pt) -- cycle;
\draw (377.000000, 45.000000) node {$+x_2$};
\end{scope}
\filldraw (377.000000, 0.000000) circle(1.500000pt);
\draw (419.000000,45.000000) -- (419.000000,0.000000);
\begin{scope}
\draw[fill=white] (419.000000, 45.000000) +(-45.000000:21.213203pt and 8.485281pt) -- +(45.000000:21.213203pt and 8.485281pt) -- +(135.000000:21.213203pt and 8.485281pt) -- +(225.000000:21.213203pt and 8.485281pt) -- cycle;
\clip (419.000000, 45.000000) +(-45.000000:21.213203pt and 8.485281pt) -- +(45.000000:21.213203pt and 8.485281pt) -- +(135.000000:21.213203pt and 8.485281pt) -- +(225.000000:21.213203pt and 8.485281pt) -- cycle;
\draw (419.000000, 45.000000) node {$\verb|neg|$};
\end{scope}
\filldraw (419.000000, 0.000000) circle(1.500000pt);
\draw[color=black] (440.000000,45.000000) node[right] {$\ket{x_3}$};
\draw[color=black] (440.000000,30.000000) node[right] {$\ket{y_3}$};
\draw[color=black] (440.000000,15.000000) node[right] {$\ket{0}$};
\draw[color=black] (440.000000,0.000000) node[right] {$\ket{ctrl}$};
\end{tikzpicture}
		\caption{Circuit for point addition. Here, $\pm a$ denotes the operation $\ket{x} \mapsto \ket{(x\pm a) \bmod{p}}$, \texttt{neg} denotes the operation $\ket{x} \mapsto \ket{(-x) \bmod{p}}$, $\texttt{mul}_F$ denotes the operation $\ket{x}\ket{y}\ket{z} \mapsto \ket{x}\ket{y}\ket{z \oplus (x\cdot y \bmod{p})}$ where $\oplus$ is the bit-wise XOR, $\texttt{mul}^+$ denotes the operation $\ket{x}\ket{y}\ket{z} \mapsto \ket{x}\ket{y}\ket{(x\cdot y + z) \bmod{p}}$, $\texttt{squ}^-$ denotes the operation $\ket{x}\ket{y} \mapsto \ket{(x-y^2)\bmod{p}}\ket{y}$, and \texttt{div} denotes the operation $\ket{x}\ket{y}\ket{0} \mapsto \ket{x}\ket{y}\ket{y/x \bmod{p}}$. The $\blacktriangleright$ in the circuit denotes the output qubit of an operation.}\label{figure:circuit-point-addition}
	\end{figure}
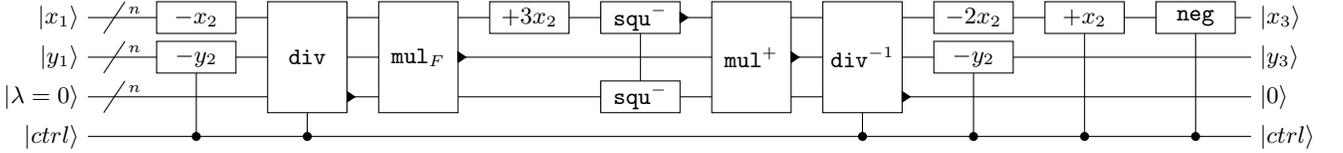

	In Shor's algorithm, one of the inputs to the elliptic-curve point addition is either $P$ or $Q$, whose values are stored classically rather than in qubits. This allows us to embed the corresponding information directly into the unitary circuit, thereby eliminating the need for qubits to store $P$ and $Q$ and simplifying the overall design. Based on Algorithm~\ref{algorithm:point-addition}, the quantum circuit for elliptic-curve point addition is shown in Fig.~\ref{figure:circuit-point-addition}. The circuit involves two divisions, two (full) multiplications, one (full) squaring, and nine integer additions (all of them are modular), of which two additions can be performed in parallel. The implementation of integer addition is described in Sec.~\ref{sec:improvedadder}. It also serves as the fundamental building block for multiplication and inversion, while division is implemented by combining inversion with multiplication. The detailed constructions of multiplication and division are provided in Appendix~\ref{subsec:multiplication} and~\ref{subsec:division}.
	
	Notably, since no in-place implementations of multiplication and division are currently available, we employ out-of-place versions, defined analogously to the out-of-place adder: $U\ket{x}\ket{y}\ket{0}=\ket{x}\ket{y}\ket{x\cdot y}$ and $U\ket{x}\ket{y}\ket{0}=\ket{x}\ket{y}\ket{x/y}$. Both multiplication and division require sequences of additions. To realize these operations within limited qubit space, we must adopt in-place addition as the basic building block, even though out-of-place addition typically achieves only about half the complexity. This necessity also motivates our exclusive focus on the in-place adder in Sec.~\ref{sec:improvedadder}.
	
	The presented circuit does not correctly handle the special case involving the identity element $O$. To avoid this issue, the quantum register is typically initialized to a state representing $kG$ (for some integer $k$) rather than $O$~\cite{proos2003shor}. Such initialization does not affect the correctness of the final result, as it only introduces a global phase shift after applying the quantum Fourier transform. Additionally, the circuit encounters difficulties when the condition $x_1 = x_2$ arises; however, the probability of this event is exponentially small~\cite{proos2003shor} and therefore has a negligible effect on the fidelity of the algorithm's output.
	
	In terms of resource complexity, point addition requires Toffoli depth $56n\log n + 40n\frac{\log n}{\log\log n}$, Toffoli count $572n^2 + 140\frac{n^2}{\log\log n}$, CNOT depth $48n\log n + 120n\frac{\log n}{\log\log n}$, and CNOT count $735n^2 + 40n^2 \tfrac{\log n}{\log\log n}$. The dominant contribution arises from division. Since division has been studied far less extensively than addition and exhibits substantially higher complexity, we expect it to provide the most promising avenue for further optimization.
	
	The circuit for point addition requires $17n$ qubits in total, matching the cost of modular division, and has gate interaction distance $5$. In each step of the procedure in Fig.~\ref{figure:circuit-point-addition}, the qubits involved must maintain a fixed positional relationship. Our qubit layout is designed to satisfy this requirement wherever possible, while minor deviations can always be corrected with a constant number of SWAP operations, whose overhead is negligible compared to the overall algorithm.

	\subsection{Shor's Algorithm}\label{subsec:full-circuit}
\begin{figure}[ht]
		\captionsetup{justification=raggedright,singlelinecheck=false}
		\centering
		\begin{tikzpicture}[scale=0.880000,x=1pt,y=1pt,transform shape]
\filldraw[color=white] (0.000000, -7.500000) rectangle (528.000000, 37.500000);
\draw[color=black] (0.000000,15.000000) -- (92.000000,15.000000);
\draw[color=black] (92.000000,14.500000) -- (117.500000,14.500000);
\draw[color=black] (92.000000,15.500000) -- (117.500000,15.500000);
\draw[color=black] (144.500000,15.000000) -- (279.000000,15.000000);
\draw[color=black] (279.000000,14.500000) -- (304.500000,14.500000);
\draw[color=black] (279.000000,15.500000) -- (304.500000,15.500000);
\draw[color=black] (358.500000,15.000000) -- (516.000000,15.000000);
\draw[color=black] (516.000000,14.500000) -- (528.000000,14.500000);
\draw[color=black] (516.000000,15.500000) -- (528.000000,15.500000);
\draw[color=black] (0.000000,15.000000) node[left] {$\ket{0}$};
\draw[color=black] (0.000000,0.000000) -- (528.000000,0.000000);
\draw[color=black] (0.000000,0.000000) node[left] {$\ket{P_0}$};
\draw (8.000000, -6.000000) -- (16.000000, 6.000000);
\draw (14.000000, 3.000000) node[right] {$\scriptstyle{2n}$};
\begin{scope}
\draw[fill=white] (12.000000, 15.000000) +(-45.000000:8.485281pt and 8.485281pt) -- +(45.000000:8.485281pt and 8.485281pt) -- +(135.000000:8.485281pt and 8.485281pt) -- +(225.000000:8.485281pt and 8.485281pt) -- cycle;
\clip (12.000000, 15.000000) +(-45.000000:8.485281pt and 8.485281pt) -- +(45.000000:8.485281pt and 8.485281pt) -- +(135.000000:8.485281pt and 8.485281pt) -- +(225.000000:8.485281pt and 8.485281pt) -- cycle;
\draw (12.000000, 15.000000) node {$H$};
\end{scope}
\draw (40.000000,15.000000) -- (40.000000,0.000000);
\begin{scope}
\draw[fill=white] (40.000000, 0.000000) +(-45.000000:14.142136pt and 8.485281pt) -- +(45.000000:14.142136pt and 8.485281pt) -- +(135.000000:14.142136pt and 8.485281pt) -- +(225.000000:14.142136pt and 8.485281pt) -- cycle;
\clip (40.000000, 0.000000) +(-45.000000:14.142136pt and 8.485281pt) -- +(45.000000:14.142136pt and 8.485281pt) -- +(135.000000:14.142136pt and 8.485281pt) -- +(225.000000:14.142136pt and 8.485281pt) -- cycle;
\draw (40.000000, 0.000000) node {$+P$};
\end{scope}
\filldraw (40.000000, 15.000000) circle(1.500000pt);
\begin{scope}
\draw[fill=white] (68.000000, 15.000000) +(-45.000000:8.485281pt and 8.485281pt) -- +(45.000000:8.485281pt and 8.485281pt) -- +(135.000000:8.485281pt and 8.485281pt) -- +(225.000000:8.485281pt and 8.485281pt) -- cycle;
\clip (68.000000, 15.000000) +(-45.000000:8.485281pt and 8.485281pt) -- +(45.000000:8.485281pt and 8.485281pt) -- +(135.000000:8.485281pt and 8.485281pt) -- +(225.000000:8.485281pt and 8.485281pt) -- cycle;
\draw (68.000000, 15.000000) node {$H$};
\end{scope}
\draw[fill=white] (86.000000, 9.000000) rectangle (98.000000, 21.000000);
\draw[very thin] (92.000000, 15.600000) arc (90:150:6.000000pt);
\draw[very thin] (92.000000, 15.600000) arc (90:30:6.000000pt);
\draw[->,>=stealth] (92.000000, 9.600000) -- +(80:10.392305pt);
\draw[color=black] (110.000000,15.000000) node[fill=white,right,minimum height=15.000000pt,minimum width=15.000000pt,inner sep=0pt] {\phantom{$\mu_0$}};
\draw[color=black] (110.000000,15.000000) node[right] {$\mu_0$};
\draw[color=black] (152.000000,15.000000) node[fill=white,left,minimum height=15.000000pt,minimum width=15.000000pt,inner sep=0pt] {\phantom{$\ket{0}$}};
\draw[color=black] (152.000000,15.000000) node[left] {$\ket{0}$};
\begin{scope}
\draw[fill=white] (170.000000, 15.000000) +(-45.000000:8.485281pt and 8.485281pt) -- +(45.000000:8.485281pt and 8.485281pt) -- +(135.000000:8.485281pt and 8.485281pt) -- +(225.000000:8.485281pt and 8.485281pt) -- cycle;
\clip (170.000000, 15.000000) +(-45.000000:8.485281pt and 8.485281pt) -- +(45.000000:8.485281pt and 8.485281pt) -- +(135.000000:8.485281pt and 8.485281pt) -- +(225.000000:8.485281pt and 8.485281pt) -- cycle;
\draw (170.000000, 15.000000) node {$H$};
\end{scope}
\draw (200.500000,15.000000) -- (200.500000,0.000000);
\begin{scope}
\draw[fill=white] (200.500000, 0.000000) +(-45.000000:17.677670pt and 8.485281pt) -- +(45.000000:17.677670pt and 8.485281pt) -- +(135.000000:17.677670pt and 8.485281pt) -- +(225.000000:17.677670pt and 8.485281pt) -- cycle;
\clip (200.500000, 0.000000) +(-45.000000:17.677670pt and 8.485281pt) -- +(45.000000:17.677670pt and 8.485281pt) -- +(135.000000:17.677670pt and 8.485281pt) -- +(225.000000:17.677670pt and 8.485281pt) -- cycle;
\draw (200.500000, 0.000000) node {$+2P$};
\end{scope}
\filldraw (200.500000, 15.000000) circle(1.500000pt);
\begin{scope}
\draw[fill=white] (231.000000, 15.000000) +(-45.000000:8.485281pt and 8.485281pt) -- +(45.000000:8.485281pt and 8.485281pt) -- +(135.000000:8.485281pt and 8.485281pt) -- +(225.000000:8.485281pt and 8.485281pt) -- cycle;
\clip (231.000000, 15.000000) +(-45.000000:8.485281pt and 8.485281pt) -- +(45.000000:8.485281pt and 8.485281pt) -- +(135.000000:8.485281pt and 8.485281pt) -- +(225.000000:8.485281pt and 8.485281pt) -- cycle;
\draw (231.000000, 15.000000) node {$R_1$};
\end{scope}
\begin{scope}
\draw[fill=white] (255.000000, 15.000000) +(-45.000000:8.485281pt and 8.485281pt) -- +(45.000000:8.485281pt and 8.485281pt) -- +(135.000000:8.485281pt and 8.485281pt) -- +(225.000000:8.485281pt and 8.485281pt) -- cycle;
\clip (255.000000, 15.000000) +(-45.000000:8.485281pt and 8.485281pt) -- +(45.000000:8.485281pt and 8.485281pt) -- +(135.000000:8.485281pt and 8.485281pt) -- +(225.000000:8.485281pt and 8.485281pt) -- cycle;
\draw (255.000000, 15.000000) node {$H$};
\end{scope}
\draw[fill=white] (273.000000, 9.000000) rectangle (285.000000, 21.000000);
\draw[very thin] (279.000000, 15.600000) arc (90:150:6.000000pt);
\draw[very thin] (279.000000, 15.600000) arc (90:30:6.000000pt);
\draw[->,>=stealth] (279.000000, 9.600000) -- +(80:10.392305pt);
\draw[color=black] (297.000000,15.000000) node[fill=white,right,minimum height=15.000000pt,minimum width=15.000000pt,inner sep=0pt] {\phantom{$\mu_1$}};
\draw[color=black] (297.000000,15.000000) node[right] {$\mu_1$};
\draw[color=black] (331.500000, 0.000000) node [fill=white] {$\cdots$};
\draw[color=black] (366.000000,15.000000) node[fill=white,left,minimum height=15.000000pt,minimum width=15.000000pt,inner sep=0pt] {\phantom{$\ket{0}$}};
\draw[color=black] (366.000000,15.000000) node[left] {$\ket{0}$};
\begin{scope}
\draw[fill=white] (384.000000, 15.000000) +(-45.000000:8.485281pt and 8.485281pt) -- +(45.000000:8.485281pt and 8.485281pt) -- +(135.000000:8.485281pt and 8.485281pt) -- +(225.000000:8.485281pt and 8.485281pt) -- cycle;
\clip (384.000000, 15.000000) +(-45.000000:8.485281pt and 8.485281pt) -- +(45.000000:8.485281pt and 8.485281pt) -- +(135.000000:8.485281pt and 8.485281pt) -- +(225.000000:8.485281pt and 8.485281pt) -- cycle;
\draw (384.000000, 15.000000) node {$H$};
\end{scope}
\draw (417.000000,15.000000) -- (417.000000,0.000000);
\begin{scope}
\draw[fill=white] (417.000000, 0.000000) +(-45.000000:21.213203pt and 8.485281pt) -- +(45.000000:21.213203pt and 8.485281pt) -- +(135.000000:21.213203pt and 8.485281pt) -- +(225.000000:21.213203pt and 8.485281pt) -- cycle;
\clip (417.000000, 0.000000) +(-45.000000:21.213203pt and 8.485281pt) -- +(45.000000:21.213203pt and 8.485281pt) -- +(135.000000:21.213203pt and 8.485281pt) -- +(225.000000:21.213203pt and 8.485281pt) -- cycle;
\draw (417.000000, 0.000000) node {$+2^nP$};
\end{scope}
\filldraw (417.000000, 15.000000) circle(1.500000pt);
\begin{scope}
\draw[fill=white] (459.000000, 15.000000) +(-45.000000:21.213203pt and 8.485281pt) -- +(45.000000:21.213203pt and 8.485281pt) -- +(135.000000:21.213203pt and 8.485281pt) -- +(225.000000:21.213203pt and 8.485281pt) -- cycle;
\clip (459.000000, 15.000000) +(-45.000000:21.213203pt and 8.485281pt) -- +(45.000000:21.213203pt and 8.485281pt) -- +(135.000000:21.213203pt and 8.485281pt) -- +(225.000000:21.213203pt and 8.485281pt) -- cycle;
\draw (459.000000, 15.000000) node {$R_{2n+1}$};
\end{scope}
\begin{scope}
\draw[fill=white] (492.000000, 15.000000) +(-45.000000:8.485281pt and 8.485281pt) -- +(45.000000:8.485281pt and 8.485281pt) -- +(135.000000:8.485281pt and 8.485281pt) -- +(225.000000:8.485281pt and 8.485281pt) -- cycle;
\clip (492.000000, 15.000000) +(-45.000000:8.485281pt and 8.485281pt) -- +(45.000000:8.485281pt and 8.485281pt) -- +(135.000000:8.485281pt and 8.485281pt) -- +(225.000000:8.485281pt and 8.485281pt) -- cycle;
\draw (492.000000, 15.000000) node {$H$};
\end{scope}
\draw[fill=white] (510.000000, 9.000000) rectangle (522.000000, 21.000000);
\draw[very thin] (516.000000, 15.600000) arc (90:150:6.000000pt);
\draw[very thin] (516.000000, 15.600000) arc (90:30:6.000000pt);
\draw[->,>=stealth] (516.000000, 9.600000) -- +(80:10.392305pt);
\draw[color=black] (528.000000,15.000000) node[right] {$\mu_{2n+1}$};
\end{tikzpicture}
		\caption{Circuit for Shor's algorithm. Each $R_{i}$ is a z-rotational gate whose rotation angle is based on the previous measurement results $\mu_0,\mu_1,\ldots,\mu_{i-1}$. The ``$+$'' in the blocks ``$+2^i P$'' stand for point addition. $P_0$ is a random multiple of $G$ to avoid adding on the identity element $O$, as discussed in~\cite{proos2003shor}.}\label{figure:qft-without-window}
	\end{figure}
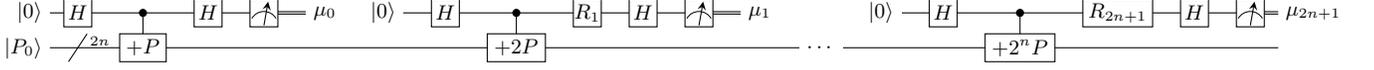

	If we directly apply the algorithm of Sec.~\ref{subsec:shor-algorithm}, implementing the full point addition  
	\begin{equation}
	\frac{1}{2^n}\sum_{x,y=0}^{2^n-1}\ket{x}\ket{y}\ket{O}\mapsto\frac{1}{2^n}\sum_{x,y=0}^{2^n-1}\ket{x}\ket{y}\ket{xG+yQ}
	\end{equation}
	requires $2n$ controlled point additions, each conditioned on the corresponding bit of $x$ or $y$. A detailed resource analysis shows that this construction has Toffoli depth $112n^2\log n + o(n^2\log n)$, Toffoli count $1144n^3 + o(n^3)$, and CNOT count $80n^3 \tfrac{\log n}{\log\log n} + 1490n^3 + o(n^3)$. The qubit cost is $17n$, the same as for a single point addition. As illustrated in Fig.~\ref{figure:qft-without-window}, dynamic-circuit techniques can be employed to compress the $2n$ qubits of the registers $\ket{x},\ket{y}$ used in the quantum Fourier transform into a single qubit, ensuring that no additional columns are introduced.
	
\begin{figure}[ht]
		\captionsetup{justification=raggedright,singlelinecheck=false}
		\centering
		\input{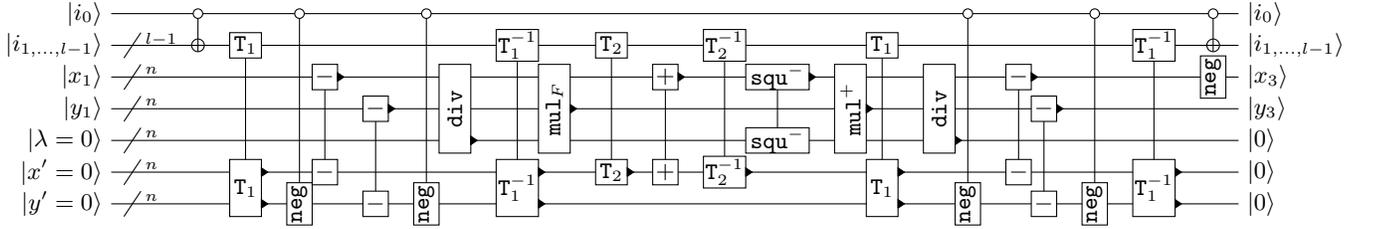}
		\caption{The circuit for windowed point addition with window size $l$ (the first window). Here, $\pm$ denotes the operation $\ket{x}\ket{y}\mapsto\ket{x}\ket{(y \pm x) \bmod{p}}$, $\texttt{T}_1$ is the quantum table with input $i$ and output ${(iP)}_x, {(iP)}_y$, where $P$ is a point on the elliptic curve, $\texttt{T}_2$ is the quantum table with input $i$ and output $3{(iP)}_x$, and the meanings of \texttt{neg}, $\texttt{mul}_F, \texttt{mul}^+, \texttt{squ}^-$, \texttt{div} are explained in the caption of Fig.~\ref{figure:circuit-point-addition}.}\label{figure:circuit-point-add-win}
	\end{figure}

	The windowed trick~\cite{haner2020improved} can further reduce the overall resource cost. Intuitively, it groups multiple blocks of Fig.~\ref{figure:qft-without-window} into a single operation, thereby decreasing the number of point additions at the expense of additional pre-computing. In the naive approach, one requires $2n$ controlled point additions of the form $+G, +2G, +4G, \ldots$ and $+Q, +2Q, +4Q, \ldots$. In contrast, with a window size $l$, $l$ control bits $\ket{b=b_{l-1}\ldots b_{0}}$ are processed simultaneously. In the $k$-th window, the operation $+bP_k$ is applied, where $P_k = 2^{l(k-1)}P$ with $P\in\{G,Q\}$. The required value of $bP_k$ is obtained from a quantum lookup table. This approach saves $l-1$ point additions per window, at the cost of maintaining a table of $2^l$ pre-computed states. The resulting circuit is shown in Fig.~\ref{figure:circuit-point-add-win}.
	
	The cost of quantum tables can be reduced by a factor of two using the signed trick. Define $b' = b - b_{l-1}2^{l-1}$, where $b_{l-1}$ is the most significant bit of the window. If $b_{l-1}=1$, we look up $b'P_k$; if $b_{l-1}=0$, we instead look up $(2^{l-1}-b')P_k$ and apply a negation. Since $-P = (P_x,-P_y)$, this negation can be implemented by a simple \verb|neg| operation on the $y$ coordinate. The procedure effectively outputs $(b-2^{l-1})P_k$, so the quantum table needs to store only $2^{l-1}$ entries rather than $2^l$. This introduces only a global phase shift in each window after the QFT, which does not affect the final result.
	
\begin{figure}[ht]
		\captionsetup{justification=raggedright,singlelinecheck=false}
		\centering
		\begin{tikzpicture}[scale=1.000000,x=1pt,y=1pt]
    \filldraw[color=white] (0.000000, -7.500000) rectangle (360.000000, 67.500000);
    \draw[color=black] (0.000000,60.000000) -- (112.000000,60.000000);
    \draw[color=black] (112.000000,59.500000) -- (137.500000,59.500000);
    \draw[color=black] (112.000000,60.500000) -- (137.500000,60.500000);
    \draw[color=black] (0.000000,60.000000) node[left] {$\ket{0}$};
    \draw[color=black] (0.000000,45.000000) -- (211.000000,45.000000);
    \draw[color=black] (211.000000,44.500000) -- (236.500000,44.500000);
    \draw[color=black] (211.000000,45.500000) -- (236.500000,45.500000);
    \draw[color=black] (0.000000,45.000000) node[left] {$\ket{0}$};
    \draw[color=black] (0.000000,30.000000) node[left] {$\vdots$};
    \draw[color=black] (0.000000,15.000000) -- (319.500000,15.000000);
    \draw[color=black] (319.500000,14.500000) -- (346.500000,14.500000);
    \draw[color=black] (319.500000,15.500000) -- (346.500000,15.500000);
    \draw[color=black] (0.000000,15.000000) node[left] {$\ket{0}$};
    \draw[color=black] (0.000000,0.000000) -- (360.000000,0.000000);
    \draw[color=black] (0.000000,0.000000) node[left] {$\ket{kG}$};
    \draw (8.000000, -6.000000) -- (16.000000, 6.000000);
    \draw (14.000000, 3.000000) node[right] {$\scriptstyle{2n}$};
    \begin{scope}
    \draw[fill=white] (12.000000, 60.000000) +(-45.000000:8.485281pt and 8.485281pt) -- +(45.000000:8.485281pt and 8.485281pt) -- +(135.000000:8.485281pt and 8.485281pt) -- +(225.000000:8.485281pt and 8.485281pt) -- cycle;
    \clip (12.000000, 60.000000) +(-45.000000:8.485281pt and 8.485281pt) -- +(45.000000:8.485281pt and 8.485281pt) -- +(135.000000:8.485281pt and 8.485281pt) -- +(225.000000:8.485281pt and 8.485281pt) -- cycle;
    \draw (12.000000, 60.000000) node {$H$};
    \end{scope}
    \begin{scope}
    \draw[fill=white] (12.000000, 45.000000) +(-45.000000:8.485281pt and 8.485281pt) -- +(45.000000:8.485281pt and 8.485281pt) -- +(135.000000:8.485281pt and 8.485281pt) -- +(225.000000:8.485281pt and 8.485281pt) -- cycle;
    \clip (12.000000, 45.000000) +(-45.000000:8.485281pt and 8.485281pt) -- +(45.000000:8.485281pt and 8.485281pt) -- +(135.000000:8.485281pt and 8.485281pt) -- +(225.000000:8.485281pt and 8.485281pt) -- cycle;
    \draw (12.000000, 45.000000) node {$H$};
    \end{scope}
    \begin{scope}
    \draw[fill=white] (12.000000, 15.000000) +(-45.000000:8.485281pt and 8.485281pt) -- +(45.000000:8.485281pt and 8.485281pt) -- +(135.000000:8.485281pt and 8.485281pt) -- +(225.000000:8.485281pt and 8.485281pt) -- cycle;
    \clip (12.000000, 15.000000) +(-45.000000:8.485281pt and 8.485281pt) -- +(45.000000:8.485281pt and 8.485281pt) -- +(135.000000:8.485281pt and 8.485281pt) -- +(225.000000:8.485281pt and 8.485281pt) -- cycle;
    \draw (12.000000, 15.000000) node {$H$};
    \end{scope}
    \draw (50.000000,60.000000) -- (50.000000,0.000000);
    \begin{scope}
    \draw[fill=white] (50.000000, 30.000000) +(-45.000000:28.284271pt and 50.911688pt) -- +(45.000000:28.284271pt and 50.911688pt) -- +(135.000000:28.284271pt and 50.911688pt) -- +(225.000000:28.284271pt and 50.911688pt) -- cycle;
    \clip (50.000000, 30.000000) +(-45.000000:28.284271pt and 50.911688pt) -- +(45.000000:28.284271pt and 50.911688pt) -- +(135.000000:28.284271pt and 50.911688pt) -- +(225.000000:28.284271pt and 50.911688pt) -- cycle;
    \draw (50.000000, 30.000000) node {\parbox{40pt}{\texttt{point}\\\texttt{addition}}};
    \end{scope}
    \begin{scope}
    \draw[fill=white] (88.000000, 60.000000) +(-45.000000:8.485281pt and 8.485281pt) -- +(45.000000:8.485281pt and 8.485281pt) -- +(135.000000:8.485281pt and 8.485281pt) -- +(225.000000:8.485281pt and 8.485281pt) -- cycle;
    \clip (88.000000, 60.000000) +(-45.000000:8.485281pt and 8.485281pt) -- +(45.000000:8.485281pt and 8.485281pt) -- +(135.000000:8.485281pt and 8.485281pt) -- +(225.000000:8.485281pt and 8.485281pt) -- cycle;
    \draw (88.000000, 60.000000) node {$H$};
    \end{scope}
    \draw[fill=white] (106.000000, 54.000000) rectangle (118.000000, 66.000000);
    \draw[very thin] (112.000000, 60.600000) arc (90:150:6.000000pt);
    \draw[very thin] (112.000000, 60.600000) arc (90:30:6.000000pt);
    \draw[->,>=stealth] (112.000000, 54.600000) -- +(80:10.392305pt);
    \draw[color=black] (130.000000,60.000000) node[fill=white,right,minimum height=15.000000pt,minimum width=15.000000pt,inner sep=0pt] {\phantom{$\mu_0$}};
    \draw[color=black] (130.000000,60.000000) node[right] {$\mu_0$};
    \begin{scope}
    \draw[fill=white] (163.000000, 45.000000) +(-45.000000:8.485281pt and 8.485281pt) -- +(45.000000:8.485281pt and 8.485281pt) -- +(135.000000:8.485281pt and 8.485281pt) -- +(225.000000:8.485281pt and 8.485281pt) -- cycle;
    \clip (163.000000, 45.000000) +(-45.000000:8.485281pt and 8.485281pt) -- +(45.000000:8.485281pt and 8.485281pt) -- +(135.000000:8.485281pt and 8.485281pt) -- +(225.000000:8.485281pt and 8.485281pt) -- cycle;
    \draw (163.000000, 45.000000) node {$R_1$};
    \end{scope}
    \begin{scope}
    \draw[fill=white] (187.000000, 45.000000) +(-45.000000:8.485281pt and 8.485281pt) -- +(45.000000:8.485281pt and 8.485281pt) -- +(135.000000:8.485281pt and 8.485281pt) -- +(225.000000:8.485281pt and 8.485281pt) -- cycle;
    \clip (187.000000, 45.000000) +(-45.000000:8.485281pt and 8.485281pt) -- +(45.000000:8.485281pt and 8.485281pt) -- +(135.000000:8.485281pt and 8.485281pt) -- +(225.000000:8.485281pt and 8.485281pt) -- cycle;
    \draw (187.000000, 45.000000) node {$H$};
    \end{scope}
    \draw[fill=white] (205.000000, 39.000000) rectangle (217.000000, 51.000000);
    \draw[very thin] (211.000000, 45.600000) arc (90:150:6.000000pt);
    \draw[very thin] (211.000000, 45.600000) arc (90:30:6.000000pt);
    \draw[->,>=stealth] (211.000000, 39.600000) -- +(80:10.392305pt);
    \draw[color=black] (229.000000,45.000000) node[fill=white,right,minimum height=15.000000pt,minimum width=15.000000pt,inner sep=0pt] {\phantom{$\mu_1$}};
    \draw[color=black] (229.000000,45.000000) node[right] {$\mu_1$};
    \begin{scope}
    \draw[fill=white] (266.000000, 15.000000) +(-45.000000:14.142136pt and 8.485281pt) -- +(45.000000:14.142136pt and 8.485281pt) -- +(135.000000:14.142136pt and 8.485281pt) -- +(225.000000:14.142136pt and 8.485281pt) -- cycle;
    \clip (266.000000, 15.000000) +(-45.000000:14.142136pt and 8.485281pt) -- +(45.000000:14.142136pt and 8.485281pt) -- +(135.000000:14.142136pt and 8.485281pt) -- +(225.000000:14.142136pt and 8.485281pt) -- cycle;
    \draw (266.000000, 15.000000) node {$R_{l-1}$};
    \end{scope}
    \begin{scope}
    \draw[fill=white] (294.000000, 15.000000) +(-45.000000:8.485281pt and 8.485281pt) -- +(45.000000:8.485281pt and 8.485281pt) -- +(135.000000:8.485281pt and 8.485281pt) -- +(225.000000:8.485281pt and 8.485281pt) -- cycle;
    \clip (294.000000, 15.000000) +(-45.000000:8.485281pt and 8.485281pt) -- +(45.000000:8.485281pt and 8.485281pt) -- +(135.000000:8.485281pt and 8.485281pt) -- +(225.000000:8.485281pt and 8.485281pt) -- cycle;
    \draw (294.000000, 15.000000) node {$H$};
    \end{scope}
    \draw[fill=white] (313.500000, 9.000000) rectangle (325.500000, 21.000000);
    \draw[very thin] (319.500000, 15.600000) arc (90:150:6.000000pt);
    \draw[very thin] (319.500000, 15.600000) arc (90:30:6.000000pt);
    \draw[->,>=stealth] (319.500000, 9.600000) -- +(80:10.392305pt);
    \draw[color=black] (312.000000,30.000000) node[fill=white,right,minimum height=15.000000pt,minimum width=15.000000pt,inner sep=0pt] {\phantom{$\ddots$}};
    \draw[color=black] (312.000000,30.000000) node[right] {$\ddots$};
    \draw[color=black] (339.000000,15.000000) node[fill=white,right,minimum height=15.000000pt,minimum width=15.000000pt,inner sep=0pt] {\phantom{$\mu_{l-1}$}};
    \draw[color=black] (339.000000,15.000000) node[right] {$\mu_{l-1}$};
    \end{tikzpicture}
    
		\caption{The first cell for point addition, following Fig.~2 in~\cite{roetteler2017quantum}. The whole circuit consists of $\left\lceil \frac{2n}{l} \right\rceil$ cells. Here, $R_k$ are rotation gates $\mathrm{diag}(1, e^{i\theta_k})$, and $\theta_k = -\pi \sum_{j=0}^{k-1} 2^{k-j} \mu_j$ where $\mu_j \in \{0,1\}$ is the measurement outcome.}\label{figure:circuit-full}
	\end{figure}
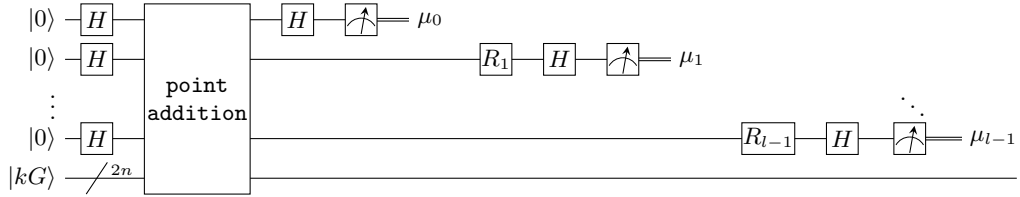

	Intuitively, the windowed trick reduces the number of point additions by a factor of $l$, but requires an additional $O(2^l)$ resources to construct the lookup table each time a point addition is performed. Since the Toffoli depth of a single point addition without windowing is $\Theta(n\log n)$, the optimal window size is $l=\log n+o(\log n)$ for sufficiently large $n$. The overall circuit still requires $17n$ qubits, as the extra $l$ ancillas introduced by the window trick can be placed at the ends of existing columns. An illustration of the complete circuit is given in Fig.~\ref{figure:circuit-full}.
	
	\subsection{Circuit Parameters}\label{subsec:circuit-parameters}
	In this section, we summarize and explain the resource costs of our algorithm. Recall that the circuit consists of Pauli gates, $H$ gates, $S$ gates, CNOT gates, SWAP gates, and Toffoli gates (with Toffoli gates further decomposed into CNOT and $H,T,T^\dag$ gates). In our estimation, however, we count only CNOT and Toffoli gates. The reason for omitting Pauli, Hadamard, $S$, and SWAP gates is explained below.  
	
	Pauli gates can be eliminated entirely by a standard postponement trick. Since $Y$ is proportional to $XZ$, it suffices to consider $X$ and $Z$. When a Pauli gate $P$ is encountered, we look ahead to the next operation. If the next gate is a $Z$-basis measurement, we simply measure first and apply the effect of $P$ to the measurement result: $X$ flips the outcome, while $Z$ has no effect. If the next gate is another Pauli, the two can be combined. If it is a Clifford gate $U$, then $P'=UPU^\dag$ is still a Pauli operator, and we may implement $U$ first while postponing $P$. Finally, if the next gate is a non-Clifford, the only possibilities in our algorithm are $T$ and $T^\dag$. These commute with $Z$, and conjugation with $X$ only changes $T$ to $T^\dag$ or vice versa, so $T$ and $T^\dag$ can also be applied first, leaving the Pauli gate postponed.  

	Hadamard and $S$ gates, outside their role in Toffoli decompositions, occur only rarely in Shor's algorithm. We therefore absorb their cost into that of the Toffoli gates.  
	
	SWAP gates are slightly different. Each SWAP can be decomposed into three CNOT gates, but we treat them separately to highlight the effect of the two-dimensional qubit layout. Specifically, our CNOT counts include only those gates required by the logical structure of the algorithm, while SWAP counts measure the additional gates needed to respect the interaction-distance constraints discussed in Sec.~\ref{subsec:2d-layout}. Under this convention, the number of SWAPs turns out to be negligible, as shown in Appendix~\ref{sec:resource-cost}.  
	
	With these conventions, our resource estimates include qubit count, CNOT depth, CNOT count, Toffoli depth, and Toffoli count. Exact costs for inputs of $25$, $192$, $256$, $384$, and $521$ bits, together with the leading-order terms for general $n$, are summarized in Table~\ref{table:resource-estimate}. A brief explanation of these items is given below, which is the same as that in Sec.~\ref{sec:intro}.  
	
	CNOT depth and count measure only the CNOT gates that arise directly from the algorithm itself, including those in addition, in the construction of quantum tables, and in the translation steps of multiplication and division. Detailed derivations are presented in Appendix~\ref{sec:circuit-construction}.  
	
	Toffoli depth and count are the most important parameters, since under error-correcting codes such as the surface code, Toffoli gates dominate both time and fidelity costs. Our adjustable parameters, namely the window sizes, are chosen specifically to minimize Toffoli depth. Almost all Toffoli gates in the algorithm are long-range gates implemented by dynamic circuits.

	\section{Discussion}\label{sec:discussion}
	Although quantum algorithms for integer factoring have received extensive attention, far fewer works have systematically addressed Shor's elliptic curve algorithm, despite its comparable importance in modern cryptography. Our study fills this gap by presenting a complete implementation and resource analysis of Shor's elliptic curve algorithm tailored to two-dimensional nearest-neighbor architectures. The key enabling technique is the use of dynamic circuits, which allow long-range Toffoli gates and unbounded GHZ states to be realized with constant overhead. These primitives provide the essential capabilities for long-distance qubit interaction and parallel execution of controlled operations. By incorporating them into the design of modular arithmetic, we demonstrate that the elliptic curve discrete logarithm problem can be realistically compiled for hardware-constrained architectures with an acceptable overhead, and give a detailed resource analysis on qubit number, CNOT depth and counts, and Toffoli depth and counts. Our results therefore establish a concrete baseline for the future experimental realization of quantum attacks on elliptic curve cryptography.

	Beyond the application of the elliptic curve discrete logarithm problem, our work makes contributions of broader relevance. A central technical advance is our improved carry-lookahead adder, which achieves best-known Toffoli depth while requiring only linear ancilla overhead. Since addition is a ubiquitous subroutine across quantum algorithms—including factoring, lattice problems, and quantum simulation—this design has potential applications well beyond elliptic curves.

	At the same time, our analysis reveals that the dominant resource cost of Shor's elliptic curve algorithm arises from modular division. As this operation is far less studied than addition and is substantially more expensive, improving quantum modular division directly promises to lower the overall cost of Shor's elliptic curve algorithm and represents a natural target for further optimization.

	It is also instructive to compare our findings with recent resource estimates for factoring-based cryptanalysis~\cite{gidney2025factor}. Classically, the security of 256-bit elliptic curve cryptography is considered comparable to that of 3072-bit RSA\@. However, our analysis shows that the quantum resources required to break these problems differ substantially. Solving the 256-bit elliptic curve cryptography requires about $2.25$ times more qubits than solving the 3072-bit RSA problem, but only about $1/13$ as many Toffoli gates. This contrast suggests that, under realistic quantum architectures, elliptic-curve-based systems may be more vulnerable to large-scale quantum attacks than RSA, despite their stronger classical security guarantees.

	Our results also highlight several avenues for future improvement. While our construction is adapted to a two-dimensional topology, this constraint inevitably imposes additional overhead. In particular, our layout arranges qubits in a long, narrow configuration, which increases the reliance on dynamic circuits to mediate long-range interactions. For example, SWAP operations used to implement long-range Toffoli gates (see Sec.~\ref{subsec:dynamic circuits}) dominate the SWAP cost of the entire algorithm. A more flexible qubit topology, featuring enhanced connectivity beyond nearest neighbors or a square rather than rectangular arrangement of qubits, can mitigate this overhead by reducing the frequency of dynamic-circuit primitives and lowering the associated resource requirements.
	
	Finally, we note that our analysis has been carried out at the logical gate level, without considering the detailed implementation of quantum error correction (QEC). Since Toffoli gates dominate both runtime and error budgets under leading error-correcting codes, a more detailed QEC-based analysis would provide valuable insights into fault-tolerant cost and overhead. Such work would be an important next step toward connecting algorithmic resource estimates with the concrete capabilities of near-future hardware.

	\begin{acknowledgements}
	This work was supported by the National Natural Science Foundation of China Grant No.~12174216 and the Innovation Program for Quantum Science and Technology Grant No.~2021ZD0300804 and No.~2021ZD0300702.
	\end{acknowledgements}
	
	\newpage
	\appendix

	\section{Toffoli Gates}\label{sec:toffoli-gate}
	As the most resource-consuming part in our circuit, the implementation of Toffoli gates involves $T$ gates. Optimizing the $T$-depth of Toffoli gate compiling is important to save the resource. In this work, we use the construction using logical AND gate and uncomputation gate given in~\cite{gidney2018halving}, as shown in Fig.~\ref{figure:toffoli}. Here, the magic resource state $\ket{T} \equiv TH\ket{0} = \frac{1}{\sqrt{2}} \qty(\ket{0} + e^{i\frac{\pi}{4}}\ket{1})$. If the output qubit of the Toffoli gate should be initially $\ket{0}$, we can use the magic resource state $\ket{T}$ to replace it and therefore need no ancilla, as shown in Fig.~\ref{figure:toffolia}. Otherwise, we need the state $\ket{T}$ as an ancilla. Note that all the magic resource states $\ket{T}$ can be prepared simultaneously with a single layer of $T$ gate, the effective $T$-depth of this circuit is actually one~\cite{dutta2025exact}.
	
	\begin{figure}[ht]
		\captionsetup{justification=raggedright,singlelinecheck=false}
		\centering
		\begin{minipage}[t]{0.9\textwidth}
			\subcaption{}
			\centering
			\begin{tikzpicture}[scale=1.000000,x=1pt,y=1pt]
\filldraw[color=white] (0.000000, -7.500000) rectangle (180.000000, 37.500000);
\draw[color=black] (0.000000,30.000000) -- (180.000000,30.000000);
\draw[color=black] (0.000000,30.000000) node[left] {$\ket{x}$};
\draw[color=black] (0.000000,15.000000) -- (180.000000,15.000000);
\draw[color=black] (0.000000,15.000000) node[left] {$\ket{y}$};
\draw[color=black] (0.000000,0.000000) -- (180.000000,0.000000);
\draw[color=black] (0.000000,0.000000) node[left] {$\ket{T}$};
\draw (9.000000,30.000000) -- (9.000000,0.000000);
\begin{scope}
\draw[fill=white] (9.000000, 0.000000) circle(3.000000pt);
\clip (9.000000, 0.000000) circle(3.000000pt);
\draw (6.000000, 0.000000) -- (12.000000, 0.000000);
\draw (9.000000, -3.000000) -- (9.000000, 3.000000);
\end{scope}
\filldraw (9.000000, 30.000000) circle(1.500000pt);
\draw (27.000000,15.000000) -- (27.000000,0.000000);
\begin{scope}
\draw[fill=white] (27.000000, 0.000000) circle(3.000000pt);
\clip (27.000000, 0.000000) circle(3.000000pt);
\draw (24.000000, 0.000000) -- (30.000000, 0.000000);
\draw (27.000000, -3.000000) -- (27.000000, 3.000000);
\end{scope}
\filldraw (27.000000, 15.000000) circle(1.500000pt);
\draw (45.000000,15.000000) -- (45.000000,0.000000);
\begin{scope}
\draw[fill=white] (45.000000, 15.000000) circle(3.000000pt);
\clip (45.000000, 15.000000) circle(3.000000pt);
\draw (42.000000, 15.000000) -- (48.000000, 15.000000);
\draw (45.000000, 12.000000) -- (45.000000, 18.000000);
\end{scope}
\filldraw (45.000000, 0.000000) circle(1.500000pt);
\draw (63.000000,30.000000) -- (63.000000,0.000000);
\begin{scope}
\draw[fill=white] (63.000000, 30.000000) circle(3.000000pt);
\clip (63.000000, 30.000000) circle(3.000000pt);
\draw (60.000000, 30.000000) -- (66.000000, 30.000000);
\draw (63.000000, 27.000000) -- (63.000000, 33.000000);
\end{scope}
\filldraw (63.000000, 0.000000) circle(1.500000pt);
\begin{scope}
\draw[fill=white] (84.000000, 30.000000) +(-45.000000:8.485281pt and 8.485281pt) -- +(45.000000:8.485281pt and 8.485281pt) -- +(135.000000:8.485281pt and 8.485281pt) -- +(225.000000:8.485281pt and 8.485281pt) -- cycle;
\clip (84.000000, 30.000000) +(-45.000000:8.485281pt and 8.485281pt) -- +(45.000000:8.485281pt and 8.485281pt) -- +(135.000000:8.485281pt and 8.485281pt) -- +(225.000000:8.485281pt and 8.485281pt) -- cycle;
\draw (84.000000, 30.000000) node {$T^\dagger$};
\end{scope}
\begin{scope}
\draw[fill=white] (84.000000, 15.000000) +(-45.000000:8.485281pt and 8.485281pt) -- +(45.000000:8.485281pt and 8.485281pt) -- +(135.000000:8.485281pt and 8.485281pt) -- +(225.000000:8.485281pt and 8.485281pt) -- cycle;
\clip (84.000000, 15.000000) +(-45.000000:8.485281pt and 8.485281pt) -- +(45.000000:8.485281pt and 8.485281pt) -- +(135.000000:8.485281pt and 8.485281pt) -- +(225.000000:8.485281pt and 8.485281pt) -- cycle;
\draw (84.000000, 15.000000) node {$T^\dagger$};
\end{scope}
\begin{scope}
\draw[fill=white] (84.000000, 0.000000) +(-45.000000:8.485281pt and 8.485281pt) -- +(45.000000:8.485281pt and 8.485281pt) -- +(135.000000:8.485281pt and 8.485281pt) -- +(225.000000:8.485281pt and 8.485281pt) -- cycle;
\clip (84.000000, 0.000000) +(-45.000000:8.485281pt and 8.485281pt) -- +(45.000000:8.485281pt and 8.485281pt) -- +(135.000000:8.485281pt and 8.485281pt) -- +(225.000000:8.485281pt and 8.485281pt) -- cycle;
\draw (84.000000, 0.000000) node {$T$};
\end{scope}
\draw (105.000000,30.000000) -- (105.000000,0.000000);
\begin{scope}
\draw[fill=white] (105.000000, 30.000000) circle(3.000000pt);
\clip (105.000000, 30.000000) circle(3.000000pt);
\draw (102.000000, 30.000000) -- (108.000000, 30.000000);
\draw (105.000000, 27.000000) -- (105.000000, 33.000000);
\end{scope}
\filldraw (105.000000, 0.000000) circle(1.500000pt);
\draw (123.000000,15.000000) -- (123.000000,0.000000);
\begin{scope}
\draw[fill=white] (123.000000, 15.000000) circle(3.000000pt);
\clip (123.000000, 15.000000) circle(3.000000pt);
\draw (120.000000, 15.000000) -- (126.000000, 15.000000);
\draw (123.000000, 12.000000) -- (123.000000, 18.000000);
\end{scope}
\filldraw (123.000000, 0.000000) circle(1.500000pt);
\begin{scope}
\draw[fill=white] (144.000000, 0.000000) +(-45.000000:8.485281pt and 8.485281pt) -- +(45.000000:8.485281pt and 8.485281pt) -- +(135.000000:8.485281pt and 8.485281pt) -- +(225.000000:8.485281pt and 8.485281pt) -- cycle;
\clip (144.000000, 0.000000) +(-45.000000:8.485281pt and 8.485281pt) -- +(45.000000:8.485281pt and 8.485281pt) -- +(135.000000:8.485281pt and 8.485281pt) -- +(225.000000:8.485281pt and 8.485281pt) -- cycle;
\draw (144.000000, 0.000000) node {$H$};
\end{scope}
\begin{scope}
\draw[fill=white] (168.000000, 0.000000) +(-45.000000:8.485281pt and 8.485281pt) -- +(45.000000:8.485281pt and 8.485281pt) -- +(135.000000:8.485281pt and 8.485281pt) -- +(225.000000:8.485281pt and 8.485281pt) -- cycle;
\clip (168.000000, 0.000000) +(-45.000000:8.485281pt and 8.485281pt) -- +(45.000000:8.485281pt and 8.485281pt) -- +(135.000000:8.485281pt and 8.485281pt) -- +(225.000000:8.485281pt and 8.485281pt) -- cycle;
\draw (168.000000, 0.000000) node {$S$};
\end{scope}
\draw[color=black] (180.000000,30.000000) node[right] {$\ket{x}$};
\draw[color=black] (180.000000,15.000000) node[right] {$\ket{y}$};
\draw[color=black] (180.000000,0.000000) node[right] {$\ket{x\cdot y}$};
\end{tikzpicture}\label{figure:toffolia}
		\end{minipage}
		\qquad
		\begin{minipage}[t]{0.9\textwidth}
			\subcaption{}
			\centering
			\begin{tikzpicture}[scale=1.000000,x=1pt,y=1pt]
\filldraw[color=white] (0.000000, -7.500000) rectangle (270.000000, 52.500000);
\draw[color=black] (0.000000,45.000000) -- (270.000000,45.000000);
\draw[color=black] (0.000000,45.000000) node[left] {$\ket{x}$};
\draw[color=black] (0.000000,30.000000) -- (270.000000,30.000000);
\draw[color=black] (0.000000,30.000000) node[left] {$\ket{y}$};
\draw[color=black] (0.000000,15.000000) -- (234.000000,15.000000);
\draw[color=black] (234.000000,14.500000) -- (258.000000,14.500000);
\draw[color=black] (234.000000,15.500000) -- (258.000000,15.500000);
\draw[color=black] (0.000000,15.000000) node[left] {$\ket{T}$};
\draw[color=black] (0.000000,0.000000) -- (270.000000,0.000000);
\draw[color=black] (0.000000,0.000000) node[left] {$\ket{z}$};
\draw (9.000000,45.000000) -- (9.000000,15.000000);
\begin{scope}
\draw[fill=white] (9.000000, 15.000000) circle(3.000000pt);
\clip (9.000000, 15.000000) circle(3.000000pt);
\draw (6.000000, 15.000000) -- (12.000000, 15.000000);
\draw (9.000000, 12.000000) -- (9.000000, 18.000000);
\end{scope}
\filldraw (9.000000, 45.000000) circle(1.500000pt);
\draw (27.000000,30.000000) -- (27.000000,15.000000);
\begin{scope}
\draw[fill=white] (27.000000, 15.000000) circle(3.000000pt);
\clip (27.000000, 15.000000) circle(3.000000pt);
\draw (24.000000, 15.000000) -- (30.000000, 15.000000);
\draw (27.000000, 12.000000) -- (27.000000, 18.000000);
\end{scope}
\filldraw (27.000000, 30.000000) circle(1.500000pt);
\draw (45.000000,30.000000) -- (45.000000,15.000000);
\begin{scope}
\draw[fill=white] (45.000000, 30.000000) circle(3.000000pt);
\clip (45.000000, 30.000000) circle(3.000000pt);
\draw (42.000000, 30.000000) -- (48.000000, 30.000000);
\draw (45.000000, 27.000000) -- (45.000000, 33.000000);
\end{scope}
\filldraw (45.000000, 15.000000) circle(1.500000pt);
\draw (63.000000,45.000000) -- (63.000000,15.000000);
\begin{scope}
\draw[fill=white] (63.000000, 45.000000) circle(3.000000pt);
\clip (63.000000, 45.000000) circle(3.000000pt);
\draw (60.000000, 45.000000) -- (66.000000, 45.000000);
\draw (63.000000, 42.000000) -- (63.000000, 48.000000);
\end{scope}
\filldraw (63.000000, 15.000000) circle(1.500000pt);
\begin{scope}
\draw[fill=white] (84.000000, 45.000000) +(-45.000000:8.485281pt and 8.485281pt) -- +(45.000000:8.485281pt and 8.485281pt) -- +(135.000000:8.485281pt and 8.485281pt) -- +(225.000000:8.485281pt and 8.485281pt) -- cycle;
\clip (84.000000, 45.000000) +(-45.000000:8.485281pt and 8.485281pt) -- +(45.000000:8.485281pt and 8.485281pt) -- +(135.000000:8.485281pt and 8.485281pt) -- +(225.000000:8.485281pt and 8.485281pt) -- cycle;
\draw (84.000000, 45.000000) node {$T^\dagger$};
\end{scope}
\begin{scope}
\draw[fill=white] (84.000000, 30.000000) +(-45.000000:8.485281pt and 8.485281pt) -- +(45.000000:8.485281pt and 8.485281pt) -- +(135.000000:8.485281pt and 8.485281pt) -- +(225.000000:8.485281pt and 8.485281pt) -- cycle;
\clip (84.000000, 30.000000) +(-45.000000:8.485281pt and 8.485281pt) -- +(45.000000:8.485281pt and 8.485281pt) -- +(135.000000:8.485281pt and 8.485281pt) -- +(225.000000:8.485281pt and 8.485281pt) -- cycle;
\draw (84.000000, 30.000000) node {$T^\dagger$};
\end{scope}
\begin{scope}
\draw[fill=white] (84.000000, 15.000000) +(-45.000000:8.485281pt and 8.485281pt) -- +(45.000000:8.485281pt and 8.485281pt) -- +(135.000000:8.485281pt and 8.485281pt) -- +(225.000000:8.485281pt and 8.485281pt) -- cycle;
\clip (84.000000, 15.000000) +(-45.000000:8.485281pt and 8.485281pt) -- +(45.000000:8.485281pt and 8.485281pt) -- +(135.000000:8.485281pt and 8.485281pt) -- +(225.000000:8.485281pt and 8.485281pt) -- cycle;
\draw (84.000000, 15.000000) node {$T$};
\end{scope}
\draw (105.000000,45.000000) -- (105.000000,15.000000);
\begin{scope}
\draw[fill=white] (105.000000, 45.000000) circle(3.000000pt);
\clip (105.000000, 45.000000) circle(3.000000pt);
\draw (102.000000, 45.000000) -- (108.000000, 45.000000);
\draw (105.000000, 42.000000) -- (105.000000, 48.000000);
\end{scope}
\filldraw (105.000000, 15.000000) circle(1.500000pt);
\draw (123.000000,30.000000) -- (123.000000,15.000000);
\begin{scope}
\draw[fill=white] (123.000000, 30.000000) circle(3.000000pt);
\clip (123.000000, 30.000000) circle(3.000000pt);
\draw (120.000000, 30.000000) -- (126.000000, 30.000000);
\draw (123.000000, 27.000000) -- (123.000000, 33.000000);
\end{scope}
\filldraw (123.000000, 15.000000) circle(1.500000pt);
\begin{scope}
\draw[fill=white] (144.000000, 15.000000) +(-45.000000:8.485281pt and 8.485281pt) -- +(45.000000:8.485281pt and 8.485281pt) -- +(135.000000:8.485281pt and 8.485281pt) -- +(225.000000:8.485281pt and 8.485281pt) -- cycle;
\clip (144.000000, 15.000000) +(-45.000000:8.485281pt and 8.485281pt) -- +(45.000000:8.485281pt and 8.485281pt) -- +(135.000000:8.485281pt and 8.485281pt) -- +(225.000000:8.485281pt and 8.485281pt) -- cycle;
\draw (144.000000, 15.000000) node {$H$};
\end{scope}
\begin{scope}
\draw[fill=white] (168.000000, 15.000000) +(-45.000000:8.485281pt and 8.485281pt) -- +(45.000000:8.485281pt and 8.485281pt) -- +(135.000000:8.485281pt and 8.485281pt) -- +(225.000000:8.485281pt and 8.485281pt) -- cycle;
\clip (168.000000, 15.000000) +(-45.000000:8.485281pt and 8.485281pt) -- +(45.000000:8.485281pt and 8.485281pt) -- +(135.000000:8.485281pt and 8.485281pt) -- +(225.000000:8.485281pt and 8.485281pt) -- cycle;
\draw (168.000000, 15.000000) node {$S$};
\end{scope}
\draw (189.000000,15.000000) -- (189.000000,0.000000);
\filldraw (189.000000, 15.000000) circle(1.500000pt);
\begin{scope}
\draw[fill=white] (189.000000, 0.000000) circle(3.000000pt);
\clip (189.000000, 0.000000) circle(3.000000pt);
\draw (186.000000, 0.000000) -- (192.000000, 0.000000);
\draw (189.000000, -3.000000) -- (189.000000, 3.000000);
\end{scope}
\begin{scope}
\draw[fill=white] (210.000000, 15.000000) +(-45.000000:8.485281pt and 8.485281pt) -- +(45.000000:8.485281pt and 8.485281pt) -- +(135.000000:8.485281pt and 8.485281pt) -- +(225.000000:8.485281pt and 8.485281pt) -- cycle;
\clip (210.000000, 15.000000) +(-45.000000:8.485281pt and 8.485281pt) -- +(45.000000:8.485281pt and 8.485281pt) -- +(135.000000:8.485281pt and 8.485281pt) -- +(225.000000:8.485281pt and 8.485281pt) -- cycle;
\draw (210.000000, 15.000000) node {$H$};
\end{scope}
\draw[fill=white] (228.000000, 9.000000) rectangle (240.000000, 21.000000);
\draw[very thin] (234.000000, 15.600000) arc (90:150:6.000000pt);
\draw[very thin] (234.000000, 15.600000) arc (90:30:6.000000pt);
\draw[->,>=stealth] (234.000000, 9.600000) -- +(80:10.392305pt);
\draw (258.000000,45.000000) -- (258.000000,30.000000);
\draw (257.500000,30.000000) -- (257.500000,15.000000);
\draw (258.500000,30.000000) -- (258.500000,15.000000);
\begin{scope}
\draw[fill=white] (258.000000, 30.000000) +(-45.000000:8.485281pt and 8.485281pt) -- +(45.000000:8.485281pt and 8.485281pt) -- +(135.000000:8.485281pt and 8.485281pt) -- +(225.000000:8.485281pt and 8.485281pt) -- cycle;
\clip (258.000000, 30.000000) +(-45.000000:8.485281pt and 8.485281pt) -- +(45.000000:8.485281pt and 8.485281pt) -- +(135.000000:8.485281pt and 8.485281pt) -- +(225.000000:8.485281pt and 8.485281pt) -- cycle;
\draw (258.000000, 30.000000) node {$Z$};
\end{scope}
\filldraw (258.000000, 45.000000) circle(1.500000pt);
\filldraw (258.000000, 15.000000) circle(1.500000pt);
\draw[color=black] (270.000000,45.000000) node[right] {$\ket{x}$};
\draw[color=black] (270.000000,30.000000) node[right] {$\ket{y}$};
\draw[color=black] (270.000000,0.000000) node[right] {$\ket{z\oplus x\cdot y}$};
\end{tikzpicture}\label{figure:toffolib}
		\end{minipage}
		\caption{Implementation of a Toffoli gate with $T$-depth $1$ and a $\ket{T}$ state.\ (a) Toffoli gate with output qubit initialized in $\ket{z}=\ket{0}$.\ (b): Toffoli gate in general.}\label{figure:toffoli}
	\end{figure}

	There are some cases where we only need the output qubit of Toffoli gate as an intermediate value rather than the final output, which is called ``Toffoli pre- and post- processing''. Intuitively, we would first use a Toffoli gate to get the intermediate value, apply the target gate, and then use another Toffoli gate to uncompute it. Such an implementation requires two Toffoli gates. However, one can again utilize the idea of logical AND gate and uncomputation gate to achieve this with a cost equivalent to only one Toffoli gate, as shown in Fig.~\ref{figure:anc-toffoli}. Therefore, in circuit construction and corresponding resource estimation, we will regard a ``Toffoli pre- and post- processing'' as a Toffoli gate.

	\begin{figure}[ht]
		\captionsetup{justification=raggedright,singlelinecheck=false}
		\centering
		\begin{minipage}[t]{0.9\textwidth}
			\subcaption{}
			\centering
\begin{tikzpicture}[scale=1.000000,x=1pt,y=1pt]
\filldraw[color=white] (0.000000, -7.500000) rectangle (279.000000, 37.500000);
\draw[color=black] (0.000000,30.000000) -- (279.000000,30.000000);
\draw[color=black] (0.000000,30.000000) node[left] {$\ket{x}$};
\draw[color=black] (0.000000,15.000000) -- (279.000000,15.000000);
\draw[color=black] (0.000000,15.000000) node[left] {$\ket{y}$};
\draw[color=black] (0.000000,0.000000) -- (243.000000,0.000000);
\draw[color=black] (243.000000,-0.500000) -- (267.000000,-0.500000);
\draw[color=black] (243.000000,0.500000) -- (267.000000,0.500000);
\draw[color=black] (0.000000,0.000000) node[left] {$\ket{T}$};
\draw (9.000000,30.000000) -- (9.000000,0.000000);
\begin{scope}
\draw[fill=white] (9.000000, 0.000000) circle(3.000000pt);
\clip (9.000000, 0.000000) circle(3.000000pt);
\draw (6.000000, 0.000000) -- (12.000000, 0.000000);
\draw (9.000000, -3.000000) -- (9.000000, 3.000000);
\end{scope}
\filldraw (9.000000, 30.000000) circle(1.500000pt);
\draw (27.000000,15.000000) -- (27.000000,0.000000);
\begin{scope}
\draw[fill=white] (27.000000, 0.000000) circle(3.000000pt);
\clip (27.000000, 0.000000) circle(3.000000pt);
\draw (24.000000, 0.000000) -- (30.000000, 0.000000);
\draw (27.000000, -3.000000) -- (27.000000, 3.000000);
\end{scope}
\filldraw (27.000000, 15.000000) circle(1.500000pt);
\draw (45.000000,15.000000) -- (45.000000,0.000000);
\begin{scope}
\draw[fill=white] (45.000000, 15.000000) circle(3.000000pt);
\clip (45.000000, 15.000000) circle(3.000000pt);
\draw (42.000000, 15.000000) -- (48.000000, 15.000000);
\draw (45.000000, 12.000000) -- (45.000000, 18.000000);
\end{scope}
\filldraw (45.000000, 0.000000) circle(1.500000pt);
\draw (63.000000,30.000000) -- (63.000000,0.000000);
\begin{scope}
\draw[fill=white] (63.000000, 30.000000) circle(3.000000pt);
\clip (63.000000, 30.000000) circle(3.000000pt);
\draw (60.000000, 30.000000) -- (66.000000, 30.000000);
\draw (63.000000, 27.000000) -- (63.000000, 33.000000);
\end{scope}
\filldraw (63.000000, 0.000000) circle(1.500000pt);
\begin{scope}
\draw[fill=white] (84.000000, 30.000000) +(-45.000000:8.485281pt and 8.485281pt) -- +(45.000000:8.485281pt and 8.485281pt) -- +(135.000000:8.485281pt and 8.485281pt) -- +(225.000000:8.485281pt and 8.485281pt) -- cycle;
\clip (84.000000, 30.000000) +(-45.000000:8.485281pt and 8.485281pt) -- +(45.000000:8.485281pt and 8.485281pt) -- +(135.000000:8.485281pt and 8.485281pt) -- +(225.000000:8.485281pt and 8.485281pt) -- cycle;
\draw (84.000000, 30.000000) node {$T^\dagger$};
\end{scope}
\begin{scope}
\draw[fill=white] (84.000000, 15.000000) +(-45.000000:8.485281pt and 8.485281pt) -- +(45.000000:8.485281pt and 8.485281pt) -- +(135.000000:8.485281pt and 8.485281pt) -- +(225.000000:8.485281pt and 8.485281pt) -- cycle;
\clip (84.000000, 15.000000) +(-45.000000:8.485281pt and 8.485281pt) -- +(45.000000:8.485281pt and 8.485281pt) -- +(135.000000:8.485281pt and 8.485281pt) -- +(225.000000:8.485281pt and 8.485281pt) -- cycle;
\draw (84.000000, 15.000000) node {$T^\dagger$};
\end{scope}
\begin{scope}
\draw[fill=white] (84.000000, 0.000000) +(-45.000000:8.485281pt and 8.485281pt) -- +(45.000000:8.485281pt and 8.485281pt) -- +(135.000000:8.485281pt and 8.485281pt) -- +(225.000000:8.485281pt and 8.485281pt) -- cycle;
\clip (84.000000, 0.000000) +(-45.000000:8.485281pt and 8.485281pt) -- +(45.000000:8.485281pt and 8.485281pt) -- +(135.000000:8.485281pt and 8.485281pt) -- +(225.000000:8.485281pt and 8.485281pt) -- cycle;
\draw (84.000000, 0.000000) node {$T$};
\end{scope}
\draw (105.000000,30.000000) -- (105.000000,0.000000);
\begin{scope}
\draw[fill=white] (105.000000, 30.000000) circle(3.000000pt);
\clip (105.000000, 30.000000) circle(3.000000pt);
\draw (102.000000, 30.000000) -- (108.000000, 30.000000);
\draw (105.000000, 27.000000) -- (105.000000, 33.000000);
\end{scope}
\filldraw (105.000000, 0.000000) circle(1.500000pt);
\draw (123.000000,15.000000) -- (123.000000,0.000000);
\begin{scope}
\draw[fill=white] (123.000000, 15.000000) circle(3.000000pt);
\clip (123.000000, 15.000000) circle(3.000000pt);
\draw (120.000000, 15.000000) -- (126.000000, 15.000000);
\draw (123.000000, 12.000000) -- (123.000000, 18.000000);
\end{scope}
\filldraw (123.000000, 0.000000) circle(1.500000pt);
\begin{scope}
\draw[fill=white] (144.000000, 0.000000) +(-45.000000:8.485281pt and 8.485281pt) -- +(45.000000:8.485281pt and 8.485281pt) -- +(135.000000:8.485281pt and 8.485281pt) -- +(225.000000:8.485281pt and 8.485281pt) -- cycle;
\clip (144.000000, 0.000000) +(-45.000000:8.485281pt and 8.485281pt) -- +(45.000000:8.485281pt and 8.485281pt) -- +(135.000000:8.485281pt and 8.485281pt) -- +(225.000000:8.485281pt and 8.485281pt) -- cycle;
\draw (144.000000, 0.000000) node {$H$};
\end{scope}
\begin{scope}
\draw[fill=white] (168.000000, 0.000000) +(-45.000000:8.485281pt and 8.485281pt) -- +(45.000000:8.485281pt and 8.485281pt) -- +(135.000000:8.485281pt and 8.485281pt) -- +(225.000000:8.485281pt and 8.485281pt) -- cycle;
\clip (168.000000, 0.000000) +(-45.000000:8.485281pt and 8.485281pt) -- +(45.000000:8.485281pt and 8.485281pt) -- +(135.000000:8.485281pt and 8.485281pt) -- +(225.000000:8.485281pt and 8.485281pt) -- cycle;
\draw (168.000000, 0.000000) node {$S$};
\end{scope}
\draw[color=black] (193.500000, 30.000000) node [fill=white] {$\cdots$};
\draw[color=black] (193.500000, 15.000000) node [fill=white] {$\cdots$};
\draw[color=black] (193.500000, 0.000000) node [fill=white] {$\cdots$};
\begin{scope}
\draw[fill=white] (219.000000, 0.000000) +(-45.000000:8.485281pt and 8.485281pt) -- +(45.000000:8.485281pt and 8.485281pt) -- +(135.000000:8.485281pt and 8.485281pt) -- +(225.000000:8.485281pt and 8.485281pt) -- cycle;
\clip (219.000000, 0.000000) +(-45.000000:8.485281pt and 8.485281pt) -- +(45.000000:8.485281pt and 8.485281pt) -- +(135.000000:8.485281pt and 8.485281pt) -- +(225.000000:8.485281pt and 8.485281pt) -- cycle;
\draw (219.000000, 0.000000) node {$H$};
\end{scope}
\draw[fill=white] (237.000000, -6.000000) rectangle (249.000000, 6.000000);
\draw[very thin] (243.000000, 0.600000) arc (90:150:6.000000pt);
\draw[very thin] (243.000000, 0.600000) arc (90:30:6.000000pt);
\draw[->,>=stealth] (243.000000, -5.400000) -- +(80:10.392305pt);
\draw (267.000000,30.000000) -- (267.000000,15.000000);
\draw (266.500000,15.000000) -- (266.500000,0.000000);
\draw (267.500000,15.000000) -- (267.500000,0.000000);
\begin{scope}
\draw[fill=white] (267.000000, 15.000000) +(-45.000000:8.485281pt and 8.485281pt) -- +(45.000000:8.485281pt and 8.485281pt) -- +(135.000000:8.485281pt and 8.485281pt) -- +(225.000000:8.485281pt and 8.485281pt) -- cycle;
\clip (267.000000, 15.000000) +(-45.000000:8.485281pt and 8.485281pt) -- +(45.000000:8.485281pt and 8.485281pt) -- +(135.000000:8.485281pt and 8.485281pt) -- +(225.000000:8.485281pt and 8.485281pt) -- cycle;
\draw (267.000000, 15.000000) node {$Z$};
\end{scope}
\filldraw (267.000000, 30.000000) circle(1.500000pt);
\filldraw (267.000000, 0.000000) circle(1.500000pt);
\draw[color=black] (279.000000,30.000000) node[right] {$\ket{x}$};
\draw[color=black] (279.000000,15.000000) node[right] {$\ket{y}$};
\draw[decorate,decoration={brace,mirror,amplitude = 4.000000pt},very thick] (3.000000,-7.500000) -- (177.000000,-7.500000);
\draw (90.000000, -11.500000) node[text width=144pt,below,text centered] {logical AND gate};
\draw[decorate,decoration={brace,mirror,amplitude = 4.000000pt},very thick] (210.000000,-7.500000) -- (276.000000,-7.500000);
\draw (243.000000, -11.500000) node[text width=144pt,below,text centered] {uncomputation gate};
\end{tikzpicture}
		\end{minipage}
		\begin{minipage}[t]{0.9\textwidth}
			\subcaption{}
			\centering
\begin{tikzpicture}[scale=1.000000,x=1pt,y=1pt]
\filldraw[color=white] (0.000000, -7.500000) rectangle (315.000000, 52.500000);
\draw[color=black] (0.000000,45.000000) -- (315.000000,45.000000);
\draw[color=black] (0.000000,45.000000) node[left] {$\ket{x}$};
\draw[color=black] (0.000000,30.000000) -- (315.000000,30.000000);
\draw[color=black] (0.000000,30.000000) node[left] {$\ket{y}$};
\draw[color=black] (0.000000,15.000000) -- (279.000000,15.000000);
\draw[color=black] (279.000000,14.500000) -- (303.000000,14.500000);
\draw[color=black] (279.000000,15.500000) -- (303.000000,15.500000);
\draw[color=black] (0.000000,15.000000) node[left] {$\ket{T}$};
\draw[color=black] (0.000000,0.000000) -- (315.000000,0.000000);
\draw[color=black] (0.000000,0.000000) node[left] {$\ket{z}$};
\draw (9.000000,45.000000) -- (9.000000,15.000000);
\begin{scope}
\draw[fill=white] (9.000000, 15.000000) circle(3.000000pt);
\clip (9.000000, 15.000000) circle(3.000000pt);
\draw (6.000000, 15.000000) -- (12.000000, 15.000000);
\draw (9.000000, 12.000000) -- (9.000000, 18.000000);
\end{scope}
\filldraw (9.000000, 45.000000) circle(1.500000pt);
\draw (27.000000,30.000000) -- (27.000000,15.000000);
\begin{scope}
\draw[fill=white] (27.000000, 15.000000) circle(3.000000pt);
\clip (27.000000, 15.000000) circle(3.000000pt);
\draw (24.000000, 15.000000) -- (30.000000, 15.000000);
\draw (27.000000, 12.000000) -- (27.000000, 18.000000);
\end{scope}
\filldraw (27.000000, 30.000000) circle(1.500000pt);
\draw (45.000000,30.000000) -- (45.000000,15.000000);
\begin{scope}
\draw[fill=white] (45.000000, 30.000000) circle(3.000000pt);
\clip (45.000000, 30.000000) circle(3.000000pt);
\draw (42.000000, 30.000000) -- (48.000000, 30.000000);
\draw (45.000000, 27.000000) -- (45.000000, 33.000000);
\end{scope}
\filldraw (45.000000, 15.000000) circle(1.500000pt);
\draw (63.000000,45.000000) -- (63.000000,15.000000);
\begin{scope}
\draw[fill=white] (63.000000, 45.000000) circle(3.000000pt);
\clip (63.000000, 45.000000) circle(3.000000pt);
\draw (60.000000, 45.000000) -- (66.000000, 45.000000);
\draw (63.000000, 42.000000) -- (63.000000, 48.000000);
\end{scope}
\filldraw (63.000000, 15.000000) circle(1.500000pt);
\begin{scope}
\draw[fill=white] (84.000000, 45.000000) +(-45.000000:8.485281pt and 8.485281pt) -- +(45.000000:8.485281pt and 8.485281pt) -- +(135.000000:8.485281pt and 8.485281pt) -- +(225.000000:8.485281pt and 8.485281pt) -- cycle;
\clip (84.000000, 45.000000) +(-45.000000:8.485281pt and 8.485281pt) -- +(45.000000:8.485281pt and 8.485281pt) -- +(135.000000:8.485281pt and 8.485281pt) -- +(225.000000:8.485281pt and 8.485281pt) -- cycle;
\draw (84.000000, 45.000000) node {$T^\dagger$};
\end{scope}
\begin{scope}
\draw[fill=white] (84.000000, 30.000000) +(-45.000000:8.485281pt and 8.485281pt) -- +(45.000000:8.485281pt and 8.485281pt) -- +(135.000000:8.485281pt and 8.485281pt) -- +(225.000000:8.485281pt and 8.485281pt) -- cycle;
\clip (84.000000, 30.000000) +(-45.000000:8.485281pt and 8.485281pt) -- +(45.000000:8.485281pt and 8.485281pt) -- +(135.000000:8.485281pt and 8.485281pt) -- +(225.000000:8.485281pt and 8.485281pt) -- cycle;
\draw (84.000000, 30.000000) node {$T^\dagger$};
\end{scope}
\begin{scope}
\draw[fill=white] (84.000000, 15.000000) +(-45.000000:8.485281pt and 8.485281pt) -- +(45.000000:8.485281pt and 8.485281pt) -- +(135.000000:8.485281pt and 8.485281pt) -- +(225.000000:8.485281pt and 8.485281pt) -- cycle;
\clip (84.000000, 15.000000) +(-45.000000:8.485281pt and 8.485281pt) -- +(45.000000:8.485281pt and 8.485281pt) -- +(135.000000:8.485281pt and 8.485281pt) -- +(225.000000:8.485281pt and 8.485281pt) -- cycle;
\draw (84.000000, 15.000000) node {$T$};
\end{scope}
\draw (105.000000,45.000000) -- (105.000000,15.000000);
\begin{scope}
\draw[fill=white] (105.000000, 45.000000) circle(3.000000pt);
\clip (105.000000, 45.000000) circle(3.000000pt);
\draw (102.000000, 45.000000) -- (108.000000, 45.000000);
\draw (105.000000, 42.000000) -- (105.000000, 48.000000);
\end{scope}
\filldraw (105.000000, 15.000000) circle(1.500000pt);
\draw (123.000000,30.000000) -- (123.000000,15.000000);
\begin{scope}
\draw[fill=white] (123.000000, 30.000000) circle(3.000000pt);
\clip (123.000000, 30.000000) circle(3.000000pt);
\draw (120.000000, 30.000000) -- (126.000000, 30.000000);
\draw (123.000000, 27.000000) -- (123.000000, 33.000000);
\end{scope}
\filldraw (123.000000, 15.000000) circle(1.500000pt);
\begin{scope}
\draw[fill=white] (144.000000, 15.000000) +(-45.000000:8.485281pt and 8.485281pt) -- +(45.000000:8.485281pt and 8.485281pt) -- +(135.000000:8.485281pt and 8.485281pt) -- +(225.000000:8.485281pt and 8.485281pt) -- cycle;
\clip (144.000000, 15.000000) +(-45.000000:8.485281pt and 8.485281pt) -- +(45.000000:8.485281pt and 8.485281pt) -- +(135.000000:8.485281pt and 8.485281pt) -- +(225.000000:8.485281pt and 8.485281pt) -- cycle;
\draw (144.000000, 15.000000) node {$H$};
\end{scope}
\begin{scope}
\draw[fill=white] (168.000000, 15.000000) +(-45.000000:8.485281pt and 8.485281pt) -- +(45.000000:8.485281pt and 8.485281pt) -- +(135.000000:8.485281pt and 8.485281pt) -- +(225.000000:8.485281pt and 8.485281pt) -- cycle;
\clip (168.000000, 15.000000) +(-45.000000:8.485281pt and 8.485281pt) -- +(45.000000:8.485281pt and 8.485281pt) -- +(135.000000:8.485281pt and 8.485281pt) -- +(225.000000:8.485281pt and 8.485281pt) -- cycle;
\draw (168.000000, 15.000000) node {$S$};
\end{scope}
\draw (189.000000,15.000000) -- (189.000000,0.000000);
\filldraw (189.000000, 15.000000) circle(1.500000pt);
\begin{scope}
\draw[fill=white] (189.000000, 0.000000) circle(3.000000pt);
\clip (189.000000, 0.000000) circle(3.000000pt);
\draw (186.000000, 0.000000) -- (192.000000, 0.000000);
\draw (189.000000, -3.000000) -- (189.000000, 3.000000);
\end{scope}
\draw[color=black] (211.500000, 45.000000) node [fill=white] {$\cdots$};
\draw[color=black] (211.500000, 30.000000) node [fill=white] {$\cdots$};
\draw[color=black] (211.500000, 15.000000) node [fill=white] {$\cdots$};
\draw[color=black] (211.500000, 0.000000) node [fill=white] {$\cdots$};
\draw (234.000000,15.000000) -- (234.000000,0.000000);
\filldraw (234.000000, 15.000000) circle(1.500000pt);
\begin{scope}
\draw[fill=white] (234.000000, 0.000000) circle(3.000000pt);
\clip (234.000000, 0.000000) circle(3.000000pt);
\draw (231.000000, 0.000000) -- (237.000000, 0.000000);
\draw (234.000000, -3.000000) -- (234.000000, 3.000000);
\end{scope}
\begin{scope}
\draw[fill=white] (255.000000, 15.000000) +(-45.000000:8.485281pt and 8.485281pt) -- +(45.000000:8.485281pt and 8.485281pt) -- +(135.000000:8.485281pt and 8.485281pt) -- +(225.000000:8.485281pt and 8.485281pt) -- cycle;
\clip (255.000000, 15.000000) +(-45.000000:8.485281pt and 8.485281pt) -- +(45.000000:8.485281pt and 8.485281pt) -- +(135.000000:8.485281pt and 8.485281pt) -- +(225.000000:8.485281pt and 8.485281pt) -- cycle;
\draw (255.000000, 15.000000) node {$H$};
\end{scope}
\draw[fill=white] (273.000000, 9.000000) rectangle (285.000000, 21.000000);
\draw[very thin] (279.000000, 15.600000) arc (90:150:6.000000pt);
\draw[very thin] (279.000000, 15.600000) arc (90:30:6.000000pt);
\draw[->,>=stealth] (279.000000, 9.600000) -- +(80:10.392305pt);
\draw (303.000000,45.000000) -- (303.000000,30.000000);
\draw (302.500000,30.000000) -- (302.500000,15.000000);
\draw (303.500000,30.000000) -- (303.500000,15.000000);
\begin{scope}
\draw[fill=white] (303.000000, 30.000000) +(-45.000000:8.485281pt and 8.485281pt) -- +(45.000000:8.485281pt and 8.485281pt) -- +(135.000000:8.485281pt and 8.485281pt) -- +(225.000000:8.485281pt and 8.485281pt) -- cycle;
\clip (303.000000, 30.000000) +(-45.000000:8.485281pt and 8.485281pt) -- +(45.000000:8.485281pt and 8.485281pt) -- +(135.000000:8.485281pt and 8.485281pt) -- +(225.000000:8.485281pt and 8.485281pt) -- cycle;
\draw (303.000000, 30.000000) node {$Z$};
\end{scope}
\filldraw (303.000000, 45.000000) circle(1.500000pt);
\filldraw (303.000000, 15.000000) circle(1.500000pt);
\draw[color=black] (315.000000,45.000000) node[right] {$\ket{x}$};
\draw[color=black] (315.000000,30.000000) node[right] {$\ket{y}$};
\draw[color=black] (315.000000,0.000000) node[right] {$\ket{z}$};
\draw[decorate,decoration={brace,mirror,amplitude = 4.000000pt},very thick] (3.000000,-7.500000) -- (195.000000,-7.500000);
\draw (99.000000, -11.500000) node[text width=144pt,below,text centered] {logical AND gate};
\draw[decorate,decoration={brace,mirror,amplitude = 4.000000pt},very thick] (228.000000,-7.500000) -- (312.000000,-7.500000);
\draw (270.000000, -11.500000) node[text width=144pt,below,text centered] {uncomputation gate};
\end{tikzpicture}
		\end{minipage}
		\caption{Implementation of ``Toffoli pre- and post- processing'' with $T$-depth $1$.\ (a) Ancillary Toffoli gate with output qubit initialized in $\ket{z}=\ket{0}$; (b) Ancillary Toffoli gate in general.}\label{figure:anc-toffoli}
	\end{figure}

	\section{Previous Works of Carry Bit Computing}\label{sec:early-carry-bit}
	In this section, we introduce two previous works of carry bit computing and their advantages and disadvantages, which serves as building blocks of our work. For convenience, we define the weight of $n$
	\begin{equation}
		w(n) = n - \sum_{i=1}^\infty \left\lfloor \frac{n}{2^i} \right\rfloor
	\end{equation}
	as the number of ones in the binary representation of $n$. We have the following lemma
	\begin{lemma}\label{lemma:weight}
		For any integer $n\geq 1$, we have
		\begin{equation}
			\sum_{i=0}^{n-1} w(i) \leq \frac{1}{2}n\log n,
		\end{equation}
	\end{lemma}
	whose proof is given in Appendix~\ref{subsec:proof-lemma-weight}.
	
	\subsection{Algorithm based on Brent-Kung tree}\label{subsec:DrapersMethod}
	The first algorithm for quantum carry-lookahead adder is given in~\cite{draper2004logarithmic}. This method uses the Brent-Kung prefix tree, as shown in Fig.~\ref{fig:carry-draper}. The detail of the corresponding classical algorithm is described in Algorithm~\ref{algorithm:carry-draper}. We need $p[i-1,i]$ and $g[i-1,i]$ as the input, both of which can be computed directly using the input. We separate each round of the carry-bit computing into four stages, as shown in Algorithm~\ref{algorithm:carry-draper}. In round $t$ of stage $P_1$, we let $P_t[m] = p[2^t m, 2^t(m+1)]$. In round $t$ of stage $G_1$, we let $G[2^t m + 2^t] = g[2^t m, 2^t(m+1)]$. In round $t$ of stage $P_2$, we let $P'_t[m] = p[0, 2^t m + 2^{t-1}]$. In round $t$ of $G_2$, we let $G[2^t m + 2^{t-1}] = g[0, 2^t m + 2^{t-1}]$. $P_1$, $P_2$ use logical AND gate with output initialized in $\ket{0}$, which requires one ancilla. $G_1$, $G_2$ use logical AND gate with nonzero output which requires two ancillas. Moreover, when we need to recover the carry bit, there are two stages in each round, denoted by $P^{-1}_1$ and $P^{-1}_2$. Each round of $P^{-1}_1$ and $P^{-1}_2$ use only uncomputation AND gate and therefore need no $T$ gate.

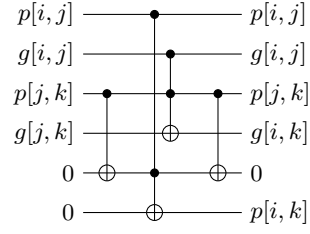
\begin{figure}[ht]
		\captionsetup{justification=raggedright,singlelinecheck=false}
		\centering
		\begin{tikzpicture}[scale=1.000000,x=1pt,y=1pt]
\filldraw[color=white] (0.000000, -7.500000) rectangle (60.000000, 82.500000);
\draw[color=black] (0.000000,75.000000) -- (60.000000,75.000000);
\draw[color=black] (0.000000,75.000000) node[left] {$p[i,j]$};
\draw[color=black] (0.000000,60.000000) -- (60.000000,60.000000);
\draw[color=black] (0.000000,60.000000) node[left] {$g[i,j]$};
\draw[color=black] (0.000000,45.000000) -- (60.000000,45.000000);
\draw[color=black] (0.000000,45.000000) node[left] {$p[j,k]$};
\draw[color=black] (0.000000,30.000000) -- (60.000000,30.000000);
\draw[color=black] (0.000000,30.000000) node[left] {$g[j,k]$};
\draw[color=black] (0.000000,15.000000) -- (60.000000,15.000000);
\draw[color=black] (0.000000,15.000000) node[left] {$0$};
\draw[color=black] (0.000000,0.000000) -- (60.000000,0.000000);
\draw[color=black] (0.000000,0.000000) node[left] {$0$};
\draw (9.000000,45.000000) -- (9.000000,15.000000);
\begin{scope}
\draw[fill=white] (9.000000, 15.000000) circle(3.000000pt);
\clip (9.000000, 15.000000) circle(3.000000pt);
\draw (6.000000, 15.000000) -- (12.000000, 15.000000);
\draw (9.000000, 12.000000) -- (9.000000, 18.000000);
\end{scope}
\filldraw (9.000000, 45.000000) circle(1.500000pt);
\draw (27.000000,75.000000) -- (27.000000,0.000000);
\begin{scope}
\draw[fill=white] (27.000000, 0.000000) circle(3.000000pt);
\clip (27.000000, 0.000000) circle(3.000000pt);
\draw (24.000000, 0.000000) -- (30.000000, 0.000000);
\draw (27.000000, -3.000000) -- (27.000000, 3.000000);
\end{scope}
\filldraw (27.000000, 75.000000) circle(1.500000pt);
\filldraw (27.000000, 15.000000) circle(1.500000pt);
\draw (33.000000,60.000000) -- (33.000000,30.000000);
\begin{scope}
\draw[fill=white] (33.000000, 30.000000) circle(3.000000pt);
\clip (33.000000, 30.000000) circle(3.000000pt);
\draw (30.000000, 30.000000) -- (36.000000, 30.000000);
\draw (33.000000, 27.000000) -- (33.000000, 33.000000);
\end{scope}
\filldraw (33.000000, 60.000000) circle(1.500000pt);
\filldraw (33.000000, 45.000000) circle(1.500000pt);
\draw (51.000000,45.000000) -- (51.000000,15.000000);
\begin{scope}
\draw[fill=white] (51.000000, 15.000000) circle(3.000000pt);
\clip (51.000000, 15.000000) circle(3.000000pt);
\draw (48.000000, 15.000000) -- (54.000000, 15.000000);
\draw (51.000000, 12.000000) -- (51.000000, 18.000000);
\end{scope}
\filldraw (51.000000, 45.000000) circle(1.500000pt);
\draw[color=black] (60.000000,75.000000) node[right] {$p[i,j]$};
\draw[color=black] (60.000000,60.000000) node[right] {$g[i,j]$};
\draw[color=black] (60.000000,45.000000) node[right] {$p[j,k]$};
\draw[color=black] (60.000000,30.000000) node[right] {$g[i,k]$};
\draw[color=black] (60.000000,15.000000) node[right] {$0$};
\draw[color=black] (60.000000,0.000000) node[right] {$p[i,k]$};
\end{tikzpicture}
		\caption{The first method to parallelize the carry computing procedure. Note that the second CNOT can be parallelized with the first one in the next round. The CNOT depth in the carry computing part is only larger than the Toffoli depth by $1$.}\label{figure:adder-parallelize}
	\end{figure}

	There are two methods to parallelize this algorithm. In the first one, we parallelize the $t$-th round of stage $P_1$ with the $t$-th round of stage $G_1$. It requires that $P_{t-1}$ serves as the control bit for stage $P_1$ and stage $G_1$ simultaneously. Thus, we need a CNOT gate to copy it, followed by another CNOT gate to recover it afterwards, as shown in Fig.~\ref{figure:adder-parallelize}. In this case, we need two ancillary qubits for parallelization of a Toffoli in stage $P_1$ and a Toffoli in stage $G_1$, one of which for copy, the other one for a general Toffoli gate implementation. Since there are $\left\lfloor n/2^t \right\rfloor$ pairs of such Toffoli gates implementing simultaneously in round $t$, we need at most $\frac{n}{2} \cdot 2 = n$ ancillas. The parallelization of stage $P_2$ and $G_2$ is exactly the same. In total, we need $n$ input qubits, $2n-w(n)-\lfloor\log n\rfloor-1$ output qubits, $n$ input and output qubits, and $n$ ancillary qubits. The Toffoli/CNOT depth/count of this method is
	\begin{equation}
		\begin{aligned}
			\text{Toffoli depth} &= \lfloor \log n \rfloor + \left\lfloor \log \frac{2n}{3} \right\rfloor, \\
			\text{Toffoli count} &= 4n - 2w(n) - 2\lfloor \log n \rfloor - 2, \\
			\text{CNOT depth} &= \lfloor \log n \rfloor + \left\lfloor \log \frac{2n}{3} \right\rfloor + 2, \\
			\text{CNOT count} &= 4n - 2w(n) - 2\lfloor \log n \rfloor - 2.
		\end{aligned}
	\end{equation}

	In the second method, we parallelize the $t$-th round of stage $P_1$ and the $(t-1)$-th round of stage $G_1$. In this case, there is no qubit serving as the control bits for both stage, and we no longer need the CNOT gates for copying and recovering. Therefore, the maximum number of ancillas we need is $\frac{n}{2}$. The Toffoli depth and number of this method is
	\begin{equation}
		\begin{aligned}
			\text{Toffoli depth} &= \lfloor \log n \rfloor + \left\lfloor \log \frac{2n}{3} \right\rfloor + 2, \\
			\text{Toffoli number} &= 4n - 2w(n) - 2\lfloor \log n \rfloor - 2, \\
		\end{aligned}
	\end{equation}
	and it need no CNOT gates.

	Comparing these two methods, we can see that the first method reduces the Toffoli depth by two at the cost of $0.5n$ more ancillas and $2\log n$ CNOT depth. The extra ancillas do not matter a lot, since the space overhead of this part is not the bottleneck of our final algorithm, as we will explain later. Whether it is worthy to replace two Toffoli depth with $2\log n$ CNOT depth, however, depends on the specific application. Since our Toffoli gate implementation have a CNOT depth of at least $6$, as shown in Fig.~\ref{figure:toffoli}, it is always better to use the first method when $n \leq 64$, if we only consider the depth. Furthermore, in many fault-tolerant quantum computing schemes, the time overhead of $T$ gates could be much larger than that for CNOT gates, making the CNOT depth less important than $T$-depth as a measure of algorithm performance. In this case, the strict superiority of the first method can be achieved for the qubit number smaller than a threshold $t$, which can be much larger than $64$. In this paper, we use the first method for the sake of simplicity.

	We should remark that the carry-bit computing introduced here is slightly more expensive than the initial one in~\cite{draper2004logarithmic}. The reason is that some of the propogation state computation (stage $P_2$ and the final round in stage $P_1$) is unnecessary if we only need the carry bits $g[0,i]$. However, the propagation bits $p[0,i]$ are also useful in our algorithm, making us retain this part. Such a difference, after all, will not affect the Toffoli depth.

	\begin{algorithm}[ht]
		\SetKwInOut{Input}{Input}
		\SetKwInOut{Output}{Output}
		\SetKw{for}{for}
		\SetKwProg{Fn}{Function}{:}{}
		\SetKwFunction{draperIn}{ComputeCarryDraperPGCompute}
		\SetKwFunction{draperOut}{ComputeCarryDraperPUncompute}
		\SetKwFunction{draper}{ComputeCarryDraper}
		
		\LinesNumbered{}
		\Fn{\draperIn{$P_0;G;P;a$}}{
			\Input{$n$-qubit $P_0$ storing $p[i-1,i]$; $n$-qubit $G$ storing $g[i-1,i]$; $(2n-w(n)-\lfloor\log n\rfloor-1)$-qubit $P$, initialized to 0; $2n$-qubit ancilla $a$, initialized to 0.}
			\Output{$P_0,a$ remains the same; $G$ holds carry bits $c_i=g[0,i]$; $P$ holds propogation states, including $p[0,i]$.}
			\For{$t \gets 1 ,\ldots, \lfloor\log n\rfloor$}{
				$P_t[m] = P_{t-1}[2m] \cdot P_{t-1}[2m+1]$ \for{} $m \gets 0 ,\ldots, \lfloor n/2^t \rfloor$ \tcp*{stage $P_1$}
				$G[2^t m + 2^t] \oplusIs G[2^t m + 2^{t-1}] \cdot P_{t-1}[2m+1]$ \for{} $m \gets 0 ,\ldots, \lfloor n/2^t \rfloor$ \tcp*{stage $G_1$}
			}
			\For{$t \gets \lfloor \log \frac{2n}{3} \rfloor ,\ldots, 1$}{
				$P'_t[m] = P_t[0] \cdot P_{t-1}[2m+1]$ \for{} $m \gets 1 ,\ldots, \lfloor (n-2^{t-1})/2^t \rfloor$ \tcp*{stage $P_2$}
				$G[2^t m + 2^{t-1}] \oplusIs G[2^t m] \cdot P_{t-1}[2m+1]$ \for{} $m \gets 1 ,\ldots, \lfloor (n-2^{t-1})/2^t \rfloor$ \tcp*{stage $G_2$}
			}
		}
		
		\Fn{\draperOut{$P_0;P;a$}}{
			\Input{$n$-qubit $P_0$ storing $p[i-1,i]$; $(2n-w(n)-\lfloor\log n\rfloor-1)$-qubit $P$ storing propogation states; $n$-bit ancilla, initialized to 0.}
			\Output{$P_0$ remains the same; $P$ recovered to 0.}
			\For{$t \gets 1 ,\ldots, \lfloor \log \frac{2n}{3} \rfloor$}{
				$P'_t[m] \oplusIs P_t[0] \cdot P_{t-1}[2m+1]$ \for{} $m \gets 1,\ldots,\lfloor(n-2^{t-1})/2^t \rfloor$ \tcp*{stage $P^{-1}_2$}
			}
			\For{$t \gets \lfloor\log n\rfloor ,\ldots, 1$}{
				$P_t[m] \oplusIs P_{t-1}[2m] \cdot P_{t-1}[2m+1]$ \for{} $m \gets 0,\ldots,\lfloor n/2^t \rfloor$ \tcp*{stage $P^{-1}_1$}
			}
		}
		
		\Fn{\draper{$P_0;G;P,a$}}{
			\Input{$n$-qubit $P_0$ storing $p[i-1,i]$; $n$-qubit $G$ storing $g[i-1,i]$; $(2n-w(n)-\lfloor\log n\rfloor-1)$-qubit ancilla $P$ and $2n$-qubit ancilla $a$, initialized to 0.}
			\Output{$P_0,P,a$ remains the same; $G$ holds carry bits $c_i=g[0,i]$.}
			\draperIn{$P_0,G,P,a$}\;
			\draperOut{$P_0,P$}\;
		}
		\caption{quantum algorithm for computing carry bits based on Brent-Kung tree.}\label{algorithm:carry-draper}
	\end{algorithm}

	\subsection{Algorithm based on Sklansky tree}
	A new method for computing the carry bits based on the Sklansky prefix tree is proposed in~\cite{wang2024optimal}, as shown in Fig.~\ref{fig:carry-wang}. This method halves the Toffoli depth of the previous method, at the cost of linear fan-out and increasing the ancilla space overhead to $O(n\log n)$. The algorithm is shown in Algorithm~\ref{algorithm:carry-wang}. We separate each round of the carry-bit computing into two stages. In the $t$-th round of stage $P$, we set $P_k[l,m] = p[2^k\cdot l, 2^k\cdot l+m]$, and in the $t$-th round of stage $G$, we set $G[2^k\cdot l+m] = g[2^k\cdot l, 2^k\cdot l+m]$. Similar to the previous method, stage $P$ and $G$ use logical AND gates, and stage $P^{-1}$ uses uncomputation gates.

	We still have two parallelization methods. The first method parallelizes round $t$ of stage $P$ and round $t$ of stage $G$, which needs CNOT gates. There are $\frac{n}{2}$ Toffoli gates in stage $P$ and stage $G$ implemented simultaneously, implying that the ancillas needed is $\frac{n}{2}\cdot(1+1)=n$. In total, we need $n$ input qubits, $\frac{n}{2}\log n$ output qubits, $n$ input and output qubits, and $n$ ancillas. The Toffoli/CNOT cost of this method is
	\begin{equation}
		\begin{aligned}
			\text{Toffoli depth} &= \left\lceil \log n \right\rceil, \\
			\text{Toffoli number} &= 2\sum_{i=0}^{n-1} w(i) \leq n\log n, \\
			\text{CNOT depth} &= \left\lceil \log n \right\rceil + 1, \\
			\text{CNOT number} &= 2\sum_{i=0}^{n-1} w(i) \leq n\log n,
		\end{aligned}
	\end{equation}
	where the inequality is from Lemma~\ref{lemma:weight}.

	The second method parallelizes round $t$ of stage $P$ and round $t-1$ of stage $G$, with resource cost
	\begin{equation}
		\begin{aligned}
			\text{Toffoli depth} &= \left\lceil \log n \right\rceil+1, \\
			\text{Toffoli number} &= 2\sum_{i=0}^{n-1} w(i) \leq n\log n, \\
		\end{aligned}
	\end{equation}
	with no CNOT gates. The ancillas needed in this method is $\frac{n}{2}$. Again, we choose the first method in this paper.

	\begin{algorithm}[ht]
		\SetKwInOut{Input}{Input}
		\SetKwInOut{Output}{Output}
		\SetKw{for}{for}
		\SetKwProg{Fn}{Function}{:}{}
		\SetKwFunction{wangIn}{ComputeCarryWangPGCompute}
		\SetKwFunction{wangOut}{ComputeCarryWangPUncompute}
		\SetKwFunction{wang}{ComputeCarryWang}
		
		\LinesNumbered{}
		\Fn{\wangIn{$P_0;G;P;a$}}{
			\Input{$n$-qubit $P_0$ storing $p[i-1,i]$; $n$-qubit $G$ storing $g[i-1,i]$; $\frac{n}{2}\log n$-qubit $P$, initialized to 0; $2n$-qubit ancilla $a$, initialized to 0.}
			\Output{$P_0,a$ remains the same; $G$ holds carry bits $c_i=g[0,i]$; $P$ holds propogation states, including $p[0,i]$.}
			\For{$k \gets 1,\ldots, \left\lceil \log n \right\rceil-1$\tcp*{stage $P$}}{
				\For{$l\gets1,\ldots,\left\lceil n/2^k \right\rceil$}{
					$P_k[l,m]=P_{k-1}[2l,2^{k-1}]\cdot P_{k-1}[2l+1,m-2^{k-1}]$ \for{} $m\gets 2^{k-1}+1,\ldots,\min\{2^k,n-2^{k-1}(2l+1)\}$
				}
			}
			\For{$k \gets 1,\ldots,\left\lceil \log n \right\rceil$\tcp*{stage $G$}}{
				\For{$l\gets0,\ldots,\left\lceil n/2^k \right\rceil$}{
					$G[2^k\cdot l+m]\oplusIs G[2^{k-1}(2l+1)]\cdot P_{k-1}[2l+1,m-2^{k-1}]$ \for{} $m\gets 2^{k-1}+1,\ldots,\min\{2^k,n-2^{k-1}(2l+1)\}$
				}
			}
		}
		
		\Fn{\wangOut{$P_0;P$}}{
			\Input{$n$-qubit $P_0$ storing $p[i-1,i]$; $\frac{n}{2}\log n$-qubit $P$ storing propogation states.}
			\Output{$P_0$ remains the same; $P$ recovered to 0.}
			\For{$k \gets \left\lceil \log n \right\rceil-1,\ldots, 1$\tcp*{stage $P^{-1}$}}{
				\For{$l\gets\left\lceil n/2^k \right\rceil,\ldots,1$}{
					$P_k[l,m]\oplusIs P_{k-1}[2l,2^{k-1}]\cdot P_{k-1}[2l+1,m-2^{k-1}]$ \for{} $m\gets 2^{k-1}+1,\ldots,\min\{2^k,n-2^{k-1}(2l+1)\}$
				}
			}
		}
		
		\Fn{\wang{$P_0;G;P,a$}}{
			\Input{$n$-qubit $P_0$ storing $p[i-1,i]$; $n$-qubit $G$ storing $g[i-1,i]$; $\frac{n}{2}\log n$-qubit ancilla $P$ and $2n$-qubit ancilla $a$, initialized to 0.}
			\Output{$P_0,P,a$ remains the same; $G$ holds carry bits $c_i=g[0,i]$.}
			\wangIn{$P_0,G,P,a$}\;
			\wangOut{$P_0,P$}\;
		}
		\caption{quantum algorithm for computing carry bits based on Sklansky tree.}\label{algorithm:carry-wang}
	\end{algorithm}

	\subsection{Proof of Lemma~\ref{lemma:weight}}\label{subsec:proof-lemma-weight}

	For simplicity, we define $S_n = \sum_{i=0}^{n-1} w(i)$, and let $S_0=0$. We have two observations of $S_n$:
	\begin{observation}\label{observation:weight-1}
		If $n=2^k$ for some non-negative integer $k$, $S_n = \frac{1}{2}n\log n$.
	\end{observation}

	\begin{observation}\label{observation:weight-2}
		For any integer $n\geq 1$, let $k=\lfloor \log n \rfloor$, then
		\begin{equation}
			\begin{aligned}
				S_n &= S_{2^k} + (n-2^k) + S_{n-2^k} \\
				&= n - 2^k + \frac{1}{2}k2^k + S_{n-2^k}, \\
			\end{aligned}
		\end{equation}
		where the second equality is from Observation~\ref{observation:weight-1}.
	\end{observation}

	Now, we begin our proof by induction on $n$. The base cases $n=1,2$ are trivial. Suppose the lemma holds for all integers smaller than $n$. If $\log n \in \mathbb{Z}$, the inequality holds by Observation~\ref{observation:weight-1}. Thus, we only need to consider the cases $\log n \notin \mathbb{Z}$ from now on. Let $k=\lfloor \log n \rfloor$. By Observation~\ref{observation:weight-2}, we have
	\begin{equation}
		S_n = n - 2^k + \frac{1}{2}k2^k + S_{n-2^k} \leq n - 2^k + \frac{1}{2}k2^k + \frac{1}{2}(n-2^k)\log(n-2^k).
	\end{equation}

	Let $f(x) = n-x + \frac{1}{2}x\log x + \frac{1}{2}(n-x)\log(n-x)$. We only need to prove $f(2^k) \leq \frac{1}{2}n\log n$. Note that $\frac{n}{2} < 2^k < n$, we can generalize the statement and try to prove that $f(x) \leq \frac{1}{2}n\log n$ for all $x \in (\frac{n}{2}, n)$. Taking the derivative of $f(x)$, we have
	\begin{gather}
		f'(x) = \frac{1}{2}\log\frac{x}{n-x} - 1, \\
		f''(x) = \frac{1}{2}\qty(\frac{1}{x} + \frac{1}{n-x}) > 0.
	\end{gather}
	Furthermore, we have $f'(\frac{n}{2}) = -1 < 0$ and $\lim_{x \to n} f'(x) = +\infty > 0$. We can conclude that $f(x)$ first decreases and then increases in the interval $(\frac{n}{2}, n)$, which implies that $f(x) \leq \max\{f\qty(\frac{n}{2}), \lim_{x \to n} f(x)\}$. Given
	\begin{gather}
		f\qty(\frac{n}{2}) = \frac{n}{2} + \frac{n}{4}\log\frac{n}{2} + \frac{n}{4}\log\frac{n}{2} = \frac{1}{2}n\log n, \\
		\lim_{x \to n} f(x) = 0 + \frac{1}{2}n\log n + 0 = \frac{1}{2}n\log n,
	\end{gather}
	we have $f(x) \leq \frac{1}{2}n\log n$ for $x \in (\frac{n}{2},n)$, which completes our proof.

	\section{Circuit Construction}\label{sec:circuit-construction}
	In this section, we introduce the circuit for the components of point addition, including modular addition, modular multiplication and modular division, followed by a detailed analysis of resource consumption, including circuit depth, width and gate number. Recall that modular addition implements $\ket{x}\ket{y}\mapsto\ket{x}\ket{(y+x) \bmod{p}}$, modular multiplication implements $\ket{x}\ket{y}\ket{z}\mapsto\ket{x}\ket{y}\ket{(z\oplus x\cdot y) \bmod{p}}$, and modular division implements $\ket{x}\ket{y}\ket{z}\mapsto\ket{x}\ket{y}\ket{(z\oplus \frac{y}{x}) \bmod{p}}$. We will only discuss Toffoli depth, Toffoli number, CNOT depth, and CNOT number as explained in Sec.~\ref{subsec:circuit-parameters}, and will mainly focus on their leading terms. Specifically, as an example, if we say ``the Toffoli depth is $4n\log n$'' for simplicity, we actually mean ``the Toffoli depth is $4n\log n + o(n\log n)$''. For equations, we will use ``$=$'' for precise numbers, and ``$\simeq$'' for asymptotic equivalence up to leading terms. That is, $x\simeq f(n)$ means $x=f(n)+o(f(n))$. Since Toffoli gates are the most time-consuming part in our circuit, our main optimization goal is to make Toffoli depth as small as possible.

	For simplicity, this section will only include the main operations, and will only include the result of the resource cost of these operations. Detailed computation of the resource cost can be found in Appendix~\ref{sec:resource-cost}. Other circuit constructions can be found in Appendix~\ref{sec:other-circuit}.
	
	\subsection{Modular Addition}\label{subsec:addition}
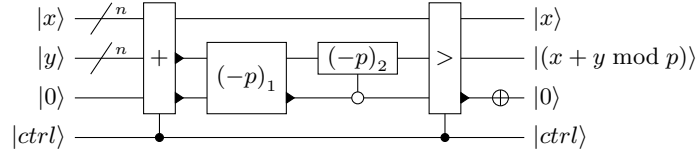
\begin{figure}[ht]
		\captionsetup{justification=raggedright,singlelinecheck=false}
		\centering
		\begin{tikzpicture}[scale=1.000000,x=1pt,y=1pt]
\filldraw[color=white] (0.000000, -7.500000) rectangle (170.000000, 52.500000);
\draw[color=black] (0.000000,45.000000) -- (170.000000,45.000000);
\draw[color=black] (0.000000,45.000000) node[left] {$\ket{x}$};
\draw[color=black] (0.000000,30.000000) -- (170.000000,30.000000);
\draw[color=black] (0.000000,30.000000) node[left] {$\ket{y}$};
\draw[color=black] (0.000000,15.000000) -- (170.000000,15.000000);
\draw[color=black] (0.000000,15.000000) node[left] {$\ket{0}$};
\draw[color=black] (0.000000,0.000000) -- (170.000000,0.000000);
\draw[color=black] (0.000000,0.000000) node[left] {$\ket{ctrl}$};
\draw (6.000000, 39.000000) -- (14.000000, 51.000000);
\draw (12.000000, 48.000000) node[right] {$\scriptstyle{n}$};
\draw (6.000000, 24.000000) -- (14.000000, 36.000000);
\draw (12.000000, 33.000000) node[right] {$\scriptstyle{n}$};
\draw (32.000000,45.000000) -- (32.000000,0.000000);
\begin{scope}
\draw[fill=white] (32.000000, 30.000000) +(-45.000000:8.485281pt and 29.698485pt) -- +(45.000000:8.485281pt and 29.698485pt) -- +(135.000000:8.485281pt and 29.698485pt) -- +(225.000000:8.485281pt and 29.698485pt) -- cycle;
\clip (32.000000, 30.000000) +(-45.000000:8.485281pt and 29.698485pt) -- +(45.000000:8.485281pt and 29.698485pt) -- +(135.000000:8.485281pt and 29.698485pt) -- +(225.000000:8.485281pt and 29.698485pt) -- cycle;
\draw (32.000000, 30.000000) node {$+$};
\end{scope}
\qOutput{38}{15}{black}
\qOutput{38}{30}{black}
\filldraw (32.000000, 0.000000) circle(1.500000pt);
\draw (65.000000,30.000000) -- (65.000000,15.000000);
\begin{scope}
\draw[fill=white] (65.000000, 22.500000) +(-45.000000:21.213203pt and 19.091883pt) -- +(45.000000:21.213203pt and 19.091883pt) -- +(135.000000:21.213203pt and 19.091883pt) -- +(225.000000:21.213203pt and 19.091883pt) -- cycle;
\clip (65.000000, 22.500000) +(-45.000000:21.213203pt and 19.091883pt) -- +(45.000000:21.213203pt and 19.091883pt) -- +(135.000000:21.213203pt and 19.091883pt) -- +(225.000000:21.213203pt and 19.091883pt) -- cycle;
\draw (65.000000, 22.500000) node {${(-p)}_1$};
\end{scope}
\qOutput{80}{15}{black}
\draw (107.000000,30.000000) -- (107.000000,15.000000);
\begin{scope}
\draw[fill=white] (107.000000, 30.000000) +(-45.000000:21.213203pt and 8.485281pt) -- +(45.000000:21.213203pt and 8.485281pt) -- +(135.000000:21.213203pt and 8.485281pt) -- +(225.000000:21.213203pt and 8.485281pt) -- cycle;
\clip (107.000000, 30.000000) +(-45.000000:21.213203pt and 8.485281pt) -- +(45.000000:21.213203pt and 8.485281pt) -- +(135.000000:21.213203pt and 8.485281pt) -- +(225.000000:21.213203pt and 8.485281pt) -- cycle;
\draw (107.000000, 30.000000) node {${(-p)}_2$};
\end{scope}
\draw[fill=white] (107.000000, 15.000000) circle(2.250000pt);
\draw (140.000000,45.000000) -- (140.000000,0.000000);
\begin{scope}
\draw[fill=white] (140.000000, 30.000000) +(-45.000000:8.485281pt and 29.698485pt) -- +(45.000000:8.485281pt and 29.698485pt) -- +(135.000000:8.485281pt and 29.698485pt) -- +(225.000000:8.485281pt and 29.698485pt) -- cycle;
\clip (140.000000, 30.000000) +(-45.000000:8.485281pt and 29.698485pt) -- +(45.000000:8.485281pt and 29.698485pt) -- +(135.000000:8.485281pt and 29.698485pt) -- +(225.000000:8.485281pt and 29.698485pt) -- cycle;
\draw (140.000000, 30.000000) node {$>$};
\end{scope}
\qOutput{146}{15}{black}
\filldraw (140.000000, 0.000000) circle(1.500000pt);
\begin{scope}
\draw[fill=white] (161.000000, 15.000000) circle(3.000000pt);
\clip (161.000000, 15.000000) circle(3.000000pt);
\draw (158.000000, 15.000000) -- (164.000000, 15.000000);
\draw (161.000000, 12.000000) -- (161.000000, 18.000000);
\end{scope}
\draw[color=black] (170.000000,45.000000) node[right] {$\ket{x}$};
\draw[color=black] (170.000000,30.000000) node[right] {$\ket{(x+y \bmod{p})}$};
\draw[color=black] (170.000000,15.000000) node[right] {$\ket{0}$};
\draw[color=black] (170.000000,0.000000) node[right] {$\ket{ctrl}$};
\end{tikzpicture}
		\caption{Circuit for controlled modular addition~\cite{haner2020improved}. Here, ``$+$'' denotes controlled integer addition, ``$-p$'' denotes controlled constant substraction of $p$ where ``${(-p)}_1$'' and ``${(-p)}_2$'' are the two parts of one controlled constant subtraction, as explained in Appendix~\ref{subsubsec:constantadder}, ``$>$'' denotes controlled comparator, and $\oplus$ means a Pauli $X$ operation. Note that the first ``$+$'' has one more qubit output, which is the highest carry bit, since $x+y$ can be an $(n+1)$-bit number.}\label{figure:circuit-mod-addition}
	\end{figure}

	Recall that we have discussed integer addition in detail in Sec.~\ref{sec:improvedadder}. The key idea of implementing modular addition with integer addition is to use a sign qubit to decide whether $x+y$ is greater than $p$ or not~\cite{roetteler2017quantum}. To be specific, one can apply the integer addition $\ket{x}\ket{y} \mapsto \ket{x}\ket{x+y}$, followed by a constant addition $-p$ controlled by whether $x+y-p$ is positive or not. By utilizing the leading carry bit of $x+y-p$ as the controlled bit, one realize the former idea with a single controlled constant addition~\cite{rines2018high}. Moreover, to recover the sign qubit, a comparator that realize $\ket{x}\ket{y}\ket{b}\mapsto\ket{x}\ket{y}\ket{b\oplus c(x,y)}$ is applied to the final state, where $c(x,y)=0$ if $x > y$ and $c(x,y)=1$ otherwise. The former procedure is shown by the circuit in Fig.~\ref{figure:circuit-mod-addition}, and detailed discussions for the constructions of constant addition and comparator are given in Appendices~\ref{subsubsec:constantadder} and~\ref{subsubsec:comparator}.

	In Shor's algorithm, the required operation is a controlled modular addition, indicating that all gates in the addition circuit must be conditioned on a control qubit, denoted as $ctrl$, as illustrated in Fig.~\ref{figure:circuit-mod-addition}~\cite{roetteler2017quantum}. It is worth noting that, except for the integer addition and comparator components, the remaining operations, labeled ``${(-p)}_1$'', ``${(-p)}_2$'', and ``$\oplus$'', mutually cancel when the circuit is executed in sequence. Consequently, the control need only be applied to the integer addition and comparator subroutines. The detailed implementations of the controlled integer adder and controlled comparator are provided in Appendices~\ref{subsubsec:constantadder} and~\ref{subsubsec:comparator}.
	
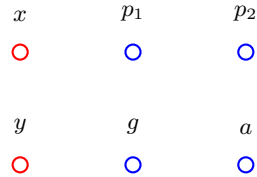
\begin{figure}[ht]
		\captionsetup{justification=raggedright,singlelinecheck=false}
		\centering
		\begin{tikzpicture}
    \draw[thick, red] (0, 1.5) circle(0.1);
    \node[above] at (0, 1.8) {$x$};
    \draw[thick, red] (0, 0) circle(0.1);
    \node[above] at (0, 0.3) {$y$};
    \draw[thick, blue] (1.5, 1.5) circle(0.1);
    \node[above] at (1.5, 1.8) {$p_1$};
    \draw[thick, blue] (1.5, 0) circle(0.1);
    \node[above] at (1.5, 0.3) {$g$};
    \draw[thick, blue] (3, 1.5) circle(0.1);
    \node[above] at (3, 1.8) {$p_2$};
    \draw[thick, blue] (3, 0) circle(0.1);
    \node[above] at (3, 0.3) {$a$};
\end{tikzpicture}
		\caption{Qubits layout for modular addition. Each point stands for a column of $n$ qubits. Red points $x$, $y$ are data qubits, and blue points $p_1$, $p_2$, $g$, and $a$ are ancillary qubits.}\label{figure:layout-addition}
	\end{figure}

	We now analyze the qubit layout for modular addition. As discussed in Sec.~\ref{sec:improvedadder}, the integer addition requires $4n$ ancillary qubits, corresponding to four columns in our layout. Beyond these, no additional column is needed to store the constant $p$, since its value is embedded directly through the simplification of constant addition. Likewise, no separate column is required for the control qubit, as it can be conveniently prepared from the ancillas of the addition circuit whenever needed. The resulting layout is illustrated in Fig.~\ref{figure:layout-addition}. The maximum gate interaction distance is $2$, occurring between the $\ket{y}$ register and the ancillary qubits $p_1$ and $g$. Overall, the modular addition circuit requires a Toffoli depth of $(5\log n + 4\log\log n)$, a Toffoli count of $33n$, a CNOT depth of $(5\log n + 4\log\log n)$, and a CNOT count of $42n$. A detailed breakdown of these resource estimates is provided in Appendix~\ref{sec:resource-cost}.
	
	\subsection{Modular Multiplication}\label{subsec:multiplication}
	
	The modular multiplication implements $\ket{x}\ket{y}\ket{z}\mapsto\ket{x}\ket{y}\ket{z\oplus (x\cdot y \bmod{p})}$. To make the multiplication more efficient, we choose the Montgomery representation~\cite{montgomery1985modular} rather than the normal binary representation. In the Montgomery representation, an integer $x$ is expressed as $x' = x2^n \bmod{p}$, where $n = \lceil \log p \rceil$. Therefore, any integer in this representation is intrinsically mod $p$. Note that the addition in Montgomery representation is the consistent to that in standard representation since $x'+y'=(x+y)2^n \bmod{p} = (x+y)'$. Moreover, to compute $x \cdot y$, we first compute $x' \cdot y' = x \cdot y \cdot 2^{2n} \bmod{p}$, then perform a $2^n$ reduction to get $(x \cdot y)' = x \cdot y \cdot 2^n \bmod{p}$.
	
	To achieve modular multiplication in the Montgomery representation, we first initialize one column of $n$ qubits to $\ket{0^n}$ as the output qubits. Denote the binary representation of $x$ as $x = x_{n-1} \ldots x_1 x_0$. In the $i$-th step, we add $y$ to the result conditioned on whether $x_i$ is $1$ or not, and perform a $2$-reduction. A $2$-reduction $\ket{x} \mapsto \ket{x/2 \bmod{p}}$ is achieved through a controlled $+p$ conditioned on whether $x$ is odd or even, followed by a right shift, which is similar to what we do in our modular addition. This method computes
	\begin{equation}
		((x_0 y / 2 + x_1 y) / 2 + \cdots +x_{n-1} y) / 2 \bmod{p}
		= (x_0 + x_1 2^1 + \cdots + x_{n-1} 2^{n-1}) \cdot y / 2^n \bmod{p}
		= (x \cdot y) / 2^n \bmod{p},
	\end{equation}
	which is the Montgomery representation of $x \cdot y$.
	
\begin{figure}[ht]
		\captionsetup{justification=raggedright,singlelinecheck=false}
		\centering
		\input{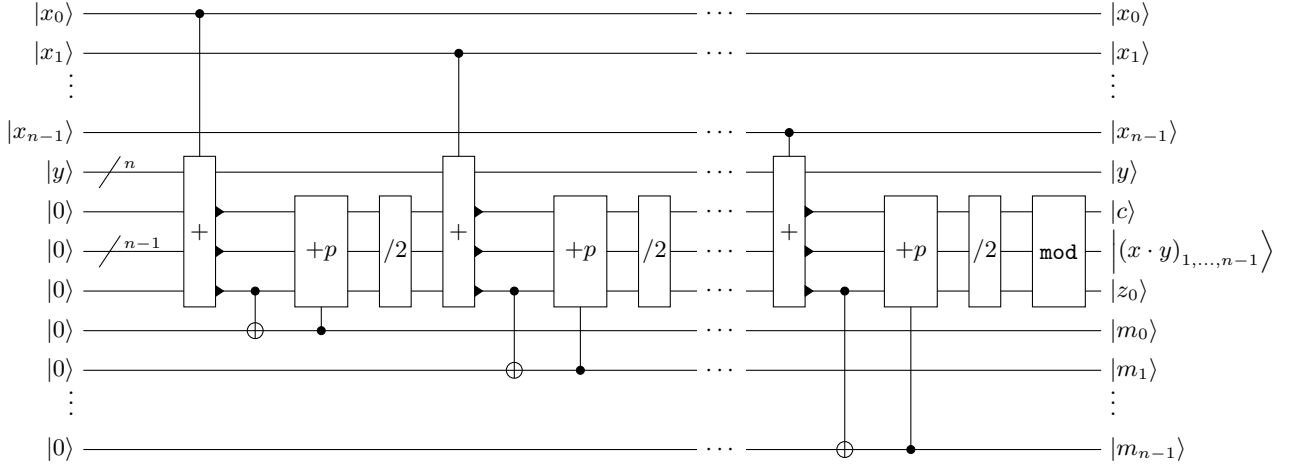}
		\caption{Quantum circuit for modular multiplication without the windowed trick. The ``$+$'' (modular addition) operation performs the transformation $\ket{y}\ket{0}\ket{z_{n-1}\cdots z_1}\ket{z_0} \mapsto \ket{y}\ket{c'}\ket{z'_{n-1}\cdots z'_1}\ket{z'_0}$, where $c'$ denotes the highest carry bit and $z' = (z + y) \bmod p$. Here, $z_0$ and $z'_0$ are the least significant bits of $z$ and $z'$, respectively. The highest carry bit $c'$ can be regarded as $z_n$, effectively extending $z$ to an $(n+1)$-bit string. The ``$+p$'' block represents a (controlled) constant integer addition, as discussed in Appendix~\ref{subsubsec:constantadder}. The ``$/2$'' block denotes a (controlled) $1$-bit right shifting operation, as discussed in Appendix~\ref{subsubsec:shifting}.\ \texttt{mod} corresponds to the modular reduction, as discussed in Appendix~\ref{subsubsec:reduction}.}\label{figure:circuit-mod-multiplication-raw}
	\end{figure}

	In total, this method requires $n$ controlled additions, $n$ controlled constant additions, and $n$ one-bit right shifts. Right shifts are achieved with SWAP gates and need no Toffoli gate. Since each addition costs $4\log n$ Toffoli depth, this method costs $8n \log n$ Toffoli depth. The circuit of this method of modular multiplication is shown in Fig.~\ref{figure:circuit-mod-multiplication-raw}.

	The windowed trick~\cite{gidney2019windowed} can be employed to further reduce the computational cost. In the previous procedure, the value of $y$ is added conditionally based on a single bit $x_i$ during the $i$-th step. Instead, we can consider a $k$-bit segment of $x$, and use it to control the $y$-addition to the output register in a single step. More precisely, the $n$-bit string $x$ can be partitioned into disjoint $k$-bit blocks $x_{\qty(\left\lfloor\frac{n}{k}\right\rfloor)}, x_{\qty(\left\lfloor\frac{n}{k}\right\rfloor - 1)}, \ldots, x_{(0)}$, where
	\begin{equation}
		x_{(0)} = x_{k-1}\ldots x_0;\ x_{(1)} = x_{2k-1}\ldots x_{k};\ \ldots;\ x_{\qty(\left\lfloor\frac{n}{k}\right\rfloor - 1)} = x_{k \left\lfloor \frac{n}{k} \right\rfloor - 1} \ldots x_{k \qty(\left\lfloor \frac{n}{k} \right\rfloor-1)};\ x_{\qty(\left\lfloor \frac{n}{k} \right\rfloor)} = x_{n-1}\ldots x_{k \left\lfloor \frac{n}{k} \right\rfloor}.
	\end{equation}
	Among them, $x_{(0)}, x_{(1)}, \ldots, x_{\qty(\left\lfloor \frac{n}{k} \right\rfloor-1)}$ are all $k$-bit numbers, and $x_{\qty(\left\lfloor \frac{n}{k} \right\rfloor)}$ are an $r$-bit number with $r = n - k \left\lfloor \frac{n}{k} \right\rfloor$.
	
	Then, we initialize a $(n+k)$-qubit register as the output qubits. In the $i$-th step, we add $x_{(i)} \cdot y$ to the result, which is achieved by a sequence of controlled addition as shown in Fig.~\ref{figure:circuit-multiplication-window}, and get the last $k$ bits $m_i$. To perform a $2^k$-reduction, we use a quantum table introduced in Appendix~\ref{subsec:Qtable} to find the appropriate multiples of $p$, $t_{m_i} p$, such that $t_{m_i} p + m_i \equiv 0 \bmod{p}$, and add them to the result. Note that after adding $t_{m_i} p$ to the register, the last $k$ bits of the result are all $0$, and we can perform a $k$-bit right shift to achieve a $2^k$-reduction. After these steps, we get the result
	\begin{equation}
		\begin{aligned}
			&\qty(\qty(\qty(x_{(0)} y / 2^k + x_{(1)} y) / 2^k + \cdots + x_{\qty(\left\lfloor\frac{n}{k}\right\rfloor - 1)} y) / 2^k + x_{\qty(\left\lfloor\frac{n}{k}\right\rfloor)} y) / 2^r \bmod{p} \\
			=& \qty(x_{(0)} + x_{(1)} 2^k + \cdots + x_{\qty(\left\lfloor\frac{n}{k}\right\rfloor - 1)} 2^{k\qty(\left\lfloor\frac{n}{k}\right\rfloor - 1)} + x_{\qty(\left\lfloor\frac{n}{k}\right\rfloor)} 2^{k \left\lfloor\frac{n}{k}\right\rfloor}) \cdot y / 2^n \bmod{p} \\
			=& (x \cdot y) / 2^n \bmod{p}.
		\end{aligned}
	\end{equation}
	The whole circuit is shown in Fig.~\ref{figure:circuit-mod-multiplication} and the construction of each building blocks \texttt{mul\_win} is shown in Fig.~\ref{figure:circuit-multiplication-window}. It is worth noting that this circuit produces one column of garbage qubits $m$, which we need to recover later by the reverse operation.
	
	\begin{figure}[ht]
		\captionsetup{justification=raggedright,singlelinecheck=false}
		\centering
		\begin{tikzpicture}[scale=1.000000,x=1pt,y=1pt]
\filldraw[color=white] (0.000000, -7.500000) rectangle (215.000000, 112.500000);
\draw[color=black] (0.000000,105.000000) -- (247.000000,105.000000);
\draw[color=black] (0.000000,105.000000) node[left] {$\ket{x_{(0)}}$};
\draw[color=black] (0.000000,90.000000) -- (247.000000,90.000000);
\draw[color=black] (0.000000,90.000000) node[left] {$\ket{x_{(1)}}$};
\draw[color=black] (0.000000,75.000000) node[anchor=mid east] {$\vdots$};
\draw[color=black] (0.000000,60.000000) -- (247.000000,60.000000);
\draw[color=black] (0.000000,60.000000) node[left] {$\ket{x_{(\lfloor n / k \rfloor)}}$};
\draw[color=black] (0.000000,45.000000) -- (247.000000,45.000000);
\draw[color=black] (0.000000,45.000000) node[left] {$\ket{y}$};
\draw[color=black] (0.000000,30.000000) -- (247.000000,30.000000);
\draw[color=black] (0.000000,30.000000) node[left] {$\ket{0}$};
\draw[color=black] (0.000000,15.000000) -- (247.000000,15.000000);
\draw[color=black] (0.000000,15.000000) node[left] {$\ket{0}$};
\draw[color=black] (0.000000,0.000000) -- (247.000000,0.000000);
\draw[color=black] (0.000000,0.000000) node[left] {$\ket{0}$};
\draw (6.000000, 99.000000) -- (14.000000, 111.000000);
\draw (12.000000, 108.000000) node[right] {$\scriptstyle{k}$};
\draw (6.000000, 84.000000) -- (14.000000, 96.000000);
\draw (12.000000, 93.000000) node[right] {$\scriptstyle{k}$};
\draw (6.000000, 54.000000) -- (14.000000, 66.000000);
\draw (12.000000, 63.000000) node[right] {$\scriptstyle{r}$};
\draw (6.000000, 39.000000) -- (14.000000, 51.000000);
\draw (12.000000, 48.000000) node[right] {$\scriptstyle{n}$};
\draw (6.000000, 24.000000) -- (14.000000, 36.000000);
\draw (12.000000, 33.000000) node[right] {$\scriptstyle{n}$};
\draw (6.000000, 9.000000) -- (14.000000, 21.000000);
\draw (12.000000, 18.000000) node[right] {$\scriptstyle{n}$};
\draw (6.000000, -6.000000) -- (14.000000, 6.000000);
\draw (12.000000, 3.000000) node[right] {$\scriptstyle{n+2k}$};
\begin{scope}[color=white]
\begin{scope}[color=white]
\begin{scope}
\draw[fill=none] (26.000000, 45.000000) +(-45.000000:0.000000pt) -- +(45.000000:0.000000pt) -- +(135.000000:0.000000pt) -- +(225.000000:0.000000pt) -- cycle;
\clip (26.000000, 45.000000) +(-45.000000:0.000000pt) -- +(45.000000:0.000000pt) -- +(135.000000:0.000000pt) -- +(225.000000:0.000000pt) -- cycle;
\filldraw (26.000000, 45.000000) circle(0.250000pt);
\end{scope}
\end{scope}
\end{scope}
\draw (58.000000,105.000000) -- (58.000000,0.000000);
\begin{scope}
\draw[fill=white] (58.000000, 105.000000) +(-45.000000:28.284271pt and 8.485281pt) -- +(45.000000:28.284271pt and 8.485281pt) -- +(135.000000:28.284271pt and 8.485281pt) -- +(225.000000:28.284271pt and 8.485281pt) -- cycle;
\clip (58.000000, 105.000000) +(-45.000000:28.284271pt and 8.485281pt) -- +(45.000000:28.284271pt and 8.485281pt) -- +(135.000000:28.284271pt and 8.485281pt) -- +(225.000000:28.284271pt and 8.485281pt) -- cycle;
\draw (58.000000, 105.000000) node {$\verb|mul_win|$};
\end{scope}
\begin{scope}
\draw[fill=white] (58.000000, 22.500000) +(-45.000000:28.284271pt and 40.305087pt) -- +(45.000000:28.284271pt and 40.305087pt) -- +(135.000000:28.284271pt and 40.305087pt) -- +(225.000000:28.284271pt and 40.305087pt) -- cycle;
\clip (58.000000, 22.500000) +(-45.000000:28.284271pt and 40.305087pt) -- +(45.000000:28.284271pt and 40.305087pt) -- +(135.000000:28.284271pt and 40.305087pt) -- +(225.000000:28.284271pt and 40.305087pt) -- cycle;
\draw (58.000000, 22.500000) node {$\verb|mul_win|$};
\end{scope}
\qOutput{78}{30}{black}
\qOutput{78}{15}{white}
\draw (110.000000,90.000000) -- (110.000000,0.000000);
\begin{scope}
\draw[fill=white] (110.000000, 90.000000) +(-45.000000:28.284271pt and 8.485281pt) -- +(45.000000:28.284271pt and 8.485281pt) -- +(135.000000:28.284271pt and 8.485281pt) -- +(225.000000:28.284271pt and 8.485281pt) -- cycle;
\clip (110.000000, 90.000000) +(-45.000000:28.284271pt and 8.485281pt) -- +(45.000000:28.284271pt and 8.485281pt) -- +(135.000000:28.284271pt and 8.485281pt) -- +(225.000000:28.284271pt and 8.485281pt) -- cycle;
\draw (110.000000, 90.000000) node {$\verb|mul_win|$};
\end{scope}
\begin{scope}
\draw[fill=white] (110.000000, 22.500000) +(-45.000000:28.284271pt and 40.305087pt) -- +(45.000000:28.284271pt and 40.305087pt) -- +(135.000000:28.284271pt and 40.305087pt) -- +(225.000000:28.284271pt and 40.305087pt) -- cycle;
\clip (110.000000, 22.500000) +(-45.000000:28.284271pt and 40.305087pt) -- +(45.000000:28.284271pt and 40.305087pt) -- +(135.000000:28.284271pt and 40.305087pt) -- +(225.000000:28.284271pt and 40.305087pt) -- cycle;
\draw (110.000000, 22.500000) node {$\verb|mul_win|$};
\end{scope}
\qOutput{130}{30}{black}
\qOutput{130}{15}{white}
\draw[color=black] (149.500000, 105.000000) node [fill=white] {$\cdots$};
\draw[color=black] (149.500000, 90.000000) node [fill=white] {$\cdots$};
\draw[color=black] (149.500000, 60.000000) node [fill=white] {$\cdots$};
\draw[color=black] (149.500000, 45.000000) node [fill=white] {$\cdots$};
\draw[color=black] (149.500000, 30.000000) node [fill=white] {$\cdots$};
\draw[color=black] (149.500000, 15.000000) node [fill=white] {$\cdots$};
\draw[color=black] (149.500000, 0.000000) node [fill=white] {$\cdots$};
\draw (189.000000,60.000000) -- (189.000000,0.000000);
\begin{scope}
\draw[fill=white] (189.000000, 30.000000) +(-45.000000:28.284271pt and 50.911688pt) -- +(45.000000:28.284271pt and 50.911688pt) -- +(135.000000:28.284271pt and 50.911688pt) -- +(225.000000:28.284271pt and 50.911688pt) -- cycle;
\clip (189.000000, 30.000000) +(-45.000000:28.284271pt and 50.911688pt) -- +(45.000000:28.284271pt and 50.911688pt) -- +(135.000000:28.284271pt and 50.911688pt) -- +(225.000000:28.284271pt and 50.911688pt) -- cycle;
\draw (189.000000, 30.000000) node {$\verb|mul_win|$};
\end{scope}
\qOutput{209}{30}{black}
\qOutput{209}{15}{white}
\begin{scope}
\draw[fill=white] (231.000000, 30.000000) +(-45.000000:14.142136pt and 8.485281pt) -- +(45.000000:14.142136pt and 8.485281pt) -- +(135.000000:14.142136pt and 8.485281pt) -- +(225.000000:14.142136pt and 8.485281pt) -- cycle;
\clip (231.000000, 30.000000) +(-45.000000:14.142136pt and 8.485281pt) -- +(45.000000:14.142136pt and 8.485281pt) -- +(135.000000:14.142136pt and 8.485281pt) -- +(225.000000:14.142136pt and 8.485281pt) -- cycle;
\draw (231.000000, 30.000000) node {$\verb|mod|$};
\end{scope}
\draw[color=black] (247.000000,105.000000) node[right] {$\ket{x_{(0)}}$};
\draw[color=black] (247.000000,90.000000) node[right] {$\ket{x_{(1)}}$};
\draw[color=black] (247.000000,75.000000) node[anchor=mid west] {$\vdots$};
\draw[color=black] (247.000000,60.000000) node[right] {$\ket{x_{(\lfloor n / k \rfloor)}}$};
\draw[color=black] (247.000000,45.000000) node[right] {$\ket{y}$};
\draw[color=black] (247.000000,30.000000) node[right] {$\ket{x \cdot y \bmod{p}}$};
\draw[color=black] (247.000000,15.000000) node[right] {$\ket{m}$};
\draw[color=black] (247.000000,0.000000) node[right] {$\ket{0}$};
\end{tikzpicture}
		\caption{Quantum circuit for modular multiplication using the windowed trick. The \texttt{mul\_win} block is shown in~\ref{figure:circuit-multiplication-window}, and the \texttt{mod} block is the modular reduction described in Appendix~\ref{subsubsec:reduction}. The $\blacktriangleright$ and $\triangleright$ represent different outputs of the \texttt{mul\_win} blocks. The white one $\triangleright$ usually represents the garbage output that needs to be uncomputed, corresponding to the output $\ket{m_i}$ in Fig.~\ref{figure:circuit-multiplication-window}.}\label{figure:circuit-mod-multiplication}
	\end{figure}
	
	\begin{figure}[ht]
		\captionsetup{justification=raggedright,singlelinecheck=false}
		\centering
		\input{picture/circuit/mul-window}
		\caption{Quantum circuit for a single \texttt{mul\_win} operation.\ \texttt{copy} denotes a bit-by-bit CNOT.\@ The ``$/2^k$'' block denotes a $k$-bit right shifting operation, as discussed in Appendix~\ref{subsubsec:shifting}. The block \texttt{table} is a quantum table that takes a $k$-bit number $m_i$ as the input and outputs an $(n + k)$-bit number $t_{m_i} p$, and $\texttt{table}^{-1}$ is its inverse, as discussed in Appendix~\ref{subsec:Qtable}.}\label{figure:circuit-multiplication-window}
	\end{figure}

	The parameter $k$ in the windowed trick should be optimized to reduce the Toffoli depth. We conclude that the optimized $k$ is given by $k = \log\log n + o(\log\log n)$ and put the details of the optimization in Appendix~\ref{sec:resource-cost}. Specifically, when $n=256$, we have $k=3$. As a result, the corresponding Toffoli depth is $2n\log n + 4n\frac{\log n}{\log\log n}$, Toffoli number is $16n^2 + 14\frac{n^2}{\log\log n}$, CNOT depth is $2n\log n + 12n \frac{\log n}{\log\log n}$, and CNOT number is $4n^2 \frac{\log n}{\log\log n} + \frac{37}{2}n^2$. Note that although the first term $n^2 \frac{\log n}{\log\log n}$ in the CNOT number is asymptotically larger than the Toffoli number, it is even smaller than the second one for practical cases such as $n=256$. Therefore, we can still regard the leading term as $O(n^2)$.
	
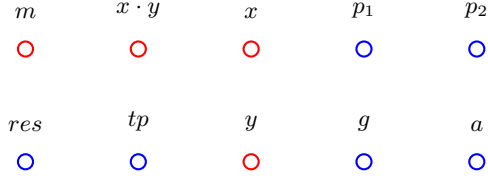
\begin{figure}[ht]
		\captionsetup{justification=raggedright,singlelinecheck=false}
		\centering
		\begin{tikzpicture}
    \draw[thick, red] (0, 1.5) circle(0.1);
    \node[above] at (0, 1.8) {$m$};
    \draw[thick, blue] (0, 0) circle(0.1);
    \node[above] at (0, 0.3) {$res$};
    \draw[thick, red] (1.5, 1.5) circle(0.1);
    \node[above] at (1.5, 1.8) {$x \cdot y$};
    \draw[thick, blue] (1.5, 0) circle(0.1);
    \node[above] at (1.5, 0.3) {$tp$};
    \draw[thick, red] (3, 1.5) circle(0.1);
    \node[above] at (3, 1.8) {$x$};
    \draw[thick, red] (3, 0) circle(0.1);
    \node[above] at (3, 0.3) {$y$};
    \draw[thick, blue] (4.5, 1.5) circle(0.1);
    \node[above] at (4.5, 1.8) {$p_1$};
    \draw[thick, blue] (4.5, 0) circle(0.1);
    \node[above] at (4.5, 0.3) {$g$};
    \draw[thick, blue] (6, 1.5) circle(0.1);
    \node[above] at (6, 1.8) {$p_2$};
    \draw[thick, blue] (6, 0) circle(0.1);
    \node[above] at (6, 0.3) {$a$};
\end{tikzpicture}
		\caption{Qubits layout for modular multiplication. $x, y$ are the input, $x\cdot y, m$ are the real and garbage outputs, respectively. $tp$ stores the number that we get by looking up the table. $p_1, p_2, g, a$ are the qubits for addition. $res$ stores the additional $5k$ qubits for the $n+k$-bit addition.}\label{figure:layout-multiplication}
	\end{figure}

	
	Finally, we show the layout of qubits used in modular multiplication in Fig.~\ref{figure:layout-multiplication}. There are totally $10n$ qubits in this figure. Note that since we need to implement an $n+k$-bit addition, there are $k$ additional qubits for $p_1, p_2, g, a, x\cdot y$, and $tp$. Furthermore, we need an additional $k$ qubits as input to the table. We store these $5k$ qubits in $res$. As long as $n \geq 10$, $n\geq 5k$, and a column of qubits is enough for them. (When $n < 10$, it is not so necessary to use this windowed trick.) Our implementation does not contain a control qubit $ctrl$, as it is not required in our overall point addition. The gate interaction distance is $4$, the distance between $\ket{res}$ and the dynamic circuit qubits.
	
\begin{figure}[ht]
		\captionsetup{justification=raggedright,singlelinecheck=false}
		\centering
		\begin{minipage}[t]{0.45\textwidth}
			\subcaption{}\label{figure:circuit-mul-full}
			\centering
			\begin{tikzpicture}[scale=1.000000,x=1pt,y=1pt]
\filldraw[color=white] (0.000000, -7.500000) rectangle (126.000000, 67.500000);
\draw[color=black] (0.000000,60.000000) -- (126.000000,60.000000);
\draw[color=black] (0.000000,60.000000) node[left] {$\ket{x}$};
\draw[color=black] (0.000000,45.000000) -- (126.000000,45.000000);
\draw[color=black] (0.000000,45.000000) node[left] {$\ket{y}$};
\draw[color=black] (0.000000,30.000000) -- (126.000000,30.000000);
\draw[color=black] (0.000000,30.000000) node[left] {$\ket{0}$};
\draw[color=black] (0.000000,15.000000) -- (126.000000,15.000000);
\draw[color=black] (0.000000,15.000000) node[left] {$\ket{0}$};
\draw[color=black] (0.000000,0.000000) -- (126.000000,0.000000);
\draw[color=black] (0.000000,0.000000) node[left] {$\ket{z}$};
\draw (21.000000,60.000000) -- (21.000000,15.000000);
\begin{scope}
\draw[fill=white] (21.000000, 37.500000) +(-45.000000:21.213203pt and 40.305087pt) -- +(45.000000:21.213203pt and 40.305087pt) -- +(135.000000:21.213203pt and 40.305087pt) -- +(225.000000:21.213203pt and 40.305087pt) -- cycle;
\clip (21.000000, 37.500000) +(-45.000000:21.213203pt and 40.305087pt) -- +(45.000000:21.213203pt and 40.305087pt) -- +(135.000000:21.213203pt and 40.305087pt) -- +(225.000000:21.213203pt and 40.305087pt) -- cycle;
\draw (21.000000, 37.500000) node {\texttt{mul}};
\end{scope}
\qOutput{36}{15}{black}
\qOutput{36}{30}{white}
\draw (63.000000,15.000000) -- (63.000000,0.000000);
\begin{scope}
\draw[fill=white] (63.000000, 7.500000) +(-45.000000:21.213203pt and 19.091883pt) -- +(45.000000:21.213203pt and 19.091883pt) -- +(135.000000:21.213203pt and 19.091883pt) -- +(225.000000:21.213203pt and 19.091883pt) -- cycle;
\clip (63.000000, 7.500000) +(-45.000000:21.213203pt and 19.091883pt) -- +(45.000000:21.213203pt and 19.091883pt) -- +(135.000000:21.213203pt and 19.091883pt) -- +(225.000000:21.213203pt and 19.091883pt) -- cycle;
\draw (63.000000, 7.500000) node {\texttt{copy}};
\end{scope}
\qOutput{78}{0}{black}
\draw (105.000000,60.000000) -- (105.000000,15.000000);
\begin{scope}
\draw[fill=white] (105.000000, 37.500000) +(-45.000000:21.213203pt and 40.305087pt) -- +(45.000000:21.213203pt and 40.305087pt) -- +(135.000000:21.213203pt and 40.305087pt) -- +(225.000000:21.213203pt and 40.305087pt) -- cycle;
\clip (105.000000, 37.500000) +(-45.000000:21.213203pt and 40.305087pt) -- +(45.000000:21.213203pt and 40.305087pt) -- +(135.000000:21.213203pt and 40.305087pt) -- +(225.000000:21.213203pt and 40.305087pt) -- cycle;
\draw (105.000000, 37.500000) node {$\texttt{mul}^{-1}$};
\end{scope}
\qOutput{120}{15}{black}
\qOutput{120}{30}{white}
\draw[color=black] (126.000000,60.000000) node[right] {$\ket{x}$};
\draw[color=black] (126.000000,45.000000) node[right] {$\ket{y}$};
\draw[color=black] (126.000000,30.000000) node[right] {$\ket{0}$};
\draw[color=black] (126.000000,15.000000) node[right] {$\ket{0}$};
\draw[color=black] (126.000000,0.000000) node[right] {$\ket{z\oplus(xy\bmod{p})}$};
\end{tikzpicture}
		\end{minipage}
		\qquad
		\begin{minipage}[t]{0.45\textwidth}
			\subcaption{}\label{figure:circuit-mul-add}
			\centering
			\begin{tikzpicture}[scale=1.000000,x=1pt,y=1pt]
\filldraw[color=white] (0.000000, -7.500000) rectangle (108.000000, 67.500000);
\draw[color=black] (0.000000,60.000000) -- (108.000000,60.000000);
\draw[color=black] (0.000000,60.000000) node[left] {$\ket{x}$};
\draw[color=black] (0.000000,45.000000) -- (108.000000,45.000000);
\draw[color=black] (0.000000,45.000000) node[left] {$\ket{y}$};
\draw[color=black] (0.000000,30.000000) -- (108.000000,30.000000);
\draw[color=black] (0.000000,30.000000) node[left] {$\ket{0}$};
\draw[color=black] (0.000000,15.000000) -- (108.000000,15.000000);
\draw[color=black] (0.000000,15.000000) node[left] {$\ket{0}$};
\draw[color=black] (0.000000,0.000000) -- (108.000000,0.000000);
\draw[color=black] (0.000000,0.000000) node[left] {$\ket{z}$};
\draw (21.000000,60.000000) -- (21.000000,15.000000);
\begin{scope}
\draw[fill=white] (21.000000, 37.500000) +(-45.000000:21.213203pt and 40.305087pt) -- +(45.000000:21.213203pt and 40.305087pt) -- +(135.000000:21.213203pt and 40.305087pt) -- +(225.000000:21.213203pt and 40.305087pt) -- cycle;
\clip (21.000000, 37.500000) +(-45.000000:21.213203pt and 40.305087pt) -- +(45.000000:21.213203pt and 40.305087pt) -- +(135.000000:21.213203pt and 40.305087pt) -- +(225.000000:21.213203pt and 40.305087pt) -- cycle;
\draw (21.000000, 37.500000) node {\texttt{mul}};
\end{scope}
\qOutput{36}{15}{black}
\qOutput{36}{30}{white}
\draw (54.000000,15.000000) -- (54.000000,0.000000);
\begin{scope}
\draw[fill=white] (54.000000, 7.500000) +(-45.000000:8.485281pt and 19.091883pt) -- +(45.000000:8.485281pt and 19.091883pt) -- +(135.000000:8.485281pt and 19.091883pt) -- +(225.000000:8.485281pt and 19.091883pt) -- cycle;
\clip (54.000000, 7.500000) +(-45.000000:8.485281pt and 19.091883pt) -- +(45.000000:8.485281pt and 19.091883pt) -- +(135.000000:8.485281pt and 19.091883pt) -- +(225.000000:8.485281pt and 19.091883pt) -- cycle;
\draw (54.000000, 7.500000) node {$+$};
\end{scope}
\qOutput{60}{0}{black}
\draw (87.000000,60.000000) -- (87.000000,15.000000);
\begin{scope}
\draw[fill=white] (87.000000, 37.500000) +(-45.000000:21.213203pt and 40.305087pt) -- +(45.000000:21.213203pt and 40.305087pt) -- +(135.000000:21.213203pt and 40.305087pt) -- +(225.000000:21.213203pt and 40.305087pt) -- cycle;
\clip (87.000000, 37.500000) +(-45.000000:21.213203pt and 40.305087pt) -- +(45.000000:21.213203pt and 40.305087pt) -- +(135.000000:21.213203pt and 40.305087pt) -- +(225.000000:21.213203pt and 40.305087pt) -- cycle;
\draw (87.000000, 37.500000) node {$\texttt{mul}^{-1}$};
\end{scope}
\qOutput{102}{15}{black}
\qOutput{102}{30}{white}
\draw[color=black] (108.000000,60.000000) node[right] {$\ket{x}$};
\draw[color=black] (108.000000,45.000000) node[right] {$\ket{y}$};
\draw[color=black] (108.000000,30.000000) node[right] {$\ket{0}$};
\draw[color=black] (108.000000,15.000000) node[right] {$\ket{0}$};
\draw[color=black] (108.000000,0.000000) node[right] {$\ket{(xy+z) \bmod{p}}$};
\end{tikzpicture}
		\end{minipage}
		\caption{The quantum circuits for (a) $\texttt{mul}_F$ and (b) $\texttt{mul}^+$.\ \texttt{copy} denotes a bit-by-bit CNOT, and $\texttt{mul}^{-1}$ is the reverse operation of $\texttt{mul}$.}\label{figure:circuit-mul}
	\end{figure}
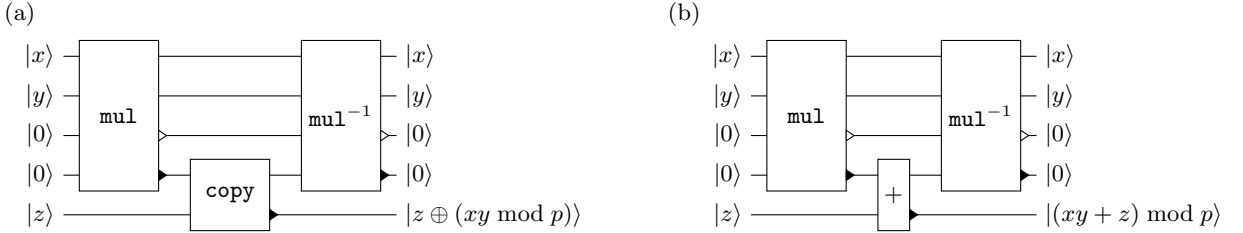

	It is worth noting that the above circuit, denoted as \texttt{mul}, produces a column of garbage qubits $m$, which must be cleaned up after the computation. This can be achieved by performing a \texttt{copy} operation followed by the inverse multiplication circuit $\texttt{mul}^{-1}$. We will denote the whole procedure as $\texttt{mul}_F$, where the subscript $F$ stands for ``full'', as illustrated in Fig.~\ref{figure:circuit-mul-full}. In our construction of elliptic-curve point addition, the multiply-and-add operation $\ket{x}\ket{y}\ket{z} \mapsto \ket{x}\ket{y}\ket{(z + xy) \bmod{p}}$ is also frequently used. This operation can be realized through a sequence of $\texttt{mul}$, $+$, and $\texttt{mul}^{-1}$ operations, and we denote it as $\texttt{mul}^+$, as shown in Fig.~\ref{figure:circuit-mul-add}. Finally, modular squaring $\ket{x}\ket{0} \mapsto \ket{x}\ket{(x^2) \bmod{p}}$ is implemented in an analogous manner to modular multiplication, as discussed in Appendix~\ref{subsubsec:squaring}, while the square-and-subtract operation $\texttt{squ}^-$ is constructed in parallel to $\texttt{mul}^+$.

	\subsection{Modular Division}\label{subsec:division}
	
	The modular division implements $\ket{x}\ket{y}\ket{0}\mapsto\ket{x}\ket{y}\ket{x^{-1}\cdot y \bmod{p}}$. The division can be implemented by computing the inversion of $y$ and the multiplying it with $x$. Thus, we will discuss the modular inversion first.

\begin{figure}[ht]
		\captionsetup{justification=raggedright,singlelinecheck=false}
		\centering
		\input{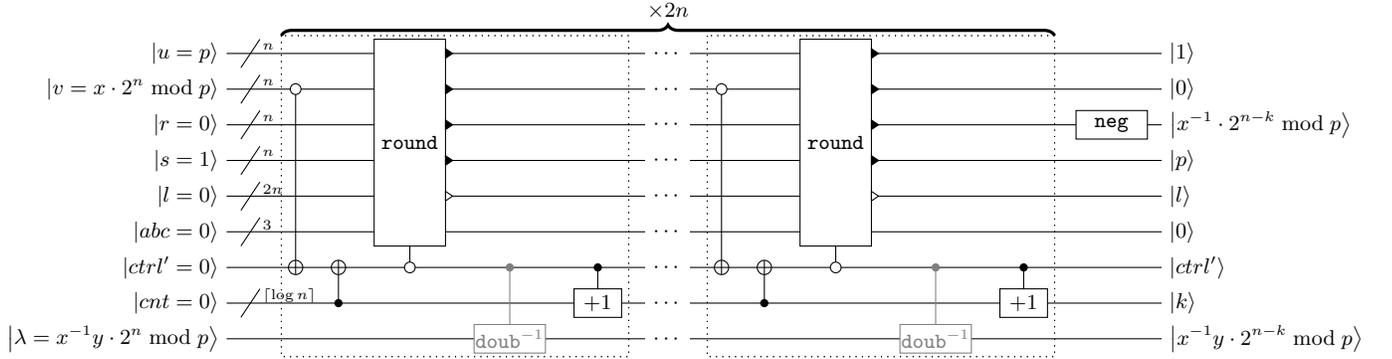}
		\caption{Quantum circuit for modular inversion. CNOT with multiple control qubits is the AND operation mentioned in Appendix~\ref{subsec:OR-AND}. The $+1$ block is a controlled constant addition of $1$.\ \texttt{round} is shown in Fig.~\ref{figure:circuit-inversion-round}. $\texttt{doub}^{-1}$ is the controlled doubling operation used to convert the result from a ``pseudo-inverse'' to a ``real-inverse''. $\ket{\lambda}$ is the column of qubit that stores the final result of the modular division, as shown in Fig.~\ref{figure:circuit-division}. This $\texttt{doub}^{-1}$ only occurs at the inverse of modular inversion, so it is denoted in gray here. Also, it functions as the modular doubling in the inverse circuit, so here we use a superscript of $-1$ to denote it.\ \texttt{neg} is modular negation $\ket{x} \mapsto \ket{p-x}$, as discussed in Appendix~\ref{subsubsec:reduction}.}\label{figure:circuit-inversion}
	\end{figure}

\begin{figure}[ht]
		\captionsetup{justification=raggedright,singlelinecheck=false}
		\centering
		\input{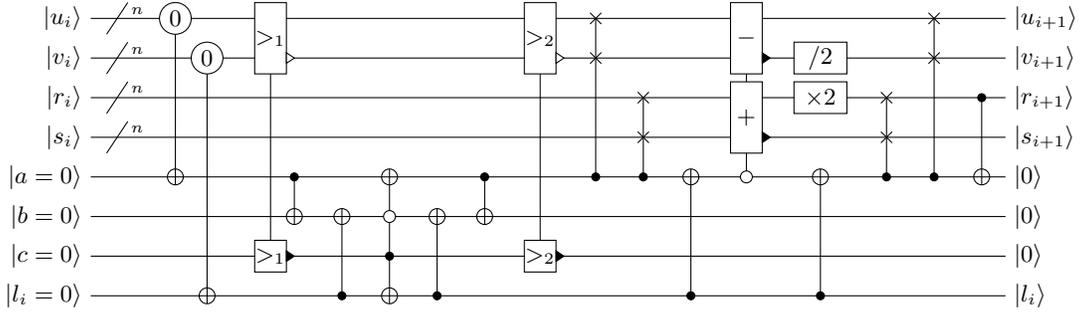}
		\caption{Quantum circuit for the \texttt{round} operation. The first two gates are CNOT gates, where the number ``$0$'' in the white circle means that they use the lowest bit of $u$ or $v$, which denotes the parity, as the control bit, conditional on the lowest bit being $\ket{0}$. $\times 2, /2$ are simply $1$-bit shifts given in Appendix~\ref{subsubsec:shifting}. $>_1, >_2$ are the half comparators, as discussed in Appendix~\ref{subsubsec:comparator}.}\label{figure:circuit-inversion-round}
	\end{figure}

	To ensure consistency with modular multiplication, modular inversion should also be performed in the Montgomery representation. Accordingly, the input and output of the inversion circuit are defined as $x' = (x \cdot 2^n) \bmod p$ and $\qty(x^{-1})' = \qty(x^{-1} \cdot 2^n) \bmod p$, respectively. The classical Montgomery inversion algorithm~\cite{kaliski1995montgomery} is provided in Algorithm~\ref{algorithm:inversion}. This algorithm is founded on the Euclidean algorithm, thereby circumventing the logically intricate trial division method, and exclusively employs addition, subtraction, and bit-shift operations, which are well-suited for quantum circuits. The algorithm we present exhibits a subtle distinction from the original algorithm in~\cite{kaliski1995montgomery}: we impose that the loop iterates exactly $2n$ times, whereas the original algorithm utilizes a while loop conditioned on $v>0$. It can be demonstrated that the number of loop iterations $k$ is bounded by $2n$. To ensure consistency in the quantum algorithm, it is imperative to execute the loop for the maximum possible number of iterations. For an effective iteration count $k$, the Montgomery inversion algorithm ultimately yields a ``pseudo-inverse'' $\qty(x^{-1} \cdot 2^{n-k}) \bmod p$, which necessitates $k$ doubling operations relative to the correct result $\qty(x^{-1} \cdot 2^n) \bmod p$. This is achieved through a sequence of modular doubling in parallel with the implementation of the Montgomery inversion algorithm. Consequently, $k$ must be stored in a register, and the result must be adjusted accordingly based on $k$. The quantum implementation~\cite{haner2020improved, roetteler2017quantum} is illustrated in Fig.~\ref{figure:circuit-inversion} and Fig.~\ref{figure:circuit-inversion-round}.

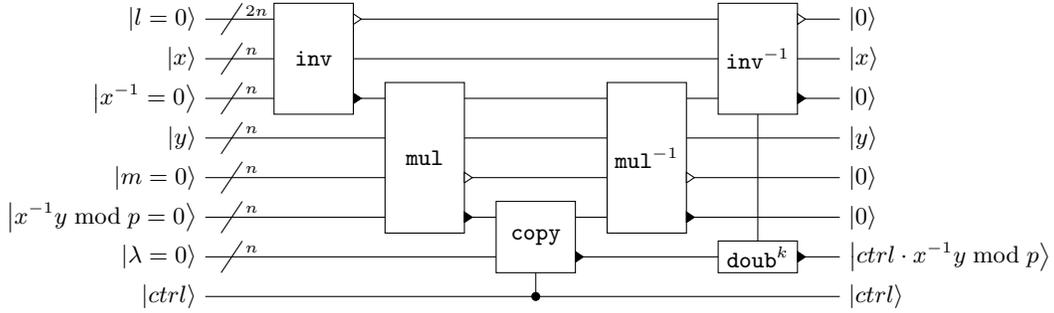
\begin{figure}[ht]
		\captionsetup{justification=raggedright,singlelinecheck=false}
		\centering
		\begin{tikzpicture}[scale=1.000000,x=1pt,y=1pt]
\filldraw[color=white] (0.000000, -7.500000) rectangle (240.000000, 112.500000);
\draw[color=black] (0.000000,105.000000) -- (240.000000,105.000000);
\draw[color=black] (0.000000,105.000000) node[left] {$\ket{l=0}$};
\draw[color=black] (0.000000,90.000000) -- (240.000000,90.000000);
\draw[color=black] (0.000000,90.000000) node[left] {$\ket{x}$};
\draw[color=black] (0.000000,75.000000) -- (240.000000,75.000000);
\draw[color=black] (0.000000,75.000000) node[left] {$\ket{x^{-1}=0}$};
\draw[color=black] (0.000000,60.000000) -- (240.000000,60.000000);
\draw[color=black] (0.000000,60.000000) node[left] {$\ket{y}$};
\draw[color=black] (0.000000,45.000000) -- (240.000000,45.000000);
\draw[color=black] (0.000000,45.000000) node[left] {$\ket{m=0}$};
\draw[color=black] (0.000000,30.000000) -- (240.000000,30.000000);
\draw[color=black] (0.000000,30.000000) node[left] {$\ket{x^{-1}y \bmod{p}=0}$};
\draw[color=black] (0.000000,15.000000) -- (240.000000,15.000000);
\draw[color=black] (0.000000,15.000000) node[left] {$\ket{\lambda=0}$};
\draw[color=black] (0.000000,0.000000) -- (240.000000,0.000000);
\draw[color=black] (0.000000,0.000000) node[left] {$\ket{ctrl}$};
\draw (6.000000, 84.000000) -- (14.000000, 96.000000);
\draw (12.000000, 93.000000) node[right] {$\scriptstyle{n}$};
\draw (6.000000, 69.000000) -- (14.000000, 81.000000);
\draw (12.000000, 78.000000) node[right] {$\scriptstyle{n}$};
\draw (6.000000, 54.000000) -- (14.000000, 66.000000);
\draw (12.000000, 63.000000) node[right] {$\scriptstyle{n}$};
\draw (6.000000, 9.000000) -- (14.000000, 21.000000);
\draw (12.000000, 18.000000) node[right] {$\scriptstyle{n}$};
\draw (6.000000, 24.000000) -- (14.000000, 36.000000);
\draw (12.000000, 33.000000) node[right] {$\scriptstyle{n}$};
\draw (6.000000, 39.000000) -- (14.000000, 51.000000);
\draw (12.000000, 48.000000) node[right] {$\scriptstyle{n}$};
\draw (6.000000, 99.000000) -- (14.000000, 111.000000);
\draw (12.000000, 108.000000) node[right] {$\scriptstyle{2n}$};
\draw (41.000000,105.000000) -- (41.000000,75.000000);
\begin{scope}
\draw[fill=white] (41.000000, 90.000000) +(-45.000000:21.213203pt and 29.698485pt) -- +(45.000000:21.213203pt and 29.698485pt) -- +(135.000000:21.213203pt and 29.698485pt) -- +(225.000000:21.213203pt and 29.698485pt) -- cycle;
\clip (41.000000, 90.000000) +(-45.000000:21.213203pt and 29.698485pt) -- +(45.000000:21.213203pt and 29.698485pt) -- +(135.000000:21.213203pt and 29.698485pt) -- +(225.000000:21.213203pt and 29.698485pt) -- cycle;
\draw (41.000000, 90.000000) node {$\verb|inv|$};
\end{scope}
\qOutput{56}{75}{black}
\qOutput{56}{105}{white}
\draw (83.000000,75.000000) -- (83.000000,30.000000);
\begin{scope}
\draw[fill=white] (83.000000, 52.500000) +(-45.000000:21.213203pt and 40.305087pt) -- +(45.000000:21.213203pt and 40.305087pt) -- +(135.000000:21.213203pt and 40.305087pt) -- +(225.000000:21.213203pt and 40.305087pt) -- cycle;
\clip (83.000000, 52.500000) +(-45.000000:21.213203pt and 40.305087pt) -- +(45.000000:21.213203pt and 40.305087pt) -- +(135.000000:21.213203pt and 40.305087pt) -- +(225.000000:21.213203pt and 40.305087pt) -- cycle;
\draw (83.000000, 52.500000) node {$\verb|mul|$};
\end{scope}
\qOutput{98}{30}{black}
\qOutput{98}{45}{white}
\draw (125.000000,30.000000) -- (125.000000,0.000000);
\begin{scope}
\draw[fill=white] (125.000000, 22.500000) +(-45.000000:21.213203pt and 19.091883pt) -- +(45.000000:21.213203pt and 19.091883pt) -- +(135.000000:21.213203pt and 19.091883pt) -- +(225.000000:21.213203pt and 19.091883pt) -- cycle;
\clip (125.000000, 22.500000) +(-45.000000:21.213203pt and 19.091883pt) -- +(45.000000:21.213203pt and 19.091883pt) -- +(135.000000:21.213203pt and 19.091883pt) -- +(225.000000:21.213203pt and 19.091883pt) -- cycle;
\draw (125.000000, 22.500000) node {$\verb|copy|$};
\end{scope}
\filldraw (125.000000, 0.000000) circle(1.500000pt);
\qOutput{140}{15}{black}
\draw (167.000000,75.000000) -- (167.000000,30.000000);
\begin{scope}
\draw[fill=white] (167.000000, 52.500000) +(-45.000000:21.213203pt and 40.305087pt) -- +(45.000000:21.213203pt and 40.305087pt) -- +(135.000000:21.213203pt and 40.305087pt) -- +(225.000000:21.213203pt and 40.305087pt) -- cycle;
\clip (167.000000, 52.500000) +(-45.000000:21.213203pt and 40.305087pt) -- +(45.000000:21.213203pt and 40.305087pt) -- +(135.000000:21.213203pt and 40.305087pt) -- +(225.000000:21.213203pt and 40.305087pt) -- cycle;
\draw (167.000000, 52.500000) node {$\verb|mul|^{-1}$};
\end{scope}
\qOutput{182}{30}{black}
\qOutput{182}{45}{white}
\draw (209.000000,105.000000) -- (209.000000,15.000000);
\begin{scope}
\draw[fill=white] (209.000000, 90.000000) +(-45.000000:21.213203pt and 29.698485pt) -- +(45.000000:21.213203pt and 29.698485pt) -- +(135.000000:21.213203pt and 29.698485pt) -- +(225.000000:21.213203pt and 29.698485pt) -- cycle;
\clip (209.000000, 90.000000) +(-45.000000:21.213203pt and 29.698485pt) -- +(45.000000:21.213203pt and 29.698485pt) -- +(135.000000:21.213203pt and 29.698485pt) -- +(225.000000:21.213203pt and 29.698485pt) -- cycle;
\draw (209.000000, 90.000000) node {$\verb|inv|^{-1}$};
\end{scope}
\begin{scope}
\draw[fill=white] (209.000000, 15.000000) +(-45.000000:21.213203pt and 8.485281pt) -- +(45.000000:21.213203pt and 8.485281pt) -- +(135.000000:21.213203pt and 8.485281pt) -- +(225.000000:21.213203pt and 8.485281pt) -- cycle;
\clip (209.000000, 15.000000) +(-45.000000:21.213203pt and 8.485281pt) -- +(45.000000:21.213203pt and 8.485281pt) -- +(135.000000:21.213203pt and 8.485281pt) -- +(225.000000:21.213203pt and 8.485281pt) -- cycle;
\draw (209.000000, 15.000000) node {$\verb|doub|^k$};
\end{scope}
\qOutput{224}{75}{black}
\qOutput{224}{105}{white}
\qOutput{224}{15}{black}
\draw[color=black] (240.000000,105.000000) node[right] {$\ket{0}$};
\draw[color=black] (240.000000,90.000000) node[right] {$\ket{x}$};
\draw[color=black] (240.000000,75.000000) node[right] {$\ket{0}$};
\draw[color=black] (240.000000,60.000000) node[right] {$\ket{y}$};
\draw[color=black] (240.000000,45.000000) node[right] {$\ket{0}$};
\draw[color=black] (240.000000,30.000000) node[right] {$\ket{0}$};
\draw[color=black] (240.000000,15.000000) node[right] {$\ket{ctrl \cdot x^{-1}y \bmod{p}}$};
\draw[color=black] (240.000000,0.000000) node[right] {$\ket{ctrl}$};
\end{tikzpicture}
		\caption{Quantum circuit for modular division.\ \texttt{inv} is the modular inversion operation and $\texttt{inv}^{-1}$ is its inverse.\ \texttt{mul} is the modular multiplication operation and $\texttt{mul}^{-1}$ is its inverse.\ \texttt{copy} is a bit-to-bit Toffoli controlled by $\ket{ctrl}$.\ \texttt{doub} is the modular doubling given in Appendix~\ref{subsubsec:doubling}. It is a sequence of $2n$ modular doublings controlled by a qubit in $\texttt{inv}^{-1}$, as shown in Fig.~\ref{figure:circuit-inversion}. Only $k$ of them will be operated, so we denote them as $\texttt{doub}^k$. Here, the ancillary qubits used for the building block operations are neglected.}\label{figure:circuit-division}
	\end{figure}

	This circuit takes $\ket{x}$ in Montgomery representation as input and produces a ``pseudo-inverse'' state $\ket{\qty(x^{-1} \cdot 2^{n-k}) \bmod p}$ as output, together with ancillary registers: a $2n$-qubit garbage register $\ket{l}$, a single-qubit flag $\ket{ctrl'}$, and a $(\lceil \log n \rceil + 1)$-qubit counter storing the value of $k$. In each iteration, the \texttt{round} operation is responsible for handling $u, v, r, s$, while the remaining operations are responsible for handling $k$. The missing $k$ doublings are later corrected in the inverse operation of the inversion procedure, conditioned on the state of $\ket{ctrl'}$, corresponding to the $\texttt{doub}^k$ operation shown in Fig.~\ref{figure:circuit-division}.
	
	\begin{algorithm}[ht]
		\SetKwInOut{Input}{Input}
		\SetKwInOut{Output}{Output}
		
		\Input{$p,x',0,1$}
		\Output{$1,0,\qty(x^{-1})',p,k$}
		\BlankLine{}
		$u\gets p, v\gets(x\cdot 2^n) \bmod{p}, r\gets0, s\gets1, k\gets0$\;
		\For{$i \gets 1 \cdots 2n$}{
			\If{$v>0$}{
				\If{$u$ is even}{$u\gets\dfrac{u}{2}, s\gets2s$\;}
				\ElseIf{$v$ is even}{$v\gets\dfrac{v}{2}, r\gets2r$\;}
				\ElseIf{$u>v$}{$u\gets\dfrac{u-v}{2},r\gets r+s,s\gets2s$\;}
				\Else{$v\gets\dfrac{v-u}{2},s\gets r+s,r\gets2r$\;}
			}
			\Else{
				$r\gets(2r)\bmod{p}$\;
				$k\gets k+1$\;
			}
		}
		$r \gets p-r$\;
		\KwRet{$u,v,r,s,k$}
		\caption{classical algorithm for computing Montgomery inversion~\cite{haner2020improved,roetteler2017quantum}}\label{algorithm:inversion}
	\end{algorithm}

	A controlled \texttt{round} costs $(3\log n + 2\log\log n)$ Toffoli depth, $45n$ Toffoli gates, $(3\log n + 2\log\log n)$ CNOT depth, and $56n$ CNOT gates. An inversion costs $(8\log n + 10\log\log n)$ Toffoli depth, $88n^2$ Toffoli gates, $(6n\log n + 8n\log\log n)$ CNOT depth, and $112n^2$ CNOT gates. Detailed counting is given in Appendix~\ref{sec:resource-cost}.

	As mentioned before, modular division consists of modular inversion and modular multiplication, including two modular multiplications, one modular inversion, one modular inversion plus modular doubling, and one copy. The whole circuit is shown in Fig.~\ref{figure:circuit-division}. The circuit design is straightforward. We first perform one inversion and one multiplication to calculate the result. Then, we copy the result to a clean register and undo the computation. Recall that we should use doubling operations to correct the result. The whole division circuit has $(22n\log n + 8n\frac{\log n}{\log\log n})$ Toffoli depth, $\qty(238n^2 + 28\frac{n^2}{\log\log n})$ Toffoli gates, $(18n\log n + 24n\frac{\log n}{\log\log n})$ CNOT depth, and $(8n^2 \frac{\log n}{\log\log n} + 309n^2)$ CNOT gates.
	
\begin{figure}[ht]
		\captionsetup{justification=raggedright,singlelinecheck=false}
		\begin{subfigure}[t]{\textwidth}
			\caption{}\label{figure:layout-inversion}
			\centering
			\begin{tikzpicture}
    \draw[thick, blue] (0, 1.5) circle(0.1);
    \node[above] at (0, 1.8) {$p_2$};
    \draw[thick, blue] (0, 0) circle(0.1);
    \node[above] at (0, 0.3) {$a_1$};
    \draw[thick, blue] (1.5, 1.5) circle(0.1);
    \node[above] at (1.5, 1.8) {$p_1$};
    \draw[thick, blue] (1.5, 0) circle(0.1);
    \node[above] at (1.5, 0.3) {$g_1$};
    \draw[thick, blue] (3, 1.5) circle(0.1);
    \node[above] at (3, 1.8) {$u$};
    \draw[thick, red] (3, 0) circle(0.1);
    \node[above] at (3, 0.3) {$x(v)$};
    \draw[thick, red] (4.5, 1.5) circle(0.1);
    \node[above] at (4.5, 1.8) {$l_1$};
    \draw[thick, red] (4.5, 0) circle(0.1);
    \node[above] at (4.5, 0.3) {$l_2$};
    \draw[thick, blue] (6, 1.5) circle(0.1);
    \node[above] at (6, 1.8) {$d$};
    \draw[thick, black] (6, 0) circle(0.1);
    \node[above] at (6, 0.3) {$\lambda$};
    \draw[thick, red] (7.5, 1.5) circle(0.1);
    \node[above] at (7.5, 1.8) {$x^{-1}(r)$};
    \draw[thick, black] (7.5, 0) circle(0.1);
    \node[above] at (7.5, 0.3) {$y$};
    \draw[thick, blue] (9, 1.5) circle(0.1);
    \node[above] at (9, 1.8) {$p_3$};
    \draw[thick, blue] (9, 0) circle(0.1);
    \node[above] at (9, 0.3) {$s$};
    \draw[thick, blue] (10.5, 1.5) circle(0.1);
    \node[above] at (10.5, 1.8) {$a_2$};
    \draw[thick, blue] (10.5, 0) circle(0.1);
    \node[above] at (10.5, 0.3) {$g_2$};
    \draw[thick, blue] (12, 1.5) circle(0.1);
    \node[above] at (12, 1.8) {$a_4$};
\end{tikzpicture}
		\end{subfigure}
		\begin{subfigure}[t]{\textwidth}
			\caption{}\label{figure:layout-div-multiplication}
			\centering
			\begin{tikzpicture}

    \draw[thick, black] (1.5, 1.5) circle(0.1);
    \draw[thick, black] (1.5, 0) circle(0.1);
    \draw[thick, black] (3, 1.5) circle(0.1);
    \node[above] at (3, 1.8) {$l_1$};
    \draw[thick, black] (3, 0) circle(0.1);
    \node[above] at (3, 0.3) {$x$};
    \draw[thick, black] (4.5, 1.5) circle(0.1);
    \node[above] at (4.5, 1.8) {$\lambda$};
    \draw[thick, black] (4.5, 0) circle(0.1);
    \node[above] at (4.5, 0.3) {$l_2$};
    \draw[thick, red] (6, 1.5) circle(0.1);
    \node[above] at (6, 1.8) {$m$};
    \draw[thick, blue] (6, 0) circle(0.1);
    \node[above] at (6, 0.3) {$res$};
    \draw[thick, red] (7.5, 1.5) circle(0.1);
    \node[above] at (7.5, 1.8) {$x^{-1}y$};
    \draw[thick, blue] (7.5, 0) circle(0.1);
    \node[above] at (7.5, 0.3) {$tp$};
    \draw[thick, red] (9, 1.5) circle(0.1);
    \node[above] at (9, 1.8) {$x^{-1}$};
    \draw[thick, red] (9, 0) circle(0.1);
    \node[above] at (9, 0.3) {$y$};
    \draw[thick, blue] (10.5, 1.5) circle(0.1);
    \node[above] at (10.5, 1.8) {$p_1$};
    \draw[thick, blue] (10.5, 0) circle(0.1);
    \node[above] at (10.5, 0.3) {$g$};
    \draw[thick, blue] (12, 1.5) circle(0.1);
    \node[above] at (12, 1.8) {$p_2$};
    \draw[thick, blue] (12, 0) circle(0.1);
    \node[above] at (12, 0.3) {$a$};
    \draw[thick, black] (13.5, 1.5) circle(0.1);
\end{tikzpicture}
		\end{subfigure}
		\caption{Qubits layout for modular division.\ (a) The inversion part. It uses two sets of qubits $a_1, p_1, p_2, g_1$ and $a_2, p_3, p_4, g_2$ to implement addition simultaneously. $u, v, r, s$ are four registers in Algorithm~\ref{algorithm:inversion}. $l_1, l_2$ are the output of $2n$ garbage qubits. $d$ is the column that stores $a, b, c, ctrl'$ and the counter $cnt$ for $k$ in the algorithm.\ (b) The multiplication part. It has the same components as the normal multiplication shown in Fig.~\ref{figure:layout-multiplication}, but with different qubit locations. To achieve a lower gate interaction distance, we change the location of $l_1$ and $l_2$ by swapping.}\label{figure:layout-division}
	\end{figure}
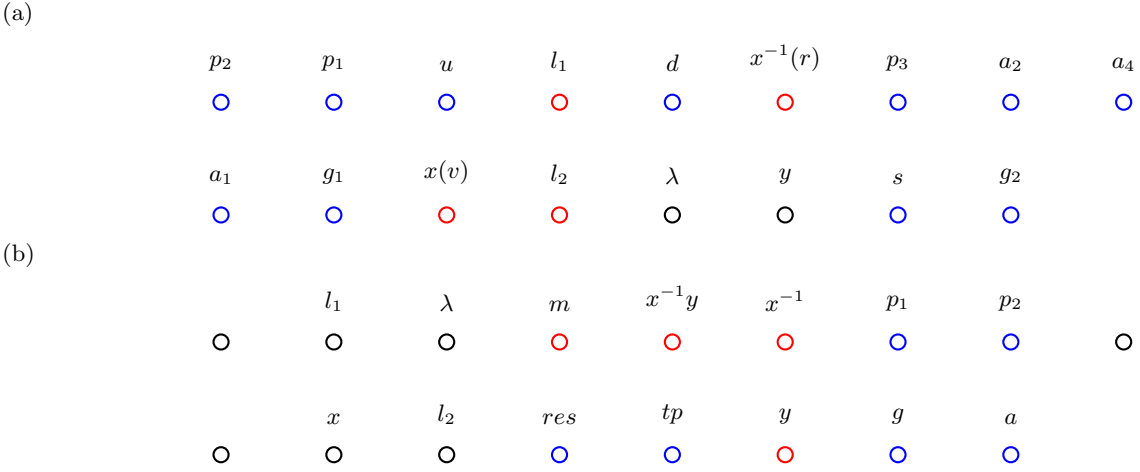

	For the space cost, our qubit layout of this operation consists of $17n$ qubits. It contains two parts, the inversion part and the multiplication part, as shown in Fig.~\ref{figure:layout-division}. The gate interaction distance is still $4$, the distance between $\ket{res}$ and the dynamic circuit qubits in the multiplication part.

	In addition, if we do not allow the parallelization of two additions, we only need $13n$ qubits. As a tradoff, the Toffoli depth will increase to $36n\log n + 8n\frac{\log n}{\log\log n}$.

	\section{Detailed Resource Cost Analysis}\label{sec:resource-cost}
	In this section, we provide a detailed analysis of the resource costs of the circuit, including the Toffoli depth, Toffoli count, CNOT depth, and CNOT count. Since many Toffoli gates in our construction are long-range and are implemented using dynamic circuits, we also evaluate the total qubit number and the corresponding gate interaction distance.

	\subsection{Modular Addition}\label{subsec:addition-cost}
	The modular addition, as shown in Fig.~\ref{figure:circuit-mod-addition}, consists of one controlled addition, one controlled constant addition, one controlled comparator, and one Pauli $X$ gate. The resource cost of all the operations above can be found in Appendix~\ref{subsec:variants-addition}. By combining them we can get the resource cost of one controlled modular addition:
	\begin{equation}
		\begin{aligned}
			\text{Toffoli depth} &\simeq 2\times(2\log n + 2\log\log n) + \log n \simeq 5\log n + 4\log\log n,\\
			\text{Toffoli number} &\simeq 16n + 13n + 4n \simeq 33n,\\
			\text{CNOT depth} &\simeq 2\times(2\log n + 2\log\log n) + \log n \simeq 5\log n + 4\log\log n,\\
			\text{CNOT number} &\simeq \frac{37}{2}n + \frac{39}{2}n + 4n \simeq 42n.
		\end{aligned}
	\end{equation}
	
	\subsection{Modular Multiplication}\label{subsec:mul-cost}
	The specific construction of modular multiplication circuit is already discussed in~\ref{subsec:multiplication}. By summing up the cost of each part, the total resource cost of one modular multiplication is
	\begin{equation}\label{equ:mod-mul-cost}
		\begin{aligned}
			\text{Toffoli depth} &\simeq \left\lfloor\frac{n}{k}\right\rfloor \qty(k \cdot (2\log n + 2\log\log n) + (2\log (n+k) + 2\log\log (n+k)) + 0 + 2 \times (2^k-1)) \\
			&+ (r \cdot (2\log n + 2\log\log n) + (2\log (n+r) + 2\log\log (n+r)) + 0 + 2^r) + (2\log n + 2\log\log n), \\
			\text{Toffoli number} &\simeq \left\lfloor\frac{n}{k}\right\rfloor \qty(k \cdot 16 n + 14(n+k) + 0 + 2 \times (2^k-1)) + (r\cdot 16n + 14(n+r) + 0 + 2^r) + 12n, \\
			\text{CNOT depth} &\simeq \left\lfloor\frac{n}{k}\right\rfloor \qty(k \cdot (2\log n + 2\log\log n) + (2\log (n+k) + 2\log\log (n+k)) + 1 + 2 \times 2^k \times 5) \\
			&+ (r \cdot (2\log n + 2\log\log n) + (2\log (n+r) + 2\log\log (n+r)) + 1 + 3(r+2) + 2^r \times 5) \\
			&+ (2\log n + 2\log\log n), \\
			\text{CNOT number} &\simeq \left\lfloor\frac{n}{k}\right\rfloor \qty(k \cdot \frac{37}{2}n + 17(n+k) + k + 2 \times 2^k \times (2(n+k)+1)) \\
			&+ \qty(r \cdot \frac{37}{2}n + 17(n+r) + r + 3n(r+1) + 2^r \times (2(n+r)+1)) + \frac{39}{2}n, \\
			\text{SWAP depth} &\simeq \left\lfloor\frac{n}{k}\right\rfloor (k+2) + (r+2), \\
			\text{SWAP number} &\simeq \left\lfloor\frac{n}{k}\right\rfloor n(k+1) + n(r+1),
		\end{aligned}
	\end{equation}
	where $k$ is the window size, which needs to be optimized, and $r = n - k \left\lfloor \frac{n}{k} \right\rfloor$ is the number of bits in the last group.
	
	In any of the above equations, the first term represents the cost associated with operations on the $k$-bit numbers, the second term corresponds to operations on the $r$-bit number, and the third term accounts for the modular reduction \texttt{mod} step. We now focus on the terms within the parentheses of the first term, as the interpretation for the second term follows analogously. Specifically, the first component in the parentheses corresponds to the $k$ controlled additions, the second to a single standard addition, the third to the \texttt{copy} operation, and the last to the two quantum table lookups. The SWAP gates originate entirely from the $k$-bit shift operations. The origin of these quantities are provided in Appendices~\ref{subsec:addition-cost} and~\ref{sec:other-circuit}.
	
	To optimize the window size $k$ and achieve the lowest Toffoli depth, we only consider the leading terms of the Toffoli depth. We have $\dv{k} \qty(2n\log n + \frac{2n}{k}\qty(\log n + 2^k)) = 0$, which gives $k = \dfrac{1}{\ln 2}\left(W \qty(\frac{\log n}{e})+1\right)$, where the Lambert W function $y=W(x)$ is the inverse function of $x=f(y)=y \cdot e^y$. When $n$ is sufficiently large, $W(x) \approx \ln x$, implying $k = \log\log n + o(\log\log n)$, and $\text{Toffoli depth}\simeq 2n\log n + 4n\frac{\log n}{\log\log n}$. To compare, normal multiplication costs $4n\log n + 4n\log\log n$ Toffoli depth, implying that our windowed skill saves about half the depth. In this case, the Toffoli number is $16n^2 + 14\frac{n^2}{\log\log n}$, CNOT depth is $2n\log n + 12n\frac{\log n}{\log\log n}$, CNOT number is $4n^2 \frac{\log n}{\log\log n} + \frac{37}{2}n^2$, SWAP depth is $n$, and SWAP number is $n^2$. Again, the additional factor $\frac{\log n}{\log\log n}$ in the CNOT depth and number is negligible in practical cases, since it is approximately a constant in such cases.
	
	\subsection{Modular Division}\label{subsec:division-cost}
	As shown in Fig.~\ref{figure:circuit-inversion} and Fig.~\ref{figure:circuit-division}, we should discuss the resource cost of \texttt{round} operation first, whose circuit is shown in Fig.~\ref{figure:circuit-inversion-round}. Note that this circuit only shows the uncontrolled version of this circuit, while all \texttt{round} operations in division are controlled operations actually. Thus, we should discuss the cost of controlled version.
	
	There are two CNOT gates conditioned on the lowest bits of $u$ and $v$, respectively, which must be promoted to Toffoli gates to incorporate the control. Second, there are four CNOT gates and one CCNOT gate acting on two output qubits, surrounded by two half-comparator segments. By letting the comparator itself be controlled, these gates no longer need explicit control, since when the control bit is $0$, all of these operations have no effect. Third, in the core part of the circuit, two pairs of controlled SWAP operations, one pair of controlled additions, and one pair of controlled bit shifts can be executed in parallel. Moreover, all parallelized controlled operations require only a single GHZ state generation for the control qubit. The two CNOT gates conditioned on $l_i$ do not need to be controlled since $l_i$ is always $0$ when $ctrl = 0$ (and the second of these CNOTs can be parallelized with the bit shift). However, the final CNOT must be controlled, thereby becoming a Toffoli gate. The overall resource cost for one controlled round operation is therefore given by
	\begin{equation}
		\begin{aligned}
			\text{Toffoli depth} &\simeq \log n + 2\times1 + (2\log n + 2\log\log n) + 3 + 5 \simeq 3\log n + 2\log\log n, \\
			\text{Toffoli number} &\simeq 3n + 4\times n + 2\times 16n + 2\times2n + 5 \simeq 43n, \\
			\text{CNOT depth} &\simeq \log n + 2\times5 + (2\log n + 2\log\log n) + 6 + 6 \simeq 3\log n + 2\log\log n, \\
			\text{CNOT number} &\simeq 4n + 2\times \qty(2n + \frac{7}{2}n) + \qty(17n + \frac{37}{2}n) + \qty(2n + \frac{7}{2}n) + 5 \simeq 56n,
		\end{aligned}
	\end{equation}
	where the first term corresponds to the comparator, the second term corresponds to the four controlled swappings, the third term corresponds to the two controlled additions, the fourth term corresponds to the two controlled bit shifting, and the last term corresponds to the other small terms.

	The whole inversion, as shown in Fig.~\ref{figure:circuit-inversion}, includes $2n$ iterations plus one modular negation. Each iteration comprises four components: an $n$-bit AND operation, a $(\lceil \log n \rceil + 1)$-bit AND operation, a controlled \texttt{round} operation, and a controlled constant addition of $(\lceil \log n \rceil + 1)$ bits. Consequently, the total resource cost of one inversion can be expressed as
	\begin{equation}\label{equ:inversion-cost}
		\begin{aligned}
			\text{Toffoli depth} &\simeq 2n\times\qty(\log n + \log (\lceil\log n\rceil+1) + (3\log n + 2\log\log n) + (2\log \qty(\lceil\log n\rceil+1) + 2\log\log \qty(\lceil\log n\rceil+1))) \\
			&+ (2\log n + 2\log\log n) \simeq 8n\log n + 10n\log\log n, \\
			\text{Toffoli number} &\simeq 2n\times\qty(n + (\lceil\log n\rceil+1) + 43n + 12(\lceil\log n\rceil+1)) + 12n \simeq 88n^2, \\
			\text{CNOT depth} &= 2n\times\qty(0 + 0 + (3\log n + 2\log\log n) + (2\log \qty(\lceil\log n\rceil+1) + 2\log\log \qty(\lceil\log n\rceil+1))) \\
			&+ (2\log n + 2\log\log n) \simeq 6n\log n + 8n\log\log n, \\
			\text{CNOT number} &\simeq 2n\times\qty(0 + 0 + 52n + \frac{41}{2}(\lceil\log n\rceil+1)) + 13n \simeq 112n^2,
		\end{aligned}
	\end{equation}
	where the first term corresponds to the $2n$ iterative rounds, and the second term corresponds to the final modular negation. Within the parentheses of the first term, the first two items represent the two AND operations, the third term corresponds to the controlled \texttt{round} operation, and the final term corresponds to the controlled constant addition.
	
	In the uncomputation of inversion, we need to do the controlled modular doubling at the same time, which can be partly parallelized with the AND function and the comparator in the controlled \texttt{round} operation. Therefore, the resource cost of one \texttt{round} plus \texttt{doub} operation is:
	\begin{equation}
		\begin{aligned}
			\text{Toffoli depth} &\simeq (2\log n + 2\log\log n) + 2\times1 + (2\log n + 2\log\log n) + 3 + 3 \simeq 4\log n + 4\log\log n, \\
			\text{Toffoli number} &\simeq (3n + 15n) + 4\times n + 2\times 16n + 2\times2n + 3 \simeq 58n, \\
			\text{CNOT depth} &\simeq (2\log n + 2\log\log n) + 2\times5 + (2\log n + 2\log\log n) + 6 + 6 \simeq 4\log n + 4\log\log n, \\
			\text{CNOT number} &\simeq (0 + 4n + 23n) + 2\times \qty(2n + \frac{7}{2}n) + \qty(17n + \frac{37}{2}n) + \qty(2n + \frac{7}{2}n) + 5 \simeq 79n,
		\end{aligned}
	\end{equation}
	where the first term corresponds to the comparator parallelizing with the modular doubling, the second term corresponds to the four controlled swappings, the third term corresponds to the two controlled additions, the fourth term corresponds to the two controlled bit shifting, and the last term corresponds to the other small terms.

	Similar to Equation~\eqref{equ:inversion-cost}, the uncomputation of inversion together with the controlled modular doubling costs $(10n\log n + 14n\log\log n)$ Toffoli depth, $118n^2$ Toffoli gates, $(8n\log n + 12n\log\log n)$ CNOT depth, and $158n^2$ CNOT gates.

	The whole division circuit, as shown in Fig.~\ref{figure:circuit-division}, consists of one modular inversion, one modular multiplication and its reverse, one bit-to-bit copy, and the uncomputation of inversion together with the controlled modular doubling. The resource cost of one modular division is thus
	\begin{equation}
		\begin{aligned}
			\text{Toffoli depth} &\simeq (8n\log n + 10n\log\log n) + (10n\log n + 14n\log\log n) + 2\times\qty(2n\log n + 4n\frac{\log n}{\log\log n}) + 1 \\
			&\simeq 22n\log n + 8n\frac{\log n}{\log\log n}, \\
			\text{Toffoli number} &\simeq 88n^2 + 118n^2 + 2\times\qty(16n^2 + 14\frac{n^2}{\log\log n}) + n \simeq 238n^2 + 28\frac{n^2}{\log\log n}, \\
			\text{CNOT depth} &\simeq (6n\log n + 8\log\log n) + (8n\log n + 12\log\log n) + 2\times\qty(2n\log n + 12n \frac{\log n}{\log\log n}) + 0 \\
			&\simeq 18n\log n + 24n\frac{\log n}{\log\log n}, \\
			\text{CNOT number} &\simeq 112n^2 + 158n^2 + 2\times\qty(4n^2 \frac{\log n}{\log\log n} + \frac{39}{2}n^2) + 0 \simeq 8n^2 \frac{\log n}{\log\log n} + 309n^2, \\
			\text{SWAP depth} &\simeq 0+0+2\times n+0 \simeq 2n, \\
			\text{SWAP number} &\simeq 0+0+2\times n^2+0 \simeq 2n^2,
		\end{aligned}
	\end{equation}
	where the first term corresponds to the modular inversion, the second term corresponds to the reverse modular inversion with controlled modular doubling, the third term corresponds to the modular multiplication and its reverse, and the last term corresponds to the bit-to-bit copy. Similar to the multiplication, the additional factor $\frac{\log n}{\log\log n}$ in the CNOT depth and number is approximately a constant in practical cases.

	\subsection{Point Addition}\label{subsec:point-addition-cost}
	A single point addition, as shown in Fig.~\ref{figure:circuit-point-addition}, includes two modular divisions, four modular multiplications, two modular squarings, five modular additions, three controlled modular additions, and one modular negation, among which two controlled modular additions can be parallelized with two modular additions. The resource cost of one point addition is
	\begin{equation}
		\begin{aligned}
			\text{Toffoli depth} &\simeq 2\times\qty(22n\log n + 8n\frac{\log n}{\log\log n}) + 6\times\qty(2n\log n + 4n\frac{\log n}{\log\log n}) \\
			&\simeq 56n\log n + 40n\frac{\log n}{\log\log n}, \\
			\text{Toffoli number} &\simeq 2\times\qty(238n^2 + 28\frac{n^2}{\log\log n}) + 6\times\qty(16n^2 + 14\frac{n^2}{\log\log n}) \\
			&\simeq 572n^2 + 140\frac{n^2}{\log\log n}, \\
			\text{CNOT depth} &\simeq 2\times\qty(18n\log n + 24n\frac{\log n}{\log\log n}) + 6\times\qty(2n\log n + 12n\frac{\log n}{\log\log n})  \\
			&\simeq 48n\log n + 120n\frac{\log n}{\log\log n}, \\
			\text{CNOT number} &\simeq 2\times\qty(8n^2 \frac{\log n}{\log\log n} + 309n^2) + 6\times\qty(4n^2 \frac{\log n}{\log\log n} + \frac{37}{2}n^2) \\
			&\simeq 40n^2 \frac{\log n}{\log\log n} + 735n^2, \\
			\text{SWAP depth} &\simeq 2\times 2n + 6\times n \simeq 10n, \\
			\text{SWAP number} &\simeq 2\times 2n^2 + 6\times n^2 \simeq 10n^2,
		\end{aligned}
	\end{equation}
	where the first term is the two modular divisions, the second term is the four modular multiplications together with the two modular squarings (since their costs are asymptotically same, as shown in Appendix~\ref{subsec:other-operations}), and we neglect the (controlled) modular additions and negation since they are negligible compared to the divisions, multiplications and squarings.
	
	\subsection{Shor's Algorithm}\label{subsec:full-circuit-cost}
	A direct implementation of point addition
	\begin{equation}
		\frac{1}{2^n}\sum_{x,y=0}^{2^n-1}\ket{x}\ket{y}\ket{O}
		\mapsto\frac{1}{2^n}\sum_{x,y=0}^{2^n-1}\ket{x}\ket{y}\ket{xG+yQ}
	\end{equation}
	requires $2n$ point additions, each of which is controlled by a corresponding bit in $x$ or $y$. Thus, the corresponding Toffoli/CNOT/SWAP depth/number equals to $2n$ times of that of point addition. Moreover, as discussed in Sec.~\ref{subsec:full-circuit}, the windowed trick~\cite{haner2020improved} can be used to reduce the resource costs. With the help of it, we can save $l-1$ point additions each window at the expense of one quantum table storing $2^l$ possible states, as shown in Fig.~\ref{figure:circuit-point-add-win}.
	
	The whole circuit has $\left\lceil \frac{2n}{l} \right\rceil$ iterations, each including two modular divisions, three modular multiplications, seven modular additions, five modular negations, and six quantum tables (four of them have $(l-1)$-bit input and $2n$-nit output, and the other two have $(l-1)$-bit input and $n$-bit output), among which two modular additions can be parallelized, as shown in Fig.~\ref{figure:circuit-point-add-win}. Notice that we ignore the fact the quantum table in the last iteration will have smaller input since it will not affect the asymptotic behavior. Thus, the resource cost of the whole circuit is given by
	\begin{equation}
		\begin{aligned}
			\text{Toffoli depth} &\simeq \left\lceil\frac{2n}{l}\right\rceil \qty(56n\log n + 40n\frac{\log n}{\log\log n} + 6\times(2^{l-1}-1)), \\
			\text{Toffoli number} &\simeq \left\lceil\frac{2n}{l}\right\rceil \qty(572n^2 + 140\frac{n^2}{\log\log n} + 6\times(2^{l-1}-1)), \\
			\text{CNOT depth} &\simeq \left\lceil\frac{2n}{l}\right\rceil \qty(48n\log n + 120n\frac{\log n}{\log\log n} + 6 \times 2^{l-1} \times 5), \\
			\text{CNOT number} &\simeq \left\lceil\frac{2n}{l}\right\rceil \qty(40n^2 \frac{\log n}{\log\log n} + 735n^2 + 4 \times 2^{l-1} \times (4n+1) + 2 \times 2^{l-1} \times (2n+1)), \\
			\text{SWAP depth} &\simeq \left\lceil\frac{2n}{l}\right\rceil 10n, \\
			\text{SWAP number} &\simeq \left\lceil\frac{2n}{l}\right\rceil 10n^2,
		\end{aligned}
	\end{equation}
	where $l$ is the window size. To get the minimum Toffoli depth up to $l$, we solve the equation $\dv{l} \qty(\frac{2n}{l} (56n\log n + 6\times2^{l-1})) = 0$ when only considering leading terms, which gives $l = \dfrac{W \qty(\frac{56}{3e}n\log n)+1}{\ln 2}$, where the Lambert W function $W(x)$ is the root of $t \cdot e^t = x$. When $n=256, l \approx12.31$. When $n$ is sufficiently large, $W(x) \approx \ln x$, giving $l = \log n + o(\log n)$. Our simulation shows that $l \approx \log n + 5$ for $128 \leq n \leq 16384$. In this case, the Toffoli depth is $112n^2 + 80\frac{n^2}{\log\log n}$, the Toffoli number is $1144\frac{n^3}{\log n} + 280\frac{n^3}{\log n\log\log n}$, the CNOT depth is $96n^2 + 240\frac{n^2}{\log\log n}$, the CNOT number is $80 \frac{n^3}{\log\log n} + 1490 \frac{n^3}{\log n}$, the SWAP depth is $20 \frac{n^2}{\log n}$, and the SWAP number is $20 \frac{n^3}{\log n}$.

	\section{Other Useful Circuits}\label{sec:other-circuit}
	
	In this section, we introduce some other useful components used in the circuit construction, including quantum table, unbounded AND and OR function, controlled and/or constant quantum adder, comparator, modular negation, modular reduction, modular doubling, (controlled) swapping and bit shifting, and modular squaring. We summarize the resource cost of all the circuits introduced in Sec.~\ref{sec:resource-cost} and Sec.~\ref{sec:other-circuit} in Table~\ref{table:operation-cost} and Table~\ref{table:operation-cost-swap}. The first table is for the cost of Toffoli and CNOT gates, while the second table is for the cost of SWAP gates.
	
	\begin{table}[ht]
		\caption{The Toffoli and CNOT cost of all the circuits introduced in Appendix.~\ref{sec:circuit-construction} and Appendix.~\ref{sec:other-circuit}.}\label{table:operation-cost}
		\begin{threeparttable}
    \begin{tabular}{c|cccc}
        \toprule
        \textbf{operation}                                                                        & \textbf{Toffoli depth} & \textbf{Toffoli number}  & \textbf{CNOT depth}                            & \textbf{CNOT number}                             \\
        \midrule
        GHZ state preparation                                                                     & 0                      & 0                        & 3                                              & $1.5n$                                           \\
        quantum table\tnote{1}                                                                   & $2^k-1$                & $2^k-1$                  & $2^k\cdot5$                                    & $2^k(2m+1)$                                      \\
        SWAP                                                                                      & 0                      & 0                        & 0                                              & 0                                                \\
        controlled SWAP                                                                           & 1                      & $n$                      & 5                                              & $3.5n$                                           \\
        $k$-bit shift                                                                             & 0                      & 0                        & 0                                              & 0                                                \\
        controlled 1-bit shift\tnote{2}                                                             & 3                      & $2n$                     & 6                                              & $3.5n$                                           \\
        AND                                                                                       & $\log n$              & $n$                     & 0                                              & 0                                                \\
        integer addition\tnote{3}                                                                 & $2\log n + 2\log\log n$              & $14n$                    & $2\log n + 2\log\log n$                        & $17n$                                             \\
        constant addition                                                                         & $2\log n + 2\log\log n$              & $12n$                     & $2\log n + 2\log\log n$                        & $16n$                                             \\
        controlled addition                                                                       & $2\log n + 2\log\log n$              & $16n$                    & $2\log n + 2\log\log n$                        & $18.5n$                                           \\
        controlled constant addition                                                              & $2\log n + 2\log\log n$              & $13n$                     & $2\log n + 2\log\log n$                        & $19.5n$                                           \\
        comparator                                                                                & $\log n$              & $3n$                     & $\log n$                                         & $4n$                                             \\
        modular negation                                                                          & $2\log n + 2\log\log n$              & $12n$                     & $2\log n + 2\log\log n$                        & $16n$                                             \\
        modular addition                                                                          & $5\log n + 4\log\log n$             & $31n$                    & $5\log n + 4\log\log n$                         & $40.5n$                                          \\
        controlled modular addition                                                               & $5\log n + 4\log\log n$             & $33n$                    & $5\log n + 4\log\log n$                         & $42n$                                           \\
        modular reduction                                                                         & $2\log n + 2\log\log n$              & $13n$                    & $2\log n + 2\log\log n$                        & $19.5n$                                           \\
        controlled modular doubling\tnote{4}                                                      & $2\log n + 2\log\log n$              & $15n$                    & $2\log n + 2\log\log n$                        & $23n$                                            \\
        modular multiplication\tnote{5}                                                             & $2n\log n + 4n\frac{\log n}{\log\log n}$             & $16n^2+14\frac{n^2}{\log\log n}$                  & $2n\log n + 12n\frac{\log n}{\log\log n}$              & $4n^2\frac{\log n}{\log\log n}+18.5n^2$           \\
        controlled round                                                                          & $4\log n + 2\log\log n$              & $45n$                    & $3\log n + 2\log\log n$                     & $56n$                                            \\
        \begin{tabular}[c]{@{}c@{}}controlled round combined with\\ controlled modular doubling\end{tabular}  & $4\log n + 4\log\log n$             & $60n$                    & $4\log n + 4\log\log n$                        & $79n$                                          \\
        modular inversion                                                                         & $10n\log n + 10n\log\log n$            & $92n^2$                  & $6\log n + 8\log\log n$                          & $112n^2$                                          \\
        \begin{tabular}[c]{@{}c@{}}modular inversion combined with\\ controlled modular doubling\end{tabular} & $10n\log n + 14n\log\log n$            & $122n^2$                 & $8\log n + 12\log\log n$                          & $158n^2$                                          \\
        modular division                                                                          & $24n\log n + 8n\frac{\log n}{\log\log n}$            & $246n^2+28\frac{n^2}{\log\log n}$                 & $18n\log n+24n\frac{\log n}{\log\log n}$            & $8n^2\frac{\log n}{\log\log n}+309n^2$           \\
        point addition                                                                            & $60n\log n + 40n\frac{\log n}{\log\log n}$           & $588n^2+140\frac{n^2}{\log\log n}$                 & $48n\log n+120n\frac{\log n}{\log\log n}$           & $40n^2\frac{\log n}{\log\log n}+735n^2$          \\
        full circuit                                                                              & $120n^2 + 80\frac{n^2}{\log\log n}$               & $1176\frac{n^3}{\log n}+280\frac{n^3}{\log n\log\log n}$ & $96n^2+240\frac{n^2}{\log\log n}$ & $80\frac{n^3}{\log\log n}+1490\frac{n^3}{\log n}$ \\
        \bottomrule
    \end{tabular}
    \begin{tablenotes}
        \footnotesize
        \item [1] With $k$ bits of input and $m$ bits of output.
        \item [2] We only provide the cost of controlled $1$-bit shift since only this controlled shift is used in our circuit.
        \item [3] The cost of integer subtraction is the same as addition.
        \item [4] There is no uncontrolled modular doubling operation in our circuit. However, one can simply remove the control qubit to achieve it.
        \item [5] The cost of modular squaring is the same as modular multiplication.
    \end{tablenotes}
\end{threeparttable}
	\end{table}
	
	\begin{table}[ht]
		\captionsetup{justification=raggedright,singlelinecheck=false}
		\caption{The SWAP cost of all the circuits introduced in Sec.~\ref{sec:resource-cost} and Sec.~\ref{sec:other-circuit}. For those operation presented in Table~\ref{table:operation-cost} but not presented in this table, their SWAP depth and SWAP number are both $0$.}\label{table:operation-cost-swap}
		\begin{threeparttable}
    \begin{tabular}{ccc}
        \toprule
        \textbf{operation}     & \textbf{SWAP depth}    & \textbf{SWAP number}   \\
        \midrule
        SWAP                   & 1                      & 1                      \\
        $k$-bit shift          & $k+2$                  & $n(k+1)$               \\
        modular multiplication & $n$                    & $n^2$                  \\
        modular division       & $2n$                   & $2n^2$                 \\
        point addition         & $10n$                  & $10n^2$                \\
        full circuit           & $20\frac{n^2}{\log n}$ & $20\frac{n^3}{\log n}$ \\
        \bottomrule
    \end{tabular}
\end{threeparttable}
	\end{table}
	
	\subsection{Quantum Table}\label{subsec:Qtable}
\begin{figure}[ht]
		\captionsetup{justification=raggedright,singlelinecheck=false}
		\centering
		\input{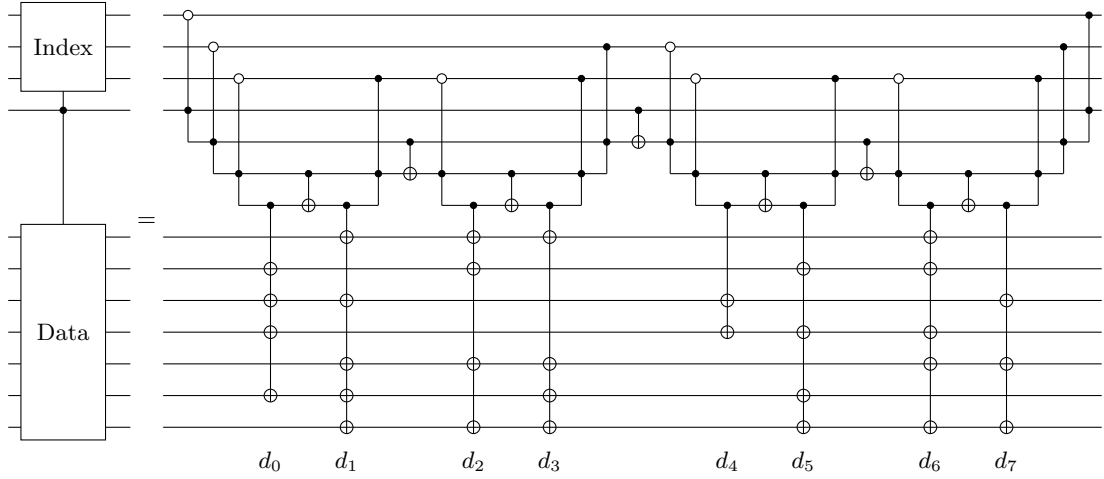}
		\caption{Quantum circuit for quantum table. On each data qubit, there are multiple controlled-$I$ gates or controlled-$X$ gates, depending on the value of $d_i$. For example, here $d_0={(0111010)}_2=58$.}\label{figure:circuit-quantum-table}
	\end{figure}

	A quantum table is a circuit that writes the corresponding data to the data register according to the contents of the index register, that is
	\begin{align}
		\sum_i\alpha_i\ket{i}\ket{0}\mapsto\sum_i\alpha_i\ket{i}\ket{a_i}.
	\end{align}
	An efficient way for implementing a quantum table is proposed in~\cite{babbush2018encoding} and shown in Fig.~\ref{figure:circuit-quantum-table}, which uses $2^k - 1$ Toffoli gates as well as depth, where $k$ is the size of the index register, and let $m$ be the bit number of the data output.
	
	Intuitively, this circuit accumulates the information in the index through ancillas, and then performs controlled gates on the data qubits using the final ancilla. For example, only when the content in the index is $\ket{000}$, can the ancilla qubit above $d_0$ be activated, thereby inputting the information of $d_0$ into the circuit. This way requires using one control qubit to control multiple $X$ gates, which can be efficiently implemented by dynamic circuits. Thus, this circuit can be modified to fit our requirements by adding a column of ancilla qubits for control.
	
	There are $2^k$ sets of CNOT gates in the ``Data'' part ($d_0, d_1, \ldots, d_7$ in Fig.~\ref{figure:circuit-quantum-table}) and $2^k-1$ disjoint CNOTs in the ``Index'' part. Each set in ``Data'' has $\frac{m}{2}$ parallel CNOT gates, and needs a GHZ state generation of $m$ bits. Since expanding a control bit into a GHZ state of $m$ bits costs $3$ CNOT depth and $\frac{3}{2}m$ CNOT number using dynamic circuit as shown in Fig.~\ref{figure:GHZ}, a quantum table with $k$-bit index and $m$-bit data has a CNOT depth of $2^k + 2^k \times 3 + 2^k = 2^k \times 5$, and a CNOT number of $2^k \frac{m}{2} + 2^k \frac{3}{2}m + 2^k = 2^k (2m+1)$.
	
	\subsection{AND and OR Function}\label{subsec:OR-AND}
	An unbounded AND gate is a generalized Toffoli gate with unbounded fan-in, i.e.,
	\begin{equation}
		\mathrm{AND}\ket{x_1}\ket{x_2}\cdots\ket{x_n}\ket{y}=\ket{x_1}\ket{x_2}\cdots\ket{x_n}\ket{y\oplus\prod_{i=1}^n x_i},
	\end{equation}
	where $x_1,x_2,\ldots,x_n,y$ are binary numbers. A naive way to implement $n$-qubit AND function is by using ancillary qubits to record temporary AND results. In the $i$-th step, we use the $i$-th data qubit and the $(i-1)$-th ancillary qubit as control, and the $i$-th ancillary qubit as target to implement a Toffoli gate $(1\leq i\leq n)$. The final result will accumulate on the $n$-th ancillary qubit.
	
	To implement an unbounded AND gate using a low-depth quantum circuit with geometric constraint, we put the data qubits and ancillary qubits stated before into two columns, and include another column of ancillary qubits used for dynamic circuits. Here, we let $n=2^t$ for convenience. Intuitively, in the $0$-th step, we divide the data qubits into adjacent groups in pairs, and accumulate the information of each group onto the adjacent ancilla qubit through Toffoli gate. After each step, the ancilla qubits obtained from the previous step will be divided into adjacent groups in pairs, and the information of each group will be accumulated through Toffoli gate onto the unused ancilla qubit located in the middle of them.
	
	More specifically, in the $0$-th step, we use the $(2j-1)$-th and the $(2j)$-th data qubits as control, and the $(2j-1)$-th ancilla qubits as target to implement a logical AND gate $(1\leq j\leq 2^{t-1})$. In the $i$-th step $(1\leq i\leq t-1)$, we use the $2^{i-1}(4j-3)$-th and the $2^{i-1}(4j-1)$-th ancilla qubits as control, and the $2^{i}(2j-1)$-th ancilla qubit as target to implement a logical AND gate $(1\leq j\leq 2^{t-1-i})$. All the long-range Toffoli gates in a step can be implemented simultaneously through disjoint dynamic circuits.
	
	Finally, we need to uncompute the ancilla qubits. This is done through uncomputation gates. Because the Toffoli depth of each step in the first phase is one, and the Toffoli number of the $i$-th step in the first phase is $n/2^{i+1}$, our AND function has a Toffoli depth $\log n$, Toffoli number $n$, and no CNOT gates. The OR function can be realized by implementing AND function after applying $X$ gate on each qubit.
	
	\subsection{Variants of addition}\label{subsec:variants-addition}
	In this subsection, we introduce several variants of quantum addition, including (controlled) out-place quantum adder, controlled (constant) in-place quantum adder, quantum negation, quantum comparator, quantum modular reduction, and (controlled) quantum modular doubling. Except the out-place adder, all these adders are based on the in-place adder shown in Algorithm~\ref{algorithm:in-place-adder}.
	
	\subsubsection{Out-place quantum adder}\label{subsubsec:outplaceadder}
	To achieve an out-place adder $\ket{x}\ket{y}\ket{0}\mapsto\ket{x}\ket{y}\ket{x+y}$, one only need to add $x_i, y_i$ onto the carry bits $c_i$ after it is computed. The algorithm is shown in Algorithm~\ref{algorithm:out-place-adder}. It has $2n$ input qubits, $n+1$ output qubits, $s_a$ ancillary qubits, Toffoli depth $d_t+1$, Toffoli number $n_t + n$, CNOT depth $d_c + 3$, and CNOT number $n_c + 3n-1$, where the parameters $s_a$, $d_t$, $n_t$, $d_c$ and $n_c$ are defined in Sec.~\ref{subsec:inplace-adder}. If we use our carry bit computing method, the leading term of these costs will be
	\begin{equation}
		\begin{aligned}
			\text{Toffoli depth} &= \log n + \log\log n + O(1), \\
			\text{Toffoli number} &= 7n + O(1), \\
			\text{CNOT depth} &= \log n + \log\log n + O(1), \\
			\text{CNOT number} &= \frac{19}{2}n + O(1), \\
			\text{Ancilla} &= 3n + O(1).
		\end{aligned}
	\end{equation}
	
	\begin{algorithm}[ht]
		\SetKwInOut{Input}{Input}
		\SetKwInOut{Output}{Output}
		\SetKw{for}{for}
		\SetKwProg{Fn}{Function}{:}{}
		\SetKwFunction{OutPlaceAdder}{OutPlaceAdder}
		\SetKwFunction{ComputeCarry}{ComputeCarry}
		
		\LinesNumbered{}
		\Fn{\OutPlaceAdder{$x,y;s;a$}}{
			\Input{$n$-qubit number $x$; $n$-qubit number $y$; $(n+1)$-qubit $s$, initialized to 0; ancilla $a$ (size depends on \ComputeCarry), initialized to 0.}
			\Output{$x, y, a$ remain the same; $s$ holds sum bits.}
			$s[i+1] \oplusIs x[i]y[i]$ \for{} $i \gets 0 ,\ldots, n-1$\tcp*{set $s[i+1] = g[i,i+1]$}
			$y[i] \oplusIs x[i]$ \for{} $i \gets 1 ,\ldots, n-1$\tcp*{set $y[i] = p[i,i+1]$ for $i\geq1$}
			\ComputeCarry{$y;s;a$}\tcp*{set $s[i] = c_i$ for $i\geq 1$}
			$s[i] \oplusIs y[i]$ \for{} $i \gets 0 ,\ldots, n-1$\tcp*{set $s[i] = x_i\oplus y_i \oplus c_i = s_i$ for $i\geq1$}
			$s[0] \oplusIs x[0]$; $y[i] \oplusIs x[i]$ \for{} $i \gets 1 ,\ldots, n-1$\tcp*{set $s[0] = s_0$ and recover $y$ to 0}
		}
		\caption{quantum algorithm for out-place adder.}\label{algorithm:out-place-adder}
	\end{algorithm}
	
	To construct a controlled out-of-place adder, we introduce $n$ additional qubits to serve as the control register. First, we prepare an $n$-qubit GHZ state to distribute the control, which requires $\frac{3}{2}n$ CNOT gates with a depth of $3$. The control qubits then condition the operations in lines 1, 4, and ``$s[0] \oplus x[0]$'' of Algorithm~\ref{algorithm:out-place-adder}. When the control bit is $0$, the sum register $s$ remains in the $\ket{0}$ state after the first line, ensuring that the carry-bit computation does not alter any values and that $s$ stays $\ket{0}$ throughout the computation. For the quantum realization of the controlled first line, the $n$ Toffoli gates are replaced by triple-controlled Toffoli ($C^3X$) gates. Each $C^3X$ gate can be decomposed into two standard Toffoli gates using one ancilla, resulting in an additional $2n$ Toffoli gates with a depth of $2$ and $n$ ancillas in total. This substitution does not increase the overall ancilla requirement. To achieve a $T$-depth of $1$ for these controlled operations, an additional $n$ ancillas are needed for GHZ-state generation of the control register; these qubits can be reused from the ancillas in the carry-bit computation. Let $d_t^o$, $n_t^o$, $d_c^o$, $n_c^o$, and $s_a^o$ denote the Toffoli depth, Toffoli count, CNOT depth, CNOT count, and ancilla space overhead of the uncontrolled out-of-place adder, respectively. Then, the controlled out-place adder has Toffoli depth $d_t^o + 2$, Toffoli number $n_t^o + 2n + 1$, CNOT depth $d_c^o + 2$, CNOT number $n_c^o +\frac{1}{2}n-1$, and ancilla space overhead $s_a^o$.
	
	If one of the input numbers is a classical constant, say $x$, all operations controlled by the qubits of $x$ can be replaced by Pauli-$X$ gates conditioned on the corresponding classical bits of $x$. Specifically, line 1 of the algorithm reduces to a single CNOT gate controlled by $y[i]$, while lines 2 and 5 simplify to unconditional $X$ gates. In this way, the constant controlled out-place adder has Toffoli depth $d_t^o - 1$, Toffoli number $n_t^o - n$, CNOT depth $d_c^o - 1$, CNOT number $n_c^o - n + 1$, and ancilla space overhead $s_a^o$.
	
	Finally, for a controlled constant out-place adder, we can just combine the idea of constant out-place adder and the controlled out-place adder. Specifically, we can replace all the operations controlled by the qubits of $x$ by Pauli-$X$ gates conditioned on the classical bits of $x$. It costs Toffoli depth $d_t^o + 1$, Toffoli number $n_t^o + n$, CNOT depth $d_c^o + 1$, CNOT number $n_c^o - \frac{1}{2}n + 1$, and ancilla space overhead $s_a^o$.

	\subsubsection{Controlled (constant) in-place quantum adder}\label{subsubsec:constantadder}
	Analogous to the construction of the controlled out-of-place adder, the controlled in-place adder can be implemented using GHZ-state preparation to expand the single control qubit into an $n$-qubit GHZ state. The corresponding circuit operations are then modified to become controlled operations. Specifically, lines 4, 5, 6, and 10 in Algorithm~\ref{algorithm:in-place-adder} are replaced by their controlled counterparts. When the control bit is $0$, the register $\ket{y}$ after line 6 remains identical to its state after line 3. Consequently, the reverse carry-bit computation restores $\ket{c}$ to $\ket{0}$, and $\ket{y}$ is recovered to its input state upon completion. To achieve a $T$-depth of $1$ for these controlled steps, an additional $n$ ancillas are required for GHZ-state generation of the control register. These ancillas can be reused from those allocated for the carry-bit computation. Let $d_t^i$, $n_t^i$, $d_c^i$, $n_c^i$, and $s_a^i$ denote the Toffoli depth, Toffoli count, CNOT depth, CNOT count, and ancilla space overhead of the uncontrolled in-place adder, respectively. Then, the controlled in-place adder has Toffoli depth $d_t^i + 2$, Toffoli number $n_t^i + 2n - 3$, CNOT depth $d_c^i$, CNOT number $n_c^i + \frac{3}{2}n + 1$, and ancilla space overhead $s_a^i$. Notice that the CNOT depth is the same as that of normal in-place adder, since GHZ preparation can be parallelized with other operations.
	
	If one of the input numbers is a classical constant, say $x$, all operations controlled by the qubits of $x$ can be replaced by Pauli-$X$ gates conditioned on the corresponding classical bits of $x$. Therefore, we can change line 1, 2, 6, 8 and 9 into simpler operations as that we do for constant out-place quantum adder. In this way, a constant in-place adder has Toffoli depth $d_t^i - 2$, Toffoli number $n_t^i - 2n + 1$, CNOT depth $d_c^i - 1$, CNOT number $n_c^i - n + 3$, and ancilla space overhead $s_a^i$.
	
	Finally, for a controlled constant in-place adder, we can just combine the idea of the constant in-place adder and the controlled in-place adder. Specifically, we can replace all the operations controlled by the qubits of $x$ by Pauli-$X$ gates conditioned on the classical bits of $x$. It costs Toffoli depth $d_t^i - 1$, Toffoli number $n_t^i - n$, CNOT depth $d_c^i + 1$, CNOT number $n_c^i + \frac{5}{2}n$, and ancilla space overhead $s_a^i$.

	It is worth noting that in our controlled addition circuit, all parts that need to be controlled comes after line 3 in Algorithm~\ref{algorithm:in-place-adder}, and by that time the carry bits have already been computed. Therefore, if we want to use the highest carry bit of an addition to control this addition itself, we can first compute the carry bits using line 1 to 3, then use the highest carry bit to control the rest of the addition process. This technique is used in our modular addition, modular reduction, and modular doubling circuits, which are discussed in Appendices~\ref{subsec:addition},~\ref{subsubsec:reduction}, and~\ref{subsubsec:doubling}, respectively, in which we use ${(-p)}_1$ to represent line 1 to 3 of a controlled $-p$ circuit, and ${(-p)}_2$ to represent line 4 to 10 of a controlled $-p$ circuit. In this way, we save a constant addition in these circuits compared with~\cite{roetteler2017quantum,haner2020improved}.
	
	\subsubsection{Comparator}\label{subsubsec:comparator}
	A comparator implements the transformation $\ket{x}\ket{y}\ket{0} \mapsto \ket{x}\ket{y}\ket{z}$, where $z = 1$ if $x >y$ and $z = 0$ otherwise. To determine whether $x < y$, it suffices to evaluate the sign bit of $x' + y$, where $x'$ denotes the bitwise complement of $x$. This task can be efficiently accomplished using our carry-bit computation method. To be specific, let $k = \lceil \log n \rceil$. We first extend both $x$ and $y$ to $2^k$ bits and compute the complementary $x'$, and then compute the most significant carry bit of $x' + y$, which indicates the comparison result. Since we only need the most significant carry bit, directly using the carry-lookahead algorithm based on Brent-Kung tree proves more efficient. This operation can be performed with a Toffoli depth of $k$, as it requires only the $P_1$ and $G_1$ stages of Algorithm~\ref{algorithm:carry-draper}. Moreover, since the bits beyond the $n$-th position are all zeros, this part of the circuit can be substantially simplified.
	
	The algorithm is shown in Algorithm~\ref{algorithm:comparator}. It has $2n$ input qubits, one output qubit, and $4n-\left\lceil \log (n-1) \right\rceil-3$ ancillas. Stage $P$ and $G$ have the same function as stage $P_1$ and $G_1$ in the previous Brent-Kung-tree-based method. Line 2, stage $P$, and stage $G$ use logical AND gates, and line 13, stage $P^{-1}$, and stage $G^{-1}$ use uncomputation gates (uncomputation gates of $G^{-1}$ costs $n$ ancillas).
	
	\begin{algorithm}[ht]
		\SetKwInOut{Input}{Input}
		\SetKwInOut{Output}{Output}
		\SetKw{for}{for}
		\SetKwProg{Fn}{Function}{:}{}
		\SetKwFunction{comparator}{Comparator}
		
		\LinesNumbered{}
		\Fn{\comparator{$x,y;G[n];G[1:n-1],P,a$}}{
			\Input{$n$-qubit number $x$; $n$-qubit number $y$; $n$-qubit string $G$, $(n-\left\lfloor \log (n-1) \right\rfloor-2)$-qubit string $P$, $2n$-qubit ancilla $a$, all initialized to 0.}
			\Output{$P_0,a$ remains the same; $G$ holds carry bits $c_i=g[0,i]$; $P$ holds propogation states, including $p[0,i]$.}
			$x[i]=-x[i]$ \for{} $i \gets 0 ,\ldots, n-1$\tcp*{set $x$ to $x'$}
			$G[i+1]\oplusIs x[i]y[i]$ \for{} $i \gets 0 ,\ldots, n-1$\tcp*{set $G[i+1] = g[i,i+1]$}
			$y[i] \oplusIs x[i]$ \for{} $i \gets 1 ,\ldots, n-1$\tcp*{set $y[i] = p[i,i+1]$}
			\For{$t \gets 1 ,\ldots, \lfloor\log n\rfloor$\tcp*{set $G=c_i=g[0,i]$ for $x'$ and $y$}}{
				$P_t[m] = P_{t-1}[2m] \cdot P_{t-1}[2m+1]$ \for{} $m \gets 0 ,\ldots, \left\lfloor 2^{\left\lceil \log n \right\rceil}/2^t \right\rfloor$ \tcp*{stage $P$; some of them may be unnecessary}
				$G[2^t m + 2^t] \oplusIs G[2^t m + 2^{t-1}] \cdot P_{t-1}[2m+1]$ \for{} $m \gets 0 ,\ldots,  \left\lfloor 2^{\left\lceil \log n \right\rceil}/2^t \right\rfloor$ \tcp*{stage $G$; some of them may be unnecessary}
			}
			\For{$t \gets \lfloor\log n\rfloor,\ldots,1$\tcp*{recover $G$ and $P$ except $G[n]$}}{
				$G[2^t m + 2^t] \oplusIs G[2^t m + 2^{t-1}] \cdot P_{t-1}[2m+1]$ \for{} $m \gets 0 ,\ldots,  \left\lfloor 2^{\left\lceil \log n \right\rceil}/2^t \right\rfloor-1$ \tcp*{stage $G^{-1}$; some of them may be unnecessary}
				$P_t[m] = P_{t-1}[2m] \cdot P_{t-1}[2m+1]$ \for{} $m \gets 0 ,\ldots, \left\lfloor 2^{\left\lceil \log n \right\rceil}/2^t \right\rfloor$ \tcp*{stage $P^{-1}$; some of them may be unnecessary}
			}
			$y[i] \oplusIs x[i]$ \for{} $i \gets 1 ,\ldots, n-1$\tcp*{recover $y$ to input}
			$G[i+1]\oplusIs x[i]y[i]$ \for{} $i \gets 0 ,\ldots, n-1$\tcp*{recover $G$ to 0 except $G[n]$}
			$x[i]=-x[i]$ \for{} $i \gets 0 ,\ldots, n-1$\tcp*{recover $x$ to input}
		}
		\caption{quantum algorithm for the comparator.}\label{algorithm:comparator}
	\end{algorithm}
	
	Stage $P$ costs $n-\left\lceil \log (n-1) \right\rceil-2$ Toffolis in depth $\left\lceil \log (n-1) \right\rceil$, and stage $G$ costs $n-1$ Toffolis in depth $\left\lceil \log (n-1) \right\rceil+1$. Parallelizing these two stages using the first method discussed in Appendix~\ref{subsec:DrapersMethod} will require $2(n-\left\lceil \log (n-1) \right\rceil-2)$ CNOT gates in depth $\left\lceil \log (n-1) \right\rceil+1$. In conclusion, the circuit parameters are
	\begin{equation}
		\begin{aligned}
			\text{Toffoli depth} &= 1 + \left\lceil \log (n-1) \right\rceil+1 = \left\lceil \log (n-1) \right\rceil+2 = \log n + O(1), \\
			\text{Toffoli number} &= n + n-\left\lceil \log (n-1) \right\rceil-2 + n-1 = 3n - \left\lceil \log (n-1) \right\rceil - 3 = 3n +O(1), \\
			\text{CNOT depth} &= 2 + \left\lceil \log (n-1) \right\rceil+1 = \left\lceil \log (n-1) \right\rceil+3 = \log n + O(1), \\
			\text{CNOT number} &= 2n-2+2(n-\left\lceil \log (n-1) \right\rceil-2) = 4n - 2\left\lceil \log (n-1) \right\rceil - 6 = 4n + O(1). \\
			\text{Ancilla} &= 4n - \left\lceil \log (n-1) \right\rceil - 3 = 4n + O(1).
		\end{aligned}
	\end{equation}
	
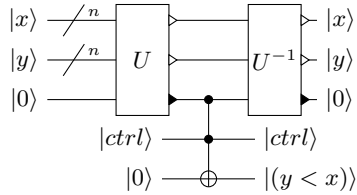
\begin{figure}[ht]
		\captionsetup{justification=raggedright,singlelinecheck=false}
		\centering
		\begin{tikzpicture}[scale=1.000000,x=1pt,y=1pt]
\filldraw[color=white] (0.000000, -7.500000) rectangle (102.000000, 67.500000);
\draw[color=black] (0.000000,60.000000) -- (102.000000,60.000000);
\draw[color=black] (0.000000,60.000000) node[left] {$\ket{x}$};
\draw[color=black] (0.000000,45.000000) -- (102.000000,45.000000);
\draw[color=black] (0.000000,45.000000) node[left] {$\ket{y}$};
\draw[color=black] (0.000000,30.000000) -- (102.000000,30.000000);
\draw[color=black] (0.000000,30.000000) node[left] {$\ket{0}$};
\draw[color=black] (36.000000,15.000000) -- (86.000000,15.000000);
\draw[color=black] (36.000000,0.000000) -- (86.000000,0.000000);
\draw (6.000000, 54.000000) -- (14.000000, 66.000000);
\draw (12.000000, 63.000000) node[right] {$\scriptstyle{n}$};
\draw (6.000000, 39.000000) -- (14.000000, 51.000000);
\draw (12.000000, 48.000000) node[right] {$\scriptstyle{n}$};
\draw (36.000000,60.000000) -- (36.000000,30.000000);
\begin{scope}
\draw[fill=white] (36.000000, 45.000000) +(-45.000000:14.142136pt and 29.698485pt) -- +(45.000000:14.142136pt and 29.698485pt) -- +(135.000000:14.142136pt and 29.698485pt) -- +(225.000000:14.142136pt and 29.698485pt) -- cycle;
\clip (36.000000, 45.000000) +(-45.000000:14.142136pt and 29.698485pt) -- +(45.000000:14.142136pt and 29.698485pt) -- +(135.000000:14.142136pt and 29.698485pt) -- +(225.000000:14.142136pt and 29.698485pt) -- cycle;
\draw (36.000000, 45.000000) node {$U$};
\end{scope}
\qOutput{46}{60}{white}
\qOutput{46}{45}{white}
\qOutput{46}{30}{black}
\draw[color=black] (43.500000,15.000000) node[fill=white,left,minimum height=15.000000pt,minimum width=15.000000pt,inner sep=0pt] {\phantom{$\ket{ctrl}$}};
\draw[color=black] (43.500000,15.000000) node[left] {$\ket{ctrl}$};
\draw[color=black] (43.500000,0.000000) node[fill=white,left,minimum height=15.000000pt,minimum width=15.000000pt,inner sep=0pt] {\phantom{$\ket{0}$}};
\draw[color=black] (43.500000,0.000000) node[left] {$\ket{0}$};
\draw (61.000000,30.000000) -- (61.000000,0.000000);
\begin{scope}
\draw[fill=white] (61.000000, 0.000000) circle(3.000000pt);
\clip (61.000000, 0.000000) circle(3.000000pt);
\draw (58.000000, 0.000000) -- (64.000000, 0.000000);
\draw (61.000000, -3.000000) -- (61.000000, 3.000000);
\end{scope}
\filldraw (61.000000, 30.000000) circle(1.500000pt);
\filldraw (61.000000, 15.000000) circle(1.500000pt);
\draw[color=black] (78.500000,15.000000) node[fill=white,right,minimum height=15.000000pt,minimum width=15.000000pt,inner sep=0pt] {\phantom{$\ket{ctrl}$}};
\draw[color=black] (78.500000,15.000000) node[right] {$\ket{ctrl}$};
\draw[color=black] (78.500000,0.000000) node[fill=white,right,minimum height=15.000000pt,minimum width=15.000000pt,inner sep=0pt] {\phantom{$\ket{(y<x)}$}};
\draw[color=black] (78.500000,0.000000) node[right] {$\ket{(y<x)}$};
\draw (86.000000,60.000000) -- (86.000000,30.000000);
\begin{scope}
\draw[fill=white] (86.000000, 45.000000) +(-45.000000:14.142136pt and 29.698485pt) -- +(45.000000:14.142136pt and 29.698485pt) -- +(135.000000:14.142136pt and 29.698485pt) -- +(225.000000:14.142136pt and 29.698485pt) -- cycle;
\clip (86.000000, 45.000000) +(-45.000000:14.142136pt and 29.698485pt) -- +(45.000000:14.142136pt and 29.698485pt) -- +(135.000000:14.142136pt and 29.698485pt) -- +(225.000000:14.142136pt and 29.698485pt) -- cycle;
\draw (86.000000, 45.000000) node {$U^{-1}$};
\end{scope}
\qOutput{96}{60}{white}
\qOutput{96}{45}{white}
\qOutput{96}{30}{black}
\draw[color=black] (102.000000,60.000000) node[right] {$\ket{x}$};
\draw[color=black] (102.000000,45.000000) node[right] {$\ket{y}$};
\draw[color=black] (102.000000,30.000000) node[right] {$\ket{0}$};
\end{tikzpicture}
		\caption{Quantum circuit for controlled comparator. Here, $U$ is the forward computation process, corresponding to Line 1--7 in Algorithm~\ref{algorithm:comparator}. An example of the details of $U$ can be found in~\cite{draper2004logarithmic}.}\label{figure:circuit-comparator}
	\end{figure}

	Changing this comparator into the controlled version is straightforward. After we computed $G[n]$, we only need to add single Toffoli gate controlled by $G[n]$ and the control bit to determine whether the final output bit is $G[n]$ or $0$. Then we do some extra uncomputation in stage $G^{-1}$ to recover $G[n]$ to $0$. This will not introduce new Toffolis, as they are all uncomputation gates. Denote the Toffoli depth, number, CNOT depth, number, and ancilla space overhead of the (uncontrolled) comparator as $d_t^c$, $n_t^c$, $d_c^c$, $n_c^c$, and $s_a^c$. The controlled comparator has Toffoli depth $d_t^c + 1$, Toffoli number $n_t^c + 1$, CNOT depth $d_c^c$, CNOT number $n_c^c$, and ancilla space overhead $s_a^c + 1$. The circuit is shown in Fig.~\ref{figure:circuit-comparator}.
	
	If one of the input numbers is a constant, say $x$, all operations controlled by the qubits of $x$ can be replaced by Pauli-$X$ gates conditioned on the corresponding classical bits of $x$. We can change line 1, 2, 3, 12, 13 and 14 into simpler operations. In this way, a constant controlled comparator has Toffoli depth $d_t^c - 1$, Toffoli number $n_t^c - n + 1$, CNOT depth $d_c^c - 2$, CNOT number $n_c^c - 2n + 2$, and ancilla space overhead $s_a^c$. A controlled constant comparator can be constructed in the same way, costing Toffoli depth $d_t^c$, Toffoli number $n_t^c -n+2$, CNOT depth $d_c^c -2$, CNOT number $n_c^c-2n+2$, and ancilla space overhead $s_a^c+1$.
	
	We denote the comparator by ``$>$''. To reduce circuit depth, it is sometimes advantageous to use only half of the (controlled) comparator, defined as 
	$\ket{x}\ket{y}\ket{0} \mapsto \ket{\bar{x}}\ket{\tilde{y}}\ket{z}$, where $\ket{\tilde{y}}$ represents an intermediate state. This allows one to perform subsequent operations before uncomputing the intermediate registers to restore the original state. We refer to these two halves of the comparator, executed in temporal order, as ``$>_1$'' and ``$>_2$'', respectively, while the full comparator is still denoted by ``$>$''.

	\subsubsection{Modular negation and reduction}\label{subsubsec:reduction}
	For modular negation $\ket{x} \mapsto \ket{-x \bmod{p}} = \ket{p-x}$, we first turn $x$ into $-x$, which consists of a bit flip and a ``$+1$''. Next, we do a ``$+p$''. To sum up, we need to do a bit flip and a ``$+(p+1)$'', thus its cost is the same as one constant addition.
	
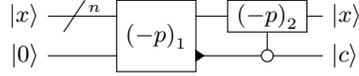
\begin{figure}[ht]
		\centering
		\begin{tikzpicture}[scale=1.000000,x=1pt,y=1pt]
\filldraw[color=white] (0.000000, -7.500000) rectangle (104.000000, 22.500000);
\draw[color=black] (0.000000,15.000000) -- (104.000000,15.000000);
\draw[color=black] (0.000000,15.000000) node[left] {$\ket{x}$};
\draw[color=black] (0.000000,0.000000) -- (104.000000,0.000000);
\draw[color=black] (0.000000,0.000000) node[left] {$\ket{0}$};
\draw (6.000000, 9.000000) -- (14.000000, 21.000000);
\draw (12.000000, 18.000000) node[right] {$\scriptstyle{n}$};
\draw (41.000000,15.000000) -- (41.000000,0.000000);
\begin{scope}
\draw[fill=white] (41.000000, 7.500000) +(-45.000000:21.213203pt and 19.091883pt) -- +(45.000000:21.213203pt and 19.091883pt) -- +(135.000000:21.213203pt and 19.091883pt) -- +(225.000000:21.213203pt and 19.091883pt) -- cycle;
\clip (41.000000, 7.500000) +(-45.000000:21.213203pt and 19.091883pt) -- +(45.000000:21.213203pt and 19.091883pt) -- +(135.000000:21.213203pt and 19.091883pt) -- +(225.000000:21.213203pt and 19.091883pt) -- cycle;
\draw (41.000000, 7.500000) node {${(-p)}_1$};
\end{scope}
\qOutput{56}{0}{black}
\draw (83.000000,15.000000) -- (83.000000,0.000000);
\begin{scope}
\draw[fill=white] (83.000000, 15.000000) +(-45.000000:21.213203pt and 8.485281pt) -- +(45.000000:21.213203pt and 8.485281pt) -- +(135.000000:21.213203pt and 8.485281pt) -- +(225.000000:21.213203pt and 8.485281pt) -- cycle;
\clip (83.000000, 15.000000) +(-45.000000:21.213203pt and 8.485281pt) -- +(45.000000:21.213203pt and 8.485281pt) -- +(135.000000:21.213203pt and 8.485281pt) -- +(225.000000:21.213203pt and 8.485281pt) -- cycle;
\draw (83.000000, 15.000000) node {${(-p)}_2$};
\end{scope}
\draw[fill=white] (83.000000, 0.000000) circle(2.250000pt);
\draw[color=black] (104.000000,15.000000) node[right] {$\ket{x}$};
\draw[color=black] (104.000000,0.000000) node[right] {$\ket{c}$};
\end{tikzpicture}
		\caption{Quantum circuit for modular reduction. ``$-p$'' denotes controlled constant substraction of $p$ where ``${(-p)}_1$'' and ``${(-p)}_2$'' are the two parts of one controlled constant subtraction, as explained in Appendix~\ref{subsubsec:constantadder}.}\label{figure:circuit-mod}
	\end{figure}

	The modular reduction operation performs $\ket{x}\ket{0} \mapsto \ket{x \bmod p}\ket{x \geq p}$ for $x < 2p$, and is denoted as \verb|mod|. This operation produces an additional garbage bit, which should be uncomputed in subsequent steps. The corresponding quantum circuit is illustrated in Fig.~\ref{figure:circuit-mod}. Similar to the modular addition circuit, only one controlled constant addition is required. That's because the control bit, i.e., the leading carry bit of $x + y - p$, is determined prior to the application of any controlled operations. Consequently, the resource cost of the modular reduction circuit is equivalent to that of a single controlled constant addition.
	
	\subsubsection{Controlled Modular Doubling}\label{subsubsec:doubling}
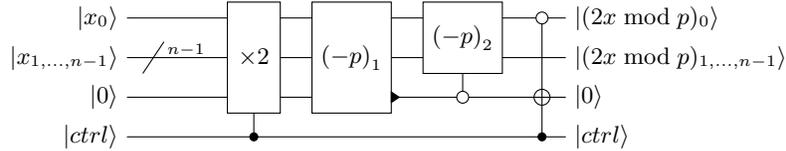
\begin{figure}[ht]
		\centering
		\begin{tikzpicture}[scale=1.000000,x=1pt,y=1pt]
\filldraw[color=white] (0.000000, -7.500000) rectangle (166.000000, 52.500000);
\draw[color=black] (0.000000,45.000000) -- (166.000000,45.000000);
\draw[color=black] (0.000000,45.000000) node[left] {$\ket{x_0}$};
\draw[color=black] (0.000000,30.000000) -- (166.000000,30.000000);
\draw[color=black] (0.000000,30.000000) node[left] {$\ket{x_{1, \ldots, n-1}}$};
\draw[color=black] (0.000000,15.000000) -- (166.000000,15.000000);
\draw[color=black] (0.000000,15.000000) node[left] {$\ket{0}$};
\draw[color=black] (0.000000,0.000000) -- (166.000000,0.000000);
\draw[color=black] (0.000000,0.000000) node[left] {$\ket{ctrl}$};
\draw (6.000000, 24.000000) -- (14.000000, 36.000000);
\draw (12.000000, 33.000000) node[right] {$\scriptstyle{n-1}$};
\begin{scope}[color=white]
\begin{scope}[color=white]
\begin{scope}
\draw[fill=none] (26.000000, 30.000000) +(-45.000000:0.000000pt) -- +(45.000000:0.000000pt) -- +(135.000000:0.000000pt) -- +(225.000000:0.000000pt) -- cycle;
\clip (26.000000, 30.000000) +(-45.000000:0.000000pt) -- +(45.000000:0.000000pt) -- +(135.000000:0.000000pt) -- +(225.000000:0.000000pt) -- cycle;
\filldraw (26.000000, 30.000000) circle(0.250000pt);
\end{scope}
\end{scope}
\end{scope}
\draw (48.000000,45.000000) -- (48.000000,0.000000);
\begin{scope}
\draw[fill=white] (48.000000, 30.000000) +(-45.000000:14.142136pt and 29.698485pt) -- +(45.000000:14.142136pt and 29.698485pt) -- +(135.000000:14.142136pt and 29.698485pt) -- +(225.000000:14.142136pt and 29.698485pt) -- cycle;
\clip (48.000000, 30.000000) +(-45.000000:14.142136pt and 29.698485pt) -- +(45.000000:14.142136pt and 29.698485pt) -- +(135.000000:14.142136pt and 29.698485pt) -- +(225.000000:14.142136pt and 29.698485pt) -- cycle;
\draw (48.000000, 30.000000) node {$\times 2$};
\end{scope}
\filldraw (48.000000, 0.000000) circle(1.500000pt);
\draw (85.000000,45.000000) -- (85.000000,15.000000);
\begin{scope}
\draw[fill=white] (85.000000, 30.000000) +(-45.000000:21.213203pt and 29.698485pt) -- +(45.000000:21.213203pt and 29.698485pt) -- +(135.000000:21.213203pt and 29.698485pt) -- +(225.000000:21.213203pt and 29.698485pt) -- cycle;
\clip (85.000000, 30.000000) +(-45.000000:21.213203pt and 29.698485pt) -- +(45.000000:21.213203pt and 29.698485pt) -- +(135.000000:21.213203pt and 29.698485pt) -- +(225.000000:21.213203pt and 29.698485pt) -- cycle;
\draw (85.000000, 30.000000) node {${(-p)}_1$};
\end{scope}
\qOutput{100}{15}{black}
\draw (127.000000,45.000000) -- (127.000000,15.000000);
\begin{scope}
\draw[fill=white] (127.000000, 37.500000) +(-45.000000:21.213203pt and 19.091883pt) -- +(45.000000:21.213203pt and 19.091883pt) -- +(135.000000:21.213203pt and 19.091883pt) -- +(225.000000:21.213203pt and 19.091883pt) -- cycle;
\clip (127.000000, 37.500000) +(-45.000000:21.213203pt and 19.091883pt) -- +(45.000000:21.213203pt and 19.091883pt) -- +(135.000000:21.213203pt and 19.091883pt) -- +(225.000000:21.213203pt and 19.091883pt) -- cycle;
\draw (127.000000, 37.500000) node {${(-p)}_2$};
\end{scope}
\draw[fill=white] (127.000000, 15.000000) circle(2.250000pt);
\draw (157.000000,45.000000) -- (157.000000,0.000000);
\draw[fill=white] (157.000000, 45.000000) circle(2.250000pt);
\filldraw (157.000000, 0.000000) circle(1.500000pt);
\begin{scope}
\draw[fill=white] (157.000000, 15.000000) circle(3.000000pt);
\clip (157.000000, 15.000000) circle(3.000000pt);
\draw (154.000000, 15.000000) -- (160.000000, 15.000000);
\draw (157.000000, 12.000000) -- (157.000000, 18.000000);
\end{scope}
\draw[color=black] (166.000000,45.000000) node[right] {$\ket{(2x \bmod{p})_0}$};
\draw[color=black] (166.000000,30.000000) node[right] {$\ket{(2x \bmod{p})_{1, \ldots, n-1}}$};
\draw[color=black] (166.000000,15.000000) node[right] {$\ket{0}$};
\draw[color=black] (166.000000,0.000000) node[right] {$\ket{ctrl}$};
\end{tikzpicture}
		\caption{Quantum circuit for modular doubling. ``$\times2$'' denotes a controlled $1$-bit left shift, and ``$-p$'' denotes controlled constant substraction of $p$ where ``${(-p)}_1$'' and ``${(-p)}_2$'' are the two parts of one controlled constant subtraction, as explained in Appendix~\ref{subsubsec:constantadder}.}\label{figure:circuit-mod-doubling}
	\end{figure}

	The modular doubling operation, $\ket{x} \mapsto \ket{(2x) \bmod p}$, can be implemented using a circuit structure similar to that of modular addition. The main difference is that the ordinary addition is replaced by a single-bit left shift, and the ancillary qubits can be recovered by checking the parity of the final result. We denote this operation as \verb|doub| to distinguish it from the simple left-shift operation ($\times 2$). This circuit requires one 1-bit left shift, one constant addition, and one controlled constant addition. Accordingly, it has a Toffoli depth of $3 + 2\log n + 2\log\log n + 1 \simeq 2\log n + 2\log\log n$, a Toffoli count of $2n + 13n = 15n$, a CNOT depth of $6 + 2\log n + 2\log\log n \simeq 2\log n + 2\log\log n$, and a CNOT count of $\tfrac{7}{2}n + \tfrac{39}{2}n = 23n$. The corresponding circuit layout is shown in Fig.~\ref{figure:circuit-mod-doubling}.
	
	\subsection{Other operations}\label{subsec:other-operations}
	
	\subsubsection{(Controlled) Swapping}\label{subsubsec:swaping}
\begin{figure}[ht]
		\centering
		\begin{tikzpicture}[scale=1.000000,x=1pt,y=1pt]
\filldraw[color=white] (0.000000, -7.500000) rectangle (99.000000, 37.500000);
\draw[color=black] (0.000000,30.000000) -- (99.000000,30.000000);
\draw[color=black] (0.000000,30.000000) node[left] {$\ket{ctrl}$};
\draw[color=black] (0.000000,15.000000) -- (99.000000,15.000000);
\draw[color=black] (0.000000,15.000000) node[left] {$\ket{x}$};
\draw[color=black] (0.000000,0.000000) -- (99.000000,0.000000);
\draw[color=black] (0.000000,0.000000) node[left] {$\ket{y}$};
\draw (9.000000,30.000000) -- (9.000000,0.000000);
\begin{scope}
\draw (6.878680, 12.878680) -- (11.121320, 17.121320);
\draw (6.878680, 17.121320) -- (11.121320, 12.878680);
\end{scope}
\begin{scope}
\draw (6.878680, -2.121320) -- (11.121320, 2.121320);
\draw (6.878680, 2.121320) -- (11.121320, -2.121320);
\end{scope}
\filldraw (9.000000, 30.000000) circle(1.500000pt);
\draw[fill=white,color=white] (24.000000, -6.000000) rectangle (39.000000, 36.000000);
\draw (31.500000, 15.000000) node {$=$};
\draw (54.000000,15.000000) -- (54.000000,0.000000);
\begin{scope}
\draw[fill=white] (54.000000, 15.000000) circle(3.000000pt);
\clip (54.000000, 15.000000) circle(3.000000pt);
\draw (51.000000, 15.000000) -- (57.000000, 15.000000);
\draw (54.000000, 12.000000) -- (54.000000, 18.000000);
\end{scope}
\filldraw (54.000000, 0.000000) circle(1.500000pt);
\draw (72.000000,30.000000) -- (72.000000,0.000000);
\begin{scope}
\draw[fill=white] (72.000000, 0.000000) circle(3.000000pt);
\clip (72.000000, 0.000000) circle(3.000000pt);
\draw (69.000000, 0.000000) -- (75.000000, 0.000000);
\draw (72.000000, -3.000000) -- (72.000000, 3.000000);
\end{scope}
\filldraw (72.000000, 15.000000) circle(1.500000pt);
\filldraw (72.000000, 30.000000) circle(1.500000pt);
\draw (90.000000,15.000000) -- (90.000000,0.000000);
\begin{scope}
\draw[fill=white] (90.000000, 15.000000) circle(3.000000pt);
\clip (90.000000, 15.000000) circle(3.000000pt);
\draw (87.000000, 15.000000) -- (93.000000, 15.000000);
\draw (90.000000, 12.000000) -- (90.000000, 18.000000);
\end{scope}
\filldraw (90.000000, 0.000000) circle(1.500000pt);
\draw[color=black] (99.000000,30.000000) node[right] {$\ket{ctrl}$};
\draw[color=black] (99.000000,15.000000) node[right] {$\ket{y}$};
\draw[color=black] (99.000000,0.000000) node[right] {$\ket{x}$};
\end{tikzpicture}
		\caption{Quantum circuit for controlled swapping.}\label{figure:circuit-swap}
	\end{figure}
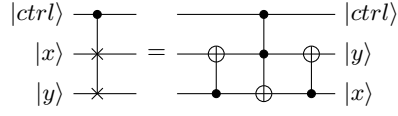

	A SWAP gate, as mentioned previously, is considered a native operation. However, a controlled SWAP cannot be implemented directly and must be decomposed into two CNOT gates and one Toffoli gate, as illustrated in Fig.~\ref{figure:circuit-swap}. For a controlled swap between two $n$-qubit registers, $\ket{x}\ket{y} \mapsto \ket{y}\ket{x}$, we first generate an $n$-qubit GHZ state from the control qubit, which requires $\tfrac{3}{2}n$ CNOT gates with a CNOT depth of $3$. The middle layer of CNOT gates in the SWAP circuit is then replaced by Toffoli gates. As a result, the total cost is $n$ Toffoli gates and $\frac{7}{2}n$ CNOT gates, with a Toffoli depth of $1$ and a CNOT depth of $5$. If $\ket{y} = \ket{0}$, the first layer of CNOT gates can be omitted, reducing the CNOT depth by $1$ and the total CNOT count by $n$. 
	
	\subsubsection{(Controlled) Shifting}\label{subsubsec:shifting}
\begin{figure}[ht]
		\captionsetup{justification=raggedright,singlelinecheck=false}
		\centering
		\begin{tikzpicture}
    \foreach \z in {0,1,2,3,4,5,6} {
        \foreach \y in {0,1,2,3,4,5,6,7} {
            \foreach \x in {0,1} {
                \draw (2*\z+0.5*\x, \y*0.5) circle (2pt);
            }
        }
    }

    \foreach \z in {0,1} {
        \foreach \y in {0,1,2,3,4} {
                \filldraw (2*\z, \y*0.5) circle (2pt);
        }
    }
    \foreach \z in {2,3,4,5} {
        \foreach \y in {0,2,4} {
                \filldraw ({2*\z}, {(\y+\z-2)*0.5}) circle (2pt);
        }
        \foreach \y in {1,3} {
                \filldraw ({2*\z+0.5}, {(\y+\z-2)*0.5}) circle (2pt);
        }
    }
    \foreach \z in {6} {
        \foreach \y in {3,4,5,6,7} {
                \filldraw (2*\z, \y*0.5) circle (2pt);
        }
    }

    \foreach \z in {0,1,2,3,4,5} {
        \draw[-implies, double equal sign distance] (1+2*\z,1.75) -- (1.5+2*\z,1.75);
    }
    \foreach \y in {1,3} {
        \draw [{Latex[length=1mm,width=1.5mm]}-{Latex[length=1mm,width=1.5mm]}] (2.1, \y*0.5) -- (2.4, \y*0.5);
        \draw [{Latex[length=1mm,width=1.5mm]}-{Latex[length=1mm,width=1.5mm]}] (10.1, \y*0.5+1.5) -- (10.4, \y*0.5+1.5);
    }
    \foreach \z in {2,3,4} {
        \foreach \y in {0,2,4} {
                \draw [{Latex[length=1mm,width=1.5mm]}-{Latex[length=1mm,width=1.5mm]}] ({2*\z}, {(\y+\z-2)*0.5+0.1}) -- ({2*\z}, {(\y+\z-2)*0.5+0.4});
        }
        \foreach \y in {1,3} {
                \draw [{Latex[length=1mm,width=1.5mm]}-{Latex[length=1mm,width=1.5mm]}] ({2*\z+0.5}, {(\y+\z-2)*0.5+0.1}) -- ({2*\z+0.5}, {(\y+\z-2)*0.5+0.4});
        }
    }

\end{tikzpicture}
		\caption{Quantum circuit for a 3-bit shifting for a 5-bit register. Black points are qubits with data, and white points are ancillas. $\leftrightarrow$ denotes a swapping.}\label{figure:circuit-shift}
	\end{figure}
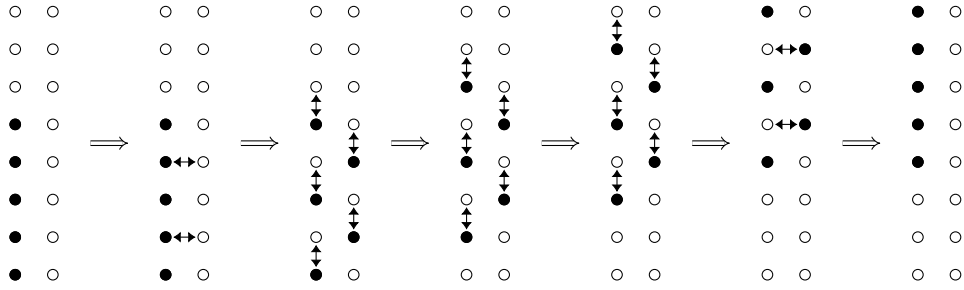

	A bit shift of $k$ positions on an $n$-qubit register can be implemented using $n(k + 1)$ SWAP gates with a SWAP depth of $k + 2$, as illustrated in Fig.~\ref{figure:circuit-shift}. For a controlled bit shift, the control qubit is first expanded into an $(n + k)$-qubit GHZ state via a dynamic circuit, which requires a CNOT depth of $3$ and $\tfrac{3}{2}(n + k)$ CNOT gates. The subsequent controlled SWAP operations are then performed using this GHZ state control. We do not provide an exact gate count for arbitrary $k$, since our circuit design only employs controlled 1-bit shifts. A controlled single-bit shift incurs a Toffoli depth of $3$, a Toffoli count of~$2n$, a CNOT depth of $1 + 2 + 3 = 6$, and a CNOT count of $2n + \tfrac{3}{2}(n + 1) \simeq \tfrac{7}{2}n$. For convenience, we denote a $k$-bit left shift as $\times 2^k$ and a $k$-bit right shift as $/ 2^k$.
	
	\subsubsection{Modular Squaring}\label{subsubsec:squaring}
\begin{figure}[ht]
		\centering
		\input{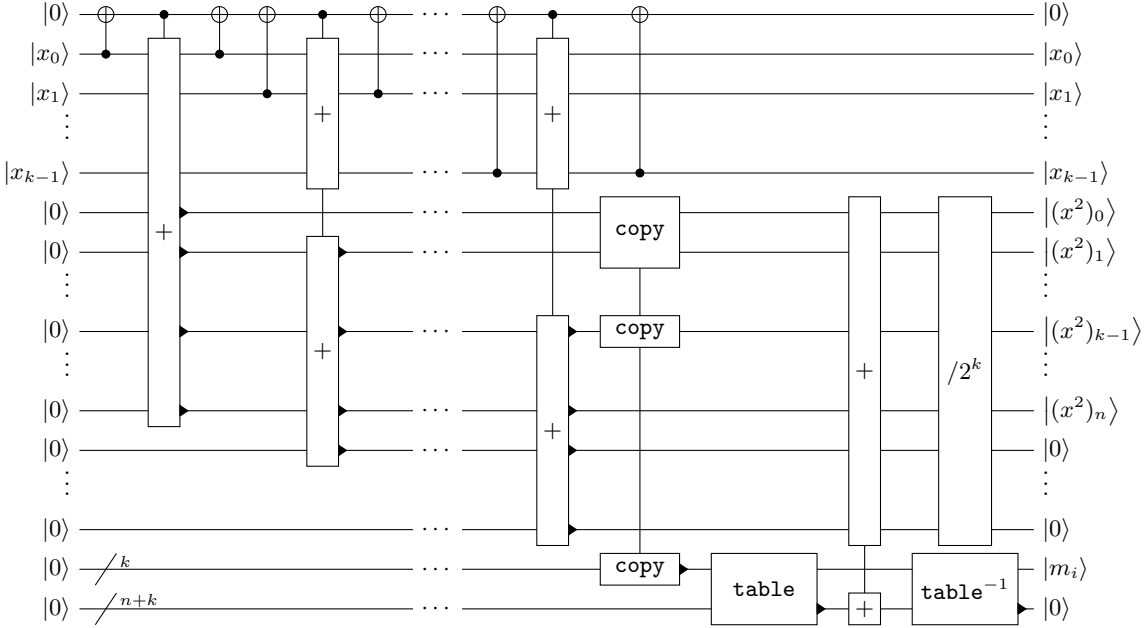}
		\caption{Quantum circuit for a single \texttt{squ\_win} operation, similar to Fig.~\ref{figure:circuit-multiplication-window}.}\label{figure:circuit-square-window}
	\end{figure}

	The overall circuit structure of modular squaring $\ket{x}\ket{0} \mapsto \ket{x}\ket{x^2 \bmod{p}}$ is the analog to that of modular multiplication, as shown in Fig.~\ref{figure:circuit-mod-multiplication}. The only difference is that we should replace the subroutine \texttt{mul\_win} by \texttt{squ\_win}, which is shown in Fig.~\ref{figure:circuit-square-window}. This modification introduces additional CNOT gates compared to modular multiplication, resulting in the following CNOT cost:
	\begin{equation}\label{equ:mod-squ-cost}
		\begin{aligned}
			\text{CNOT depth} &\simeq \left\lfloor\frac{n}{k}\right\rfloor \qty(k \cdot (2\log n + 2\log\log n) + (2\log (n+k) + 2\log\log (n+k)) + 1 + 2 \times 2^k \times 5 + (2k-1)) \\
			&+ (r \cdot (2\log n + 2\log\log n) + (2\log (n+r) + 2\log\log (n+r)) + 1 + 3(r+2) + 2^r \times 5 + (2r-1)) + 11, \\
			\text{CNOT number} &\simeq \left\lfloor\frac{n}{k}\right\rfloor \qty(k \cdot \frac{37}{2}n + 17(n+k) + k + 2 \times 2^k \times (2(n+k)+1) + 2k) \\
			&+ \qty(r \cdot \frac{37}{2}n + 17(n+r) + r + 3n(r+1) + 2^r \times (2(n+r)+1) + 2r) + \frac{39}{2}n,
		\end{aligned}
	\end{equation}
	where the meaning of all terms is the same as Equation~\ref{equ:mod-mul-cost} except for the last term in the parenthesis, which stands for the additional CNOTs. The Toffoli and SWAP costs do not change, so the optimal window size $k$ is still $\log\log n + o(\log\log n)$, and the corresponding Toffoli and SWAP costs remain the same as modular multiplication, but the CNOT depth will be $2n\log n + 12n\frac{\log n}{\log\log n}$, CNOT number is $4n^2 \frac{\log n}{\log\log n} + \frac{37}{2}n^2$ (asymptotically same with modular multiplication).
	
	where all terms have the same meaning as in Eq.~\ref{equ:mod-mul-cost}, except for the final terms in parentheses, which account for the additional CNOT gates introduced by squaring. The Toffoli and SWAP costs remain unchanged from the modular multiplication circuit. Consequently, the optimal window size is still $k = \log\log n + o(\log\log n)$, and the corresponding Toffoli and SWAP costs are identical to those of modular multiplication. The asymptotic CNOT costs, however, become $\text{CNOT depth} = 2n\log n + 12n\frac{\log n}{\log\log n}$ and $\text{CNOT count} = 4n^2 \frac{\log n}{\log\log n} + \tfrac{37}{2}n^2$, which are asymptotically equivalent to the modular multiplication results.  
	
	The modular squaring circuit requires fewer qubits than modular multiplication, as the $\ket{y}$ register is no longer needed. In total, it uses $9n$ qubits, with a gate interaction distance of $5$ for the bi-layer layout and $7$ for the single-layer layout.
	
	\section{Single-layer Layout}\label{sec:single-layer}
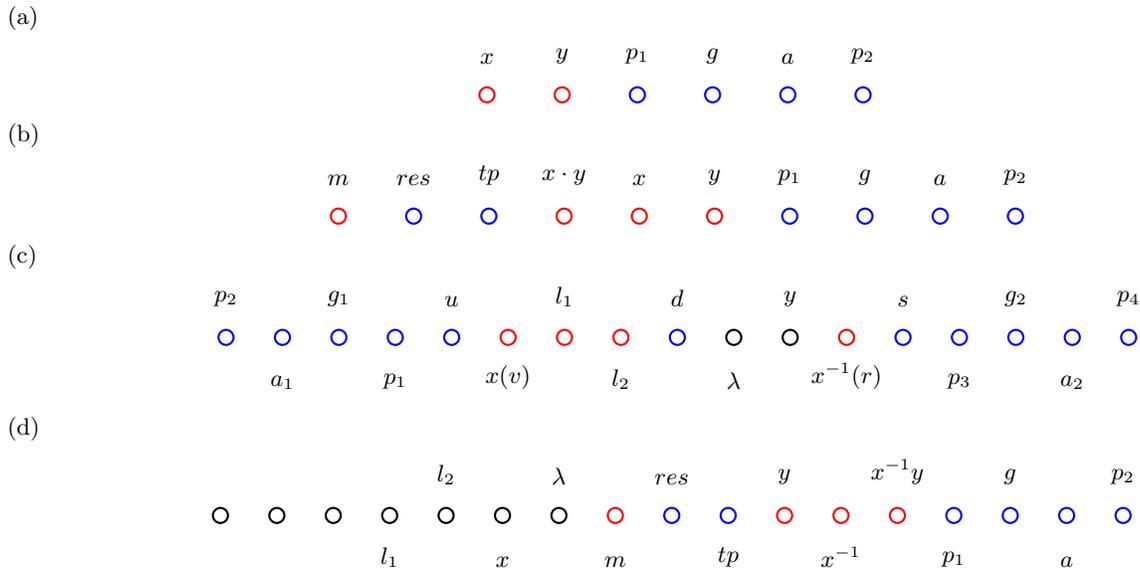
\begin{figure}[ht]
		\captionsetup{justification=raggedright,singlelinecheck=false}
		\begin{subfigure}[t]{\textwidth}
			\caption{}\label{figure:layout-addition-single}
			\centering
			\begin{tikzpicture}
    \draw[thick, red] (0, 0) circle(0.1);
    \node[above] at (0, 0.3) {$x$};
    \draw[thick, red] (1, 0) circle(0.1);
    \node[above] at (1, 0.3) {$y$};
    \draw[thick, blue] (2, 0) circle(0.1);
    \node[above] at (2, 0.3) {$p_1$};
    \draw[thick, blue] (3, 0) circle(0.1);
    \node[above] at (3, 0.3) {$g$};
    \draw[thick, blue] (4, 0) circle(0.1);
    \node[above] at (4, 0.3) {$a$};
    \draw[thick, blue] (5, 0) circle(0.1);
    \node[above] at (5, 0.3) {$p_2$};
\end{tikzpicture}
		\end{subfigure}
		\begin{subfigure}[t]{\textwidth}
			\caption{}\label{figure:layout-multiplication-single}
			\centering
			\begin{tikzpicture}
    \draw[thick, red] (1, 0) circle(0.1);
    \node[above] at (1, 0.3) {$m$};
    \draw[thick, blue] (2, 0) circle(0.1);
    \node[above] at (2, 0.3) {$res$};
    \draw[thick, blue] (3, 0) circle(0.1);
    \node[above] at (3, 0.3) {$tp$};
    \draw[thick, red] (4, 0) circle(0.1);
    \node[above] at (4, 0.3) {$x \cdot y$};
    \draw[thick, red] (5, 0) circle(0.1);
    \node[above] at (5, 0.3) {$x$};
    \draw[thick, red] (6, 0) circle(0.1);
    \node[above] at (6, 0.3) {$y$};
    \draw[thick, blue] (7, 0) circle(0.1);
    \node[above] at (7, 0.3) {$p_1$};
    \draw[thick, blue] (8, 0) circle(0.1);
    \node[above] at (8, 0.3) {$g$};
    \draw[thick, blue] (9, 0) circle(0.1);
    \node[above] at (9, 0.3) {$a$};
    \draw[thick, blue] (10, 0) circle(0.1);
    \node[above] at (10, 0.3) {$p_2$};
\end{tikzpicture}
		\end{subfigure}
		\begin{subfigure}[t]{\textwidth}
			\caption{}\label{figure:layout-inversion-single}
			\centering
			\begin{tikzpicture}
    \draw[thick, blue] (0, 0) circle(0.1);
    \node[above] at (0, 0.3) {$p_2$};
    \draw[thick, blue] (0.75, 0) circle(0.1);
    \node[above] at (0.75, -0.8) {$a_1$};
    \draw[thick, blue] (1.5, 0) circle(0.1);
    \node[above] at (1.5, 0.3) {$g_1$};
    \draw[thick, blue] (2.25, 0) circle(0.1);
    \node[above] at (2.25, -0.8) {$p_1$};
    \draw[thick, blue] (3, 0) circle(0.1);
    \node[above] at (3, 0.3) {$u$};
    \draw[thick, red] (3.75, 0) circle(0.1);
    \node[above] at (3.75, -0.8) {$x(v)$};
    \draw[thick, red] (4.5, 0) circle(0.1);
    \node[above] at (4.5, 0.3) {$l_1$};
    \draw[thick, red] (5.25, 0) circle(0.1);
    \node[above] at (5.25, -0.8) {$l_2$};
    \draw[thick, blue] (6, 0) circle(0.1);
    \node[above] at (6, 0.3) {$d$};
    \draw[thick, black] (6.75, 0) circle(0.1);
    \node[above] at (6.75, -0.8) {$\lambda$};
    \draw[thick, black] (7.5, 0) circle(0.1);
    \node[above] at (7.5, 0.3) {$y$};
    \draw[thick, red] (8.25, 0) circle(0.1);
    \node[above] at (8.25, -0.8) {$x^{-1}(r)$};
    \draw[thick, blue] (9, 0) circle(0.1);
    \node[above] at (9, 0.3) {$s$};
    \draw[thick, blue] (9.75, 0) circle(0.1);
    \node[above] at (9.75, -0.8) {$p_3$};
    \draw[thick, blue] (10.5, 0) circle(0.1);
    \node[above] at (10.5, 0.3) {$g_2$};
    \draw[thick, blue] (11.25, 0) circle(0.1);
    \node[above] at (11.25, -0.8) {$a_2$};
    \draw[thick, blue] (12, 0) circle(0.1);
    \node[above] at (12, 0.3) {$p_4$};
\end{tikzpicture}
		\end{subfigure}
		\begin{subfigure}[t]{\textwidth}
			\caption{}\label{figure:layout-div-multi-single}
			\centering
			\begin{tikzpicture}
    \draw[thick, black] (0, 0) circle(0.1);
    \draw[thick, black] (0.75, 0) circle(0.1);
    \draw[thick, black] (1.5, 0) circle(0.1);
    \draw[thick, black] (2.25, 0) circle(0.1);
    \node[above] at (2.25, -0.8) {$l_1$};
    \draw[thick, black] (3, 0) circle(0.1);
    \node[above] at (3, 0.3) {$l_2$};
    \draw[thick, black] (3.75, 0) circle(0.1);
    \node[above] at (3.75, -0.8) {$x$};
    \draw[thick, black] (4.5, 0) circle(0.1);
    \node[above] at (4.5, 0.3) {$\lambda$};
    \draw[thick, red] (5.25, 0) circle(0.1);
    \node[above] at (5.25, -0.8) {$m$};
    \draw[thick, blue] (6, 0) circle(0.1);
    \node[above] at (6, 0.3) {$res$};
    \draw[thick, blue] (6.75, 0) circle(0.1);
    \node[above] at (6.75, -0.8) {$tp$};
    \draw[thick, red] (7.5, 0) circle(0.1);
    \node[above] at (7.5, 0.3) {$y$};
    \draw[thick, red] (8.25, 0) circle(0.1);
    \node[above] at (8.25, -0.8) {$x^{-1}$};
    \draw[thick, red] (9, 0) circle(0.1);
    \node[above] at (9, 0.3) {$x^{-1}y$};
    \draw[thick, blue] (9.75, 0) circle(0.1);
    \node[above] at (9.75, -0.8) {$p_1$};
    \draw[thick, blue] (10.5, 0) circle(0.1);
    \node[above] at (10.5, 0.3) {$g$};
    \draw[thick, blue] (11.25, 0) circle(0.1);
    \node[above] at (11.25, -0.8) {$a$};
    \draw[thick, blue] (12, 0) circle(0.1);
    \node[above] at (12, 0.3) {$p_2$};
\end{tikzpicture}
		\end{subfigure}
		\caption{Single-layer layout for (a) modular addition, (b) modular multiplication, (c) modular division (inversion part), and (d) modular division (multiplication part). The meaning of each column is explained in Fig.~\ref{figure:layout-addition}, Fig.~\ref{figure:layout-multiplication} and Fig.~\ref{figure:layout-division}.}\label{figure:single-layer}
	\end{figure}

	Since bi-layer qubit layouts are not yet widely adopted in current quantum hardware architectures, we also provide an alternative single-layer realization. To minimize the gate interaction distance, the two columns of qubits in the bi-layer layout are interleaved side by side in the single-layer arrangement. The corresponding layouts for the main circuit operations are illustrated in Fig.~\ref{figure:single-layer}. A comparison of the maximum gate interaction distances for the bi-layer and single-layer layouts is summarized in Table~\ref{table:max-dist}. As shown in the table, the interaction distance in the single-layer configuration is approximately twice that of the bi-layer layout, which is consistent with our intuition and geometric expectations.
	
	\begin{table}[ht]
		\caption{Gate interaction distance of bi-layer and single-layer layouts for different operations}\label{table:max-dist}
		\begin{threeparttable}
    \begin{tabular}{ccc}
        \toprule
        \bf operation &  \bf bi-layer & \bf single-layer \\
        \midrule
        addition       & 2       & 3             \\
        multiplication & 4       & 7             \\
        division       & 4       & 7             \\
        whole circuit  & 4       & 7             \\
        \bottomrule
    \end{tabular}
\end{threeparttable}
	\end{table}

	\bibliographystyle{apsrev}

	\bibliography{./tex/bibQECDSA.bib}

\begin{thebibliography}{51}
\expandafter\ifx\csname natexlab\endcsname\relax\def\natexlab#1{#1}\fi
\expandafter\ifx\csname bibnamefont\endcsname\relax
  \def\bibnamefont#1{#1}\fi
\expandafter\ifx\csname bibfnamefont\endcsname\relax
  \def\bibfnamefont#1{#1}\fi
\expandafter\ifx\csname citenamefont\endcsname\relax
  \def\citenamefont#1{#1}\fi
\expandafter\ifx\csname url\endcsname\relax
  \def\url#1{\texttt{#1}}\fi
\expandafter\ifx\csname urlprefix\endcsname\relax\def\urlprefix{URL }\fi
\providecommand{\bibinfo}[2]{#2}
\providecommand{\eprint}[2][]{\url{#2}}

\bibitem[{\citenamefont{Shor}(1997)}]{shor1997polynomial}
\bibinfo{author}{\bibfnamefont{P.~W.} \bibnamefont{Shor}}, \bibinfo{journal}{SIAM Journal on Computing} \textbf{\bibinfo{volume}{26}}, \bibinfo{pages}{1484} (\bibinfo{year}{1997}), \eprint{https://doi.org/10.1137/S0097539795293172}, \urlprefix\url{https://doi.org/10.1137/S0097539795293172}.

\bibitem[{\citenamefont{Grover}(1996)}]{grover1996afast}
\bibinfo{author}{\bibfnamefont{L.~K.} \bibnamefont{Grover}}, in \emph{\bibinfo{booktitle}{Proceedings of the Twenty-Eighth Annual ACM Symposium on Theory of Computing}} (\bibinfo{publisher}{Association for Computing Machinery}, \bibinfo{address}{New York, NY, USA}, \bibinfo{year}{1996}), STOC '96, p. \bibinfo{pages}{212–219}, ISBN \bibinfo{isbn}{0897917855}, \urlprefix\url{https://doi.org/10.1145/237814.237866}.

\bibitem[{\citenamefont{Harrow et~al.}(2009)\citenamefont{Harrow, Hassidim, and Lloyd}}]{harrow2009quantum}
\bibinfo{author}{\bibfnamefont{A.~W.} \bibnamefont{Harrow}}, \bibinfo{author}{\bibfnamefont{A.}~\bibnamefont{Hassidim}}, \bibnamefont{and} \bibinfo{author}{\bibfnamefont{S.}~\bibnamefont{Lloyd}}, \bibinfo{journal}{Phys. Rev. Lett.} \textbf{\bibinfo{volume}{103}}, \bibinfo{pages}{150502} (\bibinfo{year}{2009}), \urlprefix\url{https://link.aps.org/doi/10.1103/PhysRevLett.103.150502}.

\bibitem[{\citenamefont{Berry et~al.}(2015)\citenamefont{Berry, Childs, and Kothari}}]{berry2015hamiltonian}
\bibinfo{author}{\bibfnamefont{D.~W.} \bibnamefont{Berry}}, \bibinfo{author}{\bibfnamefont{A.~M.} \bibnamefont{Childs}}, \bibnamefont{and} \bibinfo{author}{\bibfnamefont{R.}~\bibnamefont{Kothari}}, in \emph{\bibinfo{booktitle}{2015 IEEE 56th Annual Symposium on Foundations of Computer Science}} (\bibinfo{year}{2015}), pp. \bibinfo{pages}{792--809}.

\bibitem[{\citenamefont{Arute et~al.}(2019)\citenamefont{Arute, Arya, Babbush, Bacon, Bardin, Barends, Biswas, Boixo, Brandao, Buell et~al.}}]{arute2019quantum}
\bibinfo{author}{\bibfnamefont{F.}~\bibnamefont{Arute}}, \bibinfo{author}{\bibfnamefont{K.}~\bibnamefont{Arya}}, \bibinfo{author}{\bibfnamefont{R.}~\bibnamefont{Babbush}}, \bibinfo{author}{\bibfnamefont{D.}~\bibnamefont{Bacon}}, \bibinfo{author}{\bibfnamefont{J.~C.} \bibnamefont{Bardin}}, \bibinfo{author}{\bibfnamefont{R.}~\bibnamefont{Barends}}, \bibinfo{author}{\bibfnamefont{R.}~\bibnamefont{Biswas}}, \bibinfo{author}{\bibfnamefont{S.}~\bibnamefont{Boixo}}, \bibinfo{author}{\bibfnamefont{F.~G.} \bibnamefont{Brandao}}, \bibinfo{author}{\bibfnamefont{D.~A.} \bibnamefont{Buell}}, \bibnamefont{et~al.}, \bibinfo{journal}{Nature} \textbf{\bibinfo{volume}{574}}, \bibinfo{pages}{505} (\bibinfo{year}{2019}), \urlprefix\url{https://www.nature.com/articles/s41586-019-1666-5}.

\bibitem[{\citenamefont{Zhong et~al.}(2020)\citenamefont{Zhong, Wang, Deng, Chen, Peng, Luo, Qin, Wu, Ding, Hu et~al.}}]{zhong2020jiuzhang}
\bibinfo{author}{\bibfnamefont{H.-S.} \bibnamefont{Zhong}}, \bibinfo{author}{\bibfnamefont{H.}~\bibnamefont{Wang}}, \bibinfo{author}{\bibfnamefont{Y.-H.} \bibnamefont{Deng}}, \bibinfo{author}{\bibfnamefont{M.-C.} \bibnamefont{Chen}}, \bibinfo{author}{\bibfnamefont{L.-C.} \bibnamefont{Peng}}, \bibinfo{author}{\bibfnamefont{Y.-H.} \bibnamefont{Luo}}, \bibinfo{author}{\bibfnamefont{J.}~\bibnamefont{Qin}}, \bibinfo{author}{\bibfnamefont{D.}~\bibnamefont{Wu}}, \bibinfo{author}{\bibfnamefont{X.}~\bibnamefont{Ding}}, \bibinfo{author}{\bibfnamefont{Y.}~\bibnamefont{Hu}}, \bibnamefont{et~al.}, \bibinfo{journal}{Science} \textbf{\bibinfo{volume}{370}}, \bibinfo{pages}{1460} (\bibinfo{year}{2020}), \eprint{https://www.science.org/doi/pdf/10.1126/science.abe8770}, \urlprefix\url{https://www.science.org/doi/abs/10.1126/science.abe8770}.

\bibitem[{\citenamefont{Moses et~al.}(2023)\citenamefont{Moses, Baldwin, Allman, Ancona, Ascarrunz, Barnes, Bartolotta, Bjork, Blanchard, Bohn et~al.}}]{moses2023racetrack}
\bibinfo{author}{\bibfnamefont{S.~A.} \bibnamefont{Moses}}, \bibinfo{author}{\bibfnamefont{C.~H.} \bibnamefont{Baldwin}}, \bibinfo{author}{\bibfnamefont{M.~S.} \bibnamefont{Allman}}, \bibinfo{author}{\bibfnamefont{R.}~\bibnamefont{Ancona}}, \bibinfo{author}{\bibfnamefont{L.}~\bibnamefont{Ascarrunz}}, \bibinfo{author}{\bibfnamefont{C.}~\bibnamefont{Barnes}}, \bibinfo{author}{\bibfnamefont{J.}~\bibnamefont{Bartolotta}}, \bibinfo{author}{\bibfnamefont{B.}~\bibnamefont{Bjork}}, \bibinfo{author}{\bibfnamefont{P.}~\bibnamefont{Blanchard}}, \bibinfo{author}{\bibfnamefont{M.}~\bibnamefont{Bohn}}, \bibnamefont{et~al.}, \bibinfo{journal}{Phys. Rev. X} \textbf{\bibinfo{volume}{13}}, \bibinfo{pages}{041052} (\bibinfo{year}{2023}), \urlprefix\url{https://link.aps.org/doi/10.1103/PhysRevX.13.041052}.

\bibitem[{\citenamefont{Preskill}(2018)}]{preskill2018quantum}
\bibinfo{author}{\bibfnamefont{J.}~\bibnamefont{Preskill}}, \bibinfo{journal}{Quantum} \textbf{\bibinfo{volume}{2}}, \bibinfo{pages}{79} (\bibinfo{year}{2018}).

\bibitem[{\citenamefont{Aharonov et~al.}(1996)\citenamefont{Aharonov, {Ben-Or}, Impagliazzo, and Nisan}}]{aharonov1996limitations}
\bibinfo{author}{\bibfnamefont{D.}~\bibnamefont{Aharonov}}, \bibinfo{author}{\bibfnamefont{M.}~\bibnamefont{{Ben-Or}}}, \bibinfo{author}{\bibfnamefont{R.}~\bibnamefont{Impagliazzo}}, \bibnamefont{and} \bibinfo{author}{\bibfnamefont{N.}~\bibnamefont{Nisan}}, \emph{\bibinfo{title}{Limitations of {{Noisy Reversible Computation}}}} (\bibinfo{year}{1996}), \eprint{quant-ph/9611028}.

\bibitem[{\citenamefont{Yan et~al.}(2023)\citenamefont{Yan, Du, Chen, and Ma}}]{yan2023limitations}
\bibinfo{author}{\bibfnamefont{Y.}~\bibnamefont{Yan}}, \bibinfo{author}{\bibfnamefont{Z.}~\bibnamefont{Du}}, \bibinfo{author}{\bibfnamefont{J.}~\bibnamefont{Chen}}, \bibnamefont{and} \bibinfo{author}{\bibfnamefont{X.}~\bibnamefont{Ma}}, \emph{\bibinfo{title}{Limitations of noisy quantum devices in computational and entangling power}} (\bibinfo{year}{2023}), \eprint{2306.02836}, \urlprefix\url{https://arxiv.org/abs/2306.02836}.

\bibitem[{\citenamefont{Diffie and Hellman}(1976)}]{diffie1976new}
\bibinfo{author}{\bibfnamefont{W.}~\bibnamefont{Diffie}} \bibnamefont{and} \bibinfo{author}{\bibfnamefont{M.}~\bibnamefont{Hellman}}, \bibinfo{journal}{IEEE Transactions on Information Theory} \textbf{\bibinfo{volume}{22}}, \bibinfo{pages}{644} (\bibinfo{year}{1976}).

\bibitem[{\citenamefont{Elgamal}(1985)}]{elgamal1985public}
\bibinfo{author}{\bibfnamefont{T.}~\bibnamefont{Elgamal}}, \bibinfo{journal}{IEEE Transactions on Information Theory} \textbf{\bibinfo{volume}{31}}, \bibinfo{pages}{469} (\bibinfo{year}{1985}).

\bibitem[{\citenamefont{Johnson et~al.}(2001)\citenamefont{Johnson, Menezes, and Vanstone}}]{johnson2001elliptic}
\bibinfo{author}{\bibfnamefont{D.}~\bibnamefont{Johnson}}, \bibinfo{author}{\bibfnamefont{A.}~\bibnamefont{Menezes}}, \bibnamefont{and} \bibinfo{author}{\bibfnamefont{S.}~\bibnamefont{Vanstone}}, \bibinfo{journal}{International journal of information security} \textbf{\bibinfo{volume}{1}}, \bibinfo{pages}{36} (\bibinfo{year}{2001}).

\bibitem[{\citenamefont{Dubal and Deshmukh}(2013)}]{Dubal2013pseudorandom}
\bibinfo{author}{\bibfnamefont{M.}~\bibnamefont{Dubal}} \bibnamefont{and} \bibinfo{author}{\bibfnamefont{A.}~\bibnamefont{Deshmukh}} (\bibinfo{year}{2013}), vol. \bibinfo{volume}{377}, pp. \bibinfo{pages}{77--89}, ISBN \bibinfo{isbn}{978-3-642-40575-4}.

\bibitem[{\citenamefont{PUB}(2000)}]{pub2000digital}
\bibinfo{author}{\bibfnamefont{F.}~\bibnamefont{PUB}}, \bibinfo{journal}{Fips pub} pp. \bibinfo{pages}{186--192} (\bibinfo{year}{2000}).

\bibitem[{\citenamefont{Qu}(1999)}]{qu1999sec}
\bibinfo{author}{\bibfnamefont{M.}~\bibnamefont{Qu}}, \bibinfo{journal}{Certicom Res., Mississauga, ON, Canada, Tech. Rep. SEC2-Ver-0.6}  (\bibinfo{year}{1999}).

\bibitem[{\citenamefont{Aggarwal et~al.}(2017)\citenamefont{Aggarwal, Brennen, Lee, Santha, and Tomamichel}}]{aggarwal2017quantum}
\bibinfo{author}{\bibfnamefont{D.}~\bibnamefont{Aggarwal}}, \bibinfo{author}{\bibfnamefont{G.~K.} \bibnamefont{Brennen}}, \bibinfo{author}{\bibfnamefont{T.}~\bibnamefont{Lee}}, \bibinfo{author}{\bibfnamefont{M.}~\bibnamefont{Santha}}, \bibnamefont{and} \bibinfo{author}{\bibfnamefont{M.}~\bibnamefont{Tomamichel}}, \bibinfo{journal}{arXiv preprint arXiv:1710.10377}  (\bibinfo{year}{2017}).

\bibitem[{\citenamefont{Nakamoto}(2009)}]{Satoshi2009bitcoin}
\bibinfo{author}{\bibfnamefont{S.}~\bibnamefont{Nakamoto}}, \bibinfo{journal}{Cryptography Mailing list at https://metzdowd.com}  (\bibinfo{year}{2009}).

\bibitem[{\citenamefont{Proos and Zalka}(2004)}]{proos2004shors}
\bibinfo{author}{\bibfnamefont{J.}~\bibnamefont{Proos}} \bibnamefont{and} \bibinfo{author}{\bibfnamefont{C.}~\bibnamefont{Zalka}}, \emph{\bibinfo{title}{Shor's discrete logarithm quantum algorithm for elliptic curves}} (\bibinfo{year}{2004}), \eprint{quant-ph/0301141}, \urlprefix\url{https://arxiv.org/abs/quant-ph/0301141}.

\bibitem[{\citenamefont{Fowler et~al.}(2012)\citenamefont{Fowler, Mariantoni, Martinis, and Cleland}}]{fowler2012surface}
\bibinfo{author}{\bibfnamefont{A.~G.} \bibnamefont{Fowler}}, \bibinfo{author}{\bibfnamefont{M.}~\bibnamefont{Mariantoni}}, \bibinfo{author}{\bibfnamefont{J.~M.} \bibnamefont{Martinis}}, \bibnamefont{and} \bibinfo{author}{\bibfnamefont{A.~N.} \bibnamefont{Cleland}}, \bibinfo{journal}{Phys. Rev. A} \textbf{\bibinfo{volume}{86}}, \bibinfo{pages}{032324} (\bibinfo{year}{2012}), \urlprefix\url{https://link.aps.org/doi/10.1103/PhysRevA.86.032324}.

\bibitem[{\citenamefont{O'Gorman and Campbell}(2017)}]{gorman2017quantum}
\bibinfo{author}{\bibfnamefont{J.}~\bibnamefont{O'Gorman}} \bibnamefont{and} \bibinfo{author}{\bibfnamefont{E.~T.} \bibnamefont{Campbell}}, \bibinfo{journal}{Phys. Rev. A} \textbf{\bibinfo{volume}{95}}, \bibinfo{pages}{032338} (\bibinfo{year}{2017}), \urlprefix\url{https://link.aps.org/doi/10.1103/PhysRevA.95.032338}.

\bibitem[{\citenamefont{Gidney and Eker{\aa{}}}(2021)}]{Gidney2021howtofactorbit}
\bibinfo{author}{\bibfnamefont{C.}~\bibnamefont{Gidney}} \bibnamefont{and} \bibinfo{author}{\bibfnamefont{M.}~\bibnamefont{Eker{\aa{}}}}, \bibinfo{journal}{{Quantum}} \textbf{\bibinfo{volume}{5}}, \bibinfo{pages}{433} (\bibinfo{year}{2021}), ISSN \bibinfo{issn}{2521-327X}, \urlprefix\url{https://doi.org/10.22331/q-2021-04-15-433}.

\bibitem[{\citenamefont{Ha et~al.}(2022)\citenamefont{Ha, Lee, and Heo}}]{ha2022resource}
\bibinfo{author}{\bibfnamefont{J.}~\bibnamefont{Ha}}, \bibinfo{author}{\bibfnamefont{J.}~\bibnamefont{Lee}}, \bibnamefont{and} \bibinfo{author}{\bibfnamefont{J.}~\bibnamefont{Heo}}, \bibinfo{journal}{Quantum Information Processing} \textbf{\bibinfo{volume}{21}}, \bibinfo{pages}{60} (\bibinfo{year}{2022}), ISSN \bibinfo{issn}{1573-1332}, \urlprefix\url{https://doi.org/10.1007/s11128-021-03398-1}.

\bibitem[{\citenamefont{Gidney}(2025)}]{gidney2025factor}
\bibinfo{author}{\bibfnamefont{C.}~\bibnamefont{Gidney}}, \emph{\bibinfo{title}{How to factor 2048 bit rsa integers with less than a million noisy qubits}} (\bibinfo{year}{2025}), \eprint{2505.15917}, \urlprefix\url{https://arxiv.org/abs/2505.15917}.

\bibitem[{\citenamefont{Roetteler et~al.}(2017)\citenamefont{Roetteler, Naehrig, Svore, and Lauter}}]{roetteler2017quantum}
\bibinfo{author}{\bibfnamefont{M.}~\bibnamefont{Roetteler}}, \bibinfo{author}{\bibfnamefont{M.}~\bibnamefont{Naehrig}}, \bibinfo{author}{\bibfnamefont{K.~M.} \bibnamefont{Svore}}, \bibnamefont{and} \bibinfo{author}{\bibfnamefont{K.}~\bibnamefont{Lauter}}, in \emph{\bibinfo{booktitle}{Advances in Cryptology--ASIACRYPT 2017: 23rd International Conference on the Theory and Applications of Cryptology and Information Security, Hong Kong, China, December 3-7, 2017, Proceedings, Part II 23}} (\bibinfo{organization}{Springer}, \bibinfo{year}{2017}), pp. \bibinfo{pages}{241--270}.

\bibitem[{\citenamefont{H{\"a}ner et~al.}(2020)\citenamefont{H{\"a}ner, Jaques, Naehrig, Roetteler, and Soeken}}]{haner2020improved}
\bibinfo{author}{\bibfnamefont{T.}~\bibnamefont{H{\"a}ner}}, \bibinfo{author}{\bibfnamefont{S.}~\bibnamefont{Jaques}}, \bibinfo{author}{\bibfnamefont{M.}~\bibnamefont{Naehrig}}, \bibinfo{author}{\bibfnamefont{M.}~\bibnamefont{Roetteler}}, \bibnamefont{and} \bibinfo{author}{\bibfnamefont{M.}~\bibnamefont{Soeken}}, in \emph{\bibinfo{booktitle}{Post-Quantum Cryptography: 11th International Conference, PQCrypto 2020, Paris, France, April 15--17, 2020, Proceedings 11}} (\bibinfo{organization}{Springer}, \bibinfo{year}{2020}), pp. \bibinfo{pages}{425--444}.

\bibitem[{\citenamefont{Cuccaro et~al.}(2004)\citenamefont{Cuccaro, Draper, Kutin, and Moulton}}]{cuccaro2004new}
\bibinfo{author}{\bibfnamefont{S.~A.} \bibnamefont{Cuccaro}}, \bibinfo{author}{\bibfnamefont{T.~G.} \bibnamefont{Draper}}, \bibinfo{author}{\bibfnamefont{S.~A.} \bibnamefont{Kutin}}, \bibnamefont{and} \bibinfo{author}{\bibfnamefont{D.~P.} \bibnamefont{Moulton}}, \bibinfo{journal}{arXiv preprint quant-ph/0410184}  (\bibinfo{year}{2004}).

\bibitem[{\citenamefont{Gidney}(2018)}]{gidney2018halving}
\bibinfo{author}{\bibfnamefont{C.}~\bibnamefont{Gidney}}, \bibinfo{journal}{Quantum} \textbf{\bibinfo{volume}{2}}, \bibinfo{pages}{74} (\bibinfo{year}{2018}).

\bibitem[{\citenamefont{Draper et~al.}(2004)\citenamefont{Draper, Kutin, Rains, and Svore}}]{draper2004logarithmic}
\bibinfo{author}{\bibfnamefont{T.~G.} \bibnamefont{Draper}}, \bibinfo{author}{\bibfnamefont{S.~A.} \bibnamefont{Kutin}}, \bibinfo{author}{\bibfnamefont{E.~M.} \bibnamefont{Rains}}, \bibnamefont{and} \bibinfo{author}{\bibfnamefont{K.~M.} \bibnamefont{Svore}}, \emph{\bibinfo{title}{A logarithmic-depth quantum carry-lookahead adder}} (\bibinfo{year}{2004}), \eprint{quant-ph/0406142}, \urlprefix\url{https://arxiv.org/abs/quant-ph/0406142}.

\bibitem[{\citenamefont{Wang et~al.}(2024)\citenamefont{Wang, Mondal, and Chattopadhyay}}]{wang2024optimal}
\bibinfo{author}{\bibfnamefont{S.}~\bibnamefont{Wang}}, \bibinfo{author}{\bibfnamefont{A.}~\bibnamefont{Mondal}}, \bibnamefont{and} \bibinfo{author}{\bibfnamefont{A.}~\bibnamefont{Chattopadhyay}}, \bibinfo{journal}{ACM Transactions on Quantum Computing}  (\bibinfo{year}{2024}).

\bibitem[{\citenamefont{Montgomery}(1985)}]{montgomery1985modular}
\bibinfo{author}{\bibfnamefont{P.~L.} \bibnamefont{Montgomery}}, \bibinfo{journal}{Mathematics of computation} \textbf{\bibinfo{volume}{44}}, \bibinfo{pages}{519} (\bibinfo{year}{1985}).

\bibitem[{\citenamefont{Babbush et~al.}(2018)\citenamefont{Babbush, Gidney, Berry, Wiebe, McClean, Paler, Fowler, and Neven}}]{babbush2018encoding}
\bibinfo{author}{\bibfnamefont{R.}~\bibnamefont{Babbush}}, \bibinfo{author}{\bibfnamefont{C.}~\bibnamefont{Gidney}}, \bibinfo{author}{\bibfnamefont{D.~W.} \bibnamefont{Berry}}, \bibinfo{author}{\bibfnamefont{N.}~\bibnamefont{Wiebe}}, \bibinfo{author}{\bibfnamefont{J.}~\bibnamefont{McClean}}, \bibinfo{author}{\bibfnamefont{A.}~\bibnamefont{Paler}}, \bibinfo{author}{\bibfnamefont{A.}~\bibnamefont{Fowler}}, \bibnamefont{and} \bibinfo{author}{\bibfnamefont{H.}~\bibnamefont{Neven}}, \bibinfo{journal}{Physical Review X} \textbf{\bibinfo{volume}{8}}, \bibinfo{pages}{041015} (\bibinfo{year}{2018}).

\bibitem[{\citenamefont{Shor}(1995)}]{Shor1995code}
\bibinfo{author}{\bibfnamefont{P.~W.} \bibnamefont{Shor}}, \bibinfo{journal}{Phys. Rev. A} \textbf{\bibinfo{volume}{52}}, \bibinfo{pages}{R2493} (\bibinfo{year}{1995}), \urlprefix\url{https://link.aps.org/doi/10.1103/PhysRevA.52.R2493}.

\bibitem[{\citenamefont{Gottesman}(1997)}]{gottesman1997stabilizer}
\bibinfo{author}{\bibfnamefont{D.}~\bibnamefont{Gottesman}}, \emph{\bibinfo{title}{Stabilizer codes and quantum error correction}} (\bibinfo{publisher}{California Institute of Technology}, \bibinfo{year}{1997}).

\bibitem[{\citenamefont{Piroli et~al.}(2021)\citenamefont{Piroli, Styliaris, and Cirac}}]{Piroli2021adaptive}
\bibinfo{author}{\bibfnamefont{L.}~\bibnamefont{Piroli}}, \bibinfo{author}{\bibfnamefont{G.}~\bibnamefont{Styliaris}}, \bibnamefont{and} \bibinfo{author}{\bibfnamefont{J.~I.} \bibnamefont{Cirac}}, \bibinfo{journal}{Phys. Rev. Lett.} \textbf{\bibinfo{volume}{127}}, \bibinfo{pages}{220503} (\bibinfo{year}{2021}), \urlprefix\url{https://link.aps.org/doi/10.1103/PhysRevLett.127.220503}.

\bibitem[{\citenamefont{Yan et~al.}(2025)\citenamefont{Yan, Ma, Zhou, and Ma}}]{Yan2025Variational}
\bibinfo{author}{\bibfnamefont{Y.}~\bibnamefont{Yan}}, \bibinfo{author}{\bibfnamefont{M.}~\bibnamefont{Ma}}, \bibinfo{author}{\bibfnamefont{Y.}~\bibnamefont{Zhou}}, \bibnamefont{and} \bibinfo{author}{\bibfnamefont{X.}~\bibnamefont{Ma}}, \bibinfo{journal}{Phys. Rev. Lett.} \textbf{\bibinfo{volume}{134}}, \bibinfo{pages}{170601} (\bibinfo{year}{2025}), \urlprefix\url{https://link.aps.org/doi/10.1103/PhysRevLett.134.170601}.

\bibitem[{\citenamefont{Raussendorf et~al.}(2003)\citenamefont{Raussendorf, Browne, and Briegel}}]{Raussendorf2003MBQC}
\bibinfo{author}{\bibfnamefont{R.}~\bibnamefont{Raussendorf}}, \bibinfo{author}{\bibfnamefont{D.~E.} \bibnamefont{Browne}}, \bibnamefont{and} \bibinfo{author}{\bibfnamefont{H.~J.} \bibnamefont{Briegel}}, \bibinfo{journal}{Phys. Rev. A} \textbf{\bibinfo{volume}{68}}, \bibinfo{pages}{022312} (\bibinfo{year}{2003}), \urlprefix\url{https://link.aps.org/doi/10.1103/PhysRevA.68.022312}.

\bibitem[{\citenamefont{Briegel et~al.}(2009)\citenamefont{Briegel, Browne, D{\"u}r, Raussendorf, and Van~den Nest}}]{Briegel2009MBQC}
\bibinfo{author}{\bibfnamefont{H.~J.} \bibnamefont{Briegel}}, \bibinfo{author}{\bibfnamefont{D.~E.} \bibnamefont{Browne}}, \bibinfo{author}{\bibfnamefont{W.}~\bibnamefont{D{\"u}r}}, \bibinfo{author}{\bibfnamefont{R.}~\bibnamefont{Raussendorf}}, \bibnamefont{and} \bibinfo{author}{\bibfnamefont{M.}~\bibnamefont{Van~den Nest}}, \bibinfo{journal}{Nature Physics} \textbf{\bibinfo{volume}{5}}, \bibinfo{pages}{19} (\bibinfo{year}{2009}), \urlprefix\url{https://doi.org/10.1038/nphys1157}.

\bibitem[{\citenamefont{Takahashi and Tani}(2016)}]{Takahashi2016collapse}
\bibinfo{author}{\bibfnamefont{Y.}~\bibnamefont{Takahashi}} \bibnamefont{and} \bibinfo{author}{\bibfnamefont{S.}~\bibnamefont{Tani}}, \bibinfo{journal}{computational complexity} \textbf{\bibinfo{volume}{25}}, \bibinfo{pages}{849} (\bibinfo{year}{2016}), ISSN \bibinfo{issn}{1420-8954}, \urlprefix\url{https://doi.org/10.1007/s00037-016-0140-0}.

\bibitem[{\citenamefont{Nielsen and Chuang}(2010)}]{nielsen2010quantum}
\bibinfo{author}{\bibfnamefont{M.~A.} \bibnamefont{Nielsen}} \bibnamefont{and} \bibinfo{author}{\bibfnamefont{I.~L.} \bibnamefont{Chuang}}, \emph{\bibinfo{title}{Quantum computation and quantum information}} (\bibinfo{publisher}{Cambridge university press}, \bibinfo{year}{2010}).

\bibitem[{\citenamefont{Malz et~al.}(2024)\citenamefont{Malz, Styliaris, Wei, and Cirac}}]{Malz2024MPS}
\bibinfo{author}{\bibfnamefont{D.}~\bibnamefont{Malz}}, \bibinfo{author}{\bibfnamefont{G.}~\bibnamefont{Styliaris}}, \bibinfo{author}{\bibfnamefont{Z.-Y.} \bibnamefont{Wei}}, \bibnamefont{and} \bibinfo{author}{\bibfnamefont{J.~I.} \bibnamefont{Cirac}}, \bibinfo{journal}{Phys. Rev. Lett.} \textbf{\bibinfo{volume}{132}}, \bibinfo{pages}{040404} (\bibinfo{year}{2024}), \urlprefix\url{https://link.aps.org/doi/10.1103/PhysRevLett.132.040404}.

\bibitem[{\citenamefont{Piroli et~al.}(2024)\citenamefont{Piroli, Styliaris, and Cirac}}]{piroli2024approximatingmanybodyquantumstates}
\bibinfo{author}{\bibfnamefont{L.}~\bibnamefont{Piroli}}, \bibinfo{author}{\bibfnamefont{G.}~\bibnamefont{Styliaris}}, \bibnamefont{and} \bibinfo{author}{\bibfnamefont{J.~I.} \bibnamefont{Cirac}}, \bibinfo{journal}{Phys. Rev. Lett.} \textbf{\bibinfo{volume}{133}}, \bibinfo{pages}{230401} (\bibinfo{year}{2024}), \urlprefix\url{https://link.aps.org/doi/10.1103/PhysRevLett.133.230401}.

\bibitem[{\citenamefont{Bäumer et~al.}(2024)\citenamefont{Bäumer, Tripathi, Wang, Rall, Chen, Majumder, Seif, and Minev}}]{B_umer_2024}
\bibinfo{author}{\bibfnamefont{E.}~\bibnamefont{Bäumer}}, \bibinfo{author}{\bibfnamefont{V.}~\bibnamefont{Tripathi}}, \bibinfo{author}{\bibfnamefont{D.~S.} \bibnamefont{Wang}}, \bibinfo{author}{\bibfnamefont{P.}~\bibnamefont{Rall}}, \bibinfo{author}{\bibfnamefont{E.~H.} \bibnamefont{Chen}}, \bibinfo{author}{\bibfnamefont{S.}~\bibnamefont{Majumder}}, \bibinfo{author}{\bibfnamefont{A.}~\bibnamefont{Seif}}, \bibnamefont{and} \bibinfo{author}{\bibfnamefont{Z.~K.} \bibnamefont{Minev}}, \bibinfo{journal}{PRX Quantum} \textbf{\bibinfo{volume}{5}} (\bibinfo{year}{2024}), ISSN \bibinfo{issn}{2691-3399}, \urlprefix\url{http://dx.doi.org/10.1103/PRXQuantum.5.030339}.

\bibitem[{\citenamefont{Blelloch}(1990)}]{blelloch1990prefix}
\bibinfo{author}{\bibfnamefont{G.~E.} \bibnamefont{Blelloch}} (\bibinfo{year}{1990}).

\bibitem[{\citenamefont{Brent and Kung}(1982)}]{brent1982regular}
\bibinfo{author}{\bibnamefont{Brent}} \bibnamefont{and} \bibinfo{author}{\bibnamefont{Kung}}, \bibinfo{journal}{IEEE transactions on Computers} \textbf{\bibinfo{volume}{100}}, \bibinfo{pages}{260} (\bibinfo{year}{1982}).

\bibitem[{\citenamefont{Sklansky}(2009)}]{sklansky2009conditional}
\bibinfo{author}{\bibfnamefont{J.}~\bibnamefont{Sklansky}}, \bibinfo{journal}{IRE Transactions on Electronic computers} pp. \bibinfo{pages}{226--231} (\bibinfo{year}{2009}).

\bibitem[{\citenamefont{Proos and Zalka}(2003)}]{proos2003shor}
\bibinfo{author}{\bibfnamefont{J.}~\bibnamefont{Proos}} \bibnamefont{and} \bibinfo{author}{\bibfnamefont{C.}~\bibnamefont{Zalka}}, \bibinfo{journal}{arXiv preprint quant-ph/0301141}  (\bibinfo{year}{2003}).

\bibitem[{\citenamefont{Dutta et~al.}(2025)\citenamefont{Dutta, Wang, Baksi, Chattopadhyay, and Maitra}}]{dutta2025exact}
\bibinfo{author}{\bibfnamefont{S.}~\bibnamefont{Dutta}}, \bibinfo{author}{\bibfnamefont{S.}~\bibnamefont{Wang}}, \bibinfo{author}{\bibfnamefont{A.}~\bibnamefont{Baksi}}, \bibinfo{author}{\bibfnamefont{A.}~\bibnamefont{Chattopadhyay}}, \bibnamefont{and} \bibinfo{author}{\bibfnamefont{S.}~\bibnamefont{Maitra}}, \bibinfo{journal}{Physical Review A} \textbf{\bibinfo{volume}{111}}, \bibinfo{pages}{052611} (\bibinfo{year}{2025}).

\bibitem[{\citenamefont{Rines and Chuang}(2018)}]{rines2018high}
\bibinfo{author}{\bibfnamefont{R.}~\bibnamefont{Rines}} \bibnamefont{and} \bibinfo{author}{\bibfnamefont{I.}~\bibnamefont{Chuang}}, \bibinfo{journal}{arXiv preprint arXiv:1801.01081}  (\bibinfo{year}{2018}).

\bibitem[{\citenamefont{Gidney}(2019)}]{gidney2019windowed}
\bibinfo{author}{\bibfnamefont{C.}~\bibnamefont{Gidney}}, \bibinfo{journal}{arXiv preprint arXiv:1905.07682}  (\bibinfo{year}{2019}).

\bibitem[{\citenamefont{Kaliski}(1995)}]{kaliski1995montgomery}
\bibinfo{author}{\bibfnamefont{B.~S.} \bibnamefont{Kaliski}}, \bibinfo{journal}{IEEE transactions on computers} \textbf{\bibinfo{volume}{44}}, \bibinfo{pages}{1064} (\bibinfo{year}{1995}).

\end{thebibliography}
\end{document}